%% file: main.tex
\pdfoutput=1
\documentclass[11pt]{article}

\usepackage{heppub}
\usepackage[dvipsnames]{xcolor}
\usepackage{epsfig}
\usepackage{amssymb}
\usepackage{amsmath}
\usepackage{tikz}
\usepackage{mathrsfs}
\usepackage{hyperref}
\usepackage{multirow}
\usepackage{scalerel}
\usepackage{mathtools}
\usepackage{textcomp}
\usepackage[all]{xy}
\usepackage{dsfont}
\usepackage{amsthm}
\usepackage{blkarray}
\usepackage{bm}
\usepackage[size=footnotesize]{todonotes}
\usepackage{tikz-cd}
\usepackage{comment}
\numberwithin{equation}{section}

\usepackage{caption}
\usepackage{subcaption}
\usepackage{framed}

\usepackage[mathscr]{eucal}

\input{commands}

\newcommand{\OUE}{\Omega_{{\rm UE}}^{\figeight}}
\newcommand{\UaUE}{\Upsilon_{0{\rm UE}}^{\bndry}}
\newcommand{\UbUE}{\Upsilon_{1{\rm UE}}^{\bndry}}

\title{The Catchment Area of Groomed Jets at NNLL}

\author[a,b]{Aditya Pathak\footnote{aditya.pathak@desy.de}}
\affiliation[a]{University of Manchester, School of Physics and Astronomy, Manchester, M13 9PL, United Kingdom}
\affiliation[b]{Deutsches Elektronen-Synchrotron DESY, Notkestr. 85, 22607 Hamburg, Germany}

\abstract{Groomed jet observables have a dynamical catchment area which plays a key role in determining the leading nonperturbative power corrections and the impact of the underlying event.
Based on field-theoretic arguments, certain moments of the groomed jet radius $R_g$ capture the entirety of the kinematic and grooming parameter dependence of these effects.
These moments can be computed perturbatively in the soft drop operator expansion region where these corrections are small, but yet significant to be relevant for precision physics.
A precise determination of these moments is thus crucial to faithfully isolate the universal contributions of hadronization and the underlying event.
Building on a previously developed effective field theory framework for the doubly differential soft drop groomed jet mass and groomed jet radius measurement, we present here a calculation of these moments at next-to-next-to-leading-logarithmic (NNLL) accuracy including matching into the plain jet mass region.
We compare our predictions for these moments against parton-shower Monte Carlo simulations and find good agreement.
These results have applications for precision physics with soft drop jet mass such as determination of the strong coupling constant and the top quark mass and for improving hadronization models.}


\begin{document}
\preprint{DESY-23-003}
\maketitle

\section{Introduction}

The nonperturbative effects of quantum chromodynamics (QCD) encompass a wide array of rich phenomena ranging from low energy nuclear physics, the spectrum of hadrons, the structure of energetic protons described by parton densities, the physics of hadronization, the formation and evolution of the quark gluon plasma, and much more, each contributing to the richness of the QCD phase-diagram.
These nonperturbative effects can be probed in different experimental setups and studied via an appropriate effective field theory of QCD.
Among these, the physics of hadronization is one such complex phenomenon that has so far remained largely elusive to first principle calculations.
Studies of jets and jet substructure in high energy collisions have offered us invaluable insights about hadronization.
On the other hand, jets are also unique tools that allow us to probe fundamental aspects of QCD and in searching for physics beyond the Standard Model~\cite{Larkoski:2017jix,Kogler:2018hem}.
Hadronization impacts measurements of jet substructure observables relative to a reference parton level computation in a way that is partly unique to the observable and partly universal~\cite{Lee:2006fn,Korchemsky:1994is,Korchemsky:1997sy,Korchemsky:1999kt,Belitsky:2001ij,Dasgupta:2007wa,Stewart:2014nna}.
Because the effects of hadronization cannot be predicted from first principle calculations,
the effort has been to seek ways to eliminate or minimize hadronization corrections, such that perturbative calculations can be reliably compared to collider data.

Typically, the dominant effects of hadronization arise in the soft wide angle physics, which is challenging to bring under theoretical control even in perturbation theory. Moreover, at the Large Hadron Collider (LHC), the underlying event and pile-up also make a contribution of similar nature.
With the introduction of grooming algorithms~\cite{Ellis:2009me,Cacciari:2014gra,Krohn:2009th,Dasgupta:2013ihk,Larkoski:2014wba,Butterworth:2008iy,Frye:2017yrw,Dreyer:2018tjj} it has been possible to make first principle predictions for jet substructure observables at the LHC due to their effectiveness in removing soft wide angle contamination in the complex LHC environment and thus suppressing hadronization, underlying event and pile-up effects.
These grooming techniques were initially designed as a taggers to suppress effects of QCD in searches of multiprong decays originating from electroweak and potentially new physics,
but over the recent years they have come to be seen as tools with strong potential for applications in precision physics.\footnote{Somewhat recently, there has also been a lot of exciting progress in developing a completely complementary approach of employing energy-energy correlators (EEC) for precision collider physics~\cite{Lee:2022ige,Komiske:2022enw,Holguin:2022epo,Chen:2022swd,Chen:2022jhb,Andres:2022ovj}.}

Among various variants, the soft drop (SD) groomer~\cite{Larkoski:2014wba} (including the modified Mass
Drop Tagger~\cite{Butterworth:2008iy,Dasgupta:2013ihk} as a special case), has been extensively
studied and all-orders factorization formulae have been
established~\cite{Frye:2016aiz,Frye:2016okc,Hoang:2017kmk}.
Among an impressive list of phenomenological applications, examples of application of soft drop grooming
include measurement of the QCD splitting functions~\cite{Cal:2021fla,Larkoski:2017bvj}, study of
fragmentation structure~\cite{Gutierrez-Reyes:2019msa,Makris:2018npl}, isolation of the soft-sensitive
dynamics~\cite{Chien:2019osu,Cal:2020flh,Stewart:2022ari}, and quantification of medium
modifications~\cite{KunnawalkamElayavalli:2017hxo,Andrews:2018jcm,Mehtar-Tani:2016aco,Casalderrey-Solana:2019ubu,Brewer:2021hmh}.
The algorithm of jet grooming is intimately tied to sequential recombination jet algorithms that pair-wise cluster particles/subjets into subjets, to eventually form a jet.
The criteria for soft drop is given by a sequential test between pairs of subjets $i$ and $j$ found by de-clustering a Cambridge-Achen clustered tree (based solely on angular separation) of particles in a jet:
\begin{align}\label{eq:SDpp}
\frac{\min(p_{T_i},p_{T_j})}{p_{T_i}+p_{T_j}} > z_\cut \Big(\frac{\Delta R_{ij}}{R_0}\Big)^\beta\,.
\end{align}
If the pair collectively fails this criteria then the softer of the two (say $i$ with $p_{T_i} < p_{T_j}$) is removed and the next pair obtained by de-clustering the harder subjet ($j$) is tested for this condition. The pair that eventually satisfies this condition terminates the groomer, and the remaining particles in either of the subjets in the pair constitute the groomed jet.
The parameters $\zcut$ and $\beta$ control the strength of the groomer and are typically chosen to be $\zcut \sim 0.1$ and $\beta \sim 1$.
The careful formulation of the soft drop condition in \eq{SDpp} ensures that IRC-safe observables measured on jets remain calculable in perturbation theory even after grooming.

A classic observable studied extensively at the LHC is the jet mass. For instance,
both ungroomed and groomed jet substructure measurements have been measured by
ALICE~\cite{ALICE:2017nij},
ATLAS~\cite{ATLAS:2012am,ATLAS:2012nnf,ATLAS:2017zda,ATLAS:2019dty,ATLAS:2019mgf}, and
CMS~\cite{CMS:2017tdn,CMS:2018ypj,CMS:2019fak,CMS:2018fof} collaborations. The groomed jet
angularities (a generalization of the jet mass) and other groomed jet substructure observables have also
recently been measured by the
ALICE~\cite{ALICE:2021njq}, CMS~\cite{CMS:2018ypj,CMS:2021iwu} and ATLAS~\cite{ATLAS:2019kwg}
collaborations. Soft drop jet mass has been explored for applications
such as
precision top quark mass~\cite{Hoang:2017kmk,Bachu:2020nqn,ATLAS:2021urs} and strong coupling
constant~\cite{Marzani:2019evv,Hannesdottir:2022rsl} measurements.
State-of-the-art perturbative calculations for soft drop jet mass have reached high accuracy of
next-to-next-to-next-to-leading logarithmic accuracy (N$^3$LL) matched to next-to-next-to-leading
order (NNLO) predictions for groomed jets in the $\ee\ra q \bar q$ process~\cite{Kardos:2020gty}, and
next-to-next-to-leading-logarithmic (NNLL) accuracy~\cite{Frye:2016aiz,Hannesdottir:2022rsl} for jets at
the LHC.

To make these calculations useful for precision phenomenology, it is crucial to describe nonperturbative power corrections in the jet mass spectrum. In the region where soft drop is effective in removing soft-wide angle radiation and where perturbation theory is dominant, the nonperturbative power corrections can be as large as $\sim$ 10\%.
As was shown in a recent analysis in \Refcite{Hannesdottir:2022rsl}, these power corrections thus start to become relevant already at NNLL accuracy.
For instance, we showed in \Refcite{Hannesdottir:2022rsl} that the unconstrained effects of hadronization respectively contribute about 3\% and 8\% irreducible uncertainty for quark and gluon jets groomed with $\zcut = 0.1$ and $\beta = 1$.
Understanding these effects is also crucial for top quark mass determination using soft drop jet mass --- in \Refcite{Hoang:2017kmk}, leveraging on the resilience of soft drop against underlying event, soft drop jet mass was proposed as a candidate for precision top mass measurement, and followed up by a MC top mass calibration in \texttt{Pythia8+Powheg} by the ATLAS collaboration~\cite{ATLAS:2021urs}.
By directly comparing the theory prediction against the unfolded LHC data, a determination of the top quark mass in a short distance scheme can be envisaged.
However, the peak of the distribution where the dominant sensitivity to the top mass lies, receives significant hadronization corrections
which are important to account for in order to formulate a consistent hadron level prediction.

One of the most common method to account for non-perturbative (NP) corrections is to use hadronization
models, such as those provided with event generators such as \Pythia~\cite{Sjostrand:2007gs},
\Sherpa~\cite{Gleisberg:2008ta}, and \Herwig~\cite{Bahr:2008pv}, see for example
\Refcite{Marzani:2017mva}.
These hadronization models are extremely useful tools in guiding our intuition and allow us to study
nonperturbative effects on any arbitrarily complicated observable.
In a more recent work in \Refcite{Reichelt:2021svh}, this approach was employed to delve deeper into the
interaction between
parton showers and hadronization corrections, and account for the impact of event migration across
$p_{T,\rm jet}$ bins resulting from hadronization and underlying event.
However, it is crucial to note that these models are designed to be compatible with the respective parton
showers that are less
precise than the aforementioned analytical calculations at NNLL and beyond,
and hence are not ideal for describing NP power corrections in high precision analytical calculations.
Moreover, the parameters of event generators are tuned to a broad range of measurements that are
distinct from the target observables, thereby claiming universality. However, understanding the underlying
physics described by these models from a field theory perspective remains challenging. Due to the
intricate nature of these models, which offer multiple adjustable parameters, it is unclear how to
associate an intrinsic uncertainty with their predictions.

Another approach utilized to incorporate nonperturbative corrections involves the application of an
analytical model for
hadronization~\cite{Dasgupta:2007wa,Dasgupta:2013ihk,Marzani:2017kqd,Marzani:2019evv}. This model
is based on the dispersive representation of the strong coupling constant~\cite{Dokshitzer:1995qm},
allowing for efficient analysis of power corrections in a wide range of observables using Feynman
diagrams with `probe nonperturbative gluons'.
The dispersive approach has established a systematic framework for investigating
power corrections, addressing various limitations encountered with hadronization models in
precision phenomenology. In a first approximation, a single probe gluon is often sufficient to capture
the broad-scale kinematic features of the power corrections. For non-inclusive observables like the
jet mass, where the computation of power corrections may differ depending on whether one or
multiple probe gluons are considered, the introduction of the `Milan
factor'~\cite{Dokshitzer:1997iz,Dokshitzer:1998pt} allows for universality features within this
approach to be preserved. The Milan factor has
been computed for a wide range of $\ee$-event shapes, and studies in
\Refscite{Dasgupta:2009tm,Dasgupta:2009an} have explored the variations in the correction factor
arising from different jet clustering procedures.
Finally, in \Refcite{Salam:2001bd} has shown how naive
estimates within the dispersive framework can be further corrected to account for hadron mass
effects.
To additionally account for underlying event
effects, extensions have been made to this model in \Refscite{Dasgupta:2007wa,Marzani:2017kqd}.

However, it is important to note that the dispersive model
assumes that the effective coupling constant $\alpha_{\rm eff}(\mu^2)/\pi$, when extrapolated into
the
nonperturbative regime, remains numerically small and that higher-order effects in powers of the
effective
coupling do not significantly alter the estimation of power corrections.
Despite the considerable phenomenological success of the dispersive approach, a rigorous proof of
the universality of the infrared coupling constant and its smallness for
an expansion in the number of probe gluons is still lacking, and hence the approach fundamentally
relies on a model. To enhance the robustness and reliability
of the dispersive approach, it is important to thoroughly investigate the behavior of the effective
coupling constant in the nonperturbative regime. Such a positive proof would allow the results
obtained within this
framework to be regarded as entirely model-independent, providing more confidence in their
interpretation and applicability to high-precision phenomenology.

\subsection{A field-theoretic formalism for nonperturbative corrections}

For precision studies we need a first principles, field-theoretic paradigm for describing these nonperturbative power corrections~\cite{Lee:2006fn,Korchemsky:1994is,Korchemsky:1997sy,Korchemsky:1999kt,Belitsky:2001ij,Stewart:2014nna} that can be systematically improved and can be combined with perturbative calculations independent of their accuracy, beyond those achievable with parton showers. This has been made possible thanks to rigorous QCD factorization theorems~\cite{Collins:1981ta,Collins:1985ue,Collins:1988ig,Collins:1989gx,Collins:2011zzd}, combined with the powerful machinery of soft-collinear effective field theory (SCET)~\cite{Bauer:2000ew,Bauer:2000yr,Bauer:2001yt,Bauer:2001ct,Bauer:2002nz}. These methods have enabled a field theoretically consistent study of NP power corrections.
Although this approach cannot be applied generally to any arbitrary observable, by systematically analyzing a variety of jet substructure observables and event shapes, we have been able to draw universal and model-independent conclusions about the precise way in which NP corrections enter in these observables. By exploiting the universal properties of soft QCD, this approach allows us to identify the kinematic dependence of these NP corrections in a variety of jet observables, allowing them to be parameterized in terms of a few (a lot fewer than parameters in a hadronization model), universal $\cO(\Lambda_{\rm QCD})$ constants.
For example, an analysis along these lines reveals that nonperturbative corrections to the jet mass and jet $p_T$ take the following form~\cite{Dasgupta:2007wa,Stewart:2014nna},
\begin{align}\label{eq:NPUngroomed}
\delta m_J^2 = p_T \big(R\, \Omega_\kappa^{(1)} +R^3 \Omega_{\kappa}^{(3)} + \ldots \big) \, , \qquad
\delta p_T = \frac{1}{R}\Upsilon_\kappa^{(-1)} + R \Upsilon_\kappa^{(1)} + \ldots \, .
\end{align}
Here $\Omega_\kappa^{(i)}$ and $\Upsilon_\kappa^{(i)}$ are unknown $\cO(\Lambda_{\rm QCD})$ parameters which only depend on the flavor $\kappa = q,g$ of the parton initiating the jet, and $R$ is the jet radius. The specific scaling of these corrections with $p_T$ and the dependence on the jet radius is a model-independent statement that can be derived on general arguments of factorization of the soft physics. Here the leading terms with the smallest power of $R$ correspond to the leading hadronization effect in each of these cases. The subleading terms (including ones not shown in \eq{NPUngroomed}) arise due to contributions from the initial state radiation (ISR) and the underlying event.
Although not obtainable from a first-principle calculations, these nonperturbative parameters can be determined by making a comparison with experimental data.\footnote{Further assumptions about the hadronization physics in the dispersive approach~\cite{Dasgupta:2007wa} can be used to reduce them down to a single parameter (which also reveals that the leading hadronization correction to the jet $p_T$, $\Upsilon_1^{(-1)} < 0$).} These field theory predictions also clearly demarcate the kinematic region of validity of the proposed NP corrections as well as allow for precisely estimating the associated theory uncertainty (for example in terms of the size of higher order power corrections) and the experimental uncertainty (related to the statistics and the fitting procedure).
This approach was pursued in determination of the strong coupling constant from the LEP data in \Refscite{Abbate:2010xh,Abbate:2012jh,Hoang:2015hka}.

In this paper we focus on a factorization-based approach for analyzing the nonperturbative power corrections to groomed observables. Based on recent advancement in the understanding of the nonperturbative structure of the soft drop jet mass in \Refscite{Hoang:2019ceu,Pathak:2020iue}, the power corrections in this case can be similarly encoded into universal, $\cO(\LQCD)$ constants as in \eq{NPUngroomed}.
These power corrections in the groomed case in fact constitute a very non-trivial extension and a combination of those in \eq{NPUngroomed} for ungroomed jet mass shift and ungroomed jet $p_T$.
More specifically, it was shown in \Refcite{Hoang:2019ceu} that, in the region where soft drop is \textit{active} and where a perturbative description is valid, the leading corrections from hadronization take the following form:
\begin{align}\label{eq:NP}
\frac{1}{\sigma_\kappa} \frac{\df \sigma_\kappa}{\df m_J^2}
&= \frac{1}{\hat \sigma_\kappa} \frac{\df \hat\sigma_\kappa }{\df m_J^2} - Q \Ok \frac{\df }{\df m_J^2} \Big( \frac{1}{\hat \sigma_\kappa} \frac{\df \hat\sigma_\kappa^{\figeight} }{\df m_J^2} \Big)
+ \frac{\Uka + \beta \Ukb}{Q}
\frac{1}{\hat \sigma_\kappa} \frac{\df \hat\sigma_\kappa^{\bndry} }{\df m_J^2} + \cdots \,,
\end{align}
where the left hand side denotes the hadron-level soft drop jet mass cross section initiated by a parton $\kappa = $ quark/gluon and $Q$ is the hard scale characterizing the jet.
The first term on the right hand side with a hat, $\df \hat{\sigma}^\kappa$ is the perturbative parton level soft drop jet mass cross-section.
The $1/\hat \sigma_\kappa$ factor is included to consider normalized cross section. In the region of small jet masses that we are interested in, these functions can be unambiguously defined as jet mass distributions of jets with a specific quark or gluon flavor. See \Refcite{Hannesdottir:2022rsl} for more details on how these objects can be systematically defined in the context of inclusive or exclusive jet measurements.
The remaining pieces in \eq{NP} are the nonperturbative corrections parameterized in terms of the three $\cO(\Lambda_{\rm QCD})$ universal nonperturbative (NP) constants $\Ok$, $\Uka$ and $\Ukb$. They appear with certain jet mass dependent functions that we describe below.

\begin{figure}[t]
\centering
\includegraphics[width=.6\linewidth]{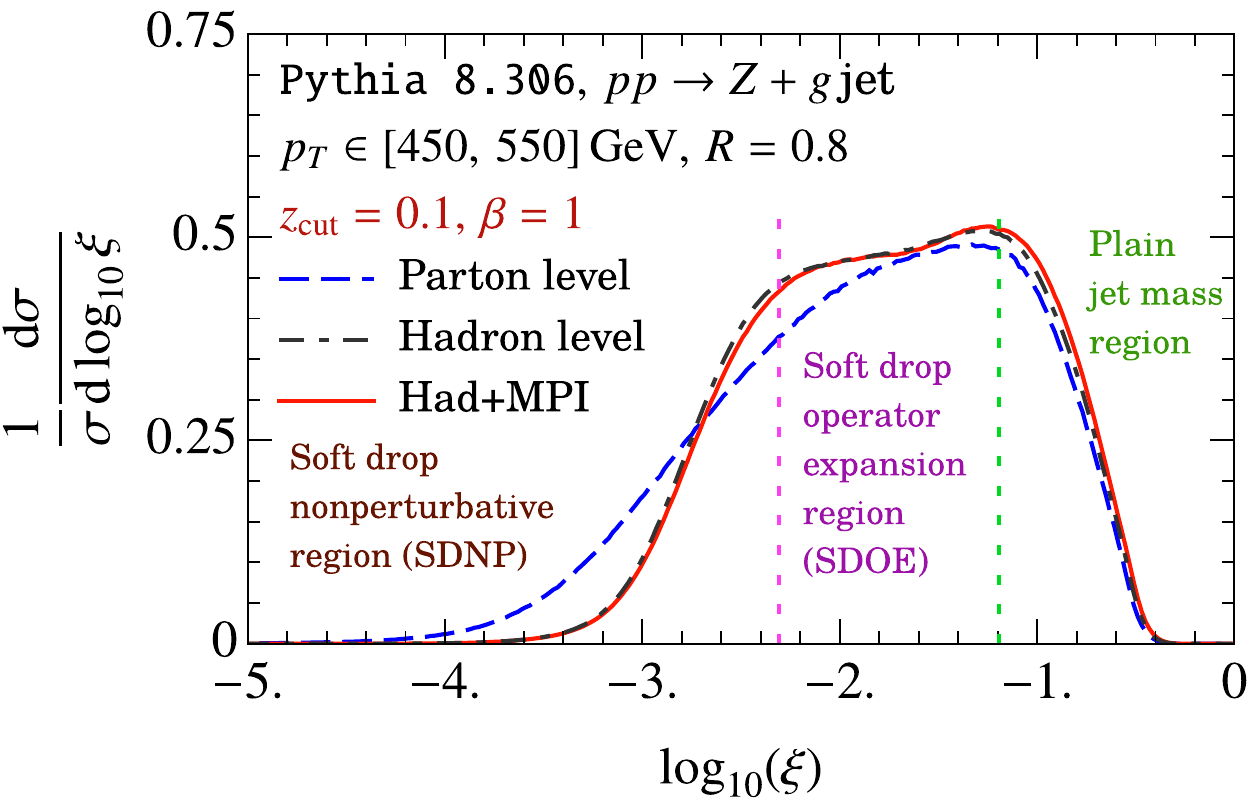}
\caption{Various regions of the soft drop jet mass spectrum that are differently affected by nonperturbative corrections. This work focuses on the middle, soft drop operator expansion region, where the hadronization corrections, while significant, can be described in a systematic expansion. Here $\xi = m_J^2/(p_TR)^2$.}
\label{fig:pythia}
\end{figure}

The form of NP power corrections in \eq{NP} holds for jet masses that satisfy the criteria
\begin{align}\label{eq:SDOE}
&\frac{Q \LQCD}{m_J^2} \Big(\frac{m_J^2}{Q\qcut}\Big)^{\frac{1}{2+\beta}} \ll 1 \,,&
&\text{(soft drop stopping emission is perturbative)} \,,&
\\
&m_J^2 < Q \qcut \, , &
&\text{(soft drop is active)} \, .& \nn
\end{align}
Here $\qcut \sim Q \zcut$ is the energy scale associated with soft drop and is precisely defined below. This region is referred to as the soft drop operator expansion (SDOE) region and is shown in \fig{pythia}. The first condition ensures that the subjet that stops the soft drop is typically perturbative, such that soft drop spectrum can be calculated in perturbation theory. This need not always be the case, but the cases where the stopping subjet is nonperturbative are power suppressed.
The second condition constraints the softer of the stopping pair to be also collinear to the jet, such that the radiation at wider angles is groomed away. For larger jet masses, the groomer can stop at wider angles and the distribution consequently looks very similar to the ungroomed jet mass spectrum,
\begin{align}\label{eq:plain}
&m_J^2\gtrsim Q \qcut \, ,& &\text{(plain jet mass resummation region)}\, .&
\end{align}
Interestingly, it is only with the LHC kinematics with jets of transverse momenta $\sim 500$ GeV that this
OPE region opens up and becomes accessible.\footnote{Soft drop jet mass distributions (and soft drop
angularities) have been measured in the legacy ALEPH data in \Refcite{Chen:2021uws}. However, at LEP
energies the above conditions cannot be satisfied and almost the entire soft drop jet mass spectrum
(including even the cusp region) is fully nonperturbative.}
We also note that in \eq{NP} we have ignored the impact of
hadronization on the hard scale $Q$. This is justified because the spectrum in
the SDOE region for large $Q$ values ($p_T \gtrsim 400$ GeV) is largely
independent of the jet $p_T$, and while hadronization and underlying event impacts the relative statistics
in different jet $p_T$ bins, the impact on normalized cross section shown in \fig{pythia} is expected to be
sub-dominant. This was also shown to be the case in \Refcite{Reichelt:2021svh} where a transfer matrix
approach was employed to investigate the effect of migration across $p_T$ bins due to hadronization and
UE.

The three parameters are grouped in two terms which have distinct physical meanings associated with them.
The first term proportional to $\Ok$ is the ``shift correction'' analogous to \eq{NPUngroomed} that captures a shift to the groomed jet mass, coming from the NP particles\footnote{We will often refer to hadrons produced at the last stage of hadronization (for example, subsequent to a parton shower in an event generator) as ``nonperturbative particles''. In the language of SCET, these corresponds to distinct modes in the effective theory that have significantly lower virtuality $\sim \Lambda_{\rm QCD}$ compared to other perturbative modes.} that survive grooming.
The second term proportional to $\Uka$ and $\Ukb$ is referred to as the ``boundary correction'' that describes how the outcome of the soft drop test (with respect to a reference parton level configuration) is altered due to hadronization.
The parameter $\Uka$ is exactly analogous to $\Upsilon_\kappa^{(-1)}$ in \eq{NPUngroomed} but now refers to the $p_T$ of the dynamically determined collinear-soft (c-soft) subjet that is found at the last stage of the soft drop groomer.
The term proportional to $\beta$ is related to the change in direction of this c-soft subjet relative to the collinear core due to hadronization.
Thus, an interesting implication of jet grooming is that both types of hadronization corrections appear in the same groomed observable.

We pause to note that `$\ldots$' in \eq{NP} refer to two types of subleading power corrections. The first kind are those that are suppressed by higher powers of $\LQCD$. These power corrections grow as the groomed jet mass is reduced and become $\cO(1)$ for small jet masses beyond the SDOE region. The second type corresponds to those where the radiation pattern of the groomed jet is more complicated than a simplest possible ``two-pronged configuration'' of a collinear and c-soft subjet. The second type of correction is a next-to-leading logarithmic (NLL) effect, as at leading-logarthmic accuracy, strong ordering of angles between the perturbative soft radiation off the jet can be assumed, such that any further radiation beyond the c-soft subjet must lie at hierarchically small angles within the collinear jet, resulting in a dipole-configuration that governs the hadronization corrections.
Equivalently, \eq{NP} can be seen as an expansion in number of identified and ordered soft subjets, analogous to the dressed gluon expansion employed in non-global logarithms (NGL) resummation~\cite{Larkoski:2015zka} and more recently for resummation of jet mass close to the cusp~\cite{Benkendorfer:2021unv}.
With this interpretation, the leading terms in \eq{NP} represent the leading single-subjet piece.
The dominance of the two-prong configuration was crucial in \Refcite{Hoang:2019ceu} for identifying universal NP constants in \eq{NP}.

Because these corrections necessarily involve kinematic properties of the c-soft subjet that cannot be unambiguously identified via the jet mass measurement, these corrections appear with certain jet mass dependent perturbative weights $\df \hat \sigma^{\figeight,\bndry}_\kappa$ that capture this dynamical effect.
Furthermore, in the SDOE region specified by \eq{SDOE}, these weights are perturbatively calculable and hence indicated with a hat.
These weights are related to moments of the groomed jet radius, $R_g$, the angular separation between the soft drop-stopping pair, at a given jet mass $m_J^2$.\footnote{Unlike \Refcite{Kang:2019prh}, we do not normalize the groomed jet radius $R_g$ by the original jet radius $R$. However, for sake of simplifying calculations we will consider a normalized variant $r_g$ defined in \eq{rgDef} below.} They are given in terms of cross sections that are differential in these kinematic properties of the c-soft subjet in addition to the jet mass:
\begin{align}\label{eq:shift}
\frac{1}{\hat \sigma_\kappa} \frac{\df \hat\sigma_\kappa^\figeight }{\df m_J^2} &\equiv \int \df r_g \: r_g \frac{1}{\hat \sigma_\kappa} \frac{\df^2 \hat \sigma_\kappa}{\df m_J^2 \df r_g} \, , \\
\label{eq:bndry}
\frac{1}{\hat \sigma_\kappa} \frac{\df \hat\sigma_\kappa^\bndry }{\df m_J^2}
&\equiv
\int \frac{\df r_g \df z_g \: \delta\big(z_g - \zcut r_g^\beta\big)}{r_g} \frac{1}{\hat \sigma_\kappa} \frac{ \df^3 \hat\sigma_\kappa }{\df m_J^2 \df r_g \df z_g} \, .
\end{align}
We have expressed here the groomed jet radius $R_g$ in terms of $r_g = R_g/R$, which we also generalize below in \eq{rgDef} to include jets in $\ee$ collisions. Here, $z_g$ is the energy (or $p_T$) fraction of the c-soft subjet.
These moments in fact appear in precisely the same fashion as the factor of jet radius $R$ does in the leading NP power correction in the ungroomed jet mass and jet $p_T$ in \eq{NPUngroomed}.
For the shift correction, the groomed jet mass shift is given by
\begin{align}
m_{J,\rm sd}^2 = \hat m_{J,\rm sd}^2 + p_T R_g \Ok \, .
\end{align}
Here $\hat m_{J,\rm \, sd}$ is the jet mass of a reference parton level configuration. Here we are implicitly considering inclusively identified jets, where, as we will discuss below, the hard scale $Q = p_TR$. Normalizing the jet mass squared by $Q^2$ we find that the shift $\delta m_{J,\rm sd}^2 = r_g \Ok/Q$. Thus, \eq{shift} indicates that $r_g$ must be averaged over all possible values allowed for a given jet mass measurement $m_J$. Lastly, because this shifts the value of the jet mass, the corresponding shift upon Taylor-expanding appears as a derivative as shown in \eq{NP}.

The term in \eq{bndry} apperaing with a $1/r_g$ factor is analogous to $1/R$ factor with the leading hadronization effect associated with the jet $p_T$ in \eq{NPUngroomed}. Instead of a shift in the jet mass, this effect modifies the normalization of the cross section. It is analogous to how the normalization of the jet mass spectrum changes from parton-level to hadron-level due to migration of events across $p_T$ bins. Just as this effect predominantly affects the events that are at the boundary of the $p_T$-bin, the analogous correction in \eq{bndry} appears in groomed jet mass when there are c-soft subjets near the ``boundary'' of passing/failing the soft drop condition, i.e. when $z_g \approx \zcut r_g^\beta$, such that effects of hadronization can lead to different outcomes at parton and hadron levels. Hence, in \eq{bndry} an additional $\delta$-function for soft drop boundary is included. The factor of $1/r_g$ can be understood by noting that the parton level values $\hat z_g$ and $\hat r_g$ upon hadronization are modified as
\begin{align}
&z_g =
\hat z_g + \frac{1}{r_g} \frac{\Uka}{Q} \, ,&
&r_g =
\hat r_g - \frac{\Ukb}{Q} \, .&
\end{align}
The two $\cO(\LQCD)$ constants $\Uka$ and $\Ukb$ respectively encode the shift in the c-soft subjet $p_T$ and the groomed jet radius. When combined together, along with the factor of $\beta$ resulting from differentiation of $r_g^\beta$, we arrive at the boundary correction in \eq{NP}.
Finally, as a technical remark, while \eq{bndry} is more intuitive to understand when expressed in terms of $z_g$, in practice, we will find it simpler to compute this correction by varying the soft drop condition itself in the doubly differential cross section in \eq{shift}, allowing us to recycle the calculations for the shift correction.

\begin{figure}[t]
\centering
\includegraphics[width=0.6\linewidth]{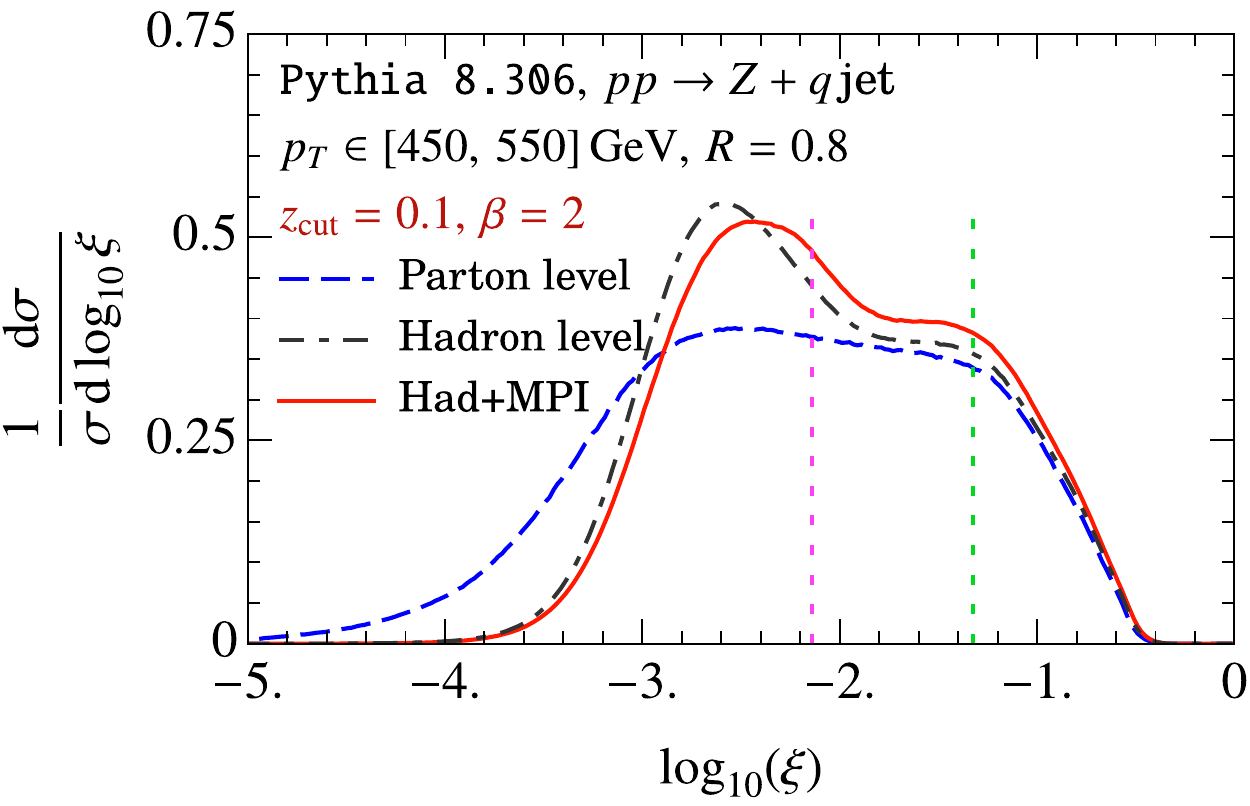}
\caption{For certain values of soft drop parameters, such as $\beta = 2$, the underlying event effects can be significant. Here $\xi = m_J^2/(p_TR)^2$.}
\label{fig:pythiaUE}
\end{figure}

Next, while typical values of the grooming parameters lead to strong suppression of underlying event (UE) and ISR effects, there are situations where less aggressive grooming is desirable. For example, in the case of groomed boosted top quark jets~\cite{Hoang:2017kmk}, a strong 10\%-level grooming invalidates a simple inclusive description of the top decay, and instead a light grooming of 1\%-level is desired. Likewise, for exclusively studying soft radiation and quark gluon discrimination using the collinear-drop~\cite{Chien:2019osu,Stewart:2022ari}, combinations of light and more aggressive soft drop is employed.
In such scenarios the effect of underlying event on the jet mass spectrum is no longer negligible. We show in \fig{pythiaUE} the spectrum for $\zcut = 0.1$ and $\beta = 2$. In the SDOE region the impact of UE is even somewhat larger than hadronization.
While it is impossible to predict the effects of UE from first principles, we can nevertheless attempt to phenomenologically describe these effects by making certain reasonable assumptions. The UE distribution is to a good approximation uniform in rapidity and independent of the hard scattering, such that it makes a contribution proportional to the jet area~\cite{Cacciari:2008gn}. Under these assumptions, in Ref.~\cite{Ferdinand:2023yyy} we show that effects of ISR and UE appear as corrections associated with higher powers of groomed jet radius:
\begin{align}\label{eq:UE}
\frac{1}{\sigma_\kappa} \frac{\df \sigma^{\rm had+UE}_\kappa}{\df m_J^2}
&= \frac{1}{ \sigma_\kappa} \frac{\df \sigma_\kappa }{\df m_J^2} - Q \OUE \frac{\df }{\df m_J^2} \Big( \frac{C_{1\kappa}^{(4)}(m_J^2)}{ \sigma_\kappa} \frac{\df \sigma_\kappa }{\df m_J^2} \Big) \\
&\quad
+ \frac{\UaUE + \beta \UbUE}{Q}
\frac{C_{2\kappa}^{(2)}(m_J^2)}{ \sigma_\kappa} \frac{\df \sigma_\kappa }{\df m_J^2} + \cdots \,, \nn
\end{align}
where $\df \sigma_\kappa$ (without a hat) is the hadron level cross section and the left hand side, $\df \sigma_\kappa^{\rm had + UE}$ is the hadron+UE level cross section. The jet mass dependent coefficients $C_{1,2\kappa}^{(n)}(m_J^2)$ are parton level $r_g$-moments of the doubly differential soft drop and boundary soft drop cross sections:
\begin{align}\label{eq:CnDef}
\frac{1}{ \hat \sigma_\kappa}\frac{\df \hat \sigma_\kappa}{\df m_J^2} C_{1\kappa}^{(n)}(m_J^2) &\equiv \int \df r_g \: r_g^n \frac{1}{\hat \sigma_\kappa} \frac{\df^2 \hat \sigma_\kappa}{\df m_J^2\df r_g} \, , \\
\frac{1}{ \hat \sigma_\kappa}\frac{\df \hat \sigma}{\df m_J^2} C_{2\kappa}^{(n)}(m_J^2)
&\equiv \int \df r_g \df z_g \: \delta(z_g - \zcut r_g^\beta) \: r_g^n\frac{1}{\hat \sigma_\kappa } \frac{\df^3 \hat \sigma_\kappa}{\df m_J^2 \df r_g \df z_g} \, . \nn
\end{align}
The appearance of $n = 4$ for the shift correction and $n = 2$ for the boundary correction is analogous to the jet radius scaling of the impact of ISR and underlying event on the jet mass and jet $p_T$ distribution respectively~\cite{Dasgupta:2007wa}.

\subsection{Precisely computing the perturbative weights of nonperturbative moments}

We thus see that \eq{NP} significantly constrains the form of the leading nonperturbative corrections, which can be parameterized for a given flavor of jet in terms of 3 constants.
However, for \eq{NP} to be useful in precision physics with soft drop jet mass, we are required to accurately calculate the jet mass dependent weights in perturbation theory. The accuracy with which these weights can be determined in turn determines the extent to which the NP parameters can be extracted or constrained in an analysis with real world collider data.
In \Refcite{Hoang:2019ceu}, a straightforward calculation of these weights at LL accuracy in the coherent branching formalism was presented.
The factorization in \eq{NP} was tested by performing a comparison between parton and hadron level jet mass spectra in the $\ee\ra q \bar q$ process in the dijet region simulated in the MC event generators using the LL-accurate predictions of the weights.
However, while a good agreement with the predictions of factorization was found,
there was no clear procedure to ascertain the perturbative uncertainty in the LL calculations.

To enable more precise predictions of these weights, it was in \Refcite{Pathak:2020iue} where their computation was first recast as moments of a multi-differential soft drop cross section.
By dissecting the kinematic phase space of $r_g$ and the $m_J$, \Refcite{Pathak:2020iue} identified the relevant set of effective field theories that are required for a precise computation of the doubly differential cross section in the SDOE region.\footnote{See also~\cite{Cal:2019gxa,Cal:2020flh,Cal:2021fla} for applications of doubly differential cross sections in the context of other groomed jet observables.} The boundary correction in \eq{bndry} was computed by considering variations in the soft drop condition in the doubly differential $\frac{\df^2 \hat \sigma_\kappa}{\df m_J^2 \df r_g}$ cross section.
The EFT formalism enabled a systematic improvement in the computation of these weights.
In \Refcite{Pathak:2020iue} these moments were computed with NLL resummation including $\cO(\as)$ singular matrix elements to achieve NLL$'$ accuracy in the SDOE region.
At the same time, the framework also enabled a systematic estimate of the perturbative uncertainty associated with these weights that was previously lacking in the LL calculations in the coherent branching framework.

In this work, with the goal of providing necessary perturbative input for \eq{NP} for precision phenomenology, we further improve the prediction of these jet mass dependent perturbative weights.
A straightforward improvement comes from employing relatively recently calculated two-loop non-cusp anomalous dimensions of certain factorization functions associated with soft drop from \Refcite{Bell:2018vaa} to extend the resummation of global logarithms in the doubly differential cross section to NNLL accuracy.
More crucially, we extend the calculation by matching the doubly differential cross section in the SDOE region in \eq{SDOE} to the ungroomed region for $m_J^2 \gtrsim Q \qcut$ which involves calculating new contributions previously not considered in \Refcite{Pathak:2020iue}. There the power corrections of the form $\big(m_J^2/(Q\qcut)\big)^{2/(2+\beta)}$ were systematically dropped, which become significant close to the soft drop cusp for $m_J^2 \lesssim Q \qcut$.
They also impact the location of the soft drop cusp which differs between NLL and NNLL estimates.
Thus, it is essential to include these power corrections in order to provide the necessary perturbative input for precision phenomenology consistent with the NNLL soft drop jet mass prediction.

In \Refcite{Hannesdottir:2022rsl}, the matching of the single differential jet mass spectrum $\df \sigma/\df m_J^2$ between soft drop and plain jet mass resummation regions was derived from demanding consistency with the doubly differential cross section $\df\Sigma(r_g)/\df m_J^2$ that in the limit $r_g = r_g^{\rm max}(m_J^2) \leq 1$ (applied as a cumulative measurement of $r_g$) reduces to the jet mass distribution, $\df\Sigma(r_g^{\rm max}(m_J^2))/\df m_J^2 = \df \sigma/\df m_J^2$. As discussed in \Refcite{Pathak:2020iue}, the calculation of the doubly differential cross section involves consideration of three different EFTs depending on whether the groomed jet radius is close to the minimum $r_g \gtrsim r_g^{\rm min}(m_J^2)$ (``min-$R_g$'') or maximum kinematic bound $r_g \lesssim r_g^{\rm max}(m_J^2)$ imposed by jet mass measurement (``max-$R_g$''), or in an intermediate region within these bounds $r_g^{\rm min}(m_J^2) \ll r_g \ll r_g^{\rm max}(m_J^2)$ (``int-$R_g$''). The regions of validity of these three EFTs are nontrivial patches in the $m_J^2$-$r_g$ plane. Consequently, the transition for the doubly differential cross section between the soft drop and plain jet mass resummation regions can be expected to be significantly more intricate than $\df \sigma /\df m_J^2$ due to nontrivial interplay of various power expansion that determine the two-dimensional region in the $m_J^2$-$r_g$ plane for the three EFTs; see \fig{regions}.

Fortunately, as we show in this paper, the results in the min-$R_g$ and int-$R_g$ regimes can be smoothly continued into the plain jet mass region. On the other hand, the max-$R_g$ regime (of which the single differential distribution $\df \sigma/\df m_J^2$ is a component) transitions \textit{discontinuously} between the soft drop and plain jet mass resummation region. This is because, different from soft drop resummation region where $r_g \leq r_g^{\rm max} \ll 1$, in the plain jet mass region the $r_g$ is close to the jet radius boundary $r_g = 1$ and the mode structure of the soft sector in the EFT changes. However, the mode and resummation structure of this EFT is identical to that of single differential jet mass (in either regions), and the dependence on $r_g$ in this EFT can be incorporated via fixed order calculations. This allows us to employ similar strategy as in \Refcite{Hannesdottir:2022rsl} to compute this matching. To achieve a smooth transition between various patches in the entire $m_J^2$-$r_g$ plane demarcating the EFTs we expand the toolbox of \Refcite{Pathak:2020iue} including new resummation kernels, generalizations of profile functions and weight functions described below. Finally, we also compute here the $\cO(\as)$ non-singular pieces that capture $\cO(\zcut)$ power corrections for the doubly differential cross section that were previously not taken into account.

As above, the computation of the boundary soft drop cross section and its moments in \eq{bndry} is also similarly extended into the plain jet mass region. In fact, we will find that including the matching to the plain jet mass region significantly impacts the $C_{2\kappa}^{(n)}$ moment defined in \eq{CnDef} in the cusp region relative to $C_{1\kappa}^{(n)}$ because this moment sharply drops to zero past the soft drop cusp. Furthermore, as shown in \Refcite{Ferdinand:2023vaf}, the dominant uncertainty in the determination of nonperturbative parameters results from that of the $C_{2\kappa}^{(-1)}$ moment because in the case of $C_{1\kappa}$, the uncertainties in the single and doubly differential cross sections are highly correlated and cancel in the ratio. The computation of $C_{2\kappa}$ is also more challenging as the tree-level result is $\cO(\as)$ and effects of resummation start at $\cO(\as^2)$ captured through cross terms of various one-loop pieces. When extending to the plain jet mass resummation region, we find that new additional $\cO(\as^2)$ cross terms arise. As a result, while conceptually straightforward, the computation of the boundary cross section quickly becomes unwieldy due to appearance of several resummation kernels associated with these new pieces in the plain jet mass region and $\cO(\as^2)$ cross terms, exhibiting intricate cancellations as we transition from groomed to ungroomed region. We overcome this and simultaneously achieve an efficient numerical implementation of these resummation kernels by recasting them as certain ``integral transforms'' involving nested plus-functions. We find a significantly faster implementation of the boundary cross section compared to \Refcite{Pathak:2020iue} where these kernels were instead expressed in terms of incomplete beta functions and their integrals.

The results of this work are used in complementary applications of \eqs{NP}{UE} and discussed in companion papers: Firstly, as already mentioned above, in \Refcite{Hannesdottir:2022rsl} the NNLL results are employed for estimating impact of the nonperturbative corrections on $\as$-determination in a completely model-independent approach. Secondly, in \Refcite{Ferdinand:2023vaf} these results are used for a precise calibration of MC hadronization models in event generators, investigating their interplay with parton showers, and rigorously testing the universality predictions of \eq{NP}.
Analysis of \Refcite{Ferdinand:2023vaf} confirms that indeed with precise calculations of the perturbative weights with reliable uncertainty estimates, a determination of these parameters with LHC data is foreseeable. Finally, in \Refcite{Ferdinand:2023yyy} the calculations of the moments relevant for UE and ISR are used for an analogous calibration of underlying event contribution in simulations.
An interesting extension of this approach will be to consider nonperturbative corrections to groomed
angularities. We expect the dependence the angularity exponent to be captured by more general
perturbative weights. However, since hadron mass effects impact jet mass and angularities (when
measured in their
usual scheme) differently, care must be taken in relating the associated nonperturbative parameters.

The organization of the paper is as follows: In \secn{ddiff} we discuss the effective field theories required for a complete prediction of the doubly differential groomed cross section and identify the relevant EFT modes for extension into the plain jet mass resummation region.
Before describing in detail the factorization formulae associated with each of these regions, we will find it advantageous to first review $\cO(\as)$ computation of various factorization functions in \secn{nlo}.
In \secn{ddifffact} we describe how the results for these functions are incorporated into the factorization formulae.
The cross sections in various regions are eventually combined together in \secn{match} where we also describe the procedure for obtaining perturbative uncertainty.
The analogous calculation for the boundary correction is described in \secn{bndry}.
Having discussed the complete calculation of the perturbative weights, in \secn{compare} we compare the results of this work with the previous calculation in \Refcite{Pathak:2020iue} as well as their impact on the jet mass spectrum and on the calibration of MC hadronization models in terms of soft drop nonperturbative moments.
In \secn{mc} we show a comparison of the NNLL resummed results and parton shower simulations for various moments of $r_g$, for both shift and boundary cross sections.
We conclude in \secn{conclusion}. In the appendices we consolidate some of the technical details.

\section{Effective theory regions and modes}
\label{sec:ddiff}

In this section we review the relevant results of previous work from Refs.~\cite{Pathak:2020iue,Hannesdottir:2022rsl} and discuss the mode structure of the doubly differential cross section in the plain jet mass region.
We first review in \secn{kin} the measurement and kinematics of inclusive jets and the various energy scales associated with the groomed jet mass measurement. In \secs{eft}{modes} we summarize the various effective field theory regions and modes derived in \Refcite{Pathak:2020iue} and the new cases from extension into the plain jet mass region.
\subsection{Measurement and kinematics}
\label{sec:kin}
For concreteness, our starting point is the inclusive jet measurement in hadron colliders. However, we will shortly generalize our notation to also simultaneously describe exclusive jets in $pp$ and inclusive jets in $\ee$ collisions. In the formal limit of jet radius $R/2 \ll 1$ the cross section for the process $pp \ra \text{jet} + X$, where $X$ includes any radiation we are inclusive over factorizes as~\cite{Ellis:1996mzs,Kang:2016mcy,Kang:2018jwa}
\begin{align}\label{eq:inclJ}
\frac{\df^3 \Sigma (R_g)}{\df p_T \df \eta_J \df m_J^2 }
= \sum_{abc}\int \frac{\df x_a \df x_b \df z}{x_a x_b z} f_a(x_a,\mu) f_b (x_b, \mu) H_{ab}^c \Big(x_a, x_b, \eta_J , \frac{p_T}{z} , \mu\Big) \cG_c (z, m_J^2 , R_g, p_T ,R, \mu) \, ,
\end{align}
Here $f_{a,b}$ are parton distribution functions which when combined with inclusive hard function $H_{ab}^c$ account for the hard process leading to production of parton $c$. This can also be generalized to describe processes with an additional vector boson $V$, such as $pp \ra \text{jet} + V +X$.
The subsequent branching of the parton $c$ to form a jet is described via the inclusive jet function $\cG_c$. In addition to the jet $p_T$ and pseudorapidity $\eta_J$, $\cG_c$ depends on $z$, the momentum fraction of original parton retained in the reconstructed jet as well as (groomed) jet mass. Finally $\df \Sigma(r_g)$ refers to the cross section that is differential in the jet kinematics and the jet mass with an additional upper bound of $R_g$ as the groomed jet radius.

In the small jet mass limit $m_J^2 \ll p_T^2 R^2$, the radiation in the jet is constrained to be predominantly soft and collinear, such that the inclusive jet function factorizes as
\begin{align}\label{eq:cGcFact}
\cG^{\rm fact.}_c \big(z, m_J^2 , R_g, p_T ,R , \mu\big) = \sum_{i} \cH_{c \ra i } \big(z, p_T R , \mu\big) \cJ_i (m_J^2, R_g, p_T , \eta_J, R , \mu) \bigg[1 + \cO\Big(\frac{m_J^2}{p_T^2R^2}\Big)\bigg] \, .
\end{align}
The factorization involves separation of hard collinear modes at scales $p_T R$ described by $\cH_{c\ra i}$, and soft and collinear modes in a multiscale function $\cJ_i$.
As discussed in \Refcite{Hannesdottir:2022rsl}, in this region, the normalized cross section can be expressed as
\begin{align}\label{eq:sigIncl}
&\frac{1}{\sigma_{\rm incl}(p_T ,\eta_J)}
\frac{\df^3 \Sigma(R_g)}{\df m_J^2 \df p_T \df \eta_J} \\
&\qquad = x_q (p_TR, \eta_J, \mu )\: \tcG_q (m_J^2, R_g,p_T R, \mu) + x_g (p_TR, \eta_J, \mu ) \tcG_g (m_J^2,R_g, p_T R, \mu) \, . \nn
\end{align}
where $x_{q,g}$ are quark/gluon fractions that are theoretically well defined, but renormalization scale $\mu$-dependent objects. These fractions depend on the underlying hard process and being normalized quantities are only mildly dependent on the parton distribution functions.
Our focus will be on computing the piece that captures the jet mass and groomed jet radius dependence $\tcGk$, which is interpreted as the normalized cross section for parton $\kappa$:
\begin{align}\label{eq:tcGDef}
\tcGk (m_J^2 ,R_g, p_TR, \mu) \equiv \frac{1}{\sigma_\kappa^{\rm incl}} \frac{\df \Sigma_\kappa(R_g)}{\df m_J^2} = N_{\rm incl}^\kappa (p_T R, \mu) \cJ_\kappa (m_J^2, p_T , \eta_J, R , \mu ) \, .
\end{align}
Here the normalization factor $N_{\rm incl}^\kappa$ arises from factoring out the piece of $\cH_{c\ra i}$ that describes the Sudakov double-logarithmic part connected with the soft-collinear factorization of $\cG_c$ in \eq{cGcFact}. The product of $N_{\rm incl}^\kappa$ and $\cJ_\kappa$ lead to an RG-invariant combination that can be independently studied.
When considering different measurements involving groomed jet mass and groomed jet radius, only the function $\cJ_\kappa$ will be modified.
We will state the all-orders factorization formulae below in terms of $\cG_c$ in \eq{cGcFact}, and employ \eq{tcGDef} for numerical implementation.

\subsubsection{Hard kinematics}

The above formulation in terms of the normalized inclusive cross section $\tcG_\kappa$ and quark/gluon fractions can be extended to exclusive, fixed number of jets with a jet veto on additional radiation, or jets in $\ee$ collisions.
Apart from different definitions of the quark gluon fractions, this involves straightforward substitutions of the hard scales and kinematic and grooming parameters. Thus, it will prove worthwhile to develop a unified notation to simultaneously treat all these cases.
We first define the hard scale for the following scenarios:
\begin{align}\label{eq:Qdef}
Q_{\rm incl}^\pp \equiv p_T R \, , \qquad Q_{\rm excl}^\pp \equiv 2 p_T \cosh \eta_J \, , \qquad Q_{\rm incl}^\ee \equiv 2 E_J \tan \frac{R}{2} \, ,
\end{align}
where the subscript `incl' (`excl') indicates that the jets are identified inclusively (exclusively) and the superscript for the incoming partons involved. This way, the function $N_{\rm incl}^\kappa$ only depends on the scale $Q$.
We now define a dimensionless variable $\xi$ as a substitute for the jet mass:
\begin{align}
\xi \equiv \frac{m_J^2}{Q^2} \, .
\end{align}
We have suppressed the superscripts and subscripts on $Q$, and will generally follow this convention, such that the above definition of $\xi$ can be adjusted for the three situations in \eq{Qdef}.

Next, to describe dynamics of the soft and collinear radiation we will work with light cone coordinates defined via the light-like vectors
\begin{align}
n^\mu \equiv \zeta^{-1} \big(1, \vec n\big) \, , \qquad \bn^\mu \equiv \zeta \big(1, - \vec n \big) \, ,
\end{align}
where the parameter $\zeta$ encodes a large boost in the jet direction given by:
\begin{align}\label{eq:zetaDef}
\zeta_{\rm incl}^\pp \equiv \frac{R}{2 \cosh \eta_J} \, , \qquad \zeta_{\rm excl}^\pp \equiv 1 \, , \qquad \zeta_{\rm incl}^\ee \equiv \tan\frac{R}{2} \, .
\end{align}
Including the boost factors in the reference light-like vectors will allow us to work with hemisphere-like coordinates and eliminate several intermediate factors involving jet radius.
In terms of these light-like vectors, we will follow the following light-cone decomposition of momentum $q^\mu$:
\begin{align}\label{eq:LC}
q^\mu =q^+ \frac{\bn^\mu}{2} + q^- \frac{n^\mu}{2} + q_\perp^\mu \, ,
\qquad
q^+ \equiv n \cdot q \, , \qquad q^- \equiv \bn \cdot q \, .
\end{align}

\subsubsection{Soft drop kinematics}

Next we define some useful variables associated with soft drop. The criteria for soft drop differs between $\ee$ and $\pp$ cases and is given by
\begin{align}\label{eq:SD}
&\frac{\min(E_i,E_j)}{E_i+E_j} > z_\cut \bigg( \sqrt{2}\frac{\sin(\theta_{ij}/2)}{\sin(R_0^\ee/2)}\bigg)^\beta\,, &
&
(\textrm{$e^+e^-$ case})\, ,&
\\
&\frac{\min(p_{T_i},p_{T_j})}{p_{T_i}+p_{T_j}} > z_\cut \Big(\frac{\Delta R_{ij}}{R_0^\pp}\Big)^\beta\,, &
&
(\textrm{$pp$ case})\, ,&
\nn
\end{align}
Application of soft drop introduces an additional scale $\qcut$ that plays a role in distinguishing the radiation that is groomed away from the radiation that is not, which is given by
\begin{align}
Q^\pp_{\rm cut,\, incl} &\equiv \zcut Q_{\rm incl}^\pp \Big(\frac{R}{R_0^\pp}\Big)^\beta \, , \qquad
Q^{\pp}_{\rm cut, \, excl} \equiv \zcut Q_{\rm excl}^\pp \Big(\frac{2 \cosh \eta_J}{R_0^\pp}\Big)^\beta \, , \\
Q^\ee_{\rm cut, \, incl} &\equiv \zcut Q_{\rm incl}^\ee \Bigg(\sqrt{2}\frac{\tan \frac{R}{2}}{\sin \frac{R_0^\ee}{2}}\Bigg)^\beta \nn \, .
\end{align}

As we probe jets with smaller masses, the emissions at wider angles become progressively softer and are groomed away.
Using the scale $\qcut$ we can identify the jet mass transition point below which soft drop becomes active. This is simply given by
\begin{align}\label{eq:xiDef}
\xi_0 \equiv \frac{\qcut}{Q} \, .
\end{align}
This point roughly defines the location of the distinct soft drop cusp, and for the reasons that will become clear below, is referred to as the ``soft-collinear transition point''. Higher order corrections slightly modify this value. We will see below that at $\cO(\as)$, the transition point appears at the location
\begin{align}\label{eq:xi0p}
\xi_0' \equiv \frac{\xi_0}{\big(1 + \zeta^2\big)^{\frac{2+\beta}{2}} } \, .
\end{align}
We will refer to $\xi_0'$ as the soft wide-angle transition point. $\xi_0'$ corresponds to the location of the cusp for the NNLL resummed groomed jet mass spectrum. For later use we also define
\begin{align}\label{eq:Qcutp}
\qcut' \equiv Q \xi_0' = \frac{\qcut}{\big(1 + \zeta^2\big)^{\frac{2+\beta}{2}} } \, .
\end{align}
Finally, the of validity of perturbative calculations is specified by \eq{SDOE}, which demarcates the extent of the SDOE region. In terms of $\xi$, we define the lower end of the perturbative region by
\begin{align}\label{eq:xiSDOE}
\xi_{\rm SDOE} \equiv \xi_0 \Big(\frac{\rho \Lambda_{\rm QCD}}{Q \xi_0} \Big)^{\frac{2+\beta}{1+\beta}} \, ,
\end{align}
where the factor $\rho > 1$ is included to approximate the strong inequality in \eq{SDOE}. Below we will take $\rho = 5$.

Finally, we now specify our choice of normalization for the groomed jet radius. We refer to $R_g$ as the groomed jet radius which can stand for rapidity invariant angular distance in $pp$ collisions or physical angular distance in $\ee$ collisions.
For simplicity, we will work with a normalized version of $R_g$ to simultaneously treat these two cases defined by
\begin{align}\label{eq:rgDef}
r_g^{\pp} \equiv \frac{R_g}{R} \, , \qquad
r_g^\ee \equiv \frac{\tan\frac{R_g}{2}}{\tan \frac{R}{2}} \, .
\end{align}
Here the same definitions apply for both inclusive and exclusive cases in $\pp$. In this normalization, $r_g$ can at most be 1, which corresponds to scenario where grooming does not eliminate any radiation from the jet.

As detailed in \Refcite{Pathak:2020iue}, the measurement of jet mass $m_J^2$, or equivalently $\xi$, puts kinematic bounds on the groomed jet radius, such that
\begin{align}
\text{Simultaneous $\xi$, $r_g$ measurement:} \qquad
r_g^{\rm min}(\xi ) \leq r_g \leq r_g^{\rm max}(\xi) \, ,
\end{align}
where
\begin{align}\label{eq:rgbounds}
r_g^{\rm min} (\xi) = \sqrt{\xi} \, , \qquad
r_g^{\rm max} (\xi) \approx \min \bigg\{\Big(\frac{\xi}{\xi_0}\Big)^{\frac{1}{2+\beta}}, 1\bigg\} \,.
\end{align}
The $r_g^{\rm min}$ arises from the kinematic bound imposed by the jet mass measurement: for a given jet mass $\xi$, it is impossible to squeeze the radiation into a jet of radius smaller than $r_g^{\rm min}(\xi)$. The maximum bound has two terms. For $\xi < \xi_0$, we are in the region where grooming is active. Here the combination of the groomed jet mass measurement and requirement that the radiation pass grooming leads to the first of the two upper bounds shown. For $\xi > \xi_0$, the radiation can be close to the jet boundary and still pass the groomer, and hence the upper bound $r_g^{\rm max}(\xi)$ saturates to 1.
Here we have stated an approximate version of $r_g^{\rm max}$ that was derived in \Refcite{Pathak:2020iue} in the limit $\xi \ll \xi_0$. We provide the formula compatible with the NNLL transition point below in \eq{rgbounds}.
We also see that for $\xi < 1$, $r_g^{\rm min}(\xi) < r_g^{\rm max} (\xi)$.

\subsection{Effective theory regions}
\label{sec:eft}
The measurement of jet mass, groomed jet radius and application of soft drop introduce several energy scales which can become hierarchical in certain regions of the $\xi$-$r_g$ phase space and consequently induce logarithmic singularities. We now list down the various regions that require a specialized effective field theory based treatment.

We first consider the single differential groomed jet mass measurement and enumerate the various factorization regimes that are relevant.
\begin{enumerate}
\item \underline{Fixed order region}:
This corresponds to the region when $\xi \lesssim 1$. Here the fixed order treatment of the (groomed) jet mass differential cross section suffices.
\item \underline{Plain jet mass resummation region}:
This is the region where $\xi_0 < \xi \ll 1$. Here $\xi \ll 1$ implies that the jet mass $m_J$ is hierarchically smaller than the hard scale $Q$. This results in dominance of the soft-collinear modes leading to Sudakov double logarithms, $\alpha_s^n \ln^k\xi$, with $k \leq 2n$. This region can be described by the standard soft-collinear Sudakov factorization for plain jet mass. The condition that $\xi > \xi_0$ implies that effects of grooming alone can be treated in fixed order perturbation theory.
\item \underline{Soft drop resummation region:}
As we discussed above, the effects of grooming on the spectrum become visible for $\xi < \xi_0$. When we move to yet lower jet masses for $\xi \ll \xi_0$, additional logarithms related to soft drop involving the ratio $\xi/\xi_0$ become large and require resummation. This results in an additional factorization of the soft radiation beyond the one mentioned above in plain jet mass resummation.
\end{enumerate}

Next, as discussed in \Refcite{Pathak:2020iue}, three different effective theory regimes related to simultaneous measurement of $r_g$ and $\xi$ arise:\footnote{In \Refcite{Pathak:2020iue} these regimes were respectively referred to as large-$R_g$, intermediate-$R_g$ and small-$R_g$ cases. We avoid using labels `small' and `large' to avoid any confusion related to numerical size of $R_g$.}

\begin{figure}[t]
\centering
\includegraphics[width=\linewidth]{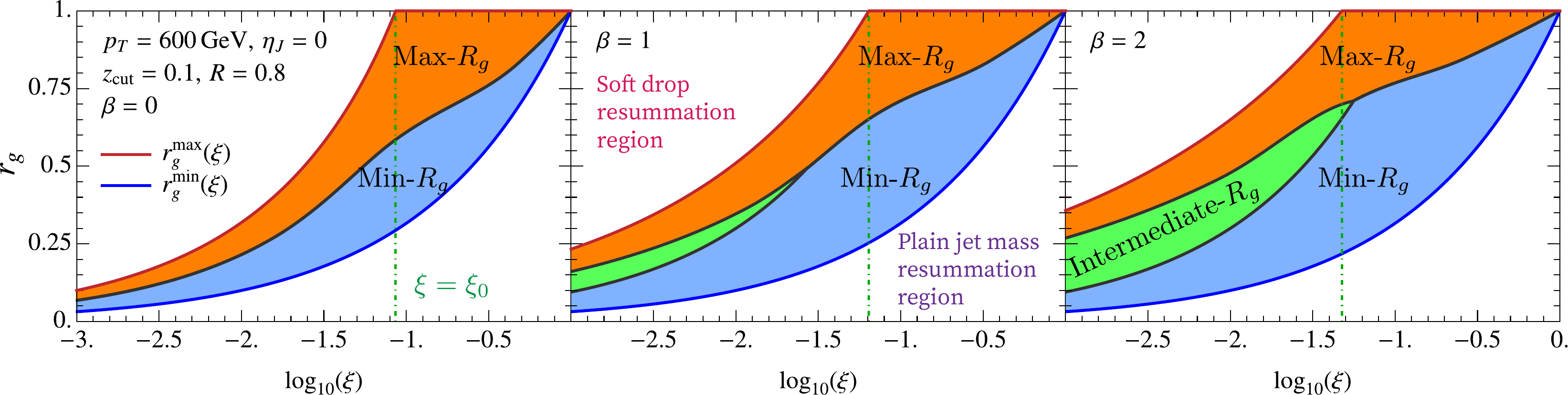}
\caption{Three regimes relevant for doubly differential groomed jet radius and groomed jet mass measurement for $\beta = 0,1,2$. The region to the left of the vertical line $\xi = \xi_0$ is the soft drop resummation region, and to the right is the plain jet mass resummation region. The boundaries of the shaded region denote the kinematic bounds imposed on groomed jet radius by the jet mass measurement. Depending on the values of grooming parameters, a third intermediate-$R_g$ regime can become relevant.}
\label{fig:regions}
\end{figure}
\begin{enumerate}
\item \underline{Max-$R_g$ regime}:
In this region the groomed jet radius is close to the maximum possible value, such that
\begin{align}
\text{Max-$R_g$ regime:} \qquad r_g^{\rm min}(\xi) \ll r_g \lesssim r_g^{\rm max}(\xi) \, .
\end{align}
As we will see below, the factorization in this regime is identical to that of the single differential groomed jet mass, and
the additional measurement of $r_g$ can be treated as a fixed order correction.
Note that unlike \Refcite{Pathak:2020iue} we do not impose an additional constraint of $r_g^{\rm max}(\xi) \ll 1$ and also include the situation $r_g^{\rm max}(\xi) \lesssim 1$ which arises in the plain jet mass region.

We also note that this regime is also relevant for the leading hadronization corrections. Here the collinear-soft modes that pass grooming lie at maximal possible angular separation from the collinear core, and hence the leading two-pronged configurations relevant for \eq{NP} arise specifically in this regime.
\item \underline{Intermediate-$R_g$ regime}:
This regime is relevant when the groomed jet radius is hierarchically separated from both minimum and maximum kinematic bounds:
\begin{align}
\text{Intermediate-$R_g$ regime:} \qquad
r_g^{\rm min} (\xi) \ll r_g \ll r_g^{\rm max} (\xi) \, .
\end{align}
Depending on the precise values of the hard scale and grooming parameters, this regime may or may not be relevant. However, because of the two hierarchies, describing this regime leads to the most factorized version of the doubly differential cross section. As a result, the intermediate regime cross section can be used as an efficient tool to subtract the singular pieces and define a stable prescription for incorporating fixed order power corrections in the entire double differential spectrum.
\item \underline{Min-$R_g$ regime:}
This regime arises when the groomed jet radius is close to the lower bound:
\begin{align}
\text{Min-$R_g$ regime:} \qquad r_g^{\rm min} (\xi) \lesssim r_g \ll r_g^{\rm max} (\xi) \, .
\end{align}
This physically corresponds to a scenario where the jet is filled uniformly with hard collinear radiation, with a haze of soft radiation at the same angular scales. If we fix the $r_g$ and vary the jet mass, this region in fact corresponds to the end-point of the jet mass spectrum. For this reason, the factorization in this regime has resemblance with the fixed-order region of the singly differential jet mass spectrum. This feature of this regime was utilized in \Refcite{Hannesdottir:2022rsl} for incorporating jet mass related power corrections in the single differential jet mass spectrum.
\end{enumerate}

We show the various regimes discussed above in \fig{regions} for jets with LHC kinematics and $\beta = 0$ and $\beta = 1$. The vertical line at $\xi = \xi_0$ separates the soft drop and plain jet mass resummation regions. The boundary of the colored region is the kinematic bound on $r_g$ given in \eq{rgbounds}.
We see that for $\beta = 0$, the intermediate-$R_g$ regime is absent whereas this regime covers a substantial phase space for $\beta = 2$ in the soft drop resummation region. The precise formulae for demarcating these regions are discussed in \secn{match} and \app{weight}.

\subsection{Effective theory modes}
\label{sec:modes}
\begin{figure}[t]
\centering
\includegraphics[width=\linewidth]{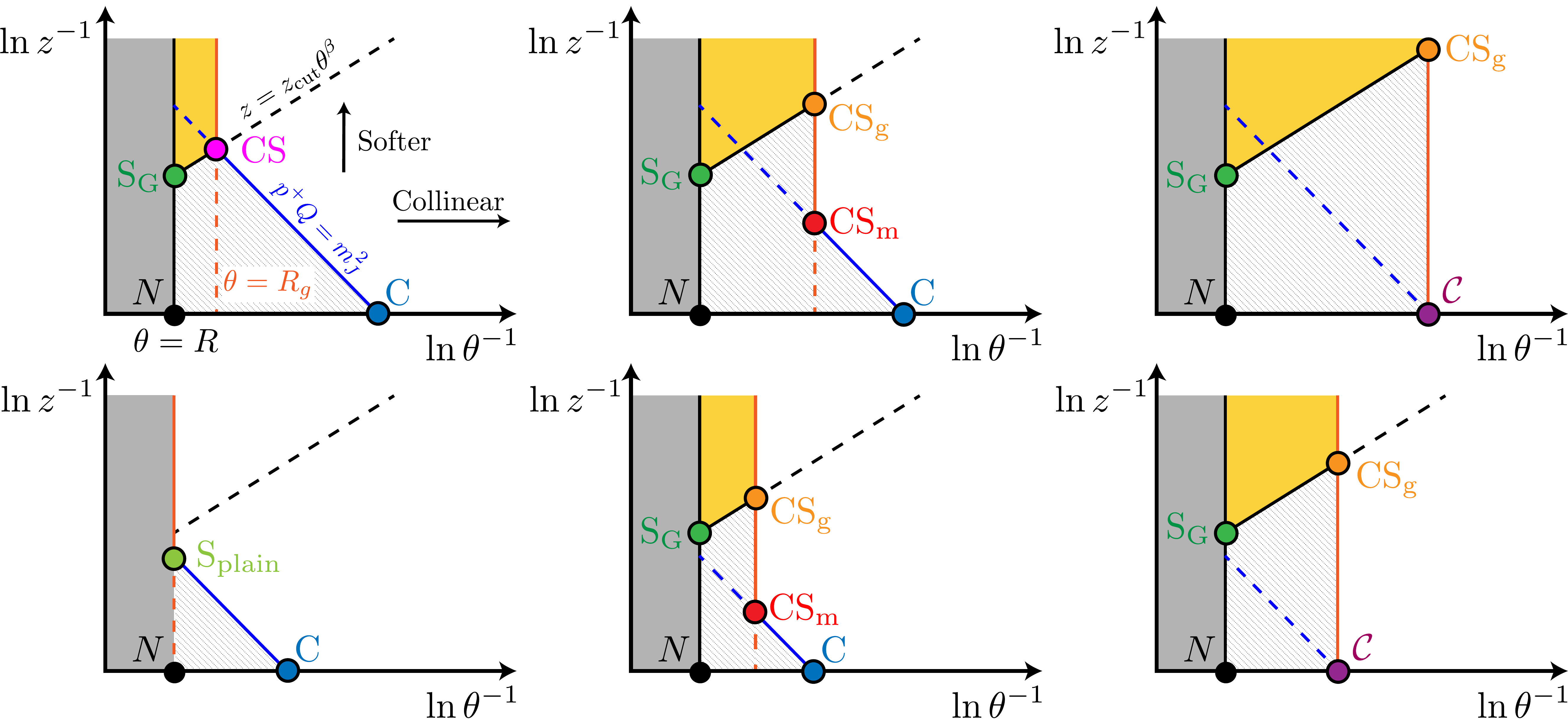}
\caption{Effective theory modes appearing in various regimes for double differential measurement. The solid yellow shaded region is where radiation is groomed away. The rectangular gray region corresponds to radiation falling outside the jet. The hatched region is the radiation vetoed by combination of grooming, jet mass and groomed jet radius measurements. The top row corresponds to the jet masses in the soft drop resummation region where as the bottom row describes scenarios in the plain jet mass resummation region.}
\label{fig:modes}
\end{figure}

From above discussion we learn that there are two jet mass regions and three regimes corresponding to groomed jet radius measurements that require resummation.
It is instructive to represent the measurements and the modes in the (primary) Lund plane of a soft/collinear emission off the jet-initiating fast parton as shown in \fig{modes}.
Here $z$ is the momentum fraction of the emission and $\theta$ is the angle it makes relative to the jet axis.
The choice of axes in \fig{modes} implies that emissions to the right are increasingly collinear and those higher are increasingly softer. These schematic figures are extremely useful in identifying the relevant effective theory modes that appear at the intersections of various measurements and constraints imposed on the jet.

We first describe how various measurements are displayed in \fig{modes}. The black vertical line labeled $\theta = R$ denotes the boundary of the jet and the gray region corresponds to radiation outside the
jet.
Emissions on the blue line with negative slope labeled $p^+ Q = m_J^2$ contribute (in combination with the fast massless jet-initiating parton) jet mass of $m_J^2$.
Thus, increasing the jet mass corresponds to moving this line downwards.
The soft drop condition is given by the dashed black line with positive slope.
The groomed jet radius measurement is given by the orange vertical line labeled $\theta = R_g$.
Emissions that are vetoed by soft drop are the ones that are encountered at angles larger than $R_g$ and are shown in yellow shaded region.
Hence, the columns from left to right correspond to max-$R_g$, intermediate-$R_g$ and min-$R_g$ regimes respectively.
Finally, a given jet mass and groomed jet radius measurement (along with jet radius and soft drop constraints) excludes any emissions that are harder shown in the hatched region.
The cases in the top row where the jet mass lies in the soft drop resummation region were already considered in \Refcite{Pathak:2020iue}. The cases in the bottom row are new and correspond to larger jet masses in the plain jet mass resummation region.
In the bottom left plot we show the completely ungroomed case when the $R_g = R$. However, this plot also includes the scenario where $R_g \lesssim R$, and where the effects of soft drop do not require any further factorization.

We notice that the relevant EFT modes appear at the intersections of two or more measurement or veto conditions.
\begin{enumerate}
\item \underline{Collinear modes}: Modes on the $x$-axis represent collinear radiation with $z \sim 1$ emitted by the fast parton at the core of the jet. In the scenarios there is a softer radiation at wider angles that stops the groomer, the jet mass measurement imposed on the collinear radiation at the center of the jet is essentially inclusive denoted by \textcolor{blue}{$C$}. On the other hand, (hard-)collinear modes denoted by $N$ and \textcolor{Purple}{$\cC$} respectively see the jet radius and groomed jet radius boundary.
\item \underline{Wide-angle soft modes}:
Modes at wider angles on the $y$-axis will naturally encounter the groomer first. For jet masses in the top row of \fig{modes}, the radiation at the jet-boundary must necessarily be groomed away, else we would have found a larger value of the jet mass. The physics associated with this radiation can be factorized in the soft drop resummation region when $\xi \ll \xi_0$ described by the global soft mode $\mSG$. On the other hand, in the plain jet mass region, for $r_g \lesssim 1$, the wide angle radiation is energetic enough to pass soft drop and thus we do not encounter this mode, as shown in the bottom left case. Here, the wide angle mode $\mSp$ is the same as in the plain jet mass resummation. These two modes only differ in their relative energy and the combination of measurements and vetoes they see.
\item \underline{Collinear-soft modes}:
Finally, we have modes that have simultaneously $z \ll 1$ and $\theta \ll 1$. These collinear soft modes are distinguished from each other via the role they play in the entire measurement. The \textcolor{Magenta}{CS} mode is the same as that appears in the single differential jet mass measurement in the soft drop resummation region. It has the largest possible angle and energies to saturate the soft drop condition. When $r_g < r_g^{\rm max}(\xi)$, we denote the soft radiation that stops soft drop as \textcolor{BurntOrange}{CS$_g$}. However, as we can see \textcolor{BurntOrange}{CS$_g$} mode does not carry sufficient energy to result in the jet mass shown. Thus we include another mode \textcolor{Red}{CS$_m$} which lies at similar angular scales but is more energetic as required by the jet mass measurement.
\end{enumerate}

\begin{table}[t]
\renewcommand{\arraystretch}{1.2}
\centering
\begin{tabular}{c | l | c}
\hline \hline
Mode & Description & Scaling \\
\hline
$N$ & Hard-collinear mode at jet boundary & $Q \big(1, \, 1, \, 1\big)$ \\
\hline
$\mcC$ & Hard-collinear mode within $R_g$ & $Qr_g \big(r_g,\, \frac{1}{r_g}, \, 1\big) $ \\
\hline
$\mC$ & Collinear mode for inclusive jet mass & $Q\sqrt{\xi} \big(\sqrt{\xi}, \, \frac{1}{\sqrt{\xi}} ,\, 1\big)$ \\
\hline
$\mSp$ & Wide-angle soft & $Q \xi \big(1, \, 1, \,1 \big)$ \\
\hline
$\mSG$ & Global soft & $\qcut \big(1, \, 1, \, 1\big) $ \\
\hline
$\mCSg$ & Collinear-soft mode for groomed jet radius & $\qcut r_g^{1+\beta} \big(r_g,\,\frac{1}{r_g},\, 1\big)$ \\
\hline
$\mCSm$ & Collinear-soft mode for jet mass& $\frac{Q\xi}{r_g} \big(r_g,\, \frac{1}{r_g}, \, 1\big)$ \\
\hline
$\mCS$ & Collinear-soft mode at maximum $R_g$ & $\frac{Q\xi}{r_g^{\rm max}(\xi)} \bigg(r_g^{\rm max}(\xi),\, \frac{1}{r_g^{\rm max}(\xi)}, \, 1\bigg) $ \\
\hline
\hline
\end{tabular}
\caption{Scalings of various modes for soft drop double differential cross section following the light cone decomposition in \eq{LC}. Including jet radius related factors in the reference light cone vectors results in hemisphere-jets like kinematics. The third argument also indicates the virtuality of these modes.}
\label{tab:modes}
\end{table}
The momentum scaling of various modes in the light cone coordinates in \eq{LC} are shown in \tab{modes}.

\section{One-loop results of factorization functions}

\label{sec:nlo}

Before we state the factorization formulae for each of the cases discussed above, we first perform one-loop caluclations of the corresponding soft and collinear factorization functions to gain familiarity with the details of measurements associated with each of the modes in \fig{modes}. We first set up a convenient notation to express one-loop phase space integrals in \secn{matrix} and review the known factorization functions results in \secn{known}. In \secn{newNLO} we describe computation of new factorization functions associated with modes in max-$R_g$ and also compute the fixed order non-singular corrections.

\subsection{Matrix elements and measurement functions}
\label{sec:matrix}
At $\cO(\as)$, the jet initating parton $i^*$ splits as $i^* \ra jk$. It will prove useful parameterize the light cone coordinates in the decomposition in \eq{LC} in terms of dimensionless numbers:
\begin{align}
\text{Soft functions}: \qquad y \equiv \frac{q_j^+}{Q} \, , \qquad x \equiv \frac{q_j^-}{Q} \, , \\
\text{Collinear functions}: \qquad y \equiv \frac{p_{i^*}^2}{Q^2} \, , \qquad x \equiv \frac{q_j^-}{Q} \, , \nn
\end{align}
where $j$ is the softer of the two final state partons. Here $p_{i^*}^2 = s = (p_j + p_k)^2$.
All the soft functions associated with the wide-angle soft and collinear-soft modes involve the same eikonal matrix element with a single-particle phase space, whereas those associated with collinear modes involve the full splitting functions and two-particle phase space.
For a generic observable $\cO$ including any veto conditions $\Theta$, the corresponding soft and collinear functions renormalized in $\MS$-scheme at $\cO(\as)$ are then given by
\begin{align} \label{eq:JSNLO}
\cJ^{[1]}_\kappa \big(\cO, \delta_\cO , \mu \big) &\equiv
\frac{\as }{2\pi} \frac{e^{\eps \gamma_E}}{\Gamma(1-\eps)}
\Big(\frac{\mu^2}{Q^2}\Big)^\eps
\int_0^1 \frac{\df x \, \df y }{y^{1+\eps}\big[x(1-x)\big]^\eps} \: \delta_\cO \big[ \hat \cO (x,y),\hat \Theta (x,y)\big] \sum_{\kappa'} \hat P_{\kappa' \kappa} (x)
\nn
\, ,
\\
\cS^{[1]}_\kappa \big(\cO, \delta_\cO , \mu \big) &\equiv \frac{\as C_\kappa}{\pi} \frac{e^{\eps \gamma_E}}{\Gamma(1-\eps)} \Big(\frac{\mu^2}{Q^2}\Big)^\eps \int_0^\infty \frac{\df x \, \df y}{(xy)^{1+\eps}} \: \delta_\cO \big[ \hat \cO^{(s,cs)} (x,y),\hat \Theta^{(s,cs)} (x,y)\big] \, ,
\end{align}
where $\delta_\cO$ combines the measurement and veto conditions on the two parton system and also includes the virtual contribution. We have suppressed dependence on kinematic and grooming parameters for simplicity.
The splitting functions are given by
\begin{align}
\hat P_{gq}(x) &= C_F \bigg[\frac{1 + (1-x)^2}{x} - \eps x \bigg] \, , \nn \\
\hat P_{gg} (x) &=C_A \bigg[\frac{x}{1-x} + \frac{1-x}{x} + x(1-x)\bigg]\, , \nn \\
\hat P_{qg} (x) &= n_f T_F \bigg[x^2 + (1-x)^2 - 2 \eps x(1-x)\bigg] \, .
\end{align}
In the soft matrix elements, we additionally take soft or collinear-soft limit of the observable and the veto condition as indicated by the superscript $(s,cs)$:
\begin{align}
&\text{Soft limit}:& &x \ra 0 \, ,& &x \sim y \, ,& \\
&\text{Collinear-soft limit}:& &x \ra 0 \, ,& &\frac{y}{x} \ra 0 \, .& \nn
\end{align}

Thus, factorization functions associated with various modes differ only in the details of the measurement and veto conditions and some of the modes may only involve veto conditions. For the cases we are interested in, we will only consider the jet mass measurement, which is the same for both collinear and soft matrix elements, and is simply given by
\begin{align}
\hat \xi (x,y) = y \, .
\end{align}
Next, we have the jet radius, groomed jet radius and soft drop constraints. The jet radius constraint corresponds to the $k_T$ clustering condition:
\begin{align}\label{eq:Thetakt}
\hat \Theta_{k_T} (x,y) \equiv \Theta \big(x(1-x) - y\big) \, .
\end{align}
Note that the jet radius does not explicitly appear in the above constraint as we have absorbed the factor $\zeta$ defined in \eq{zetaDef} in the definitions of the light cone coordinates. Similarly, it is easy to check the condition that the two partons are within the groomed jet radius is given by
\begin{align}\label{eq:Thetarg}
\hat \Theta_{r_g} (x,y,r_g) \equiv \Theta \big(r_g^2 x(1-x) - y\big) \, .
\end{align}
Finally, the full soft drop condition in \eq{SD} in terms of the variables $x$ and $y$ is given by
\begin{align}\label{eq:Thetasd}
\hat \Theta_{\rm sd} (x,y,\xi_0,\zeta) &= \Theta \bigg(\frac{\min \{x_1 ,x_2\}}{x_1 + x_2} - \xi_0 \Big(\frac{y}{4x_1x_2}\Big)^\frac{\beta}{2} \bigg) \, ,
\end{align}
where
\begin{align}
x_1(x,y) = \frac{xy \zeta^2}{2} + \frac{1-x}{2} \, , \qquad
x_2(x,y) = \frac{(1-x) y \zeta^2}{2} + \frac{x}{2} \, ,
\end{align}
and we have suppressed dependence on $\beta$ for simplicity.

The corresponding soft and collinear-soft limit of these results read:
\begin{align}\label{eq:measurements}
&\text{soft particle clustered with the jet:}&
&\hat \Theta_{k_T}^{(s,cs)}(x,y)& &\equiv& &\Theta (x - y)\,,& \\
&\text{soft emission at an angle $<r_g$:}&
&\hat \Theta_{r_g}^{(s,cs)} (x,y,r_g) & &\equiv& & \Theta (r_g^2 x - y) \, ,& \nn \\
&\text{collinear-soft emission passes soft drop:}&
&\hat \Theta^{(cs)}_{\rm sd} (x,y,\xi_0)& &\equiv& & \Theta\big (x - y^{\frac{\beta}{2+\beta}} \xi_0^{\frac{2}{2+\beta}} \big) \, .& \nn \\
&\text{soft emission passes soft drop:}&
&\hat \Theta_{\rm sd}^{(s)} (x,y,\xi_0,\zeta)& &\equiv& & \Theta\big (x + y\zeta^2 - y^{\frac{\beta}{2+\beta}} \xi_0^{\frac{2}{2+\beta}} \big) \, ,& \nn
\end{align}

Having compiled the various measurement and veto functions, we now describe how they are combined in the functions $\delta_{\cO}$ appearing in \eq{JSNLO}.
\subsection{Compilation of known results}
\label{sec:known}
We first review the known results of factorization functions.

\subsubsection{Hard-collinear radiation outside the jet}
We first state the normalization contribution of the hard-collinear near the boundary of the jet as calculated in \Refscite{Cal:2019hjc,Hannesdottir:2022rsl}:
\begin{align}
N_{\rm incl}^q(Q, \mu) &= 1 + \frac{\alpha_s(\mu) C_F}{2\pi} \Bigg\{-\frac{1}{2} \ln^2 \frac{\mu^2}{Q^2} - \frac{3}{2}\ln\frac{\mu^2}{Q^2}
-\frac{13}{2} + \frac{3\pi^2}{4} \Bigg\} \, , \\
N_{\rm incl}^g (Q, \mu) &= 1 + \frac{\alpha_s(\mu)}{2\pi}
\Bigg\{-\frac{C_A}{2} \ln^2 \frac{\mu^2}{Q^2} - \frac{\beta_0}{2}\ln\frac{\mu^2}{Q^2}
+ C_A \Big(-\frac{5}{12} + \frac{3\pi^2}{4}\Big) - \frac{23}{12} \beta_0\Bigg\} \nn \, .
\end{align}

\subsubsection{Inclusive jet mass measurement on collinear radiation}
We now consider the simplest case of inclusive ungroomed jet mass, which is relevant for the collinear mode $\mC$ in the max- and intermediate-$R_g$ cases.
Here the measurement function is given by
\begin{align}
\delta_\xi^{\mC} (x, y) = \delta (\xi - y) - \delta (y) \, .
\end{align}
such that using \eq{JSNLO} the inclusive jet function at NLO is given by
\begin{align}
J^{[1]}_\kappa (m_J^2, \mu) = \delta(m_J^2) + \frac{1}{Q^2} \cJ_\kappa^{[1]} \bigg( \xi = \frac{m_J^2}{Q^2}, \delta_\xi^{\mC} , \mu\bigg) + \cO(\as^2)\, .
\end{align}
We have expressed the jet function in terms of its natural argument, and will do the same for all the factorization functions below.
The results read~\cite{Bauer:2003pi,Bosch:2004th}
\begin{align}\label{eq:Jmom}
J_q(m_J^2 ,\mu ) &= \delta(m_J^2) + \frac{\alpha_s(\mu) C_F}{\pi} \bigg( {\cal L}_1(m_J^2, \mu^2)- \frac{6}{8}{\cal L}_0(m_J^2, \mu^2)
-\frac{\delta (m_J^2)}{4} \Big(\pi^2 - 7\Big)
\bigg)
\, , \\
J_g(m_J^2, \mu) &= \delta (m_J^2) + \frac{\alpha_s(\mu)}{\pi} \bigg(\frac{C_A}{\pi} {\cal L}_1(m_J^2, \mu^2)- \frac{\beta_0}{4}{\cal L}_0(m_J^2, \mu^2) \nn \\
&\qquad \qquad\qquad\qquad+
\frac{\delta(m_J^2) }{36}\Big(C_A\big(67 - 9\pi^2\big) - 20 n_F T_R \Big)
\bigg) \nn \, ,
\end{align}
where $\cL_n$ is the standard plus function:
\begin{align}
{\cal L}_n (\ell^+, \mu ) \equiv \frac{1}{\mu} {\cal L}_n\bigg(\frac{\ell^+}{\mu} \bigg) \, .
\end{align}
For later use we also state the result for the Laplace transform defined by
\begin{align}
\tilde J_\kappa (x, \mu) &\equiv \int_0^\infty \df m_J^2 \: e^{-x m_J^2} \, J_\kappa (m_J^2, \mu) \, ,
\end{align}
such that
\begin{align}
\tilde J_q (x,\mu) & = 1 + \frac{\alpha_s C_F}{\pi} \left[ \frac{1}{2}
\log^2 \left( x \, \mu^2 e^{\gamma_E} \right)
+ \frac{3}{4}
\log \left( x \, \mu^2 e^{\gamma_E} \right)
+ \frac{7}{4}
-
\frac{\pi^2}{6}
\right] \,, \\
\tilde J_g (x,\mu) & = 1 + \frac{\alpha_s}{\pi} \left[ \frac{C_A}{2}
\log^2 \left( x \, \mu^2 e^{\gamma_E} \right)
+ \frac{\beta_0}{4}
\log \left( x \mu^2 e^{\gamma_E} \right)
+ C_A
\left(\frac{67}{36} - \frac{\pi^2}{6} \right) -
n_F T_R \, \frac{5}{9}
\right] \,. \nn
\end{align}
\subsubsection{Hard-collinear radiation within groomed jet radius}
Next, we turn to the hard-collinear mode \textcolor{Purple}{$\cC$} which involves an additional constraint of groomed jet radius with jet mass measurement:
\begin{align}\label{eq:deltacC}
\delta_\xi^{\mcC} ( x,y,r_g) \equiv \hat \Theta_{r_g}(x,y,r_g) \big[\delta (\xi - y) - \delta (\xi) \big] \,,
\end{align}
where $\hat \Theta_{r_g}$ was defined in \eq{Thetarg}. The corresponding collinear function is then given by
\begin{align}\label{eq:CkappaDef}
\cC^\kappa\big(\xi, r_g, Q, \mu\big) \equiv \frac{1}{r_g^2}\cC^\kappa\bigg( \frac{\xi}{r_g^2}, Q r_g, \mu \bigg) = \delta ( \xi)+ \cJ_\kappa^{[1]} \big(\xi ,\delta_\xi^{\mcC} , \mu \big)+ \cO(\as^2) \, .
\end{align}
with explicit expressions of the function with three arguments being~\cite{Pathak:2020iue,Hannesdottir:2022rsl},
\begin{align}\label{eq:CollFunc}
\cC^q \big(\xi , Q, \mu \big) &= \delta(\xi ) + \frac{\alpha_s (\mu) C_F}{\pi} \Bigg\{\delta (\xi )\bigg(
\frac{1}{4} \ln^2 \frac{\mu^2}{Q^2}+ \frac{3}{4} \ln \frac{\mu^2}{Q^2}+ \frac{7}{4} - \frac{5\pi^2}{24} \bigg) \\
&\qquad
+ \Theta \big(1-4 \xi \big) \Bigg[ - {\cal L}_1 (\xi )+ {\cal L}_0 (\xi )\bigg(
2 \ln \Big(\frac{1 + \sqrt{1 - 4\xi }}{2}\Big) -\frac{3}{4} \sqrt{1-4\xi } \bigg)
\Bigg]
\Bigg\} \, ,\nn
\\
\cC^g \big(\xi , Q , \mu \big) &=
\delta (\xi) + \frac{\alpha_s(\mu)}{\pi} \Biggl\{ \delta (\xi) \bigg(
\frac{C_A}{4} \ln^2 \frac{\mu^2}{Q^2}+ \frac{\beta_0}{4} \ln \frac{\mu^2}{Q^2}+
C_A \Big(\frac{67}{36} - \frac{5\pi^2}{24}\Big) - \frac{5}{9}n_f T_F\bigg) \nn \\
&\qquad+ \Theta \big(1-4 \xi\big)
\Bigg[ - C_A {\cal L}_1 (\xi) + {\cal L}_0(\xi) \bigg(2 C_A \log \Big(\frac{1 + \sqrt{1-4 \xi}}{2}\Big) \nn \\
&\qquad \qquad -\frac{\beta_0}{4}\sqrt{1-4\xi } + \frac{C_A - 2 n_f T_F}{6} \xi \sqrt{1-4 \xi}\bigg)
\Bigg]
\Biggr\} \nn \, .
\end{align}

\subsubsection{Collinear-soft radiation within groomed jet radius}
We now consider soft functions and consider the simplest case of $\mCSm$ mode with jet mass measurement and groomed jet radius boundary. The measurement function is simply a soft limit of the previous case and is defined as:
\begin{align}\label{eq:deltaCSm}
\delta_\xi^{\mCSm} ( x,y,r_g) \equiv \hat \Theta^{(cs)}_{r_g} \big(x,y,r_g\big) \big[\delta(x - y ) - \delta (\xi)\big] \, ,
\end{align}
with the corresponding function given by
\begin{align}\label{eq:ScmDef}
S_{c_m}^{\kappa} \bigg(\frac{\ell^+}{r_g}, \mu \bigg) = \delta \Big(\frac{\ell^+}{r_g} \Big) + \frac{1}{Q}\cS^{[1]}_\kappa \bigg(\xi = \frac{\ell^+}{Q}, \delta_\xi^{\mCSm} , \mu \bigg)+ \cO(\as^2) \, .
\end{align}
which yields
\begin{align}\label{eq:Scm1L}
S_{c_m}^{\kappa} \bigg(\frac{\ell^+}{r_g}, \mu \bigg)&=
\delta \bigg(\frac{\ell^+}{r_g}\bigg) + \frac{\alpha_s C_\kappa}{2\pi} \Bigg[-4 {\cal L}_1 \bigg(\frac{\ell^+}{r_g}, \mu\bigg) + \frac{\pi^2}{12} \delta \bigg(\frac{\ell^+}{r_g}\bigg) \Bigg] \, .
\end{align}

We next define the Laplace transform:
\begin{align}
\tilde S_{c_m } (u,\mu) &\equiv \int_0^\infty \df \ell^+ \: e^{-u\ell^+} \, S_{c_m}^\kappa (\ell^+,\mu) \, ,
\end{align}
which yields
\begin{align}\label{eq:Lap}
\tilde S_{c_m}^\kappa\bigl[u,\mu \bigr]
&=
1 + \frac{\alpha_s C_\kappa}{\pi}
\bigg[
-
\log^2 \big( u e^{\gamma_E}\mu \big)
-\frac{\pi^2}{8}
\bigg]
\, .
\end{align}

\subsubsection{Wide-angle soft radiation failing soft drop}
The next case we consider is the global soft modes $\mSG$ that fail soft drop. Here we only have a veto condition and no measurement:
\begin{align}
\hat \Theta^{\mSG} (x,y,\xi_0,\zeta) \equiv \hat \Theta_{k_T}^{(s)} (x,y)\big(1 - \hat \Theta_{\rm sd}^{(s)} (x,y,\xi_0,\zeta)\big) \, ,
\end{align}
such that
\begin{align}\label{eq:SGDef}
S_G^{\kappa} \big(\qcut, \zeta, \beta, \mu \big) = 1 + \cS_\kappa^{[1]} \big(\cdot \, , \hat \Theta^{\mSG} , \mu \big) + \cO(\as^2)\, ,
\end{align}
Here the empty slot in the first argument simply denotes that the function has no differential measurement applied on it and only contributes to the normalization.
At one-loop the result reads~\cite{Frye:2016aiz,Kang:2018jwa}
\begin{align}\label{eq:SG1loop}
S_G^\kappa\big(\qcut,\zeta, \beta, \mu\big)&=
1 + \frac{\alpha_s(\mu)C_\kappa}{\pi}
\bigg[
\frac{1}{(1+\beta)} \log^2\Big(\frac{\mu}{\qcut}\Big)
-\frac{\pi^2}{24}
\Big(
\frac{1}{1+\beta}
\Big)
\\
&\qquad \qquad \qquad \qquad
-\frac{(2+\beta)}{4}
\Big(
2{\rm Li}_2\Big[ \frac{\zeta^2}{1 + \zeta^2} \Big]
+ \log^2\big [1 + \zeta^2\big]
\Big)
\bigg] \nn \, .
\end{align}
Note that the above global soft function is really only valid for inclusive jets measurement. For exclusive measurements, one additionally needs to include contributions from the beam region and other trigger jets. However, as detailed in \Refcite{Hannesdottir:2022rsl}, the above result is nevertheless useful in the case of exclusive jets with an appropriate treatment of the quark-gluon fractions.

\subsubsection{Collinear-soft radiation at intermediate groomed jet radius}
Analogous to above, we consider the case of $\mCSg$ modes that pass soft drop and are within the required groomed jet radius, but do not contribute to the jet mass measurement:
\begin{align}\label{eq:ThetaCSg}
\hat \Theta^{\mCSg} (x,y,r_g,\xi_0) \equiv \hat \Theta^{(cs)}_{r_g} (x,y,r_g) \hat \Theta_{\rm sd}^{(cs)} (x,y,\xi_0) \,.
\end{align}
The corresponding function being
\begin{align}
S_{c_g}^\kappa \big(\qcut r_g^{1+\beta} , \beta\big) = 1 + \cS_\kappa^{[1]} \big (\cdot \, , \hat \Theta^{\mCSg} , \mu \big) + \cO(\as^2) \, .
\end{align}
While formally in \eq{ThetaCSg} we are required to take the collinear-soft limit of the soft drop constraint, we will also find it useful to employ the results of intermediate-$R_g$ regime for implementing fixed-order subtractions. To this end, it is helpful to evaluate the above function in the soft-wide angle limit, for which we can directly recycle the computation of the previous result of global soft function (while being careful about minus signs):
\begin{align}\label{eq:ScGNLO}
S_{c_g}^\kappa \big(\qcut r_g^{1+\beta} , \zeta_g, \beta\big)
&=
1 - \frac{\alpha_s(\mu)C_\kappa}{\pi}
\bigg[
\frac{1}{(1+\beta)} \log^2\Big(\frac{\mu}{\qcut r_g^{1+\beta} }\Big)
-\frac{\pi^2}{24}
\Big(
\frac{1}{1+\beta}
\Big) \nn \\
&\qquad \qquad
-\frac{(2+\beta)}{4}
\Big(
2{\rm Li}_2\Big[ \frac{\zeta_g^2}{1 + \zeta_g^2} \Big]
+ \log^2\big [1 + \zeta_g^2\big]
\Big)
\bigg]
\, .
\end{align}
Here we have defined a variable $\zeta_g$ analogous to $\zeta$ defined above in \eq{zetaDef}
\begin{align}\label{eq:zetagDef}
\zeta_{g,\rm incl}^\pp \equiv \frac{R_g}{2 \cosh \eta_J} \, , \qquad \zeta_{g,\rm excl}^\pp \equiv 1 \, , \qquad \zeta_{g,\rm incl}^\ee \equiv \tan\frac{R_g}{2} \, .
\end{align}

\subsection{One-loop results in max-$R_g$ and fixed-order regime}
\label{sec:newNLO}

We now state results for the remaining functions that are relevant for max-$R_g$ regimes. Some of the results below have already been calculated elsewhere but only after expanding in $r_g \ll1$ limit, and we reinstate the full $r_g$ dependence at $\cO(\as)$.
\subsubsection{Widest angle collinear soft radiation passing soft drop}
\label{sec:Srg}
We consider the $\mCS$ mode that saturates the kinematic constraints imposed by jet mass measurement and soft drop passing condition. Here we have
\begin{align}\label{eq:deltaCS}
\delta_\xi^{\mCS} (x,y,r_g,\xi_0) \equiv \hat \Theta_{\rm sd}^{(cs)}(x,y,\xi_0) \big[\hat \Theta_{r_g}^{(cs)}(x,y,r_g) \delta (\xi - y) - \delta (\xi)\big]
\end{align}
This condition can be straightforwardly obtained by demanding that modes that pass soft drop are also required to satisfy the groomed jet radius constraint. On the other hand, modes that fail soft drop and the virtual piece, however, do not see this constraint.
The collinear-soft function is then given by
\begin{align}\label{eq:ScDef}
S_c^\kappa \big(\tilde k , r_g,\qcut, \beta, \mu \big) \equiv \delta (\tilde k ) + \frac{1}{Q \qcut^{\frac{1}{1+\beta}}} \cS_\kappa^{[1]} \bigg(\xi = \frac{\tilde k}{Q \qcut^{\frac{1}{1+\beta}}} , \delta_\xi^{\mCS}, \mu \bigg) + \cO(\as^2)\, .
\end{align}
Note that we have made use of a $\frac{2+\beta}{1+\beta}$ dimensional variable which appears as the natural argument for this function.
It will be helpful to split the measurement in \eq{deltaCS} as
\begin{align}\label{eq:deltaCS2}
\delta_\xi^{\mCS} (x,y,r_g,\xi_0) = \hat \Theta_{\rm sd}^{(cs)} (x,y,\xi_0) \big[ \delta (\xi - y) - \delta (\xi)\big] + \Delta \delta_{\xi,\,r_g}^{\mCS} (x,y,r_g,\xi_0)\, ,
\end{align}
where
\begin{align}
\Delta \delta_{\xi,\,r_g}^{\mCS} (x,y,r_g,\xi_0)\equiv - \big(1- \hat \Theta_{r_g}^{(cs)}(x,y,r_g) \big) \hat \Theta_{\rm sd}^{(cs)} (x,y,\xi_0)\delta (\xi - y) \, ,
\end{align}
such that the first term simply results in the standard collinear-soft function for single differential jet mass, and the second piece in a finite fixed order correction:
\begin{align}
S_c^\kappa \big(\tilde k , r_g, \qcut, \beta, \mu \big) = S_c^\kappa \big(\tilde k , \beta, \mu\big) + \Delta S_{r_g}^\kappa \big(\tilde k , r_g ,\qcut, \beta, \as(\mu) \big) \, ,
\end{align}
where~\cite{Frye:2016aiz} (see also \Refcite{Hannesdottir:2022rsl})
\begin{align}
S_c^\kappa \big(\tilde k,\beta, \mu \big)
&= \delta(\tilde k)+ \frac{\alpha_s C_\kappa}{\pi} \Biggl[
\frac{-2(1+\beta)}{2 + \beta}
\,{\cal L}_1 \left(\tilde k , \mu^{\frac{2+\beta}{1+\beta}} \right)
+ \frac{\pi^2}{24} \frac{2 + \beta}{1 + \beta}
\delta(\tilde k)
\Biggr] \, .
\end{align}
Here the dependence on $\qcut$ drops out as it is a high scale from the perspective of low energy collinear-soft modes.
The Laplace transform is defined by
\begin{align}
\tilde S_c^\kappa (s,\beta,\mu) &\equiv \int \df \tilde k \: e^{-s \tilde k } S_c^{\kappa} \big (\tilde k , \beta ,\mu \big ) \, ,
\end{align}
such that
\begin{align}
\tilde S_c^\kappa (s)
&=
1 + \frac{\alpha_s C_\kappa}{\pi}
\bigg[
- \Big(
\frac{1+\beta}{2+\beta}
\Big)
\log^2 \big( s e^{\gamma_E}\mu^{\frac{2+\beta}{1+\beta}} \big)
-\frac{\pi^2}{24}
\frac{\beta(3\beta+ 4)}{(1+\beta)(2+\beta)}
\bigg]
\, .
\end{align}

Next we turn to the finite correction $\Delta S_{r_g}^\kappa$ that describes fixed-order corrections due to $r_g$ measurement and re-introduces dependence on $\qcut$. This correction was evaluated in \Refcite{Pathak:2020iue} in the $r_g \ll 1$ limit. Since we are also interested in covering the region close to the soft drop cusp, where $r_g \lesssim1 $, we will find it helpful for the purposes of matching to include the full jet radius dependence by employing the soft-wide angle limit of the soft drop constraint $\hat \Theta_{\rm sd}^{(s)}$ in \eq{deltaCS} and including jet radius constraint $\hat \Theta_{k_T}^{(s)}$, such that we use in \eq{deltaCS2}
\begin{align}\label{eq:DeltadeltaCS}
\Delta \delta_{\xi,\, r_g}^{\mCS,\, \rm full}(x,y,r_g,\xi_0,\zeta)\equiv - \hat \Theta_{k_T}^{(s)} (x,y)\big(1- \hat \Theta_{r_g}^{(s)}(x,y,r_g) \big) \hat \Theta_{\rm sd}^{(s)} (x,y,\xi_0,\zeta)\delta (\xi - y) \, .
\end{align}
which yields the correction piece
\begin{align}\label{eq:DeltaScg}
&\qcut^{\frac{1}{1+\beta}} \Delta S_{r_g}^\kappa \bigg( \ell^+ \qcut^{\frac{1}{1+\beta}} , r_g ,\qcut, \zeta, \beta, \as(\mu) \bigg)
= \frac{1}{Q } \cS_\kappa^{[1]} \bigg(\xi = \frac{\ell^+}{Q} , \Delta \delta_{\xi,\, r_g}^{\mCS,\, \rm full}, \mu \bigg) \\
& \quad=
\frac{\alpha_s C_\kappa}{\pi} \frac{\Theta (\ell^+ -\qcut' v(r_g))}{ \ell^+ } \Bigg[
\ln(r_g^2)
+ \Theta (\qcut' - \ell^+)
\ln\bigg( \Big(\frac{\qcut}{\ell^+}\Big)^\frac{2}{2+\beta} - \zeta^2\bigg)
\Bigg]
\, , \nn
\end{align}
where we have defined
\begin{align}\label{eq:vDef}
v(r_g) \equiv\Bigg( \frac{1 + \zeta^2}{ \frac{1}{r_g^2} + \zeta^2 }\Bigg)^{\frac{2+\beta}{2}} \, ,
\end{align}
and have expressed the result in terms of $\ell^+ \qcut^{\frac{1}{1+\beta}}$ due to explicit $\qcut$ dependence from constraining $r_g$. We had defined $\qcut'$ in \eq{Qcutp}.
We can check that by setting $\zeta$ to zero, we recover the result in the soft drop resummation region ($\xi < \xi_0$) calculated in \Refcite{Pathak:2020iue}:
\begin{align}
&\qcut^{\frac{1}{1+\beta}}\Delta S_{r_g}^\kappa \bigg(\ell^+ \qcut^{\frac{1}{1+\beta}} < \qcut^{\frac{2+\beta}{1+\beta}} , r_g ,\qcut, \zeta = 0, \beta, \as(\mu) \bigg) \\
&\qquad = \frac{-2}{2+\beta} \frac{\alpha_s C_\kappa}{\pi}\Theta \bigg(\frac{\ell^+}{\qcut} - r_g^{2+\beta}\bigg) \frac{1}{ \ell^+ }
\log\bigg(\frac{\ell^+}{\qcut r_g^{2+\beta}}\bigg) \, . \nn
\end{align}
We thus see that including soft-wide angle effects modify the transition point from $\qcut$ to $\qcut'$.

\subsubsection{Wide-angle soft radiation in plain jet mass region}
\label{sec:Ssd}
We now consider the final case where the wide-angle soft mode $\mSp$ is tested for soft drop, jet radius constraints with jet mass measurement. This is relevant for the max-$R_g$ regime in the plain jet mass resummation region. The measurement function is given by
\begin{align}\label{eq:deltaSplain}
\delta_\xi^{\mSp} (x,y,r_g,\xi_0,\zeta)\equiv \hat \Theta^{(s)}_{k_T}(x,y) \hat \Theta_{\rm sd}^{(s)}(x,y,\xi_0,\zeta) \big[ \hat \Theta_{r_g}^{(s)} (x,y,r_g)\delta (\xi - y) - \delta (\xi)\big]
\end{align}
This expression is a simple extension of \eq{deltaCS} where we have also included the jet radius constraint. We will find it helpful to split the measurement into chunks that we have already encountered:
\begin{align}\label{eq:deltaSplainSplit}
\delta_\xi^{\mSp} (x,y,r_g,\xi_0,\zeta) &= \delta_\xi^{\mCSm} (x,y,1) + \Delta \delta_{\xi,\, r_g}^{\mCS,\, \rm full} (x,y,r_g,\xi_0,\zeta) + \Delta \delta_{\xi,\, \rm sd}^{\mSp} (x,y,\xi_0,\zeta)
\, .
\end{align}
In the first term on the right hand side, setting $r_g = 1$ in \eq{deltaCSm} results in the same jet radius constraint, and hence is simply the familiar ungroomed soft function. The second piece accounts for the cumulative measurement of groomed jet radius using \eq{DeltadeltaCS}. Finally, the new piece accounts for effects of soft drop on wide-angle soft modes:
\begin{align}\label{eq:deltaSD0}
\Delta \delta_{\xi,\,\rm sd}^{\mSp} (x,y,\xi_0,\zeta) = -\hat \Theta_{k_T}^{(s)}(x,y) \big(1 - \hat \Theta_{\rm sd}^{(s)} (x,y,\xi_0,\zeta)\big) \big[\delta(\xi - y) - \delta(\xi)\big] \, .
\end{align}
Hence, the soft function in the plain jet mass region is given by
\begin{align}\label{eq:SplainFull}
&S_{\rm plain}^{\kappa} \big(\ell^+, r_g , \qcut, \zeta, \beta, \mu \big) = \delta (\ell^+) + \frac{1}{Q} \cS_{\kappa}^{[1]} \bigg(\xi = \frac{\ell^+}{Q}, \delta_\xi^{\mSp}, \mu\bigg) + \cO(\as^2) \\
&\qquad = S_{c_m}^\kappa \big(\ell^+,\mu\big) + \qcut^{\frac{1}{1+\beta}} \Delta S_{r_g}^\kappa \big(\ell^+ \qcut^{\frac{1}{1+\beta}} , r_g ,\qcut, \zeta, \beta, \as(\mu) \big)
+ \Delta S_{\rm sd}^\kappa \big(\ell^+, \qcut, \beta, \zeta, \as(\mu) \big) \nn
\, ,
\end{align}
where we have written the result in terms of previous results in \eqs{Scm1L}{DeltaScg} and a new piece given by~\cite{Hannesdottir:2022rsl}
\begin{align}\label{eq:Splain1}
&\Delta S_{\rm sd}^\kappa \big(\ell^+, \qcut, \zeta, \beta, \as(\mu) \big) = \frac{1}{Q} \cS_{\kappa}^{[1]} \bigg(\xi = \frac{\ell^+}{Q}, \Delta \delta_{\xi,\, \rm sd}^{\mSp}, \mu\bigg)\\
&\qquad =
-\frac{\alpha_s (\mu)C_\kappa}{\pi} \Biggl[\frac{\Theta(\ell^+) \Theta (\qcut' - \ell^+)}{\ell^+} \ln \bigg( \Big(\frac{\qcut}{\ell^+}\Big)^\frac{2}{2+\beta} - \zeta^2\bigg) \Biggr]^{[\qcut']}_+ \, . \nn
\end{align}
Here the plus-function with a non-standard boundary condition is defined as
\begin{align}
\big[\Theta(x)q(x) \big]_+^{[x_0]}
\equiv \lim_{\eps\ra 0}
\Big( \Theta(x-\eps) q(x)
- \delta (x- \eps) \int_{\eps}^{x_0} \df x'\: q(x')
\Big) \, .
\end{align}
We also note that for functions satisfying $q(x) = \lambda^\alpha g(\lambda^{-1}x)$ we have,
\begin{align}
\big[\Theta(x)q(x) \big]_+^{[x_0]} = \lambda^\alpha \big[\Theta(\lambda^{-1}x)g (\lambda^{-1}x) \big]_+^{[\lambda^{-1}x_0]} \, .
\end{align}
This property proves useful in simplifying expressions involving integrals of such plus-functions.

\subsubsection{Fixed-order cross section}
\label{sec:FO}
We now turn to the fixed order cross section. Here the measurement function is same as \eq{deltaSplain} without any expansions:
\begin{align}\label{eq:deltaFO}
\delta_\xi^{\rm FO} (x,y,r_g,\xi_0,\zeta) &\equiv \hat \Theta_{k_T}(x,y) \hat \Theta_{\rm sd} (x,y,\xi_0,\zeta) \big[ \hat \Theta_{r_g} (x,y,r_g)\delta (\xi - y) - \delta (\xi)\big] \, ,
\end{align}
and the fixed order cross section is given by
\begin{align}
\cG^{\rm FO}_{\kappa, \rm sd} \big(\xi, r_g, \xi_0, \zeta, \alpha_s(\mu)\big) = \cJ^{[1]}\big(\xi, \delta_\xi^{\rm FO}, \mu\big) + \cO(\as^2) \, .
\end{align}
Because of the complicated form of the full soft drop condition in \eq{Thetasd} we will evaluate this numerically. To this end, we define a subtraction term with the measurement function,
\begin{align}
\delta_\xi^{\rm FO(0)} (x,y,r_g,\xi_0,\zeta) &\equiv \hat \Theta_{k_T}(x,y) \hat \Theta^{(s)}_{\rm sd} (x,y,\xi_0,\zeta) \big[ \hat \Theta_{r_g} (x,y,r_g)\delta (\xi - y) - \delta (\xi)\big] \, .
\end{align}
where we have now replaced full soft drop constraint by its soft limit in \eq{measurements}, and evaluate the following function numerically by implementing subtraction at the level of the integrand:
\begin{align}
\Delta \cG^{\rm FO[1]}_{\kappa, \rm sd} \big(\xi, r_g, \xi_0, \zeta, \alpha_s(\mu)\big) = \cG^{\rm FO}_{\kappa, \rm sd} \big(\xi, r_g, \xi_0, \zeta, \alpha_s(\mu)\big) -\cG^{\rm FO(0)}_{\kappa, \rm sd} \big(\xi, r_g, \xi_0, \zeta, \alpha_s(\mu)\big) \, .
\end{align}
The soft matrix element corresponding to this measurement is given by
\begin{align}
\tcG_{\kappa,\, \rm sd}^{\rm FO(0)} (\xi, r_g , \xi_0, \zeta , \as(\mu)) &\equiv \cS_\kappa^{[1]} \big(\xi,\delta_\xi^{\rm FO(0)}, \mu\big) + \cO(\as^2) \, ,
\end{align}
such that we have
\begin{align}
\xi\tcG_{q,\, \rm sd}^{\rm FO(0)} (\xi, r_g , \xi_0, \zeta , \as(\mu)) &= \frac{\alpha_s(\mu)C_F}{\pi} \ln \bigg(\frac{\frac{1}{2}(1 + \sqrt{1+4 \xi/r_g^2})}{\max\big\{\frac{1}{2}(1 - \sqrt{1+4 \xi/r_g^2}) , \, \xi_0^{\frac{2}{2+\beta}} y^{\frac{\beta}{2+\beta}} - y\zeta^2\big \}}\bigg) \, ,\nn \\
\xi\tcG_{g,\, \rm sd}^{\rm FO(0)} (\xi, r_g , \xi_0, \zeta , \as(\mu)) &= \frac{\alpha_s(\mu)C_A}{\pi} \ln \bigg(\frac{1/2}{\max\big\{\frac{1}{2}(1 - \sqrt{1+4 \xi/r_g^2}) , \, \xi_0^{\frac{2}{2+\beta}} y^{\frac{\beta}{2+\beta}} - y\zeta^2\big \}}\bigg) \, .\nn \\
\end{align}
Since we are interested in the differential cross section we can restrict to $\xi> 0$ by multiplying by $\xi$ and avoid considering the zero-bin terms.

\section{Factorization and resummation}
\label{sec:ddifffact}
Having discussed the mode structure, we now state the factorization formulae for the three regimes discussed here~\cite{Pathak:2020iue}. In \secn{MaxResum} we describe the factorization formula for max-$R_g$ regime, and review min and intermediate $R_g$ regimes in \secn{MinIntResum}.

\subsection{Max-$R_g$ regime}
\label{sec:MaxResum}
We first consider the max-$R_g$ regime in the plain jet mass region and identify the resummation kernels associated with the one-loop results calculated in \secn{newNLO}.
\subsubsection{Max-$R_g$ in plain jet mass resummation region}
\label{sec:maxPlain}

We recall the discussion in \secn{kin} where in \eq{cGcFact} we showed how in the small jet mass region the inclusive jet function factorizes, which can be formulated in terms of an RG invariant jet mass distribution for a jet flavor $\kappa$ in \eq{tcGDef}. We will see that the various cases we consider below for $\xi \ll 1$ will differ only in the details of the multi-scale $\cJ_\kappa$ function describing soft collinear dynamics.
We will treat the non-global logarithms at NLL accuracy where they can be factorized.
Next, in the ungroomed region the soft drop condition and measurement of the groomed jet radius are accounted for via fixed order corrections, and hence the factorization and resummation proceeds precisely the same way as the ungroomed jet mass case. This is given by the bottom left scenario in \fig{modes} involving the hard collinear modes $N$, the (inclusive) collinear modes $\mC$ and the wide angle soft modes $\mSp$. The factorized cross section is given by
\begin{align}\label{eq:factPlainNGL}
\cG_{\kappa}^{\rm plain} (z, \xi , r_g ,\mu ) &=\sum_{\kappa'} \cH_{\kappa' \ra \kappa} (z, Q, \mu) \frac{\df}{\df\xi}\Bigg[ \cS_{\rm NGL}^\kappa \big( t [\xi Q, Q] \big) \Sigma_{\rm plain}^\kappa (\xi,r_g, Q, \qcut, \zeta, \mu) \Bigg] \, ,
\end{align}
where we have suppressed dependence on kinematic and grooming parameters in $\cG_\kappa^{\rm plain}$.
As shown in \eq{cGcFact}, this factorization involves the same hard collinear function $\cH_{\kappa'\ra \kappa}$. The jet mass measurement with cumulative $r_g$ cut off is described by the derivative of the cumulative cross section:
\begin{align}\label{eq:GPlain}
\Sigma_{\rm plain}^\kappa (\xi,r_g, Q, \qcut, \zeta, \mu)
= \int_0^\infty \df s \, \df \ell_c^+\: J_\kappa \big( s , \mu \big)\: \crS_{\rm plain}^{\kappa}\big ( \ell_c^+, r_g, \qcut, \zeta, \beta, \mu \big ) \: \delta \bigg(\xi - \frac{s}{Q^2} - \frac{\ell_c^+}{Q}\bigg)
\, .
\end{align}
where $\crS_{\rm plain}^{\kappa}$ is the cumulative version of the wide-angle soft function in \eq{SplainFull}.
\begin{align}\label{eq:Ssd}
\crS_{\rm plain}^{\kappa}\big( \ell_c^+, r_g , \qcut, \zeta, \beta, \mu \big) \equiv
\int_0^{\ell_c^+} \df \ell^+\: S_{\rm plain}^{\kappa} \big(\ell^+, r_g , \qcut, \zeta, \beta, \mu \big) \, ,
\end{align}
The $\cS_{\rm NGL}^\kappa$ accounts for non-global logarithms up to NLL accuracy, and the argument is defined as
\begin{align}
t[\mu_0,\mu_1] \equiv \frac{1}{2\pi} \int_{\mu_0}^{\mu_1} \frac{d \mu'}{\mu'} \: \alpha_s (\mu') \,.
\end{align}
From \eq{GPlain} we see that the NGLs depend on the integral of running coupling between wide-angle soft and hard-collinear scales.

Finally, we point out that factorizing the cross section in \eq{GPlain} amounts to dropping the following power corrections:
\begin{align}\label{eq:MaxPC1}
\cG_{\kappa} \big(z, \xi , r_g, \alpha_s (\mu)\big) &= \cG_{\kappa}^{\rm plain} (z, \xi , r_g, \mu)\Big(1 + \cO( \xi_0, \xi) \Big) \,,
\end{align}
where the left hand side is the full QCD cross section which only depends on running coupling at a single scale $\mu$. In the resummed version of the factorized cross section we will employ separate jet mass and groomed jet radius dependent profile scales for each of the factorization functions. In addition to the $\cO(\xi)$ jet mass power corrections mentioned above, we have also dropped the finite-$\zcut$ terms of $\cO(\xi_0)$.

We now describe the resummation using the renormalization group evolution of the functions appearing in factorization formula above.
We will consider the normalized cross section $\tcG_{\kappa}^{\rm plain}$ defined in \eq{tcGDef} after stripping off the DGLAP evolution.
We will find it helpful to isolate the fixed-order corrections in $S_{\rm plain}^\kappa$ in \eq{SplainFull} and decompose $\tcG_{\kappa}^{\rm plain}$ as
\begin{align}\label{eq:GplainSplit}
&\tcG_{\kappa}^{\rm plain} ( \xi , r_g ,Q, \qcut, \zeta, \mu_{\rm plain} ) =\tcG_{\kappa\, , \rm no\,sd}^{\rm plain} (\xi ,Q, \mu_{\rm plain} ) +\Delta \tcG_{\kappa}^{\rm plain} ( \xi ,r_g, Q, \qcut, \zeta, \mu_{\rm plain} ) \, .
\end{align}
Instead of a single scale $\mu$, we employ here a set of scales $\mu_{\rm plain}$ for plain jet mass resummation that minimize the logs in each factorization function:
\begin{align}\label{eq:muplain}
\mu_{\rm plain} \equiv \{\mu_N, \mu_J (\xi), \mu_{s}(\xi) \} \, .
\end{align}
Precise implementation of these scales was discussed extensively in \Refscite{Pathak:2020iue,Hannesdottir:2022rsl}. We summarize the formulae for these scales and their variations in \app{prof}.

The first of these is the resummed ungroomed cross section given by
\begin{align}\label{eq:FactPlain}
\tilde \cG^{\rm plain}_{\kappa,{\rm no\,sd}} (\xi , Q, \mu_{\rm plain} ) =
N^\kappa_{\rm incl}( Q, \mu_N)
e^{ K_N }\Big(\frac{\mu_N}{Q}\Big)^{\omega_N}
\cJ^{\rm plain}_{\kappa,{\rm no\,sd}}(\xi, Q, \mu_{\rm plain}) \, ,
\end{align}
where $K_N$ and $\omega_N$ are resummation kernels associated with the hard collinear function $N_{\rm incl}^\kappa$.
We follow a shorthand throughout this paper
\begin{align}
K_{\cF} \equiv j_\cF K \big(\Gamma_\cF[\as], \mu,\mu_\cF\big) + \eta \big(\gamma_\cF [\as], \mu, \mu_\cF \big)\, , \qquad \omega_\cF \equiv \eta \big(\Gamma_\cF[\as], \mu,\mu_\cF \big) \, ,
\end{align}
where $K_\cF$ and $\omega_\cF$ are resummation kernels associated with any factorization function $\cF$, $j_\cF$ is the dimension of the argument of the function such as $m_J^2 (j_\cF = 2)$, $\ell^+ (j_\cF = 1)$, $\tilde k (j_\cF = \frac{2+\beta}{1+\beta})$, $\mu_\cF$ is the choice of scale used to minimize the logs and $\mu$ is the final scale up to which the function is RG evolved (which we will leave unspecified).
The functions $K(\Gamma, \mu,\mu_\cF)$ and $\eta(\Gamma , \mu ,\mu_\cF)$ are responsible for implementing single and double logarithmic resummation associated with cusp $\Gamma_\cF$ and non-cusp $\gamma_\cF$ anomalous dimensions of the functions.
The formulae for these kernels were described in detail in App.~A of \Refcite{Pathak:2020iue}. In \app{anomdim} we state the anomalous dimensions required for NNLL resummation.

The function $\cJ^{\rm plain}_{\kappa,{\rm no\,sd}}$ accounts for the remaining soft and collinear pieces:
\begin{align}\label{eq:FactPlainKappa}
\cJ^{\rm plain}_{\kappa,{\rm no\,sd}}(\xi,Q, \mu_{\rm plain}) = \frac{\df}{\df\xi} \Big( \cS_{\rm NGL}^\kappa \big( t [Q \xi, Q] \big) \,
\Sigma_{\rm no\, sd}^\kappa(\xi,Q,\mu_{\rm plain})
\Big) \, .
\end{align}
where $\Sigma_{\rm no\, sd}^\kappa$ is defined analogously to $\Sigma_{\rm plain}$ in \eq{GPlain}:
\begin{align}\label{eq:SigNoSD}
\Sigma_{\rm no\,sd}^{\kappa}\big(\xi, Q, \mu\big) =
\int \df s \int \df \ell_c^+ \: J_\kappa \big(s, \mu\big) \, \crS^\kappa_{c_m}(\ell_c^+, \mu) \, \delta \bigg(\xi - \frac{s}{Q^2} -\frac{\ell^+}{Q} \bigg) \,,
\end{align}
and similar to \eq{Ssd}, $\crS^\kappa_{c_m}$ is the cumulative version of the $S_{c_m}$ soft function defined in \eq{ScmDef}.
Making the RG evolution in \eq{FactPlainKappa} explicit, we have
\begin{align}\label{eq:FactPlainResum}
\cJ^{\rm plain}_{\kappa,{\rm no\,sd}}(\xi, Q, \mu_{\rm plain}) &=\Bigg( \cS_{\rm NGL}^\kappa \big( t [Q \xi, Q] \big) \cJ^{\rm plain}_\kappa [\partial_\Omega ; \xi, Q, \mu_{\rm plain}] \frac{e^{\gamma_E\Omega}}{\Gamma(-\Omega)} \\
&\quad + \Big(\frac{\df}{\df \ln \xi} \cS_{\rm NGL}^\kappa \big( t [Q \xi, Q] \big) \Big) \cJ^\kappa_{\rm plain} [\partial_\Omega ; \xi, Q, \mu_{\rm plain}] \frac{e^{\gamma_E\Omega}}{\Gamma(1-\Omega)}\Bigg)
\Bigg|_{\Omega = \tilde \omega_{cs_m}(\mu_s, \mu_J)} \, , \nn
\end{align}
where $\Omega$ is evaluated at
\begin{align}\label{eq:omegaPlain}
\tilde \omega_{cs_m} (\mu_s, \mu_J) \equiv \omega_{cs_m} (\mu, \mu_{s} ) + \omega_J (\mu, \mu_J) \, ,
\end{align}
and the function of the derivative operator is given by
\begin{align}\label{eq:PlainOperator}
\cJ^{\rm plain}_\kappa [\partial_\Omega; \xi, Q, \mu_{\rm plain}]& \equiv \frac{1}{\xi} e^{K_J + K_s} \frac{ \big(Q \mu_s\big)^{\omega_s} \big(\mu_J^2\big)^{\omega_J} }{(\xi Q^2)^{\Omega}}
\:
\nn
\\
&\times
\,
\tilde J_\kappa\Big [ \partial_{\Omega} + \log\Big(\frac{\mu_J^2}{Q^2 \xi}\Big),\, \alpha_s(\mu_J) \Big] \,
\tilde S_{c_m}^\kappa\Big [\partial_{\Omega} + \log\Big(\frac{\mu_{s} }{Q \xi}\Big) , \alpha_s(\mu_{s})\Big ]
\,,
\end{align}
Here $\tilde J_\kappa$ and $\tilde S_{c_m}^\kappa$ are the Laplace transforms of the jet and ungroomed soft function, and we have written them in a notation that makes the logarithms explicit:
\begin{align}\label{eq:LapAlternative}
\tilde J_\kappa \big [ \log (e^{\gamma_E} x \mu_J^2), \alpha_s (\mu_J)\big] &\equiv \tilde J_\kappa (x, \mu_J) \, ,
\\
\tilde S_{c_m}^\kappa \Big[\log \big( u e^{\gamma_E}\mu_s \big) , \alpha(\mu_{s})\Big] &\equiv
\tilde S_{c_m}^\kappa\bigl(u,\mu_{s} \bigr)\,.
\nn
\end{align}

We now turn to the remaining piece in \eq{GplainSplit}. Since this term involves fixed order terms that are not related to a boundary condition of the RG evolution, they have to be treated differently, and result in the formula
\begin{align}\label{eq:DeltaFactPlainX}
\Delta &\tilde \cG^{\rm plain}_{\kappa} (\xi ,r_g, Q, \qcut, \zeta, \mu_{\rm plain} ) = N^\kappa_{\rm incl}( Q, \mu_N)
e^{ K_N }\Big(\frac{\mu_N}{Q}\Big)^{\omega_N} \\
& \times
\Bigg( \cS_{\rm NGL}^\kappa \big( t [Q \xi, Q] \big) \cJ^{\rm plain}_\kappa[\partial_\Omega ; \xi, Q, \mu_{\rm plain}]
\cQ^{\rm plain}_{\kappa} (\Omega,\xi,r_g,\as(\mu_s))
\nn \\
& \quad+ \Big(\frac{\df}{\df \ln \xi} \cS_{\rm NGL}^\kappa \big( t [Q \xi, Q] \big) \Big)\cJ^{\rm plain}_\kappa [\partial_\Omega ; \xi, Q, \mu_{\rm plain}] \cQ^{\rm plain}_{\kappa} (\Omega - 1,\xi,r_g,\as(\mu_s)) \Bigg)\bigg|_{\Omega = \tilde \omega(\mu_{s}, \mu_J)}
\,.\nn
\end{align}
where the kernel $\cQ_\kappa^{\rm plain}$ is defined as Laplace transform of the fixed order soft function terms and convolved with RG resummation kernels:
\begin{align}\label{eq:Qplain}
\cQ_\kappa^{\rm plain}
&\equiv \cQ_\kappa^{\rm sd}\big(\Omega, \xi, \as(\mu)\big)
+ \cQ_\kappa^{r_g}\big(\Omega, \xi, r_g, \as(\mu); a_{20}^{\rm max}\big)
+ \cQ_\kappa^{({\rm sd}, r_g)}\big(\Omega, \xi,r_g,\as(\mu)\big) \, .
\end{align}
The first two terms in \eq{Qplain} involve the $\cO(\as)$ soft function pieces in \eqs{DeltaScg}{Splain1}:
\begin{align}\label{eq:Qplain1}
\cQ^{\rm sd}_{\kappa} \big(\Omega, \xi, \as(\mu)\big) &\equiv \frac{e^{\gamma_E\Omega}}{\Gamma(-\Omega)}
\int_0^\infty \df \ell^+ \: \cL_0^{-\Omega } \bigg(1 - \frac{\ell^+}{Q \xi}\bigg) \, \Delta S_{\rm sd}^\kappa \big(\ell^+, \qcut, \zeta, \beta, \as(\mu)\big) \, , \nn\\
\cQ^{r_g}_\kappa \big(\Omega, \xi, r_g, \as(\mu); a_{20}^{\rm max}\big) &\equiv
\frac{e^{\gamma_E\Omega}}{\Gamma(-\Omega)}
\int_0^\infty \df \ell^+ \: \cL_0^{-\Omega } \bigg(1 - \frac{\ell^+}{Q \xi}\bigg) \Big(1+\frac{\as(\mu_{cs})}{\pi}a_{20}^{\rm max}\Big)\nn \\
&\qquad \times \qcut^{\frac{1}{1+\beta}}\Delta S_{r_g}^\kappa \bigg(\qcut^{\frac{1}{1+\beta}}\ell^+, r_g, \qcut, \zeta, \beta, \as(\mu)\bigg) \, ,
\end{align}
and the third is an $\cO(\as^2)$ cross term:
\begin{align}\label{eq:Qplain2}
\cQ_{\kappa}^{({\rm sd}, r_g)}&\big(\Omega, \xi, r_g,\qcut, \zeta, \beta, \as(\mu)\big) \equiv
\frac{e^{\gamma_E\Omega}}{\Gamma(-\Omega)}
\int_0^\infty \df \ell_1^+\,\df \ell_2^+ \: \cL_0^{-\Omega } \bigg(1 - \frac{\ell_1^++\ell_2^+}{Q \xi}\bigg) \\
&\qquad \times \Delta S_{\rm sd}^\kappa \big(\ell_1^+, \qcut, \zeta, \beta, \as(\mu)\big)
\, \qcut^{\frac{1}{1+\beta}}\Delta S_{r_g}^\kappa \bigg(\qcut^{\frac{1}{1+\beta}}\ell_2^+, r_g, \qcut, \zeta, \beta, \as(\mu)\bigg) \, . \nn
\end{align}
Here we have defined
\begin{align}
{\cal L}_0^{a} (x) \equiv {\cal L}^a (x)+ \frac{1}{a} \delta (x)
\, , \qquad {\cal L}^{a}(x) \equiv \Big [\frac{\Theta(x)}{x^{1-a}}\Big]_+ \, , \qquad a\neq 0 \, ,
\end{align}
and case with $\Omega = 0$ corresponds to turning off resummation:
\begin{align}
\lim\limits_{\Omega\ra0} \frac{e^{\gamma_E\Omega}}{\Gamma(-\Omega)}
{\cal L}_0^{-\Omega}(x) = \delta (x) \, .
\end{align}
We note that the $r_g$-dependence in the cross section in the plain jet mass region arises at $\cO(\as)$ through the fixed order correction $\Delta S_{r_g}^\kappa$ in \eq{DeltaScg}. However, as mentioned above, $\Delta S_{r_g}^\kappa$ itself cannot provide the $\cO(\as)$ boundary condition for NLL evolution (and for NNLL accuracy).
Hence, as discussed in \Refcite{Pathak:2020iue}, we are required to consider cross terms with between $\cO(\as)$ pieces in $\cJ_\kappa^{\rm plain}$ and the normalization factor $N_{\rm incl}^\kappa$, and the $r_g$-dependent piece in \eq{DeltaScg}. We will see that the same applies to the cross section in the soft drop resummation region.
On the other hand, the role of the piece $\Delta S_{\rm sd}^\kappa$ in \eq{Splain1} is to account for effects of grooming on the soft drop jet mass which factorize into global-soft and collinear-soft pieces in the soft drop resummation region.
As we will see below,
we are required to include the cross term $\cQ_\kappa^{({\rm sd},r_g)}$ in \eq{Qplain2} in order to ensure that the $\cO(\as^2)$ terms in the plain jet mass region consistently match with $\cO(\as^2)$ pieces in the soft drop resummation region.
Finally, the computation of these kernels and others below is detailed in \secn{kernel}.

We note that unlike \Refcite{Pathak:2020iue}, we have chosen not to include additional $\cO(\as^2)$ terms proportional to $\beta_0$ in order to cancel running coupling effects in these kernels and render them $\mu$-independent. We have retained for simplicity only the minimal set of $\cO(\as^2)$ terms required to achieve NNLL accuracy, and parameterized uncertainty due to the missing $\cO(\as^2)$ corrections in terms of the nuisance parameter $a_{20}^{\rm max}$ in \eq{Qplain1}. We assume that the missing two-loop pieces in \eq{Qplain} have the same functional form as the one-loop kernels with an unknown normalization parameterized by $a_{20}^{\rm max} \in [-2\pi, 2\pi]$. Additionally, we will separately consider below the effects of two-loop logarithmic terms that arise from RG evolution.

\subsubsection{Max-$R_g$ in soft drop resummation region}
In the soft drop resummation region, the relevant modes are shown in the top left case in \fig{modes}. The factorized cross section reads
\begin{align}\label{eq:SDFact}
\cG^{\rm sd\,res.}_{\kappa } \big(z, \xi, r_g , \mu \big) &= \sum_{\kappa'} \cH_{\kappa' \ra \kappa} (z, Q, \mu) S_{G}^\kappa \big(\qcut, \zeta, \beta,\mu\big) \cS_{\rm NGL}^\kappa \big(t \big[\qcut , Q\big] \big) \\
&\quad \times \int \df \tilde k \int \df s \: J_\kappa \big(s, \mu\big) S_c^\kappa \big(\tilde k , r_g, \qcut, \beta, \mu \big)
\delta \bigg(\xi - \frac{s}{Q^2} -\frac{\tilde k(\qcut)^{\frac{-1}{1+\beta}}}{Q} \bigg)
\nn \, .
\end{align}
This involves the global-soft and c-soft functions we discussed above in \eqs{SGDef}{ScDef}. The NGLs are now independent of jet mass involving constant scales $\qcut$ and $Q$, such that we are able to write it directly as differential in groomed jet mass. The power corrections associated with this factorization are given by
\begin{align}\label{eq:MaxPC2}
\cG_{\kappa} \big(z, \xi, r_g, \alpha_s(\mu) \big) = \cG^{\rm sd\, res.}_{\kappa} \big(z, \xi,r_g,\mu \big) \Bigg[
1 + \cO \bigg( \xi_0, \frac{\xi}{r_g^2}, \Big(\frac{\xi}{ \xi_0}\Big)^{\frac{2}{2+\beta}}\bigg) \Bigg] \, ,
\end{align}
We see that the jet mass related $\cO(\xi)$ power corrections in the plain jet mass region are now replaced by $\cO(\xi/r_g^2)$, due to modification of the effective jet radius seen by collinear modes. In the plain jet mass region with $r_g = 1$, these corrections match with those in \eq{MaxPC1}. Additionally, we have new power corrections related to soft drop that have resulted in the factorization of the wide-angle soft mode into a global soft and c-soft mode.

Next we state the resummed formula for this regime for the jet mass dependent part of $\cG_{\kappa}$:
\begin{align}\label{eq:GsdFactResum}
\tilde \cG^{\rm sd\, res.}_{\kappa}& (\xi, r_g, Q, \qcut, \zeta, \beta, \mu_{\rm sd})= N_\kappa^{\rm evol} \big(\mu_N, \mu_{gs}, Q, \qcut, \zeta, \beta\big)
\cS_{\rm NGL}^\kappa\big(t \big[\qcut , Q\big] \big)\\
& \times
\cJ_\kappa^{\rm sd\,res} [\partial_\Omega ; \xi, Q, \qcut, \mu_{\rm sd}] \bigg( \frac{e^{\gamma_E\Omega}}{\Gamma(- \Omega)} + \cQ^{r_g}_\kappa \big(\Omega, \xi, r_g, \as(\mu_{cs}); a_{20}^{\rm max}\big)
\bigg)
\Bigg|_{\Omega = \tilde \omega_{cs}(\mu_{cs}, \mu_J)} \nn
\, .
\end{align}
Here $\mu_{\rm sd}$ stands for the set of scales in the max-$R_g$ regime in the soft drop resummation region:
\begin{align}\label{eq:musd}
\mu_{\rm sd}(\xi) \equiv \{\mu_N, \mu_{gs}, \mu_J (\xi), \mu_{cs}(\xi) \} \, .
\end{align}
The normalization factor $N_\kappa^{\rm evol}$ includes resummation of logarithms between the hard-collinear and global-soft scales:
\begin{align}
N_\kappa^{\rm evol} \big(\mu_N, \mu_{gs}, Q, \qcut , \zeta, \beta\big)
&\equiv N^\kappa_{\rm incl}( Q, \mu_N)
S_G^\kappa \big(\qcut, \zeta, \beta, \mu_{gs} \big)
\\
&\quad\times
e^{\big [ K_N + K_{gs} \big] }\Big(\frac{\mu_N}{Q}\Big)^{\omega_N}
\Big(\frac{\mu_{gs}}{\qcut}\Big)^{\omega_{gs}} \, . \nn
\end{align}
The function of the derivative operator is given by
\begin{align}\label{eq:SDOperator}
&\cJ_\kappa^{\rm sd\,res} [\partial_\Omega; \xi, Q, \qcut, \mu_{\rm sd}]
\equiv \frac{1}{\xi}
\:
e^{ K_{cs} + K_{J} }
\frac{\big(\mu_J^2\big)^{ \omega_J} \big(Q \mu_{cs}\big)^{\omega_{cs} }}{(\xi Q^2)^\Omega}
\Bigg(\Big(\frac{\mu_{cs}}{\qcut}\Big)^\frac{1}{1+\beta}\Bigg)^{\omega_{cs} }
\\
&\quad\times
\,
\tilde J_\kappa\Big [ \partial_{\Omega} + \log\Big(\frac{\mu_J^2}{Q^2 \xi}\Big),\, \alpha_s(\mu_J) \Big] \,
\tilde S_{c}^\kappa\Bigg [\partial_{\Omega} + \log\Bigg(\frac{ \mu_{cs}}{Q \xi}\Big(\frac{\mu_{cs}}{\qcut}\Big)^\frac{1}{1+\beta}\Bigg) , \alpha_s(\mu_{cs})\Bigg ]
\, ,\nn
\end{align}
where the Laplace transforms are written analogously to \eq{LapAlternative}. Similar to \eq{omegaPlain}, the derivatives are evaluated at $\tilde \omega_{cs}(\mu_{cs},\mu_J)$ defined as
\begin{align}\label{eq:omegaSD}
\tilde \omega_{cs} (\mu_{cs}, \mu_J) \equiv \omega_{cs} (\mu, \mu_{cs} ) + \omega_J (\mu, \mu_J) \, .
\end{align}
Finally, as in \eq{DeltaFactPlainX}, we expand the above equation to $\cO(\as^2)$ including cross terms between $\cJ_\kappa^{\rm sd\,res}$ and the same kernel $\cQ^{r_g}_\kappa$ defined in \eq{Qplain1} required for NNLL accuracy.

\subsection{Min and Intermediate $R_g$ regimes}
\label{sec:MinIntResum}

For completeness we review the factorization formulae for the remaining min and intermediate $R_g$ regimes derived in \Refcite{Pathak:2020iue}.
\subsubsection{Min-$R_g$ regime}
We now turn to the min-$R_g$ regime. As seen in the rightmost column in \fig{modes}, the cross section in the min-$R_g$ regime involves combination of the hard collinear $\mcC$ mode, the collinear soft mode $\mCSg$ and the global soft mode $\mSG$, and is given by
\begin{align}\label{eq:FactSmallRg}
\cG^{\rm min}_{\kappa}\big(z, \xi, r_g ,\mu \big)&= \sum_{\kappa'} \cH_{\kappa' \ra \kappa} (z, Q, \mu) S_{G}^\kappa \big(\qcut, \zeta, \beta,\mu\big) \cS_{\rm NGL}^\kappa \big(t \big[\qcut , Q\big] \big) \\
&\quad \times S_{c_g}^{\kappa} \big(\qcut r_g^{1+\beta}, \beta, \mu\big) \cS_{\rm NGL}^\kappa \Big(t \big[\qcut r_g^{1+\beta}, Q r_g \big] \Big)\frac{1}{r_g^2} \cC^\kappa \bigg(\frac{\xi}{r_g^2}, Qr_g, \mu \bigg)
\nn \, .
\end{align}
Here we notice appearance of new NGLs between the scales associate with $\mCSg$ and $\mcC$ modes due to an additional boundary of the groomed jet radius. The power corrections that are dropped in this formula are given by
\begin{align}\label{eq:MinPC}
\cG_{\kappa} \big(z, \xi, r_g, \alpha_s(\mu) \big) = \cG^{\rm min}_{\kappa} \big(z, \xi,r_g ,\mu \big) \Bigg[
1 + \cO \bigg( \xi_0, \Big(\frac{\xi}{ \xi_0}\Big)^{\frac{2}{2+\beta}}, r_g^{2+\beta}\frac{\xi_0}{ \xi} \bigg) \Bigg] \, ,
\end{align}
As before, the soft drop factorization proceeds by dropping the $\cO((\xi/\xi_0)^{\frac{2}{2+\beta}})$ power corrections. In contrast with the max-$R_g$ regime in \eqs{MaxPC1}{MaxPC2} collinear function now includes the $\cO(\xi/r_g^2)$ terms which become $\cO(1)$ in this regime. However, in order to resum logarithms between scales associated with $\mCSg$ and $\mcC$ modes, the power corrections of the form $r_g^{2+\beta}\frac{\xi_0}{ \xi}$ are dropped.

The resummed result for the normalized cross section is given by
\begin{align}\label{eq:MinResum}
\tilde \cG^{\rm min}_{\kappa} (\xi, r_g, Q, \qcut, \zeta, \beta, \mu_{\rm min})&= N_\kappa^{\rm evol} \big(\mu_N, \mu_{gs}, Q, \qcut, \zeta, \beta\big) \cS_{\rm NGL}^\kappa\big(t \big[\qcut , Q\big] \big)\\
& \quad \times
\cS_{\rm NGL}^\kappa \Big(t \big[\qcut r_g^{1+\beta}, Q r_g \big] \Big)
e^{K_\cC + K_{cs_g}} \Big(\frac{\mu_{gs}}{\qcut r_g^{1+\beta}}\Big)^{\omega_{cs_g}}
\Big(\frac{\mu_\cC}{Qr_g} \Big)^{\omega_\cC} \nn \\
&\quad \times
S_{c_g}^{\kappa} \big(\qcut r_g^{1+\beta}, \beta, \mu_{cs_g}\big) \frac{1}{r_g^2} \cC^\kappa \bigg(\frac{\xi}{r_g^2}, Qr_g, \mu_{\cC}; a_{20}^{\rm min} \bigg)
\, , \nn
\end{align}
with the set of profiles for this regimes being:
\begin{align}\label{eq:mumin}
\mu_{\rm min}(r_g) \equiv \{\mu_N, \mu_{gs}, \mu_\cC (r_g), \mu_{cs_g}(r_g) \} \, .
\end{align}
Here we have included additional $\cO(\as^2)$ terms in the collinear function precisely as described in \Refcite{Pathak:2020iue}:
\begin{align}\label{eq:Cmin}
\frac{\xi}{r_g^2}\cC^q \bigg(\frac{\xi}{r_g^2}, Qr_g, \mu; a_{20}^{\rm min} \bigg)
&= \frac{\as}{\pi} a_{10}^{\cC_q}\bigg(\frac{\xi}{r_g^2}\bigg) \bigg[1 + \frac{\as C_F}{\pi} \bigg(\frac{1}{4} \ln^2 \frac{\mu^2}{Q^2r_g^2}+ \frac{3}{4} \ln \frac{\mu^2}{Q^2r_g^2}\bigg)\bigg] \\
&\quad +\frac{\as}{\pi} a_{20}^{\rm min} \frac{\as}{\pi} a_{10}^{\cC_q}\bigg(\frac{16}{25}\frac{\xi}{r_g^2}\bigg) \nn \, ,\\
\frac{\xi}{r_g^2}\cC^g \bigg(\frac{\xi}{r_g^2}, Qr_g, \mu; a_{20}^{\rm min} \bigg)
&= \frac{\as}{\pi} a_{10}^{\cC_g}\bigg(\frac{\xi}{r_g^2}\bigg) \bigg[1 + \frac{\as}{\pi} \bigg(\frac{C_A}{4} \ln^2 \frac{\mu^2}{Q^2r_g^2}+ \frac{\beta_0}{4} \ln \frac{\mu^2}{Q^2r_g^2}\bigg)\bigg] \nn \\
&\quad +\frac{\as}{\pi} a_{20}^{\rm min} \frac{\as}{\pi} a_{10}^{\cC_g}\bigg(\frac{16}{25}\frac{\xi}{r_g^2}\bigg) \, .\nn
\end{align}
Here $a_{10}^{\cC_\kappa} \frac{\as(\mu)}{\pi}$ is the $\cO(\as)$ result of the collinear function in \eq{CollFunc} after including $\xi/r_g^2$ factor that sets the $\delta(\xi/r_g^2)$ terms to zero. The new $\cO(\as^2)$ pieces in the square brackets provide the boundary condition for NLL resummation, as we saw above in the max-$R_g$ case. In the second line we have included parameterized the uncertainty from missing $\cO(\as)$ pieces in terms of the parameter $a_{20}^{\rm min} \in [-2\pi, 2\pi]$, while shifting the argument appropriately to take into account the different end-point of the jet mass spectrum at NLO~\cite{Pathak:2020iue} at $r_g = 8/5 \sqrt{\xi}$ instead of $r_g = 2\sqrt{\xi}$ at LO.
Additionally, we also include $\cO(\as^2)$ cross terms from other pieces in \eq{MinResum}.

\subsubsection{Intermediate-$R_g$ regime}
Finally, we describe the factorization in the intermediate $R_g$ regime shown in the middle column in \fig{modes} which represents the most factorized scenario:
\begin{align}\label{eq:FactIntRg}
\cG^{\rm int}_{\kappa} (z, \xi , r_g ,\mu) &=
\sum_{\kappa'} \cH_{\kappa' \ra \kappa} (z, Q, \mu) S_{G}^\kappa \big(\qcut, \zeta, \beta,\mu\big)
\cS_{\rm NGL}^\kappa \big(t \big[\qcut , Q\big] \big)S_{c_g}^{\kappa} \big(\qcut r_g^{1+\beta}, \beta, \mu\big) \nn \\
& \quad\times \frac{\df }{\df \xi} \Bigg[\cS_{\rm NGL}^\kappa \bigg( t\bigg[\qcut r_g^{1+\beta}, \frac{Q \xi}{r_g} \bigg]\bigg) \Sigma_{\rm int}^{\kappa}\bigg(\frac{\xi}{r_g^2}, Qr_g, \mu\bigg) \Bigg] \, ,
\end{align}
Here the $r_g$-dependent NGLs are analogous to the previous case but involve running between the scales associated with $\mCSg$ and $\mCSm$ modes. Since they do depend on the jet mass, we have written them as a derivative of the cumulative jet mass cross section. Here $\Sigma_{\rm int}^{\kappa}$ is defined in terms of $\Sigma_{\rm no\,sd}^\kappa$ in \eq{SigNoSD}:
\begin{align}
\Sigma_{\rm int}^{\kappa}\bigg(\frac{\xi}{r_g^2}, Qr_g, \mu\bigg) =
\Sigma_{\rm no\,sd}^\kappa \bigg(\xi \ra \frac{\xi}{r_g^2}, Q \ra Q r_g , \mu\bigg)
\, .
\end{align}
The power corrections that are dropped in this regime are combinations of the previous cases:
\begin{align}\label{eq:IntPC}
\cG_{\kappa} \big(z, \xi, r_g, \alpha_s(\mu) \big) = \cG^{\rm int}_{\kappa} \big(z, \xi,r_g ,\mu \big) \Bigg[
1 + \cO \bigg( \xi_0, \Big(\frac{\xi}{ \xi_0}\Big)^{\frac{2}{2+\beta}}, r_g^{2+\beta}\frac{\xi_0}{ \xi}, \frac{\xi}{r_g^2}\bigg) \Bigg] \, .
\end{align}

Finally, we state the formula for the resummed cross section in this regime:
\begin{align}\label{eq:IntResum}
&\tilde \cG^{\rm int}_{\kappa} (\xi, r_g, Q, \qcut, \zeta, \beta, \mu_{\rm int})= N_\kappa^{\rm evol} \big(\mu_N, \mu_{gs}, Q, \qcut, \zeta, \beta\big) \cS_{\rm NGL}^\kappa\big(t \big[\qcut , Q\big] \big)\\
& \times \Bigg( \cS_{\rm NGL}^\kappa \bigg( t\bigg[\qcut r_g^{1+\beta}, \frac{Q \xi}{r_g} \bigg]\bigg) \cJ^{\rm int}_\kappa [\partial_\Omega ; \xi, r_g, Q, \qcut, \zeta, \beta , \mu_{\rm int}] \frac{e^{\gamma_E\Omega}}{\Gamma(-\Omega)} \nn \\
& + \frac{\df}{\df \ln \xi} \cS_{\rm NGL}^\kappa \bigg( t\bigg[\qcut r_g^{1+\beta}, \frac{Q \xi}{r_g} \bigg]\bigg) \times \cJ^{\rm int}_\kappa [\partial_\Omega ; \xi, r_g, Q, \qcut, \zeta, \beta , \mu_{\rm int}]\frac{e^{\gamma_E\Omega}}{\Gamma(1-\Omega)}\Bigg)
\Bigg|_{\Omega = \tilde \omega (\mu_s, \mu_J)} \, . \nn
\end{align}
where
\begin{align}
&\cJ^{\rm int}_\kappa [\partial_\Omega ; \xi, r_g, Q, \qcut, \zeta, \beta , \mu_{\rm int}] \equiv \frac{e^{K_J + K_s + K_{cs_g}} }{\xi} \Big(\frac{\mu_{cs_g}}{\qcut r_g^{1+\beta}}\Big)^{\omega_{cs_g}} \frac{ \big(Q r_g\mu_{cs_m}\big)^{\omega_{cs_m}} \!(\mu_J^2)^{\omega_J} }{(\xi Q^2)^{\Omega}} \nn
\\
&\quad \times
S_{c_g}^{\kappa} \big(\qcut r_g^{1+\beta}, \beta, \mu_{cs_g}\big)
\tilde J_\kappa\Big [ \partial_{\Omega} + \log\Big(\frac{\mu_J^2}{Q^2 \xi}\Big),\, \alpha_s(\mu_J) \Big] \,
\tilde S_{c_m}^\kappa\Big [\partial_{\Omega} + \log\Big(\frac{\mu_{cs_m}r_g }{Q \xi}\Big) , \alpha_s(\mu_{cs_m})\Big ]
\,,
\end{align}
and the intermediate-$R_g$ profiles being
\begin{align}
\mu_{\rm int}(\xi, r_g) \equiv \{\mu_N, \mu_{gs}, \mu_J(\xi), \mu_{cs_m}(\xi, r_g), \mu_{cs_g}(r_g)\} \, .
\end{align}
Unlike the previous two cases, here every function contributes to the NLL boundary condition, and hence we do not need to include any additional $\cO(\as^2)$ pieces.

\section{Matched cross section}
\label{sec:match}

We now combine the cross sections in the three regimes and the two jet mass regions to obtain the complete matched cross section. In \secn{match_max} we describe how the results of \Refcite{Hannesdottir:2022rsl} are extended to match the max-$R_g$ regime between soft drop and plain jet mass resummation regions. In \secn{match_all} we match this result to min- and int-$R_g$ regimes, including the plain jet mass region. In \secn{pertuncert} we compute the perturbative uncertainty due to scale variation and nuisance parameters. The impact of $\cO(\as^2)$ cross terms is assessed in \secn{twoloop} and that of non-singular fixed order corrections in \secn{ns}. The computation of $C_{1\kappa}^{(n)}$ weights and the impact of NGLs is discussed in \secn{c1}.

\subsection{Matching in the max-$R_g$ regime}
\label{sec:match_max}

We first begin with combining the max-$R_g$ cross sections
in \eqs{GplainSplit}{GsdFactResum} to obtain a complete matched result in this regime. Our prescription for matching the two results parallels that of the single differential jet mass distribution derived in \Refcite{Hannesdottir:2022rsl}, and is given by
\begin{align}\label{eq:MatchMax}
\tcG^{{\rm max}}_{\kappa} (\xi,r_g ) &\equiv
\tcG_\kappa^{\rm sd\,res.} (\xi, r_g , \mu_{\rm sd\ra plain})
+ \big[ \tcG_\kappa^{\rm plain} (\xi, r_g , \mu_{\rm plain})
- \tcG_\kappa^{\rm sd\,res.} (\xi, r_g , \mu_{\rm plain}) \big ]
\, .
\end{align}
For simplicity we have suppressed dependence on kinematic and grooming parameters. The profile $\mu_{\rm sd\ra plain}$ is designed such that it transitions from $\mu_{\rm sd}$ scales in \eq{musd} to $\mu_{\rm plain}$ scales in \eq{muplain} by merging the c-soft $\mu_{cs}$ and global-soft $\mu_{gs}$ scales for $\xi \geq \xi_0$. As a result the two $\tcG_\kappa^{\rm sd\,res.}$ terms cancel each other for $\xi \geq \xi_0$ leaving behind the correct $\tcG_\kappa^{\rm plain}$ cross section in this region. In the soft drop resummation region for $\xi < \xi_0$, the term $\tcG_\kappa^{\rm sd\,res.} (\xi, r_g , \mu_{\rm plain})$ acts as subtraction piece for the $\tcG_\kappa^{\rm plain}$ cross section evaluated with the same scale. By evaluating the collinear-soft and global-soft pieces at the same $\mu_s$ scale, the difference between the two amounts to $(\xi/\xi_0)^{\frac{2}{2+\beta}}$ soft drop related power corrections lacking in the $\tcG_\kappa^{\rm sd\,res.}$ in \eq{MaxPC2}.
Furthermore, since we have chosen to employ the same kernel $\cQ^{r_g}_\kappa$ in \eq{Qplain1} in all the three pieces, the matching of $r_g$-related piece simply amounts to choosing the right soft scale in the argument of $\as$ and the resummation kernels $\omega_{cs_m,s}$ in \eqs{omegaPlain}{omegaSD}.
Finally, as remarked above, including the cross term $\cQ_{\kappa}^{({\rm sd},r_g)}$ defined in \eq{Qplain2} in the plain jet mass cross section, \eq{MatchMax} seamlessly implements matching of $\cO(\as^2)$ terms as well.
\begin{figure}[t]
\centering
\includegraphics[width=.48\linewidth]{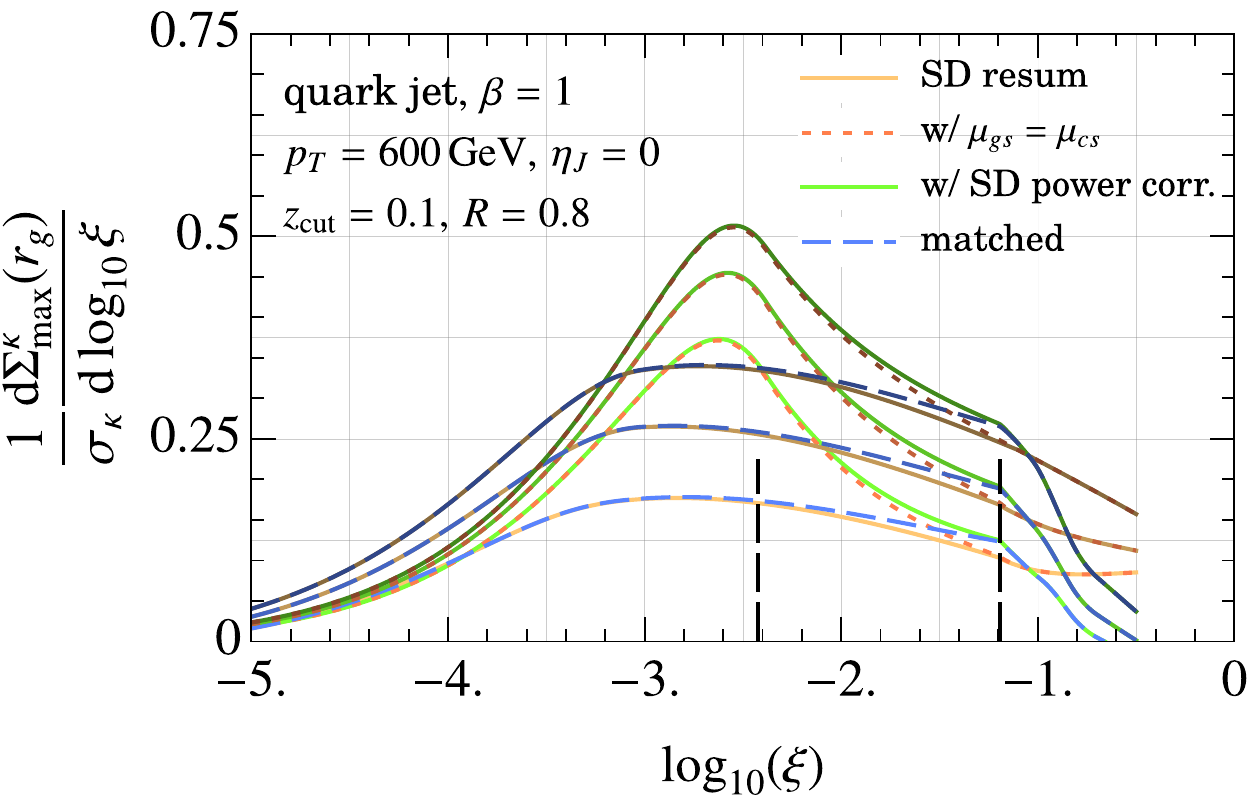}
\includegraphics[width=.48\linewidth]{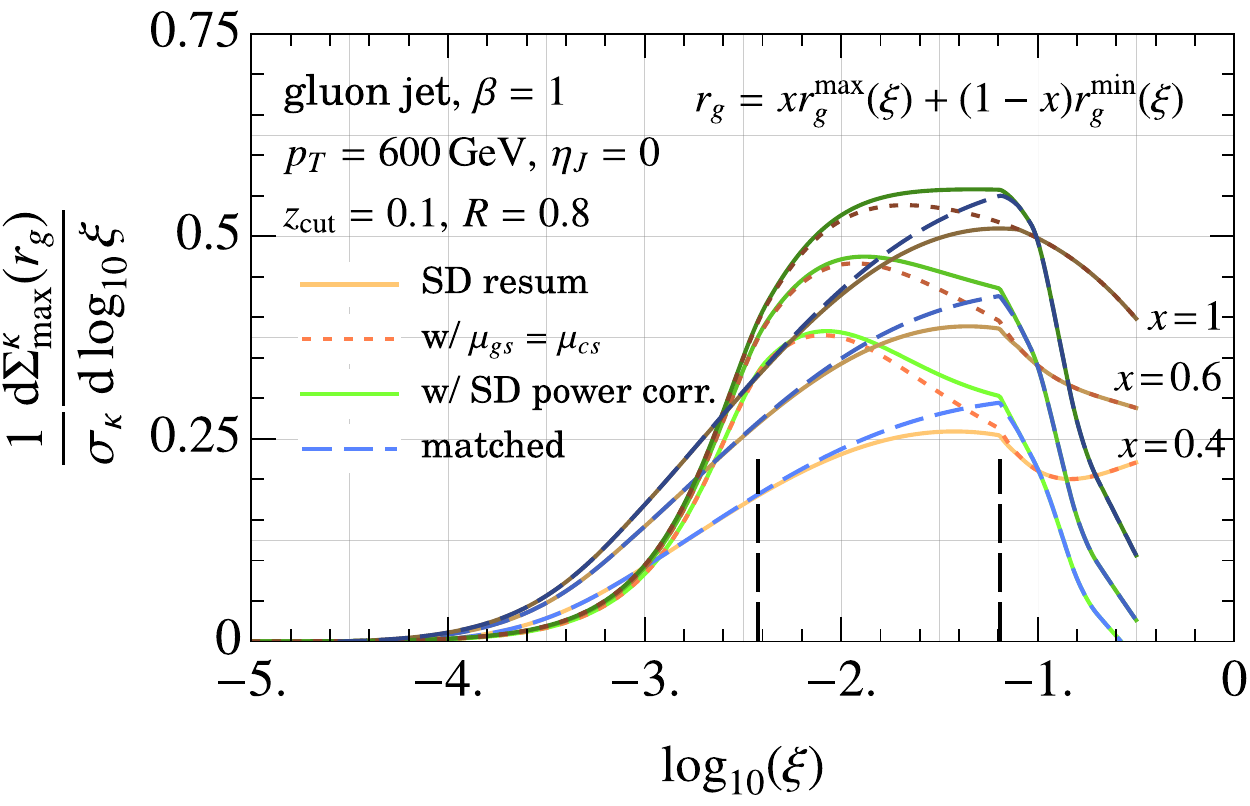}
\caption{Matching for max-$R_g$ cross section. The vertical lines denote the extent of the soft drop operator expansion region. The three curves correspond to choosing different intermediate values of groomed jet radius as a function of the jet mass.}
\label{fig:maxmatch}
\end{figure}

We show the result of matching for gluon and quark jets in \fig{maxmatch}. For now we will only include terms up to $\cO(\as)$ and discuss the effects of including $\cO(\as^2)$ cross terms below. Here we have taken the groomed jet radius cut $r_g$ to lie somewhere between the maximum and minimum value of $r_g$ for a given jet mass:
\begin{align}\label{eq:rglimits}
r_g^{\rm min}(\xi) \equiv \sqrt{\xi} \, , \qquad
r_g^{\rm max} (\xi) \equiv \min \Bigg\{1, \, \bigg[ \Big(\frac{\xi_0}{\xi}\Big)^{\frac{2}{2+\beta}} - \zeta^2 \bigg]^{-\frac{1}{2}} \Bigg \} \, .
\end{align}
In \fig{maxmatch} the vertical lines denote the extent of the soft drop operator expansion region, $\xi \in (\xi_{\rm SDOE}, \xi_0')$. As explained above, the overlap curve $\cG_\kappa^{\rm sd\,res.}(\xi, r_g, \mu_{\rm plain})$ merges with the un-factorized soft drop curve in for small jet masses and with the factorized one for $\xi > \xi_0$. We see that the effect of matching close to the cusp is particularly noticeable for gluon jets.

\subsection{Matched resummed cross section}
\label{sec:match_all}

Having defined the max-$R_g$ cross section, we now follow the same strategy as in \Refcite{Pathak:2020iue} to obtain the matched resummed cross section valid across all three regimes:
\begin{align}\label{eq:MatchDDiff}
\tcG_\kappa^{\rm match} (\xi, r_g)
&\equiv \tcG_\kappa^{{\rm int}} (\xi,r_g ,\mu_{\rm hyb})
+ \big [\tcG^{{\rm max}}_{\kappa} (\xi,r_g)
-\tcG_\kappa^{{\rm int}} (\xi,r_g ,\mu_{\rm sd\ra plain})
\big]\\
&\quad
+ \big[
\tcG_\kappa^{{\rm min}} (\xi,r_g ,\mu_{\rm min})
- \tcG_\kappa^{{\rm int}} (\xi,r_g ,\mu_{\rm min})
\big] \, . \nn
\end{align}
The new hybrid profile $\mu_{\rm hyb}$ that appears in the first term interpolates between the three sets of profiles $\mu_{\rm sd\ra plain}$, $\mu_{\rm min}$ and $\mu_{\rm int}$, and is given by
\begin{align}\label{eq:weight}
& \mu_{\rm hyb} \equiv \big(\mu_{\rm min}\big)^{w_{\rm min}}
\big(\mu_{\rm int}\big)^{w_{\rm int}}
\big (\mu_{\rm sd\ra plain}\big)^{w_{\rm sd\ra plain}}
\, ,&
&w_{\rm min} + w_{\rm int} + w_{\rm sd\ra plain} = 1 \, .&
\end{align}
where $w_i$ are weight functions that depend on both $\xi$ and $r_g$ and provide a prescription for demarcating boundaries between the three regimes, as shown in \fig{regions}.
The construction of the weight functions for soft drop resummation region was discussed in \Refcite{Pathak:2020iue}. In this work we extend the weight functions into the plain jet mass region, with their final result given by:
\begin{align}
&\text{2-EFT:}& &w_{\rm max} = X \big(r_g, r_{g,t} (\xi)\big)\,, & &w_{\rm int} = 0 \, ,& \\
&\text{3-EFT:}& &w_{\rm max } = X \big(r_g, r_{g,\rm PC}^{\rm max} (\xi)\big) \, ,&
&w_{\rm int} = \big[1 - X\big(r_g, r_{g,\rm PC}^{\rm max}(\xi )\big)\big] X\big(r_g, r_{g,\rm PC}^{\rm min}(\xi)\big) \, ,& \nn
\end{align}
and $w_{\rm min} = 1 - w_{\rm int} - w_{\rm max}$. Here $X(r_g, r_{g,t}) \ra 0$ for $r_g \ll r_{g,t}$ and $X(r_g, r_{g,t}) \ra 1$ for $r_g\gg r_{g,t}$ transitioning around $r_{g,t}$. The transition points $r_{g,\rm PC}^{\rm min}(\xi)$ and $r_{g,\rm PC}^{\rm min}(\xi)$ are derived from power expansion parameters that determine validity of the EFT.
We describe the details of the implementation in \app{weight}.

\begin{figure}[t]
\centering
\includegraphics[width=.48\linewidth]{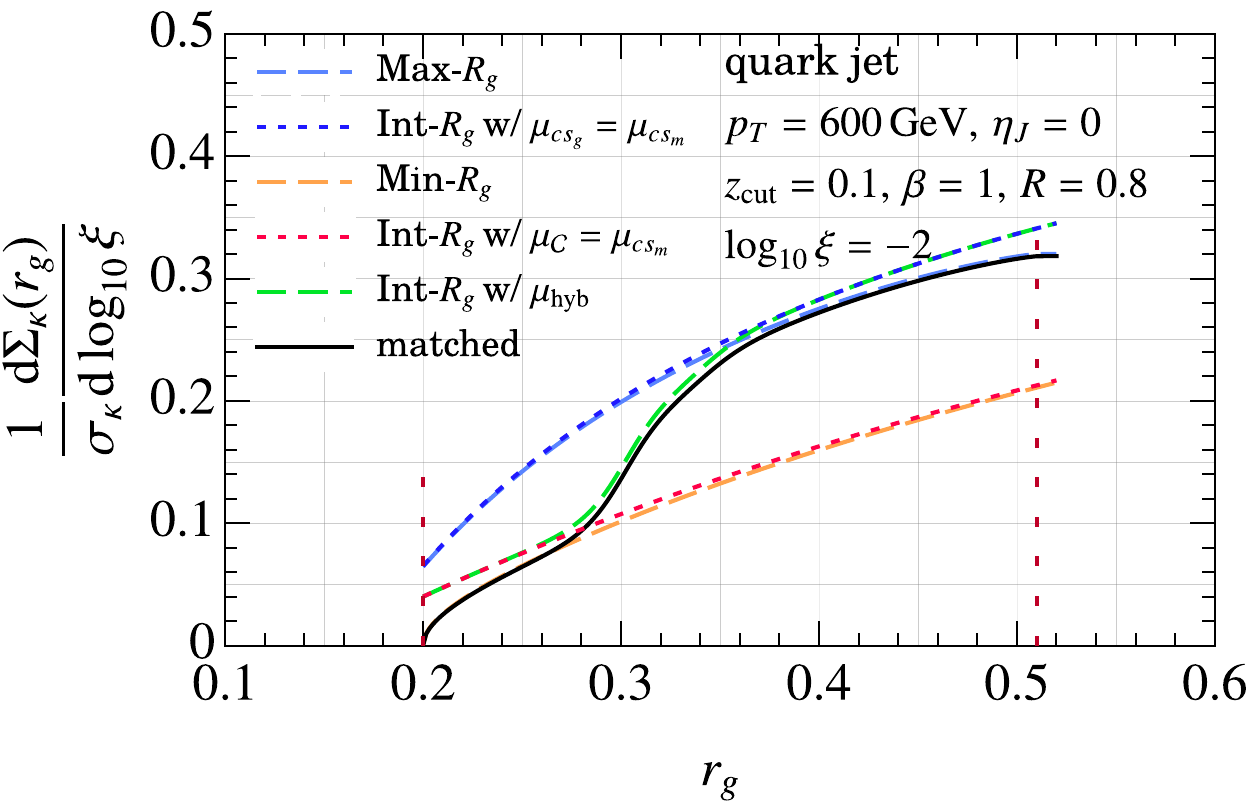}
\includegraphics[width=.48\linewidth]{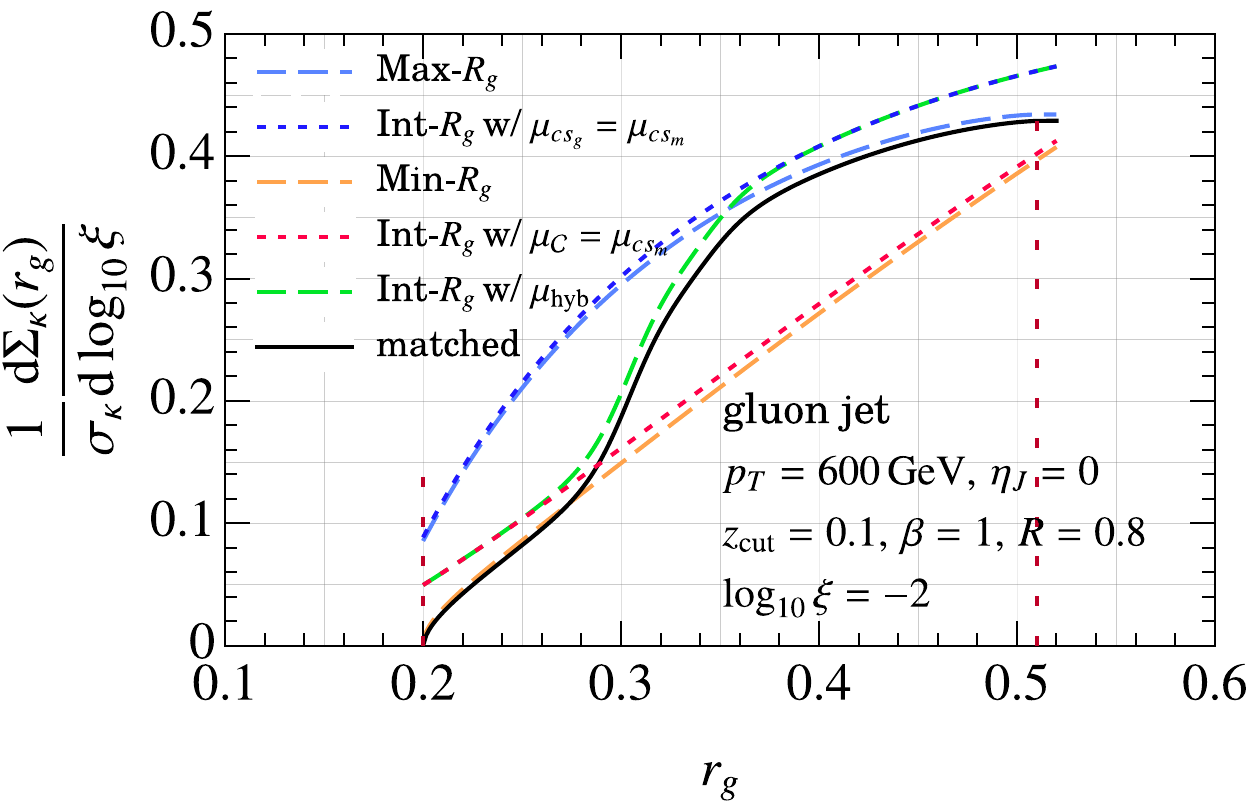}
\caption{Matching across the three $R_g$ regimes. The vertical lines denote the minimum and maximum $r_g$ allowed for the specified $\xi = 10^{-2}$ value.}
\label{fig:match}
\end{figure}

In \fig{match} we show the matching across the three regimes for quark and gluon jets. We can see that the overlap pieces obtained from the intermediate-$R_g$ regime (shown as dotted lines) cancel the max-$R_g$ cross section for $r_g \ll r_g^{\rm max}(\xi)$ and likewise the min-$R_g$ cross section for $r_g \gg r_g^{\rm min}$. The intermediate-$R_g$ cross section with hybrid profiles (dashed) interpolates between the two overlap pieces. The matched curve thus obtained agrees with the NNLL jet mass cross section at the end point $r_g = r_g^{\rm max}$, such that $\cG_\kappa^{\rm match} (\xi, r_g^{\rm max}(\xi)) = \cG_\kappa^{\rm NNLL}(\xi)$. However, note that $\cG_\kappa^{\rm match} (\xi, r_g^{\rm max}(\xi)) \neq \cG_\kappa^{\rm max}(\xi, r_g^{\rm max}(\xi))$. This is because the $\cG_\kappa^{\rm max}$ cross section is derived from factorizing the jet mass measurement and lacks jet mass related power corrections which are supplied by the min-$R_g$ cross section.

\subsection{Perturbative uncertainty}
\label{sec:pertuncert}

\begin{figure}[t]
\centering
\includegraphics[width=.48\linewidth]{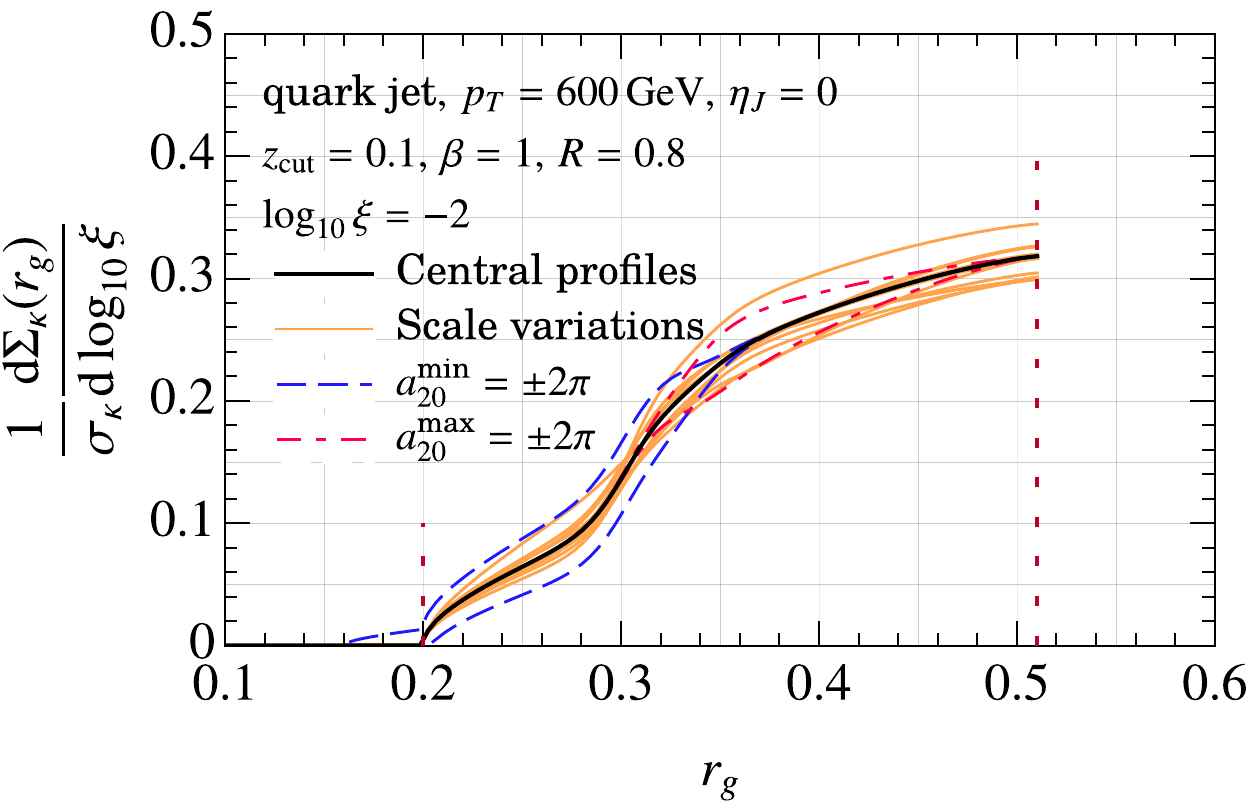}
\includegraphics[width=.48\linewidth]{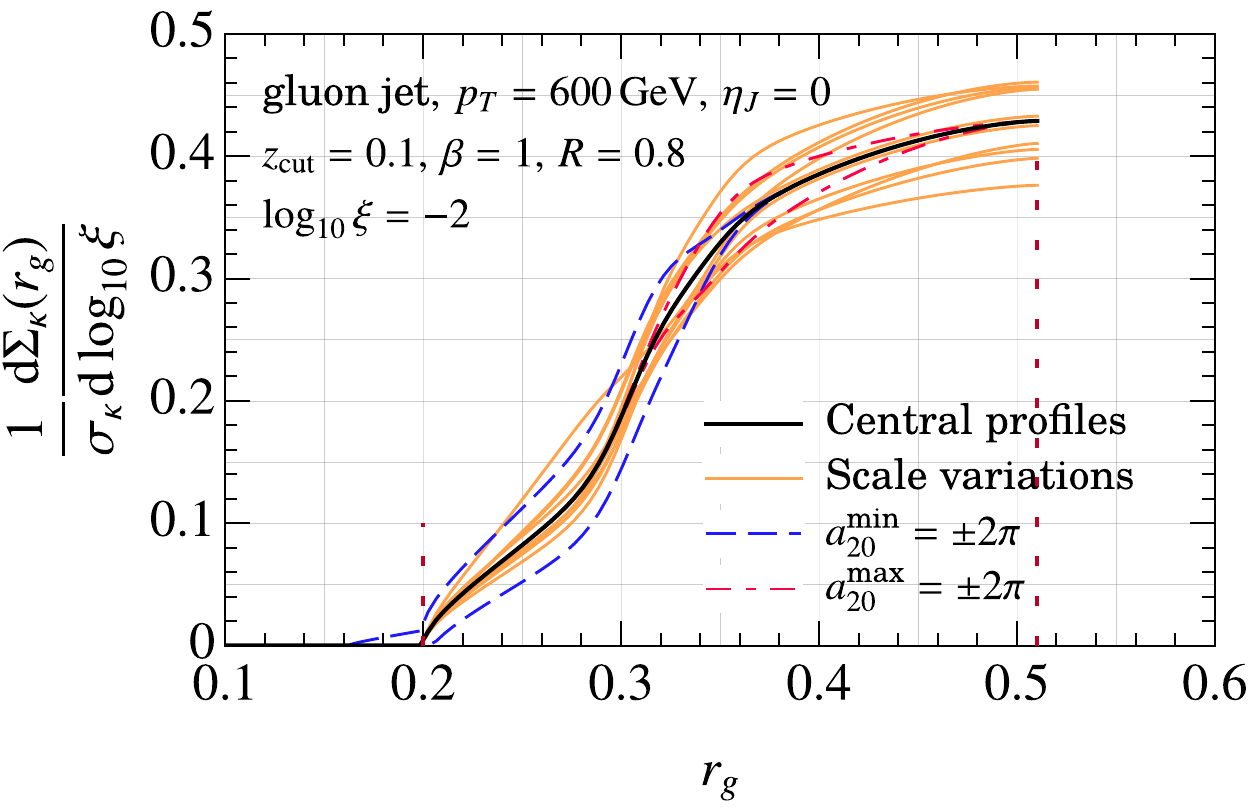}
\includegraphics[width=.48\linewidth]{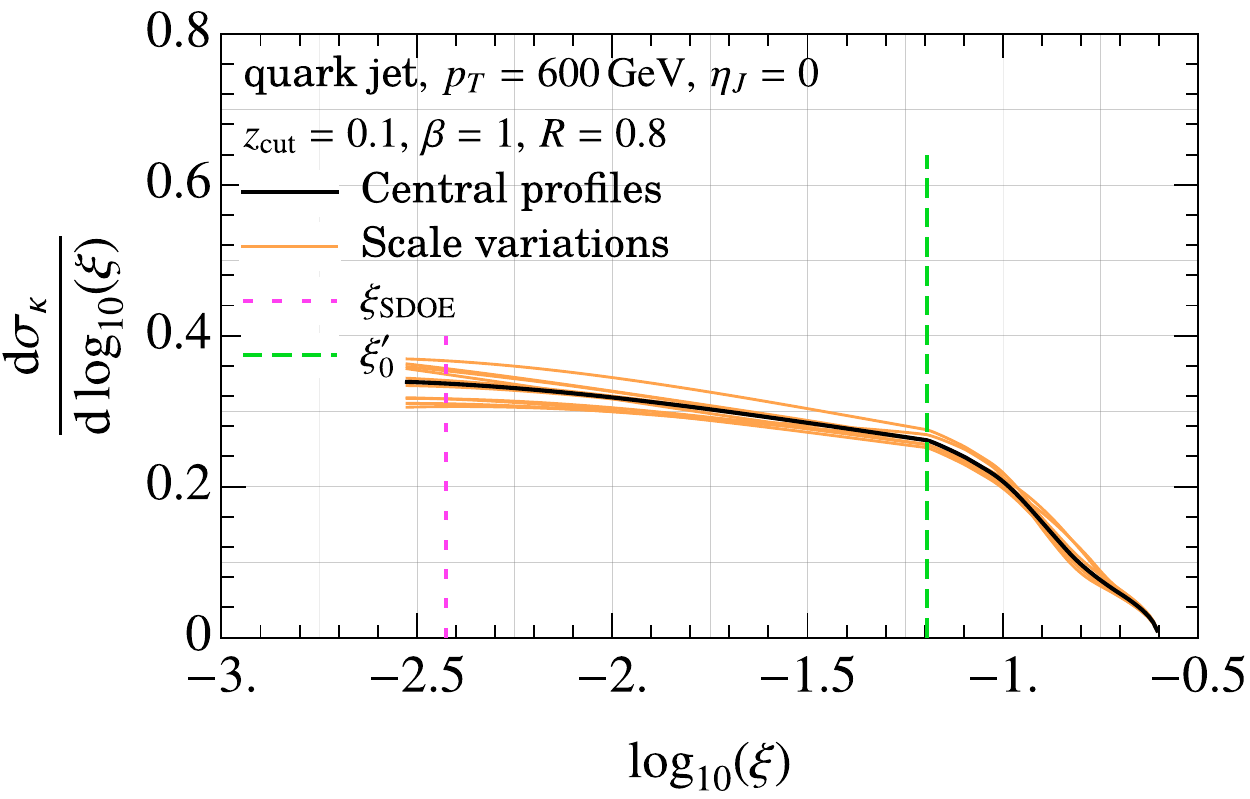}
\includegraphics[width=.48\linewidth]{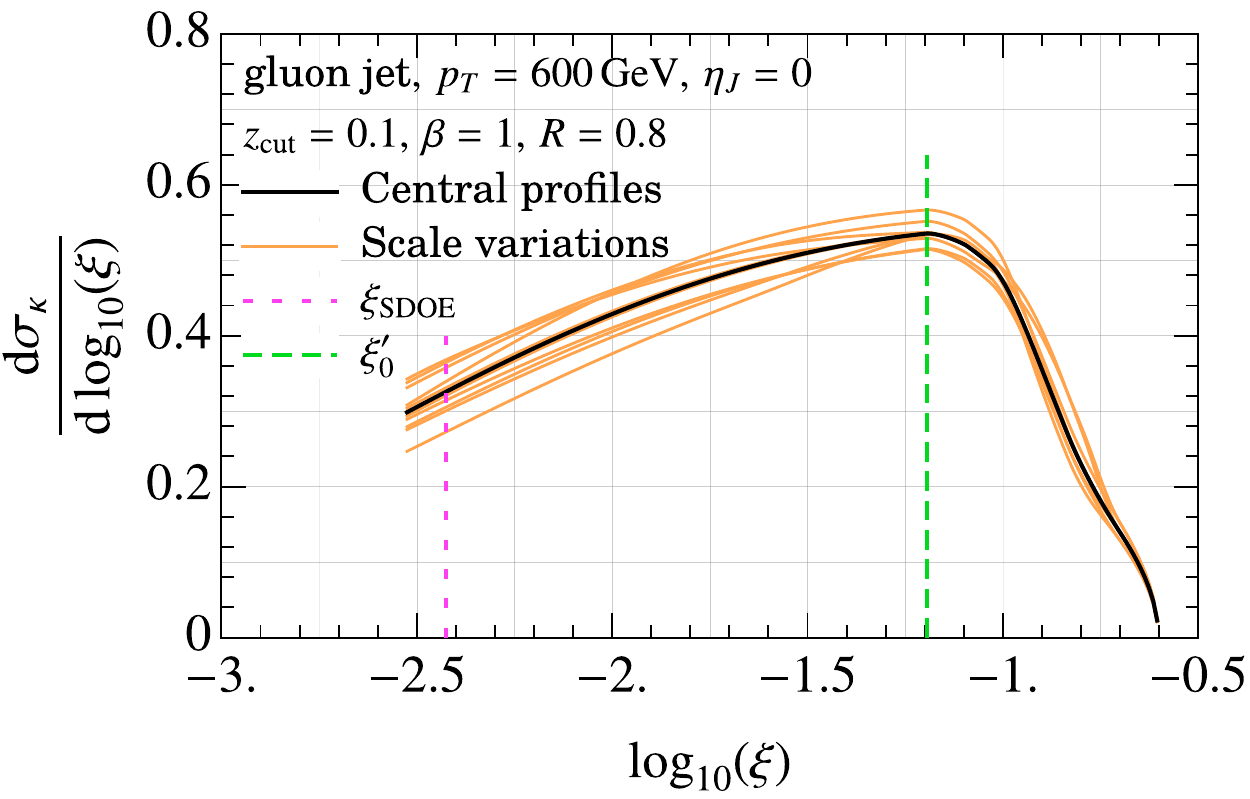}
\caption{Profile and two-loop constant terms variation of the cumulative $r_g$ cross section (top row) and the single differential jet mass cross section (bottom row) for quarks (left) and gluon (right) jets. The vertical lines in the top row denote the minimum and maximum $r_g$ allowed for the specified $\xi = 10^{-2}$ value.}
\label{fig:vary}
\end{figure}

In \fig{vary} we show the estimate of perturbative uncertainty at NNLL through scale variations and nuisance parameters. In \app{prof} we summarize the implementation of profile scales and their variation discussed in Refs.~\cite{Pathak:2020iue,Hannesdottir:2022rsl}. The variation involves six parameters that test sensitivity of the cross section to a) a uniform up/down variation of all scales, b) trumpet variation of $\xi$- and $r_g$-dependent scales in the resummation region, c) the scale where the coupling is deemed non-perturbative and frozen to a constant value, and d) breaking of three canonical relations which are used to derive the jet and (hard-)collinear scales from the soft scales.

In addition to scale variations, we also show uncertainty induced from lack of knowledge of two-loop non-logarithmic terms in the max-$R_g$ and min-$R_g$ cross sections via parameters $a_{20}^{\rm max/min}$ in \eqs{Qplain1}{Cmin}. As discussed in Ref.~\cite{Pathak:2020iue} we limit these variations within their respective regime by multiplying them by the corresponding weight functions. We see that the $a_{20}^{\rm min}$ variation allows us to probe sensitivity to the order-dependent minimum $r_g$ end point of the cross section, and that this variation also is the dominant uncertainty in the min-$R_g$ regime. On the other hand, the variation due to $a_{20}^{\rm max}$ is subdominant compared to scale variations in the max-$R_g$ region. We note that both these variations vanish for $r_g = r_g^{\rm max}$ and they do not impact the single differential jet mass cross section, whereas the scale variations are correlated between both single and doubly differential cross sections at each value of $m_J$.

\subsection{Effect of two-loop pieces for NNLL resummation}
\label{sec:twoloop}

\begin{figure}[t]
\centering
\includegraphics[width=.48\linewidth]{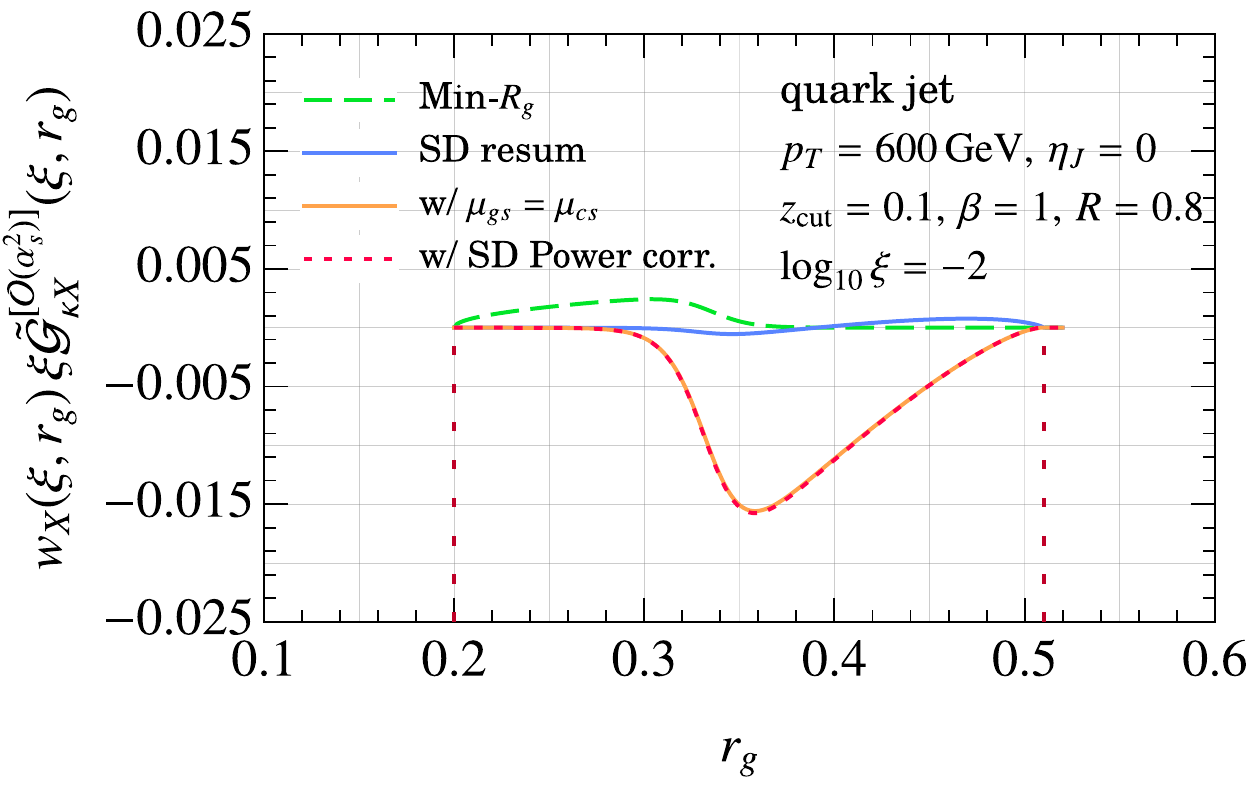}
\includegraphics[width=.48\linewidth]{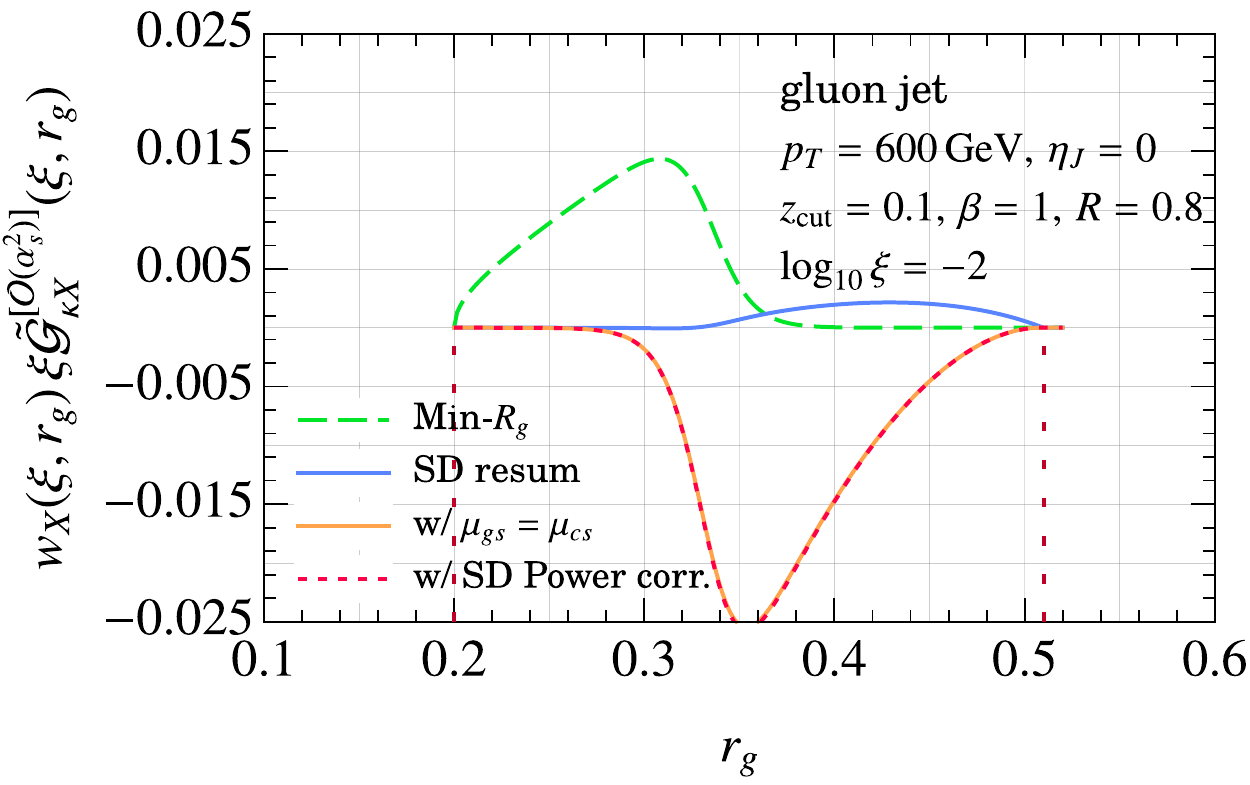}
\caption{Size of two-loop pieces of the resummed cross section required for NLL$'$ boundary condition for quark and gluon jets. Here we have multiplied each piece in a given regime by corresponding weight function.}
\label{fig:nllp}
\end{figure}

Next, we inspect the effect of including $\cO(\as^2)$ cross-terms to implement the boundary condition for NNLL resummation. As noted above, there are two places where this is needed: firstly in the max-$R_g$ cross sections in \eqs{DeltaFactPlain}{GsdFactResum} and secondly in the collinear function in the min-$R_g$ regime as shown in \eq{Cmin}. Since these two loop pieces do not match straightforwardly as the one-loop pieces above in \eq{MatchDDiff} we include them in the cross section by multiplying these terms by the corresponding weight function. We show this in \fig{nllp} where we plot
\begin{align}
w_X (\xi, r_g) \, \xi\tilde \cG_{\kappa X}^{[\cO(\as^2)]} (\xi , r_g ) \equiv w_X (\xi , r_g) \Big[ \xi \tilde \cG_{\kappa X} (\xi , r_g ) - \xi\tilde \cG_{\kappa X}^{[\cO(\as)]} (\xi , r_g )\Big] \, ,
\end{align}
where $X$ refers to the regime considered and in the second term we truncate all the fixed-order terms to $\cO(\as)$.
In \fig{nllp} we show the size of the two-loop pieces in the min-$R_g$ cross section multiplied by $w_{\rm min}(\xi,r_g)$, and those
from each of the three terms in \eq{MatchMax} with a common weight factor $w_{\rm max}(\xi,r_g)$. Note that despite the fact that corrections to the $\tilde \cG_\kappa^{\rm plain}(\xi, r_g, \mu_{\rm plain} )$ are relatively large, after including the two loop cross term in \eq{Qplain2} it cancels almost entirely against the correction from $\tilde \cG_\kappa^{\rm sd\, res.}(\xi, r_g, \mu_{\rm plain} )$ and only those remaining in the $\tilde \cG^{\rm sd\,res.}_{\kappa}$ with soft drop resummation survive. These corrections are overall small and they turn out to be smaller than the perturbative uncertainty at NNLL from scale variations and nuisance parameters in \fig{vary}. We will investigate them again when considering $r_g$-moments of the doubly differential cross section.

\subsection{Non-singular corrections}
\label{sec:ns}

Finally, we can include the non-singular pieces in the resummed, matched cross section as
\begin{align}\label{eq:MatchDDiff1}
\tcG_\kappa (\xi, r_g) &\equiv \tcG_\kappa^{{\rm match}} (\xi,r_g ) + \Delta \tcG_\kappa^{\rm [n.s.]} \big(\xi, r_g, \alpha_s (\mu_N)\big)\, ,
\end{align}
where
\begin{align}
\Delta \tcG_\kappa^{\rm [n.s.]} \big(\xi, r_g, \alpha_s (\mu_N)\big) &\equiv \tcG_\kappa^{{\rm FO}} (\xi,r_g , \alpha_s(\mu_{N}) ) - \tcG_\kappa^{{\rm match}} (\xi,r_g ,\mu_i \ra \mu_{N} )
\end{align}
In $\tcG_\kappa^{{\rm match}} (\xi,r_g ,\mu_i \ra \mu_{N} )$ all the $\mu_i$ scales are set to the hard-collinear $\mu_N$ scale, turning off all resummation and non-global logarithms.
To see how the cancellation of singular pieces between the fixed order and the matched cross section takes place, we note that the measurement function in \eq{deltaFO} can be written as
\begin{align}
\xi \delta_\xi^{\rm FO} (x,y,r_g,\xi_0, \zeta ) &\simeq \xi \Big[ \delta_\xi^{\mcC}(x,y) + \delta_{\xi}^{\mCSm}(x,y,1) - \delta_{\xi}^{\mCSm}(x,y,r_g) \\
&\qquad + \Delta \delta_{\xi, r_g}^{\mCS, \, \rm full} (x,y,r_g, \xi_0, \zeta )+ \Delta \delta_{\xi,\,\rm sd}^{\mSp} (x,y,\xi_0, \zeta)
\Big] \, , \nn
\end{align}
where the approximate equality results from dropping power suppressed terms in the soft approximations on the right hand side. The measurement functions on the right hand side were defined above in Eqs.~(\ref{eq:deltacC}), (\ref{eq:deltaCSm}), (\ref{eq:DeltadeltaCS}) and (\ref{eq:deltaSD0}) respectively. It is not hard to see that setting all the scales in \eq{MatchDDiff} to $\mu_N$ will result in the above combination of factorization functions at one-loop.

\begin{figure}[t]
\centering
\includegraphics[width=.6\linewidth]{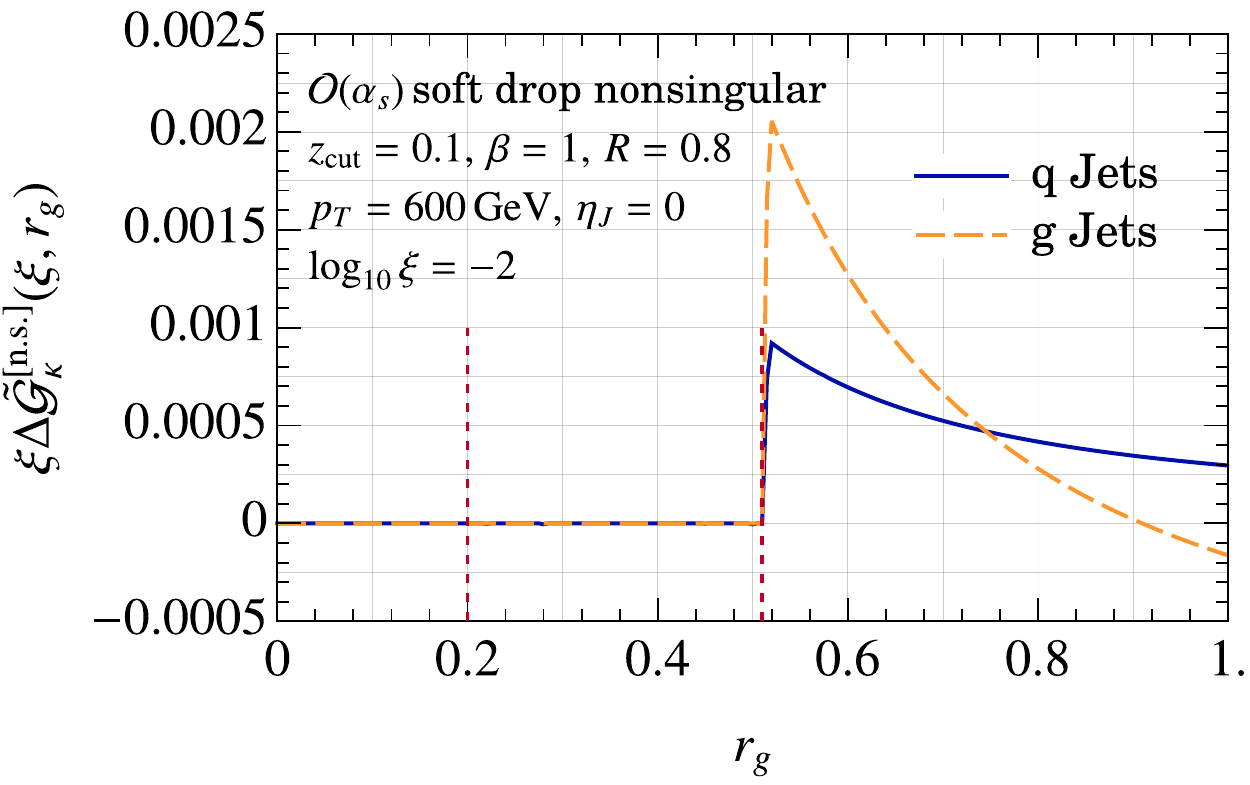}
\caption{Non-singular correction related to approximations in the soft drop constraint. The vertical lines denote the range of groomed jet radius allowed for the specified jet mass value of $\xi = 10^{-2}$. The non-singular correction related to soft drop is non-zero only beyond this range since it is where the soft drop constraint is a stronger condition than $r_g$-constraint.}
\label{fig:ns}
\end{figure}

In \fig{ns} we show the non-singular correction for quark and gluon jets. In fact, we find a complete cancellation between the fixed order and the matched cross-section for $r_g \leq r_g^{\rm max}$ defined in \eq{rglimits}. This is because we have already taken into account the mass-power corrections in the collinear function in \eq{CollFunc} and the remaining power corrections in \eq{MatchDDiff1} arise from only approximations made in the soft drop condition. For $r_g \leq r_g^{\rm max}$ the $r_g$-constraint is stronger than soft drop constraint. This also serves as a strong cross check of the numerical implementation.
On the other hand for $r_g > r_g^{\rm max}$, the resummed-cross section saturates to single differential jet mass cross section governed by the soft drop constraint. It is here there is a non-trivial non-singular correction in \eq{MatchDDiff1}. However, as can be seen in \fig{ns}, this correction is numerically negligible. It is of the same order as the non-singular correction in single differential jet mass which was also shown to be numerically negligible in \Refcite{Hannesdottir:2022rsl}. Hence, we will ignore the non-singular corrections from soft drop constraints entirely in our numerical analysis.

\subsection{$R_g$-weighted jet mass cross section}
\label{sec:c1}
\begin{figure}[t]
\centering
\includegraphics[width=.48\linewidth]{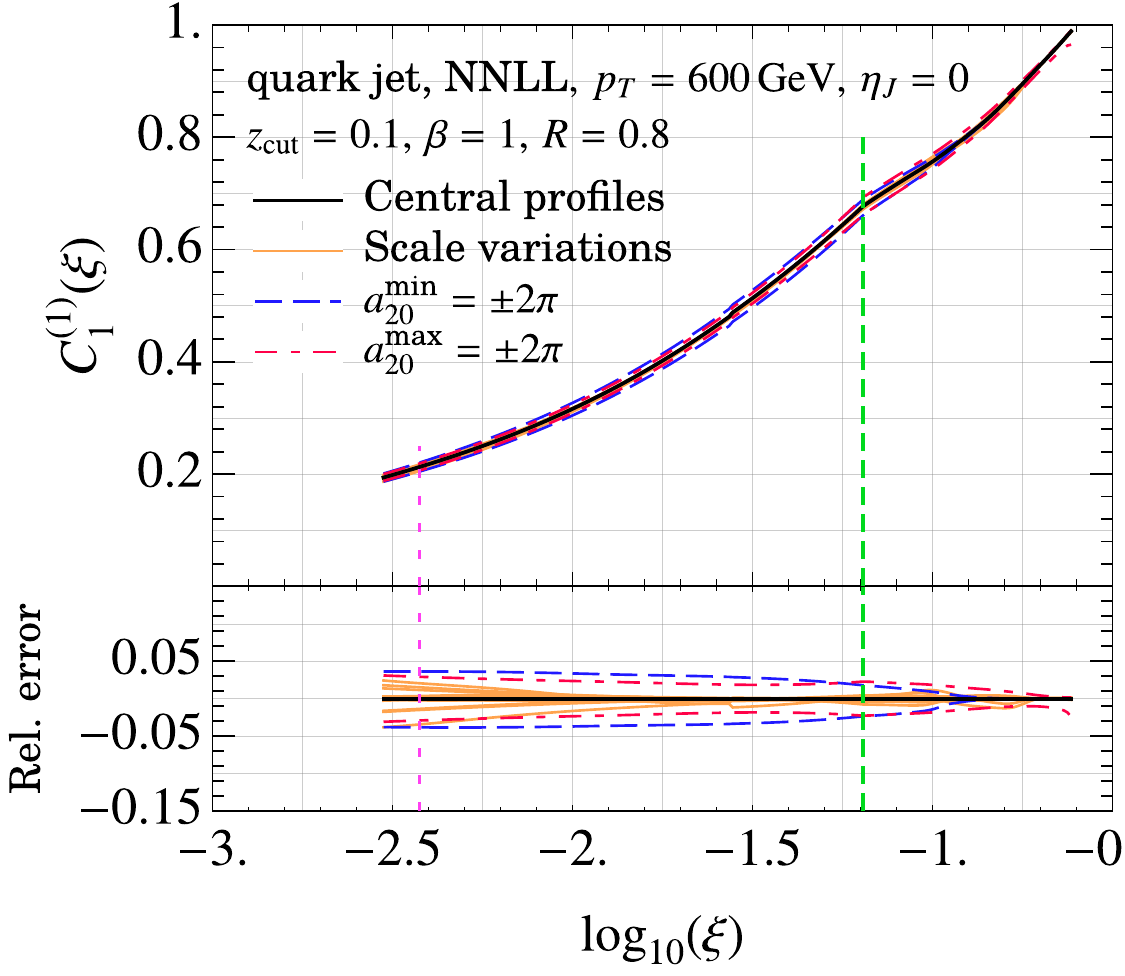}
\includegraphics[width=.48\linewidth]{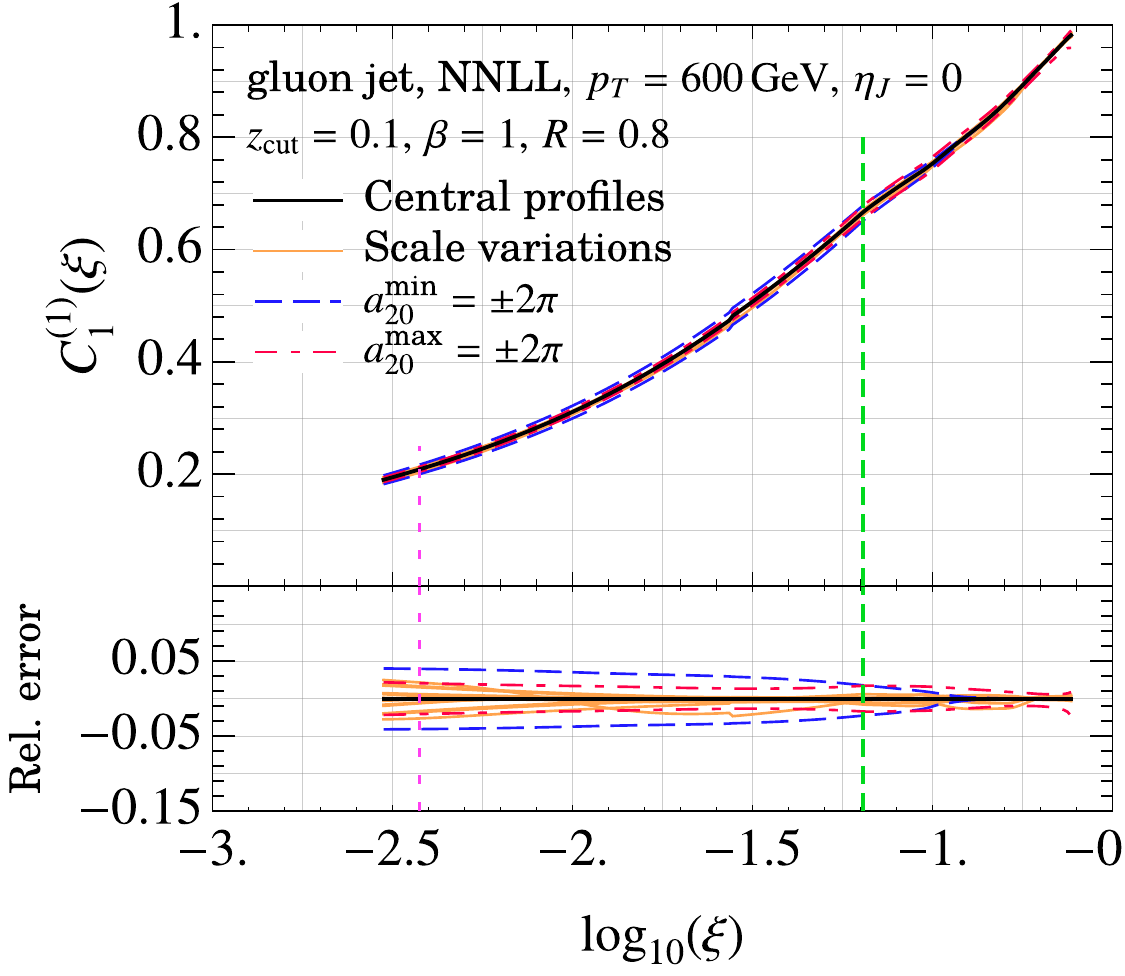}
\\
\includegraphics[width=.48\linewidth]{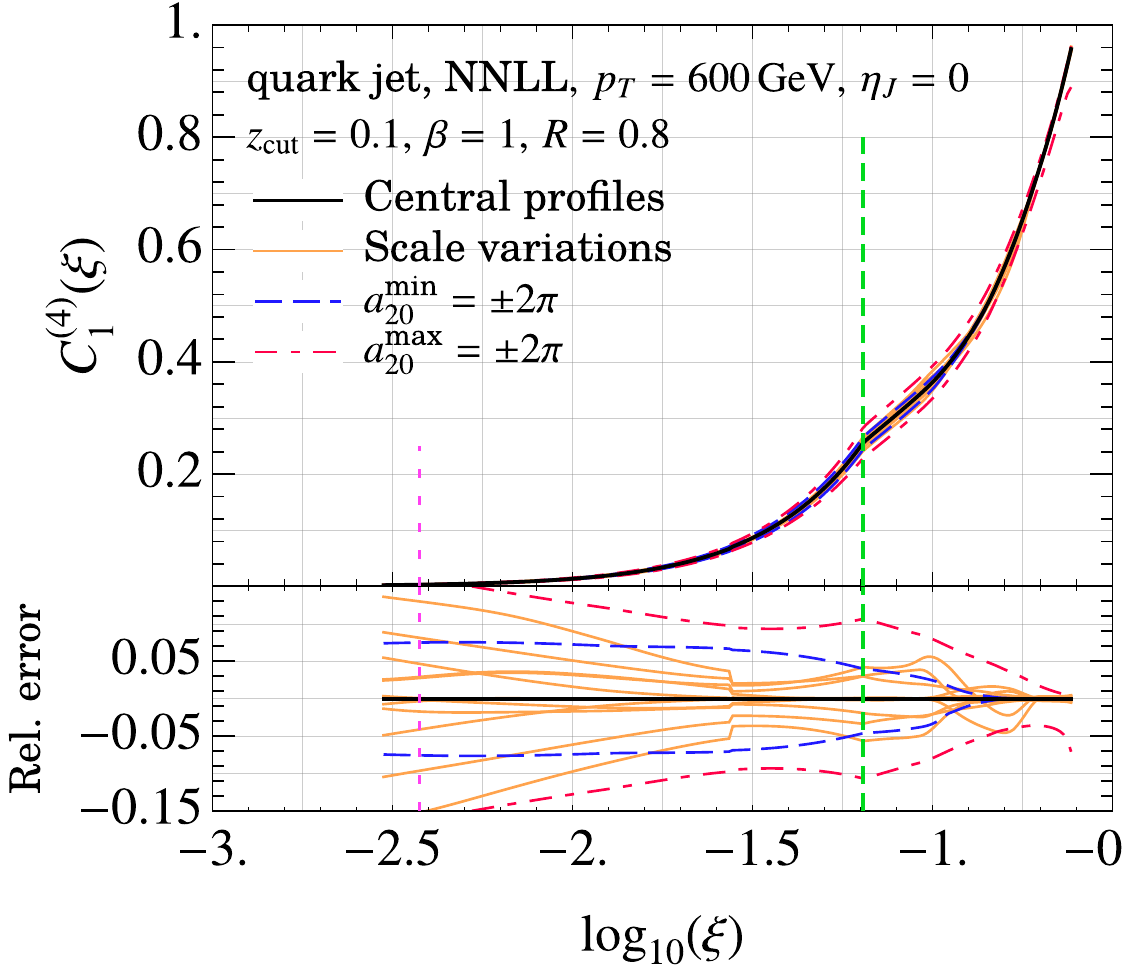}
\includegraphics[width=.48\linewidth]{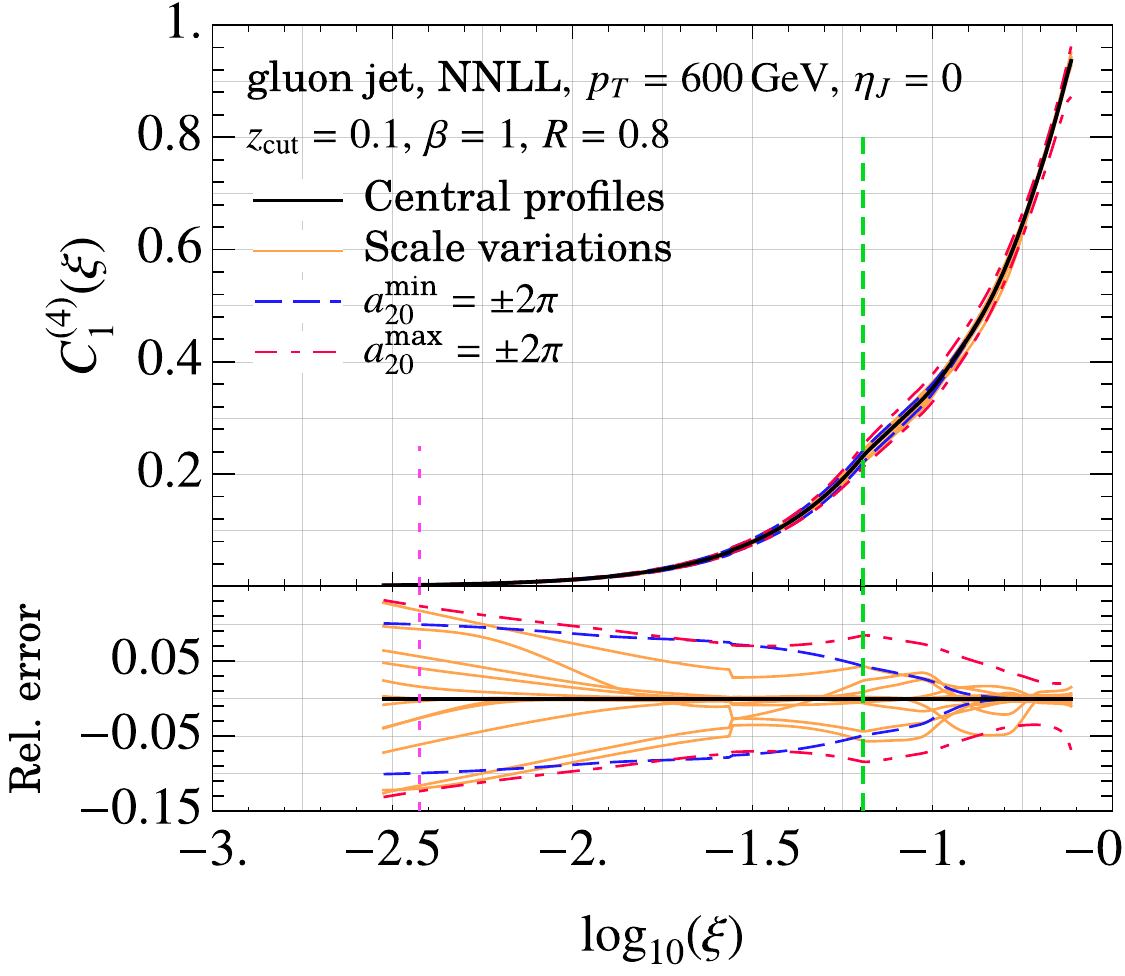}
\caption{$C_{1\kappa}^{(n)}$ for $n = 1$ (top) and $n =4$ (bottom) for quark and gluon jets. The vertical lines denote the extent of the SDOE region.}
\label{fig:c1}
\end{figure}

Having set up the matched cross section we can compute the $r_g$-weighted cross section differential in jet mass that are relevant phenomenologically as perturbative weights of hadronization and underlying event corrections. To facilitate comparison with Monte Carlo event generators we will normalize the weighted cross section with the usual jet mass cross section. To this end we define
\begin{align}
C_{1\kappa}^{(n)}(\xi) \frac{1}{\hat \sigma_\kappa}\frac{\df \hat \sigma_\kappa}{\df \xi} \equiv \int_0^1 \df r_g \: r_g^n \frac{1}{\hat \sigma_\kappa} \frac{\df^2 \hat \sigma_\kappa }{\df r_g \df \xi}
\end{align}
Since we have access to the cross section cumulative in $r_g$, we have
\begin{align}
&\frac{1}{\hat \sigma_\kappa} \frac{\df^2 \hat \sigma_\kappa }{\df r_g \df \xi} = \frac{\df \tcG_\kappa^{\rm match}(\xi,r_g)}{\df r_g} \, ,&
&\frac{1}{\hat \sigma_\kappa}\frac{\df \hat \sigma_\kappa}{\df \xi} = \tcG_\kappa^{\rm match} (\xi , 1) \equiv \tcG_\kappa^{\rm match}(\xi) \, .
\end{align}
Hence, the moment $C^{(n)}_1(\xi)$ becomes
\begin{align}
C_{1\kappa}^{(n)}(\xi)
&= \frac{1}{\tcG_\kappa^{\rm match}(\xi)}
\int_{r_g^{\rm min}(\xi)}^{r_g^{\rm max}(\xi)} \df r_g \:r_g^{n} \frac{\df}{\df r_g} \,\tcG_\kappa^{\rm match}(\xi, r_g)
\nn \\
&= \frac{1}{\tcG_\kappa^{\rm match}(\xi)}
\bigg (
\big[r_g^{\rm max}(\xi)\big]^n \tcG_\kappa^{\rm match}\big(\xi, r_g^{\rm max}(\xi)\big) - n \int_{r_g^{\rm min}(\xi)}^{r_g^{\rm max}(\xi)} \df r_g \:r_g^{n - 1} \,\tcG_\kappa^{\rm match}(\xi, r_g)
\bigg) \nn \\
&=
\big[r_g^{\rm max}(\xi)\big]^n
-\frac{n}{\tcG_\kappa^{\rm match}(\xi)}
\int_{r_g^{\rm min}(\xi)}^{r_g^{\rm max}(\xi)} \df r_g \:r_g^{n - 1} \,\tcG_\kappa^{\rm match}(\xi, r_g)
\, ,
\end{align}
where we have used the fact that the differential cross section only has support between $r_g^{\rm min}(\xi)$ and $r_g^{\rm max}(\xi)$.

In \fig{c1} we show $C_{1\kappa}^{(n)}$ for $n = 1,4$ cases, relevant for hadronization and underlying event corrections. Following the same color scheme as in \fig{vary} we find that the effect of scale variations on the normalized moment is very small. This is because the scale variation uncertainties are correlated between the doubly differential and the singly differential cross sections and cancel in the ratio. On the other hand, the dominant uncertainty in $C_{1\kappa}^{(n)}$ results from variation of the $a_{20}^{\rm min}$ and $a_{20}^{\rm max}$ nuisance parameters.

\begin{figure}[t]
\centering
\includegraphics[width=.48\linewidth]{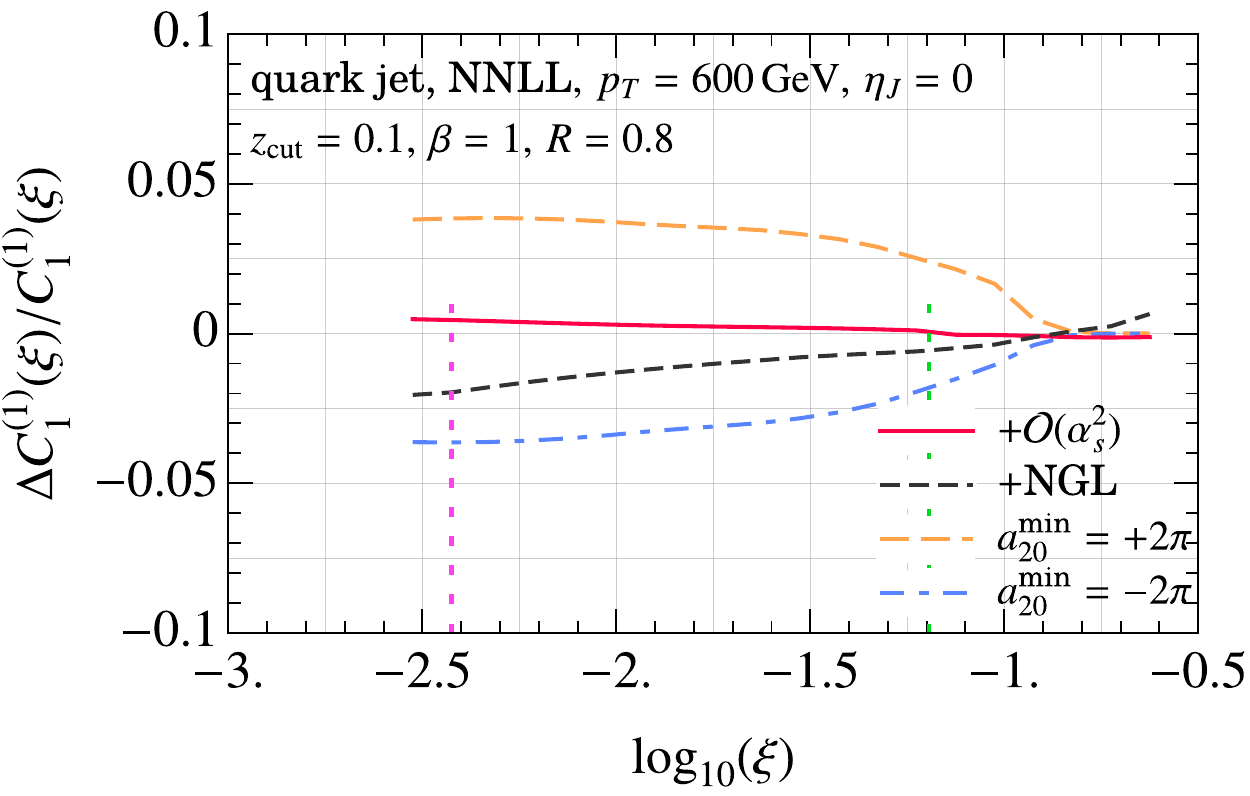}
\includegraphics[width=.48\linewidth]{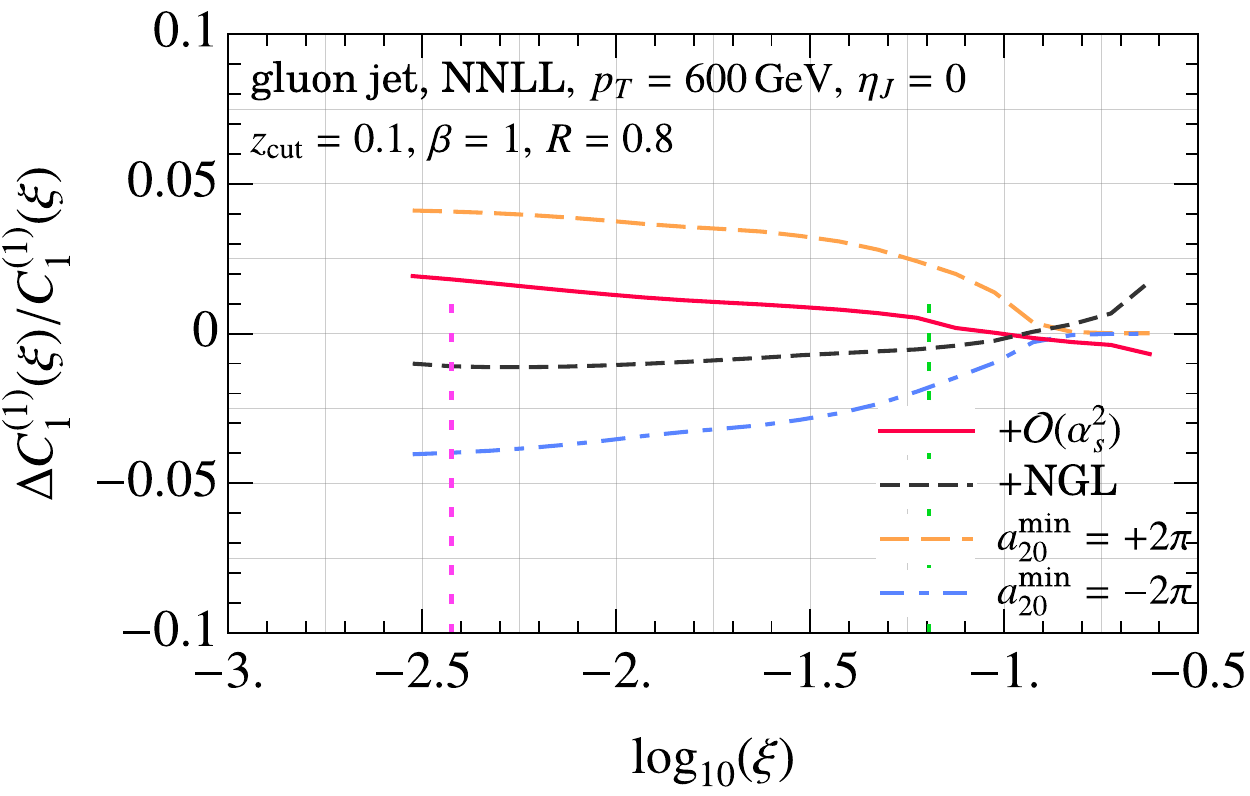}
\caption{Fractional deviation of $C_1^{(1)}(\xi)$ upon including $\cO(\as^2)$ pieces and non-global logarithms compared with that from varying the $a_{20}^{\rm min}$ nuisance parameter.}
\label{fig:c1nllp}
\end{figure}
In \fig{c1nllp} we inspect the effect of adding two-loop logarithmic pieces, that are required to implement consistently the NNLL boundary condition, and non-global logs at NLL (both separately) by considering the fractional deviation in the moment. We also show for comparison the fractional deviation from adding two loop non-logarithmic terms in the min-$R_g$ cross section by varying the nuisance parameter $a_{20}^{\rm min}$, which dominates both these effects. The $a_{20}^{\rm max}$ variation (not shown) in the ungroomed region will similarly dominate the effect of two-loop logarithmic terms. Hence, in our final numerical analysis we can safely ignore the $\cO(\as^2)$ logarithmic pieces and non-global logarithms.

\section{Soft drop boundary cross section}
\label{sec:bndry}

We now consider the computation of the soft drop boundary cross section stated in \eq{bndry}. After describing the measurement function in \secn{bndrydelta}, we discuss the computation in the plain jet mass and soft drop resummation regions in \secs{plainBndry}{sdBndry}. As we will see below, the one-loop fixed order results for soft functions are transformed into resummation kernels whose computation is discussed in \secn{kernel}. In \secn{foBndry} we discuss the fixed order non-singular corrections and show them to be small as seen above. In \secn{c2} we finally describe the computation of the $C_{2\kappa}^{(n)}$ moment.

\subsection{The measurement function}
\label{sec:bndrydelta}
We are specifically interested in the projection of the triply differential cross section on the boundary of the soft drop:
\begin{align}
\frac{1}{\hat \sigma_\kappa}\frac{\df \hat\sigma_\kappa^{\bndry}}{\df r_g \df \xi} \equiv \int \df z_g \:
\delta (z_g - \xi_0 r_g^\beta)\,\frac{1}{\hat\sigma_\kappa}\frac{\df \hat\sigma_\kappa}{\df z_g \df r_g \df \xi }
\end{align}
However, instead of directly computing the triply-differential cross section we can equivalently calculate the doubly differential cross section while including a soft drop condition small shift in the soft drop condition:
\begin{align}\label{eq:SDExp}
\hat \Theta_{\rm sd} (x, y, \xi_0,\zeta,\veps)
&= \Theta \bigg(\frac{\min \{x_1 ,x_2\}}{x_1 + x_2} - \xi_0 \Big(\frac{y}{4x_1x_2}\Big)^\frac{\beta}{2} + \veps \bigg) \nn \\
&= \hat \Theta_{\rm sd} (x,y,\xi_0, \zeta) + \veps \, \delta \bigg(\frac{\min \{x_1 ,x_2\}}{x_1 + x_2} - \xi_0 \Big(\frac{y}{4x_1x_2}\Big)^\frac{\beta}{2} \bigg) + \ldots
\, ,
\end{align}
and differentiate it with respect to $\veps$:
\begin{align}
\frac{1}{\hat \sigma_\kappa}\frac{\df \hat\sigma_\kappa^{\bndry}}{\df r_g \df \xi} =
\frac{\df}{\df \veps} \bigg( \frac{1}{\hat \sigma_\kappa} \frac{\df^2 \hat \sigma_{\kappa,\veps}}{\df r_g \df \xi}\bigg) \bigg|_{\veps \ra 0}\,,
\end{align}
where $\df \sigma_{\kappa,\veps}$ is calculated with the shifted soft drop condition.

Additionally, as explained in \Refcite{Pathak:2020iue}, we only need to consider the max-$R_g$ regime which captures the correct two-pronged geometry, whereas the contributions from the intermediate and min-$R_g$ enter beyond the LL nonperturbative factorization in \eq{NP}. In the discussion above for doubly differential cross section we did not drop contributions from these regions because effects of hadronization (and of ISR and underlying event) involve positive powers of $r_g$-moments, and hence they are naturally suppressed. Whereas for the boundary cross section we require $1/r_g$ moment, which receives large power corrections from the intermediate and min-$R_g$ region, which must be suppressed by hand when calculating using shifted soft drop condition.

As above, in practice, we will find it easier to compute first the soft matrix elements, and use them in the factorization formulae as well as subtractions for fixed order pieces.
In the soft limit, the shifted soft drop condition reads
\begin{align}\label{eq:ThetaSDPlainEps}
\hat \Theta_{\rm sd}^{(s)} (x, y, \xi_0,\zeta,\veps)
&= \Theta \Bigg(
y\zeta^2 + x - \xi_0^{\frac{2}{2+\beta}} y^{\frac{\beta}{2+\beta}} + \frac{2\veps}{2+\beta}
\Big(\frac{y}{\xi_0}\Big)^{\frac{\beta}{2+\beta}}
\Big(\zeta^2 + \frac{x}{y}\Big)^{\frac{\beta}{2}} + \ldots
\Bigg) \nn\\
&= \hat \Theta_{\rm sd}^{(s)} (x, y, \xi_0,\zeta) + \frac{2\veps}{2+\beta} \: \delta \Big(y \zeta^2 + x - \big( \xi_0\big)^{\frac{2}{2+\beta}} (y)^{\frac{\beta}{2+\beta}}\Big) + \ldots \, .
\end{align}
The same in collinear-soft limit is given by
\begin{align}
\hat \Theta_{\rm sd}^{(cs)} (x,y,\xi_0,\veps) &=
\Theta \Bigg(
x - \xi_0^{\frac{2}{2+\beta}} y^{\frac{\beta}{2+\beta}} + \frac{2\veps}{2+\beta}
\Big(\frac{y}{\xi_0}\Big)^{\frac{\beta}{2+\beta}}
\Big(\frac{x}{y}\Big)^{\frac{\beta}{2}} + \ldots
\Bigg) \nn\\
&= \hat \Theta_{\rm sd}^{(cs)} (x,y,\xi_0) +
\frac{2\veps}{2+\beta} \: \delta
\big( x - \xi_0^{\frac{2}{2+\beta}} y^{\frac{\beta}{2+\beta}}\big)+ \ldots \, .
\end{align}
Thus, we define:
\begin{align}
\hat \delta_{\rm sd}^{(s)} (x,y,\xi_0,\zeta) &\equiv \frac{2}{2+\beta} \: \delta \Big(y \zeta^2 + x - \big( \xi_0\big)^{\frac{2}{2+\beta}} (y)^{\frac{\beta}{2+\beta}}\Big) \,, \\
\hat \delta_{\rm sd}^{(cs)} (x,y,\xi_0) &\equiv \frac{2}{2+\beta} \: \delta
\big( x - \xi_0^{\frac{2}{2+\beta}} y^{\frac{\beta}{2+\beta}}\big)\nn \, .
\end{align}

\subsection{Plain jet mass region}
\label{sec:plainBndry}
We first consider the plain jet mass region. Here the modification to the soft drop condition only affects the $\mSp$ mode. Thus, the measurement function analogous to \eq{deltaSplain} is given by
\begin{align}\label{eq:deltaSplaineps}
\delta_{\xi,\veps}^{\mSp} (x,y,r_g,\xi_0,\zeta)\equiv \hat \Theta^{(s)}_{k_T}(x,y) \hat \delta_{\rm sd}^{(s)}(x,y,\xi_0,\zeta) \big[ \hat \Theta_{r_g}^{(s)} (x,y,r_g)\delta (\xi - y) - \delta (x)\big] \, ,
\end{align}
where here, and everywhere below, we will include a subscript $\veps$ to distinguish functions relevant to the soft drop boundary cross section.
Similar to \eq{deltaSplainSplit} we can isolate the piece corresponding to $r_g$ measurement:
\begin{align}\label{eq:deltaSplainSpliteps}
\delta_{\xi,\veps}^{\mSp} (x,y,r_g,\xi_0,\zeta) = \Delta \delta_{\xi,\, r_g,\,\veps}^{\mCS,\, \rm full} (x,y,r_g,\xi_0,\zeta) + \Delta \delta_{\xi,\, \rm sd,\,\veps}^{\mSp} (x,y,\xi_0,\zeta) \, .
\end{align}
Unlike \eq{deltaSplainSplit}, there is no term corresponding to ungroomed measurement as we have differentiated with respect to $\veps$. These terms are given by
\begin{align}\label{eq:deltaEps}
\Delta \delta_{\xi,\, r_g,\,\veps}^{\mCS,\, \rm full} (x,y,r_g,\xi_0,\zeta) &\equiv - \hat \Theta_{k_T}^{(s)} (x,y)\big(1- \hat \Theta_{r_g}^{(s)}(x,y,r_g) \big) \hat \delta_{\rm sd}^{(s)} (x,y,\xi_0,\zeta)\delta (\xi - y) \, , \\
\Delta \delta_{\xi,\, \rm sd,\,\veps}^{\mSp} (x,y,\xi_0,\zeta) &\equiv +\hat \Theta_{k_T}^{(s)}(x,y) \hat \delta_{\rm sd}^{(s)} (x,y,\xi_0,\zeta) \big[\delta(\xi - y) - \delta(\xi)\big] \, . \nn
\end{align}

The fixed order correction that captures the dependence on the $r_g$ measurement is given by
\begin{align}\label{eq:DeltaScgeps}
\qcut^{\frac{1}{1+\beta}} \Delta S_{r_g,\,\veps}^\kappa &\bigg( \ell^+ \qcut^{\frac{1}{1+\beta}} , r_g ,\qcut, \zeta, \beta, \as(\mu) \bigg)
= \frac{1}{Q } \cS_\kappa^{[1]} \bigg(\xi = \frac{\ell^+}{Q} , \Delta \delta_{\xi,\, r_g,\,\veps}^{\mCS,\, \rm full}, \mu \bigg) \\
& \quad=
\frac{\alpha_s C_\kappa}{\pi} \frac{\Theta (\ell^+ -\qcut' v(r_g))\Theta (\qcut' - \ell^+) }{ (\ell^+)^2/Q }
\bigg( \Big(\frac{\qcut}{\ell^+}\Big)^\frac{2}{2+\beta} - \zeta^2\bigg)^{-1}
\, , \nn
\end{align}
and analogously, the soft drop related fixed-order piece is given by
\begin{align}\label{eq:Splain1eps}
&\Delta S_{\rm sd,\,\veps}^\kappa \big(\ell^+, \qcut, \zeta, \beta, \as(\mu) \big) = \frac{1}{Q} \cS_{\kappa}^{[1]} \bigg(\xi = \frac{\ell^+}{Q}, \Delta \delta_{\xi,\, \rm sd,\, \veps}^{\mSp}, \mu\bigg)\\
&\qquad =
\frac{2}{2+\beta}\frac{\alpha_s (\mu)C_\kappa}{\pi} \Biggl[\frac{\Theta(\ell^+) \Theta (\qcut' - \ell^+)}{(\ell^+)^2/Q} \bigg( \Big(\frac{\qcut}{\ell^+}\Big)^\frac{2}{2+\beta} - \zeta^2\bigg) ^{-1} \Biggr]^{[\qcut']}_+ \, . \nn
\end{align}

These fixed order corrections are combined with the rest of the factorized cross section in precisely the same manner as in \eq{DeltaFactPlainX} above
\begin{align}\label{eq:DeltaFactPlain}
&\tilde \cG^{\rm plain}_{\kappa,\,\veps} (\xi , Q, \qcut, \zeta, \mu_{\rm plain} ) = N^\kappa_{\rm incl}( Q, \mu_N)
e^{ K_N }\Big(\frac{\mu_N}{Q}\Big)^{\omega_N} \\
&\quad \times
\Bigg( \cS_{\rm NGL}^\kappa \big( t [Q \xi, Q] \big) \cJ^{\rm plain}_\kappa[\partial_\Omega ; \xi, Q, \mu_{\rm plain}]
\cQ^{\rm plain}_{\kappa,\veps} (\Omega,\as(\mu_s))
\nn \\
& \quad \qquad + \Big(\frac{\df}{\df \ln \xi} \cS_{\rm NGL}^\kappa \big( t [Q \xi, Q] \big) \Big)\cJ^{\rm plain}_\kappa [\partial_\Omega ; \xi, Q, \mu_{\rm plain}]
\cQ^{\rm plain}_{\kappa,\veps} (\Omega - 1,\as(\mu_s))
\Bigg)\bigg|_{\Omega = \tilde \omega(\mu_{s}, \mu_J)}
\,.\nn
\end{align}

Similar to \eq{Qplain}, the kernels $\cQ_{\kappa,\veps}^{\rm plain}$ include Laplace transforms of the soft functions in \eqs{DeltaScgeps}{Splain1eps}, but also $\cO(\as^2)$ cross terms that are required to consistently carry out NNLL resummation~\cite{Pathak:2020iue}.
In \tab{KernelsEps} we summarize the various $\cO(\as^2)$ cross terms that constitute the kernel $\cQ^{\rm plain}_{\kappa,\veps}$.
We have included cross terms involving $\Delta S_{\rm sd}^{\kappa}$ pieces for consistent matching with the cross section in the soft drop resummation regime.
Finally, there are also cross terms arising from $\cO(\as)$ parts of $\cJ^{\rm plain}_\kappa$ and $\cO(\as)$ soft functions that we have not shown.

{\renewcommand{\arraystretch}{1.6}
\begin{table}[t!]
\begin{center}
\scalebox{1}{\begin{tabular}{ c | c | c }
\hline
\hline
& $\cO(\alpha_s)$ & $\cO(\alpha_s^2)$
\\[0.5ex]
\hline
\hline
Ungroomed region & $\partial_{r_g} \Delta S_{r_g,\,\veps}^{\kappa[1]}$
& $\Delta S_{\rm sd}^{\kappa[1]}\partial_{r_g} \Delta S_{r_g,\,\veps}^{\kappa[1]}$, \: $ \Delta S_{\rm sd,\,\veps}^{\kappa[1]}\partial_{r_g}\Delta S_{r_g}^{\kappa[1]}$, \: $\partial_{r_g}\big[\Delta S_{r_g}^{\kappa[1]} \Delta S_{r_g,\,\veps}^{\kappa[1]}\big]$
\\[0.5ex] \hline
SD res., $\beta = 0$ & $\partial_{r_g} \Delta S_{r_g,\,\veps}^{\kappa[1]}$
& $\cJ_{\kappa,\,\veps}^{\rm sd\,res} [\partial_\Omega ] \,\partial_{r_g} \Delta S_{r_g}^{\kappa[1]}$, \:$\partial_{r_g}\big[\Delta S_{r_g}^{\kappa[1]} \Delta S_{r_g,\,\veps}^{\kappa[1]}\big]$
\\[0.5ex] \hline
SD res., $\beta > 0$ & $\partial_{r_g} \Delta S_{r_g,\,\veps}^{\kappa[1]}$
& $\Delta S_{\rm sd,\,\veps}^{\kappa[1]} \partial_{r_g}\Delta S_{r_g}^{\kappa[1]} $, \: $\partial_{r_g}\big[\Delta S_{r_g}^{\kappa[1]} \Delta S_{r_g,\,\veps}^{\kappa[1]}\big]$
\\[0.5ex] \hline\hline
\end{tabular}}
\end{center}
\caption{Fixed order corrections to soft functions that appear in soft drop boundary resummation kernels $\cQ_{\kappa,\, \veps}^{\rm plain}$ and $\cQ_{\kappa,\veps}^{\rm sd\,res.}$ for plain and soft drop resummation regions respectively. We have not shown the additional $\cO(\as^2)$ terms that result from action of $\cJ^{\rm plain}_\kappa[\partial_\Omega]$ and $\cJ_\kappa^{\rm sd\,res.}[\partial_\Omega]$ on $\cO(\as)$ pieces}
\label{tab:KernelsEps}
\end{table}}

\subsection{Soft drop resummation region}
\label{sec:sdBndry}
We now turn to the boundary cross section in the soft drop resummation region. Here, including shift in the boundary cross section modifies the measurement for the global soft and collinear-soft functions:
\begin{align}\label{eq:deltaSDeps}
\hat \Theta^{\mSG}_{\veps} (x,y,\xi_0,\zeta) &\equiv -\hat\Theta_{k_T}^{(s)} (x,y) \hat \delta_{\rm sd}^{(s)} (x,y,\xi_0,\zeta) + \hat \delta_{\rm sd}^{(s)} (x,y,\xi_0,\zeta) \, , \\
\delta_{\xi,\,\veps}^{\mCS} (x,y,r_g,\xi_0) &\equiv \hat \delta_{\rm sd}^{(cs)} (x,y,\xi_0) \big[ \delta (\xi - y) - \delta (\xi)\big] + \Delta \delta_{\xi,\,r_g,\,\veps}^{\mCS} (x,y,r_g,\xi_0)\, , \nn
\end{align}
The corresponding soft functions are given by
\begin{align}
S_{G,\veps}^{\kappa} \big(\qcut, \zeta, \beta, \mu \big) &\equiv \cS_\kappa^{[1]} \big(\cdot \, , \hat \Theta^{\mSG}_{\veps} , \mu \big) + \cO(\as^2)\, , \\
S_{c,\veps}^\kappa \big(\tilde k , r_g,\qcut, \beta, \mu \big) &\equiv \frac{1}{Q \qcut^{\frac{1}{1+\beta}}} \cS_\kappa^{[1]} \bigg(\xi = \frac{\tilde k}{Q \qcut^{\frac{1}{1+\beta}}} , \delta_{\xi,\,\veps}^{\mCS}, \mu \bigg) + \cO(\as^2)\, .\nn
\end{align}
Here, the second term in \eq{deltaSDeps} in the global soft measurement function $\hat \Theta^{\mSG}_{\veps}$ is required to appropriately cancel the UV divergences in of the virtual piece. In the collinear-soft measurement function, the measurement $\Delta \delta_{\xi,\,r_g,\,\veps}^{\mCS}$ is simply the collinear limit of the one in \eq{deltaEps}, which yields
\begin{align}
S_{c,\veps}^\kappa \big(\tilde k , r_g, \qcut, \beta, \mu \big) = S_{c,\veps}^\kappa \big(\tilde k , \beta, \mu\big) + \Delta S_{r_g,\,\veps}^\kappa \big(\tilde k , r_g ,\qcut,\beta, \as(\mu) \big) \, ,
\end{align}
However, as we did above for the usual soft drop case, we will continue to include the power suppressed terms and employ the full measurement in \eq{deltaEps}.

It is helpful to consider the $\beta = 0$ and $\beta > 0$ cases separately. For $\beta = 0$, the soft functions are given by
\begin{align}\label{eq:SGScb0}
S_{G,\veps}^{\kappa,\rm bare}\big(\qcut,\zeta, \beta = 0 , \mu\big) &=
\frac{1}{\xi_0}\frac{\alpha_s C_\kappa}{\pi} \bigg(\frac{1}{\eps} + 2 \ln \Big(\frac{\mu}{\qcut}\Big) \bigg) \, , \\
S_{c,\veps}^{\kappa,\rm bare} \big(\tilde k,\beta = 0, \mu \big)
&= \frac{\alpha_s C_\kappa}{\pi} \frac{1}{\xi_0}\biggl[
-\frac{\delta(\tilde k)}{\veps}
+
\cL_0 (\tilde k , \mu^2 )
\biggr] \, , \nn
\end{align}
and the same $r_g$-dependent correction as in \eq{Splain1eps}. Here we find an extra log-divergence~\cite{Pathak:2020iue} which is unrelated to the soft drop factorization we considered above.
The extra divergence leads to a nontrivial non-cusp anomalous dimension $\veps \: \gamma_0^{\veps}(\zcut)$ where
\begin{align}\label{eq:gamma0eps}
\gamma_{0}^{\veps, S_G^\kappa} = - \gamma_0^{\veps, S_c^\kappa} \equiv \gamma_0^\veps(\zcut) = \frac{8 C_\kappa}{\zcut} \,, \qquad (\beta = 0) \, .
\end{align}

We can verify that for $\mu_{cs} = \mu_{gs} = \mu$ the fixed order pieces add to yield the correct fixed order result in \eq{Splain1eps} in the limit $\xi \ll \xi_0$ by noting that
\begin{align}
\qcut^{\frac{1}{1+\beta}} S_{c,\veps}^\kappa \big(\qcut^{\frac{1}{1+\beta}} \ell^+, \beta = 0, \mu\big) &= \frac{1}{\xi_0}\frac{\as C_\kappa}{\pi} \qcut\cL_0\big(\qcut \ell^+, \mu^2) \nn \\
&= \frac{1}{\xi_0}\frac{\as C_\kappa}{\pi} \bigg[\cL_0 \big(\ell^+, \mu\big) - \ln\Big(\frac{\mu}{\qcut}\Big) \delta (\ell^+)\bigg]
\end{align}
adding to this the contribution from global soft function, we find
\begin{align}
&\delta (\ell^+) S_{G,\veps}^\kappa \big(\qcut, \zeta, \beta = 0 ,\mu \big) + \qcut^{\frac{1}{1+\beta}} S_{c,\veps}^\kappa \big(\qcut^{\frac{1}{1+\beta}} \ell^+, \beta = 0, \mu\big)
\\
&\qquad = \frac{1}{\xi_0}\frac{\as C_\kappa}{\pi} \bigg[\cL_0 \big(\ell^+, \mu\big) + \ln\Big(\frac{\mu}{\qcut}\Big) \delta (\ell^+)\bigg]
\nn
\\
&\qquad = \frac{1}{\xi_0}\frac{\as C_\kappa}{\pi} \bigg[\frac{\Theta(\ell^+)}{\ell^+}\bigg]_+^{[\qcut]}
\nn \, .
\end{align}
which is precisely the limit $\ell^+/\qcut \ll \zeta$ of $\Delta S_{\rm sd,\veps}^\kappa$ above for $\beta = 0$.

On the other hand, for $\beta > 0$, there is no straightforward way to split $\Delta S_{\rm sd,\veps}^\kappa$ in \eq{Splain1eps} into two pieces as the divergence is a power law and we simply include the piece as in the case of plain jet mass region. Thus, the resummed formula in soft drop resummation region is given by
\begin{align}\label{eq:GsdFactResumeps}
\tilde \cG^{\rm sd\, res}_{\kappa,\,\veps}& (\xi, r_g, Q, \qcut, \zeta, \beta, \mu_{\rm sd})= N_\kappa^{\rm evol} \big(\mu_N, \mu_{gs}, Q, \qcut, \zeta, \beta\big)
\cS_{\rm NGL}^\kappa\big(t \big[\qcut , Q\big] \big)\\
& \times
\cJ_{\kappa,\,\veps}^{\rm sd\,res.} [\partial_\Omega ; \xi, Q, \qcut, \mu_{\rm sd}] \: \cQ_{\kappa,\, \veps}^{\rm sd\, res.} \big(\Omega, \xi, r_g, \qcut, \zeta, \beta, \as(\mu_{cs})\big)
\Big|_{\Omega = \tilde \omega(\mu_{cs}, \mu_J)} \nn
\, .
\end{align}

\begin{align}\label{eq:SDepsOperator}
\cJ_{\kappa,\,\veps}^{\rm sd\,res.} [\partial_\Omega ] &= \bigg ( 1
+ \delta_{\beta,0} \eta \big(\gamma_0^\veps(\zcut) , \mu_{cs}, \mu_{gs} \big) \bigg) \cJ_{\kappa}^{\rm sd\,res} [\partial_\Omega ] \\
&\quad+ \frac{\delta_{\beta,0}}{\xi\xi_0} \:
e^{ K_{cs} + K_{J} }
\frac{\big(\mu_J^2\big)^{ \omega_J} \big(Q \mu_{cs}\big)^{\omega_{cs} }}{(\xi Q^2)^\Omega}
\Big(\frac{\mu_{cs}}{\qcut}\Big)^\frac{\omega_{cs}}{1+\beta}
\nn \\
&\quad\quad \times \Bigg[
\frac{2\alpha_s(\mu_{gs})C_\kappa}{\pi} \ln \Big(\frac{\mu_{gs}}{\qcut }\Big)
- \frac{\alpha_s(\mu_{cs})C_\kappa}{\pi} \:\Bigg( \partial_\Omega + \log\bigg(\frac{ \mu_{cs}}{Q \xi}\Big(\frac{\mu_{cs}}{\qcut}\Big)^\frac{1}{1+\beta}\bigg) \Bigg)
\Bigg]
\, ,\nn
\end{align}
Here $\cJ_{\kappa}^{\rm sd\,res} [\partial_\Omega ] $ is the same function that appeared above for the usual soft drop cross section in \eq{SDOperator}. The additional $\cO(\veps)$ terms arise only for $\beta = 0$ case. The first of these involves $\eta(\gamma_0^\veps(\zcut),\mu_{cs},\mu_{gs})$ which is the single log non-cusp resummation kernel arising from the new non-cusp anomlous dimension in \eq{gamma0eps}:
\begin{align}
\eta\big(\gamma_0^\veps(\zcut),\mu_{cs},\mu_{gs}\big) = - \frac{\gamma_0^\veps(\zcut)}{2 \beta_0} \ln \Big(\frac{\alpha_s(\mu_{cs})}{\alpha_s(\mu_{gs})}\Big)
\, .
\end{align}
In the third line of \eq{SDepsOperator} we have included the $\cO(\as)$ logarithms associated with the factorized $\beta = 0$ boundary global-soft and collinear-soft functions in \eq{SGScb0}. For this reason these pieces are not included in the $\cQ_{\kappa,\veps}^{\rm sd\,res.}$ kernel for $\beta = 0$. For $\beta > 0$, the non-factorized $\Delta S_{\rm sd,\veps}^{\kappa}$ contribution is included directly in $\cQ_{\kappa,\veps}^{\rm sd\, res.}$ as indicated in \tab{KernelsEps}.
\subsection{Computing the resummation kernels}
\label{sec:kernel}
We now turn to the computation of the resummation kernels that appear due to non-logarithmic functions in the cross sections summarized in \tab{KernelsEps}.
For compactness, we write all the relevant soft functions as
\begin{align}\label{eq:GijDef}
&a_\kappa \cS_\kappa^{({\rm sd})} (\xi)
\equiv Q \Delta S_{\rm sd}^{\kappa[1]} \big(Q \xi, \as (\mu)\big)
\, ,&
&a_\kappa\cS^{({r_g})}_\kappa(\xi)
\equiv Q\qcut^{\frac{1}{1+\beta}}\Delta S_{r_g}^{\kappa[1]} \Big(\xi Q\qcut^{\frac{1}{1+\beta}}, r_g , \as(\mu)\Big)
\, ,& \nn \\
&a_\kappa\cS_{\kappa, \veps}^{({\rm sd})} (\xi)
\equiv Q \Delta S_{\rm sd,\,\veps}^{\kappa[1]} \big(Q \xi, \as (\mu)\big)\, ,&
&a_\kappa\cS^{({r_g})}_{\kappa,\veps}(\xi)
\equiv
Q\qcut^{\frac{1}{1+\beta}}\Delta S_{r_g,\,\veps}^{\kappa[1]} \Big(\xi Q\qcut^{\frac{1}{1+\beta}}, r_g , \as(\mu)\Big) \, .&
\end{align}
where we have defined $a_\kappa \equiv \alpha_s(\mu)C_\kappa/\pi$, and have left the $r_g$-dependence of $\cS_{\kappa}^{(r_g)}$ and $\cS_{\kappa,\veps}^{(r_g)}$ implicit. Next, since we do not intend to match the max-$R_g$ cross section with other regimes for boundary cross section, we can directly consider the $r_g$-derivative. This will help us avoid considering some of the more complicated cross terms not shown in \tab{KernelsEps}.
The $r_g$-derivatives of the soft functions are given by
\begin{align}\label{eq:DelPsiG1}
\qcut^{\frac{1}{1+\beta}}\partial_{r_g} \Delta S_{r_g}^{\kappa[1]} \Big(\ell^+\qcut^{\frac{1}{1+\beta}}, r_g , \as(\mu)\Big)
&= \frac{2\alpha_s C_\kappa}{\pi} \frac{\Theta (\ell^+ -\qcut' v(r_g))}{ \ell^+ } \, ,
\\
\qcut^{\frac{1}{1+\beta}}\partial_{r_g} \Delta S_{r_g,\,\veps}^{\kappa[1]} \Big(\ell^+\qcut^{\frac{1}{1+\beta}}, r_g , \as(\mu)\Big)
&=
\frac{2}{2+\beta} \frac{\alpha_s C_\kappa}{\pi}\frac{1}{\xi_0'} \Big(\frac{{r_g}}{ v({r_g})}\Big)^2 \partial_{r_g} v({r_g})
\delta (\ell^+ - \qcut' v({r_g}))
\nn
\, .
\end{align}

Next, we write
\begin{align}
\cQ^{[1]X}_\kappa (\Omega)&\equiv
a_\kappa \frac{e^{\gamma_E \Omega} }{\Gamma(-\Omega)} \Big( \delta_{X,{\rm plain}} \cR_\kappa^{({\rm sd})} \big(\Omega\big) + \cR_\kappa^{(r_g)} \big(\Omega\big) \Big) \, , \\
\cQ^{[2]X}_\kappa (\Omega)&\equiv a_\kappa^2
\frac{e^{\gamma_E \Omega} }{\Gamma(-\Omega)} \delta_{X,{\rm plain}}\cR_\kappa^{({\rm sd},r_g)} \big(\Omega\big) \, ,\nn \\
\partial_{r_g} \cQ^{[1]X}_{\kappa, \veps} \big(\Omega\big)
&\equiv a_\kappa
\frac{2}{2+\beta} \frac{e^{\gamma_E \Omega} }{\Gamma(-\Omega)}
\frac{\Psi(\xi,{r_g})}{\xi}
\cR_{\kappa,\veps}^{\prime (r_g)} \big(\Omega\big)
\, , \nn \\
\partial_{r_g} \cQ^{[2]X}_{\kappa, \veps} \big(\Omega\big)
&\equiv a_\kappa^2
\frac{2}{2+\beta} \frac{e^{\gamma_E \Omega} }{\Gamma(-\Omega)}
\frac{\Psi(\xi,{r_g})}{\xi}
\sum_{AB \in \{{\rm sd},\, r_g \}_X} \cR^{\prime AB}_{\kappa,\veps} \big(\Omega\big)
\, , \nn
\end{align}
where $X \in \{{\rm plain, \, sd\,res.}\}$ and the set $\{{\rm sd}, r_g\}_X$ refers to the combinations of soft functions in \eq{GijDef} relevant for plain jet mass and soft drop jet mass resummation region, as summarized in \tab{KernelsEps}.
In the last two lines, for the soft drop boundary cross section we have factored out the pre-factor,
\begin{align}\label{eq:PsiPreDef}
\Psi(\xi,{r_g}) \equiv \frac{\xi}{\xi_0'} \Big(\frac{{r_g}}{ v({r_g})}\Big)^2 \partial_{r_g} v({r_g})
\,.
\end{align}

In terms of the rescaled-soft functions defined in \eq{GijDef}, from \eq{Qplain1} we have
\begin{align}\label{eq:cRdef}
\cR_\kappa^{({\rm sd}/r_g)} \big(\Omega\big) &= \int_0^\infty \df y \:
{\cal L}_0^{-\Omega} \big(1 - y \big) \, \xi \cS^{({\rm sd}/{r_g})}_{\kappa}(\xi y ) \, , \\
\cR_\kappa^{({\rm sd},r_g)} \big(\Omega\big) &= \int_0^\infty \df y_A \df y_B \:
{\cal L}_0^{-\Omega} \big(1 -(y_A + y_B)\big) \, \xi \cS^{(\rm sd)}_{\kappa,\veps}(\xi y_A)\, \xi \cS^{(r_g)}_{\kappa}(\xi y_B) \, , \nn \\
\frac{\Psi(\xi,{r_g})}{\xi}
\frac{2}{2+\beta} \cR^{\prime(r_g)}_{\kappa,\veps} (\Omega)
&=
\int_0^\infty \df y \:
{\cal L}_0^{-\Omega} \big(1 - y \big)\, \xi \partial_{r_g} \cS^{{r_g}}_{\kappa,\veps}(\xi y) \, , \nn\\
\frac{\Psi(\xi,{r_g})}{\xi}
\frac{2}{2+\beta} \cR^{\prime AB}_{\kappa,\veps}(\Omega)
&=
\int_0^\infty \df y_A \df y_B \:
{\cal L}_0^{-\Omega} \big(1 -(y_A + y_B)\big) \partial_{r_g} \Big[ \xi \cS^{A}_{\kappa,\veps}(\xi y_A)\, \xi \cS^{B}_{\kappa}(\xi y_B) \Big]\, . \nn
\end{align}

To proceed further we define the functions,
\begin{align}\label{eq:fhgDef}
&\gamma (\xi,{r_g}) \equiv \frac{\xi_0' v({r_g})}{\xi}\,,&
&f(w) \equiv \frac{\ln ({r_g}^2)}{w} \, ,&
&h(w) \equiv \frac{1}{w}
\ln \bigg( (1+\zeta^2) \Big(\frac{\gamma_{\rm max}}{w}\Big)^{\frac{2}{2+\beta}} -\zeta^2 \bigg) \, ,&
\\
&\gamma_{\rm max}(\xi) \equiv \frac{\xi_0^\prime}{\xi} \, ,&
&f'(w) \equiv \frac{2}{w {r_g}} \, ,&
&g(w) \equiv \frac{1}{w^2}\bigg( (1+\zeta^2)\Big(\frac{\gamma_{\rm max}}{w}\Big)^{\frac{2}{2+\beta}} -\zeta^2 \bigg)^{-1} \, .&
\nn
\end{align}

We see that in terms of $\gamma(\xi,r_g)$ the prefactor $ \Psi(\gamma,r_g)$ using \eq{PsiPreDef} becomes
\begin{align}\label{eq:PsiPreGamma}
\Psi(\gamma,r_g) = \Big(\frac{r_g}{\gamma}\Big)^2 \partial_{r_g} \gamma(\xi,r_g) \, .
\end{align}
We will find it convenient later to switch variables $r_g \ra \gamma$ when considering $r_g$-moments.

In terms of these functions we have
\begin{align}
\xi \cS_\kappa^{(\rm sd)} (\xi w)&=- \big[ \Theta(\gamma_{\rm max} - w) h(w) \big]_+^{[\gamma_{\rm max}]} \, ,\nn \\
\xi \cS_\kappa^{(r_g)} (\xi w )
&=\Theta(w - \gamma) \Big[
f(w) + \Theta(\gamma_{\rm max} -w ) h(w)
\Big]\,,
\\
\xi \cS_{\kappa,\veps}^{(\rm sd)}(\xi w)
&= \frac{2}{2+\beta}\big[ \Theta(\gamma_{\rm max} - w) g(w) \big]_+^{[\gamma_{\rm max}]}
\,, \nn \\
\xi \cS_{\kappa,\veps}^{ (r_g)}(\xi w )
&= - \frac{2}{2+\beta}\Theta(\gamma_{\rm max} - w) \Theta(w - \gamma) g(w)
\,, \nn
\end{align}
such that from \eq{cRdef}, the one-loop kernels are given by
\begin{align}
\cR^{(\rm sd)}_{\kappa} (\Omega) &= - \int_0^\infty \df w \:
{\cal L}_0^{-\Omega} \big(1 - w\big)\big[\Theta(\gamma_{\rm max} - w)\Theta(w) h(w) \big]^{[\gamma_{\rm max}]}_+
\, , \\
\cR^{(r_g)}_{\kappa} (\Omega) &= \int_0^\infty \df w \:
{\cal L}_0^{-\Omega} \big(1 - w\big)\Theta(w - \gamma)\bigg[ f (w) + \Theta(\gamma_{\rm max} - w) h (w) \bigg]
\, , \nn \\
\cR_{\kappa,\veps}^{\prime(r_g)} (\Omega)
&=
{\cal L}_0^{-\Omega} \big(1 -\gamma \big)
\, , \nn
\end{align}
and the two-loop kernels $\cR_\kappa^{({\rm sd}, r_g)}$ and $\cR_{\kappa,\veps}^{\prime AB}$ being
\begin{align}
\cR_\kappa^{({\rm sd}, r_g)} &= -\int_0^{\infty} \df w\,\df u \: \cL_0^{-\Omega}(1-w-u) \big[\Theta(\gamma_{\rm max} - u)\Theta(u) h(u) \big]^{[\gamma_{\rm max}]}_+ \\
&\quad\qquad \times\Theta(w-\gamma) \Big[f(w) +\Theta(\gamma_{\rm max} - w)h(w )\Big] \nn
\, , \\
\cR_{\kappa,\veps}^{\prime({\rm sd}, r_g)} (\Omega)
&=
-
\int_0^\infty \df w \:
{\cal L}_0^{-\Omega} \big(1 - w - \gamma \big)\big[\Theta(\gamma_{\rm max} - w)\Theta(w) h(w) \big]^{[\gamma_{\rm max}]}_+
\nn \, ,
\\
\cR_{\kappa,\veps}^{\prime(r_g,{\rm sd})} (\Omega)
&=
+\Psi^{-1}
\int_0^\infty \df w \: \df u \:
{\cal L}_0^{-\Omega} \big(1 - u -w \big)
f'(w) \Theta(w - \gamma ) \big[\Theta(\gamma_{\rm max} - u)\Theta(u) g(u) \big]^{[\gamma_{\rm max}]}_+ \, ,\nn\\
\cR_{\kappa,\veps}^{\prime(r_g,r_g)} (\Omega) &=
-\Psi^{-1}\int_0^\infty \df w \: \df u \:
{\cal L}_0^{-\Omega} \big(1 - u - w \big)
f'(w) \Theta(w - \gamma)
\Theta(\gamma_{\rm max} - u) \Theta (u - \gamma ) g (u) \nn
\\
& \quad +
\int_0^\infty \df w \:
{\cal L}_0^{-\Omega} \big(1 - w - \gamma \big)\Theta(w - \gamma)\bigg[ f (w) + \Theta(\gamma_{\rm max} - w) h (w) \bigg]
\, . \nn
\end{align}
Note that we have not included $\cR_{\kappa,\veps}^{({\rm sd,sd})}$ in the list above as it does not carry any $r_g$ dependence and vanishes upon taking $r_g$-derivative.

To further compactify these expressions we introduce the following transforms acting on a function or a distribution $F(w)$:
\begin{align}
q \{ F \} (u, u_{\rm min} ) &\equiv
\Theta(1 - u_{\rm min} - u )
\int_0^\infty
\df w \:
{\cal L}_0^{-\Omega}
\big(1 - u -w \big) F(w) \Theta (w - u_{\rm min}) \, ,
\\
q_+ \{ F \} ( u, \, u_{\rm max} ) &\equiv
\Theta (1- u)
\int_0^\infty
\df w \:
{\cal L}_0^{-\Omega}
\big(1 - u -w\big) \Big[\Theta(u_{\rm max} - w)\Theta(w)F(w)\Big]^{[u_{\rm max}]}_+ \, . \nn
\end{align}
In terms of these transforms the above expressions become
\begin{align}\label{eq:Rij}
\cR^{(\rm sd)}_{\kappa} (\Omega) &=
-q_+\{h\}(0,\gamma_{\rm max}) \, ,
\\
\cR^{(r_g)}_{\kappa} (\Omega) &= q\{f +h\}( 0,\gamma ) - q\{h\}(0, \gamma_{\rm max})
\, , \nn\\
\cR_\kappa^{({\rm sd}, r_g)} (\Omega)&=- \int_\gamma^{\min\{\gamma_{\rm max} , 1\}} \df w \: q_+\{h\}\big(w,\gamma_{\rm max}\big)\Big[f(w) + \Theta(\gamma_{\rm max} -w) h(w) \Big]
\nn \, ,
\\
\cR_{\kappa,\veps}^{\prime({\rm sd },r_g)} (\Omega)
&= -
q_+\{h\}( \gamma, \gamma_{\rm max}) \, , \nn \\
\cR_{\kappa,\veps}^{\prime(r_g,{\rm sd})} (\Omega) &=\Psi^{-1} \int_0^{\gamma_{\rm max} } \df u \:
g(u) \Big [
q\{f'\}(u , \gamma) - q\{f'\}( 0 , \gamma)
\Big ] \, , \nn
\\
\cR_{\kappa,\veps}^{\prime(r_g,r_g)}(\Omega)
&= -\Psi^{-1} \int_{\gamma }^{\gamma_{\rm max}} \df u \: g(u) \, q\{f'\} (u, \gamma)
+
\Big[
q\{f + h \} (\gamma, \gamma)
- q\{h\} (\gamma,\gamma_{\rm max})
\Big]
\nn \,.
\end{align}
The transform on the combination $f+h$ in the second and last line is necessary to stabilize the numerical integration.

The transforms simplify to the following expressions:
\begin{align}
q \{ F \} (u,\, u_{\rm min} )
&=
\Bigg[
\frac{F(1 - u)}{-\Omega\big(1 - u_{\rm min} - u \big)^{\Omega}}
+
\int_{u_{\rm min}}^{1 - u} \df w \:
\frac{F(w) - F(1 - u)}{\big(1 - w-u\big)^{1 + \Omega}}
\Bigg] \, , \nn \\
q_+ \{ F \} ( u, \, u_{\rm max} )
&= \Theta (1- u - u_{\rm max}) \int_0^{u_{\rm max}} \df w \:F(w)\Big[\big(1 - w-u\Big)^{-1 - \Omega}
-\big(1 - u\big)^{-1 - \Omega}
\Big]
\nn \\
&\quad+ \Theta (u_{\rm max} - 1 + u)
\Bigg[ \int_0^{1 - u}
\df w \:
\bigg (\frac{\big(F(w) - F(1 - u) \big)}{\big(1 - w-u\big)^{1 + \Omega}}
- \frac{F(w)}{\big(1 -u\big)^{1 + \Omega}} \bigg)
\nn \\
&\qquad
-\int_{1-u}^{u_{\rm max}} \df w\: \frac{F(w)}{\big(1 - u\big)^{1 + \Omega}}
+
\frac{ F(1 - u)}{-\Omega\big(1 - u \big)^{\Omega}}
\Bigg] \,.
\end{align}
In the first expression, the $w$ integration always sees the zero of the argument of ${\cal L}_0^{-\Omega}$ for every value of $u$. In the second expression, we see that only in the second and third line is the plus prescription of ${\cal L}_0^{-\Omega}$ necessary. In writing the integrals in the form shown above, we regulate singularities in $F(w)$ at $w \ra 0$ and in ${\cal L}_0^{-\Omega}(1-u)$ for $u\ra 1$.
Finally, for $\Omega \ra 0$ limit, the resummation is turned off and we find
\begin{align}
\lim\limits_{\Omega \ra 0} \frac{e^{\gamma_E \Omega}}{\Gamma(-\Omega)} q \{ F \} (u,\, u_{\rm min} ) &= \Theta (1 - u_{\rm min} - u) F(1 - u) \, ,
\\
\lim\limits_{\Omega \ra 0}
\frac{e^{\gamma_E \Omega}}{\Gamma(-\Omega)} q_+ \{ F \} ( u,\, u_{\rm max} )
&= \Big[\Theta\big(u_{\rm max} - (1 - u)\big)\Theta(1 - u) F(1-u)\Big]^{[u_{\rm max}]}_+ \, .
\nn
\end{align}

In \Refcite{Pathak:2020iue}, the results for some of the resummation kernels above were presented in terms of incomplete beta functions and their integrals. We have checked that the results there agree with the ones above when $\zeta \ra 0$ in \eq{fhgDef}. The numerical implementation associated with this work follows the above presentation of these kernels in terms of integral transforms, and in fact results in a significantly faster and reliable code.
\subsection{Fixed order cross section}
\label{sec:foBndry}
Finally, we consider the non-singular corrections from using the full soft drop condition without expansions in the soft limit. In analogy with \eq{deltaFO} the measurement function is given by
\begin{align}\label{eq:deltaFOeps}
\delta_{\xi,\veps}^{\rm FO} (x,y,r_g,\xi_0,\zeta) &\equiv \hat \Theta_{k_T}(x,y) \delta_{\rm sd} (x,y,\xi_0,\zeta) \big[ \hat \Theta_{r_g} (x,y,r_g)\delta (\xi - y) - \delta (\xi)\big] \, ,
\end{align}
Again restricting to $\xi > 0$ for the differential cross section we can ignore the $\delta(\xi)$ term. The two delta-functions fix both $x$ and $y$, such that the fixed order cross section is given by
\begin{align}\label{eq:GFOeps}
\xi \tcG_{\kappa,\veps}^{\rm FO[1]} (\xi, r_g) &= (1 + \delta_{\kappa,g} )j(x^*,\xi)\frac{\alpha_s}{2\pi}\hat \Theta_{k_T}(x^*, \xi) \hat \Theta_{r_g} \big(x^*, \xi, r_g\big)\sum_{\kappa'} \hat P_{\kappa'\kappa}(x^*) \\
&\quad + \delta_{\kappa,q} | j(1-x^*, \xi) | \frac{\alpha_s}{2\pi}\hat \Theta_{k_T}(x^*, \xi) \hat \Theta_{r_g} \big(x^*, \xi, r_g\big)\sum_{\kappa'} \hat P_{\kappa'\kappa}(1-x^*) \nn
\, ,
\end{align}
where $x^*$ is the solution of the soft drop constraint in \eq{SDExp} for $y = \xi$ and $j(x,y)$ is the Jacobian given by
\begin{align}
j(x,y) =\bigg(\frac{1+ y \zeta^2}{1- y\zeta^2}\bigg)\bigg(1 + \frac{\beta}{2} \frac{(1- 2x)(1 - y\zeta^2)}{(1-x) + xy \zeta^2}\bigg)^{-1} \, .
\end{align}
In the soft limit ($x \sim y, x \ra 0$), $j(x,y) = 2/(2+\beta)$. The $\delta_{\rm sd}$ constraint can be easily solved for $\beta = 0$, for which we find
\begin{align}
x^*(y, \xi_0, \beta = 0) = 1 - \frac{1- \xi_0(1 + y\zeta^2)}{1-y\zeta^2} \, .
\end{align}
For other values of $\beta$ we employ the following approximation which fairly well reproduces the exact solution for the range of $\zcut$ and $\beta$ that we consider:
\begin{align}
x^*(y, \xi_0, \beta > 0) \approx y^{\frac{\beta}{2+\beta}}\xi_0^{\frac{2}{2+\beta}} -y \zeta^2 + \frac{1}{4} y\zeta^2\bigg(\frac{\zcut}{0.1}\bigg) \, .
\end{align}

\begin{figure}[t]
\centering
\includegraphics[width=.6\linewidth]{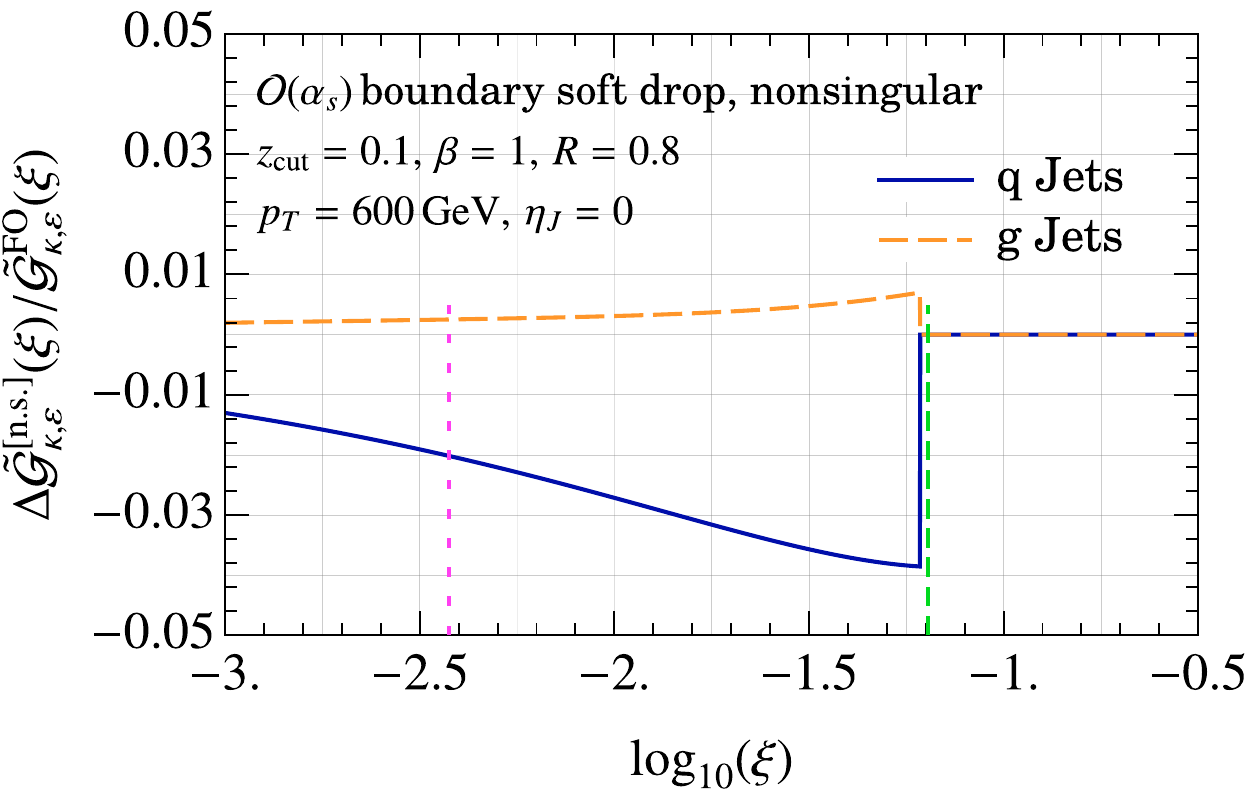}
\caption{Relative size of non-singular corrections to the boundary soft drop cross section. Using the full soft drop condition shifts the cusp slightly to the left of $\xi_0'$ (green-dashed vertical line).}
\label{fig:epsns}
\end{figure}

Since the $r_g$-dependence in the fixed order cross section in \eq{GFOeps} only arises from the measurement theta-function and collinear matrix elements, to gauge the size of non-singular corrections we can simply consider the boundary soft drop cross section for $r_g = 1$. To this end we define
\begin{align}
\Delta \tilde \cG^{\rm n.s.[1]}_{\kappa, \, \rm sd,\veps}\big(\xi, \alpha_s(\mu_N)\big) \equiv \tcG^{\rm FO}_{\kappa,\veps}\big(\xi, r_g = 1\big) - Q \Delta S_{\rm sd,\veps}^{\kappa} \big(Q \xi, \qcut, \zeta, \beta, \alpha_s(\mu_N)\big) \, .
\end{align}
In \fig{epsns} we show the size of the non-singular correction relative to the fixed-order result. We see that the deviation is below 5\% which we will find to be well within the perturbative uncertainties.

\subsection{$R_g$-weighted boundary soft drop cross section}

\label{sec:c2}

Having discussed the construction of the soft drop boundary cross section we now use it to compute the $r_g$ moments. The first step is to combine the results in the plain jet mass and soft drop resummation regions to compute the matched cross section. This is straightforwardly obtained by repeating the steps for the max-$R_g$ cross section in \eq{MatchMax}:
\begin{align}\label{eq:MatchEps}
\tcG^{\rm match}_{\kappa,\veps} (\xi, r_g )
&\equiv
\tcG_{\kappa,\veps}^{\rm sd} (\xi, r_g,\mu_{\rm sd\ra plain})
+
\big[ \tcG_{\kappa,\veps}^{\rm plain} (\xi,r_g ,\mu_{\rm plain})
- \tcG_{\kappa,\veps}^{\rm sd} (\xi,r_g ,\mu_{\rm plain}) \big ]
\, .
\end{align}
Using this result, we can compute the $C_{2\kappa}^{(n)}(\xi)$ moments:
\begin{align}\label{eq:C2Def}
C_{2\kappa}^{(n)} (\xi) = \frac{\xi}{\tcG_\kappa^{\rm match}(\xi)} \int_{r_g^{\rm min}(\xi)}^{r_g^{\rm max}(\xi)} \df r_g \:r_g^n \: w_{\rm sd \ra plain} (\xi,r_g) \, \frac{\df \tcG_{\kappa,\veps}^{\rm match}(\xi ,r_g) }{\df r_g} \, .
\end{align}
Next, we note that the cross section $\partial_{r_g} \tcG_{\kappa,\veps}^{\rm match}(\xi ,r_g) $ is proportional to $\Psi(\xi ,r_g)$ defined in \eq{PsiPreDef}:
\begin{align}
\xi \frac{\df \tcG_{\kappa,\veps}^{\rm match}(\xi ,r_g) }{\df r_g} = \Psi(\xi,r_g) \: \cF_{\kappa,\veps}^{\rm match}(\xi ,r_g) \, ,
\end{align}
and have factored out the pre-factor in \eq{DelPsiG1},
such that using \eq{PsiPreGamma} $C_{2\kappa}^{(n)}(\xi)$ in \eq{C2Def} becomes
\begin{align}
C_{2\kappa}^{(n)} (\xi) = \frac{1}{\tcG^{\rm match} (\xi)} \int_{\gamma_{\rm min}(\xi)}^{\min\{1, \gamma_{\rm max}(\xi)\}} \df \gamma\: \big[r_g(\gamma)\big]^n \Big(\frac{r_g^{\rm max}(\xi \gamma)}{\gamma}\Big)^2 \cF_{\kappa,\veps}^{\rm match}\big(\xi ,r_g^{\rm max}(\xi\gamma)\big) \, .
\end{align}
where
\begin{align}
&r_g(\xi,\gamma) \equiv r_g^{\rm max}(\xi \gamma) \, ,&
&\gamma_{\rm min} (\xi) \equiv \frac{\xi_0' v(\sqrt{\xi})}{\xi} \, ,&
&\gamma_{\rm max} (\xi) = \frac{\xi_0'}{\xi} \, .&
\end{align}

\begin{figure}[t]
\centering
\includegraphics[width=.48\linewidth]{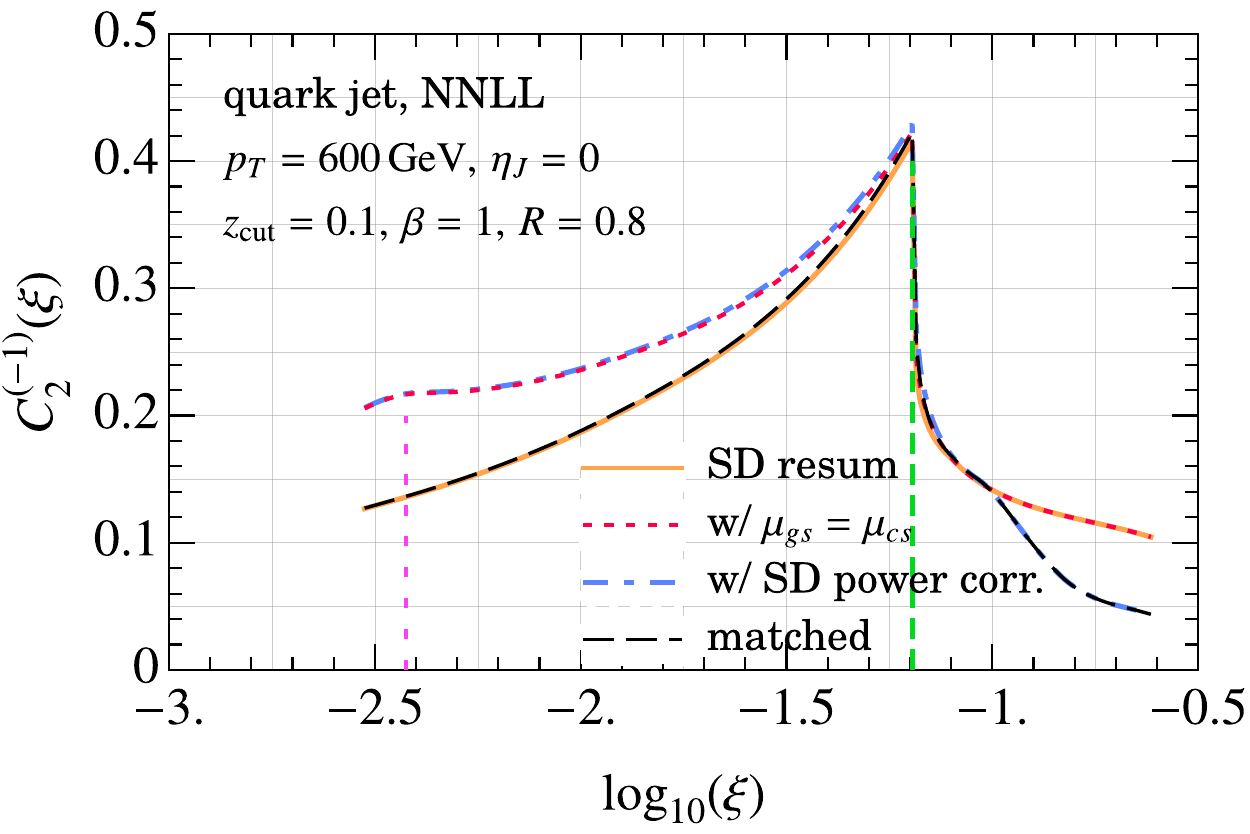}
\includegraphics[width=.48\linewidth]{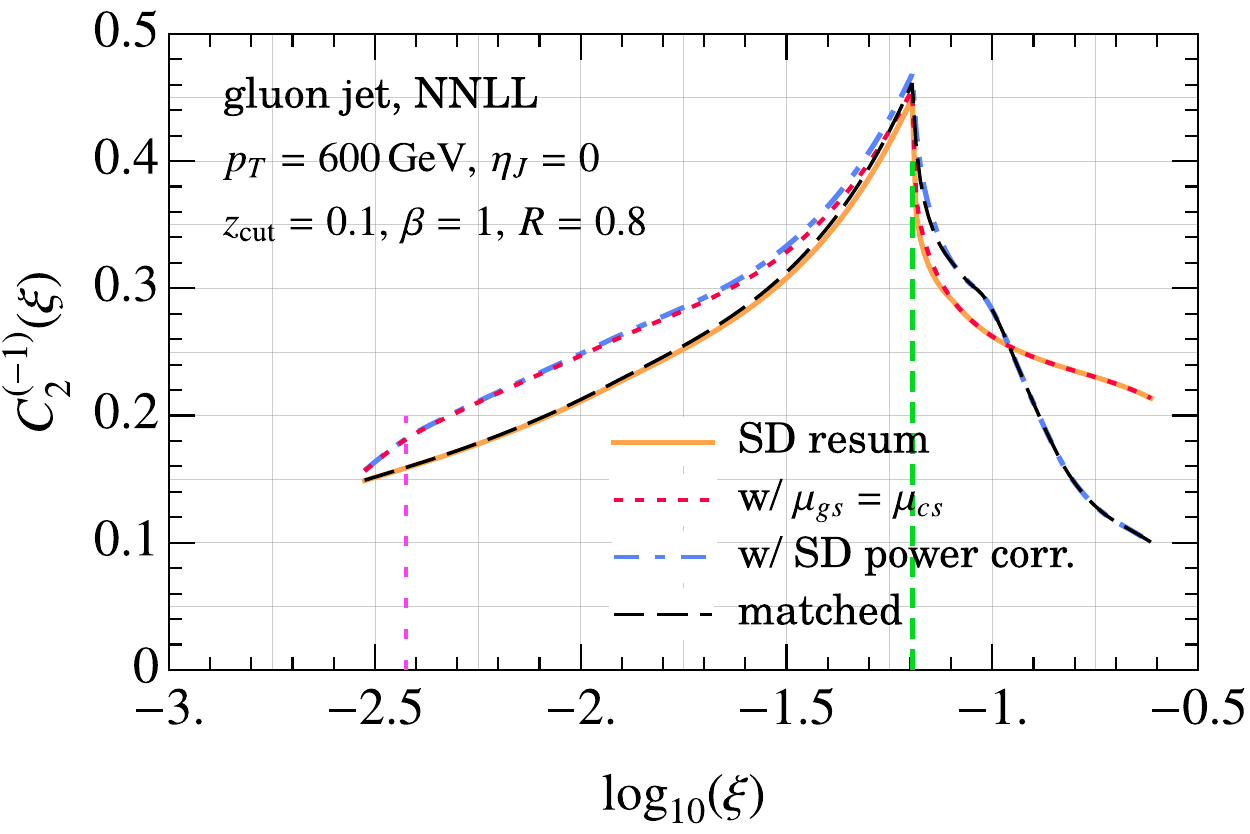}
\caption{Matching the result for $C_{2\kappa}^{(n)}$ for $n = -1$ for quark and gluon jets across the soft drop resummation and plain jet mass regions. The vertical lines denote the extent of the SDOE region.}
\label{fig:c2match}
\end{figure}
\begin{figure}[t]
\centering
\includegraphics[width=.48\linewidth]{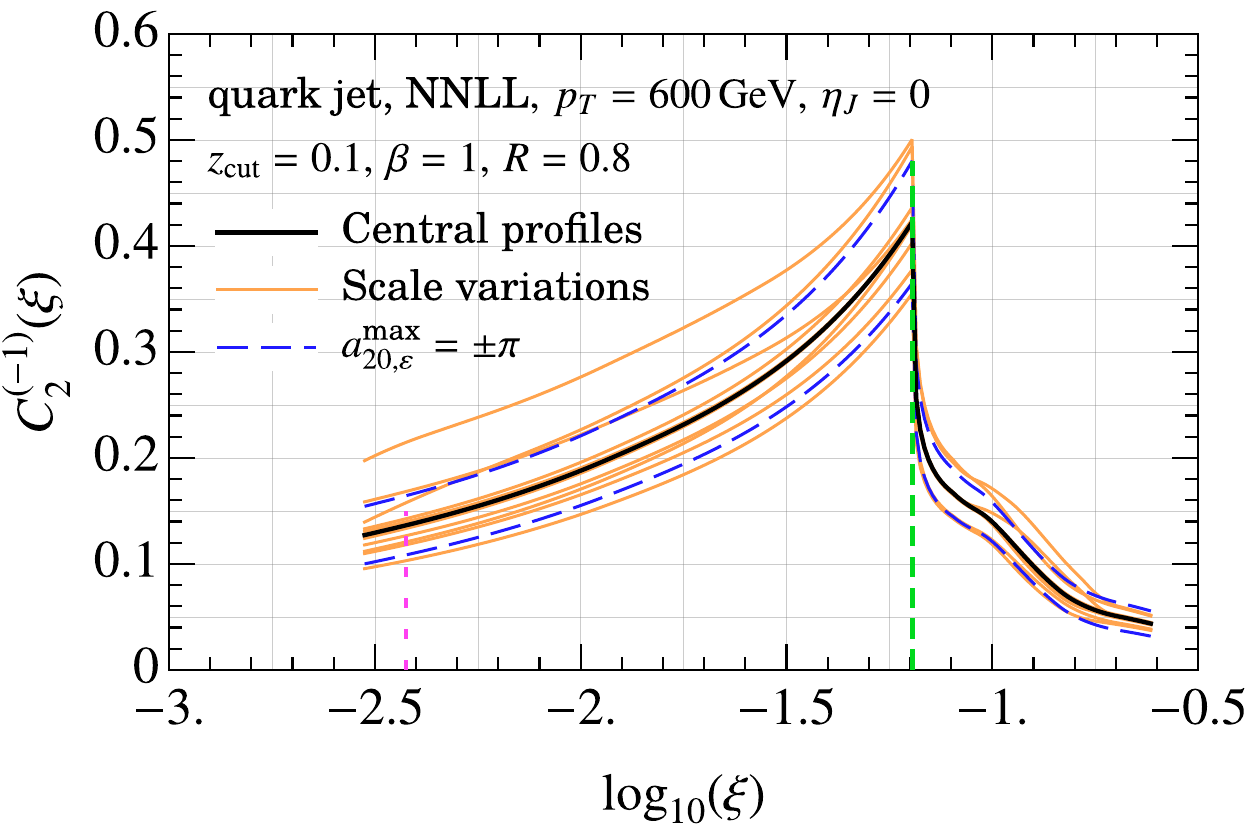}
\includegraphics[width=.48\linewidth]{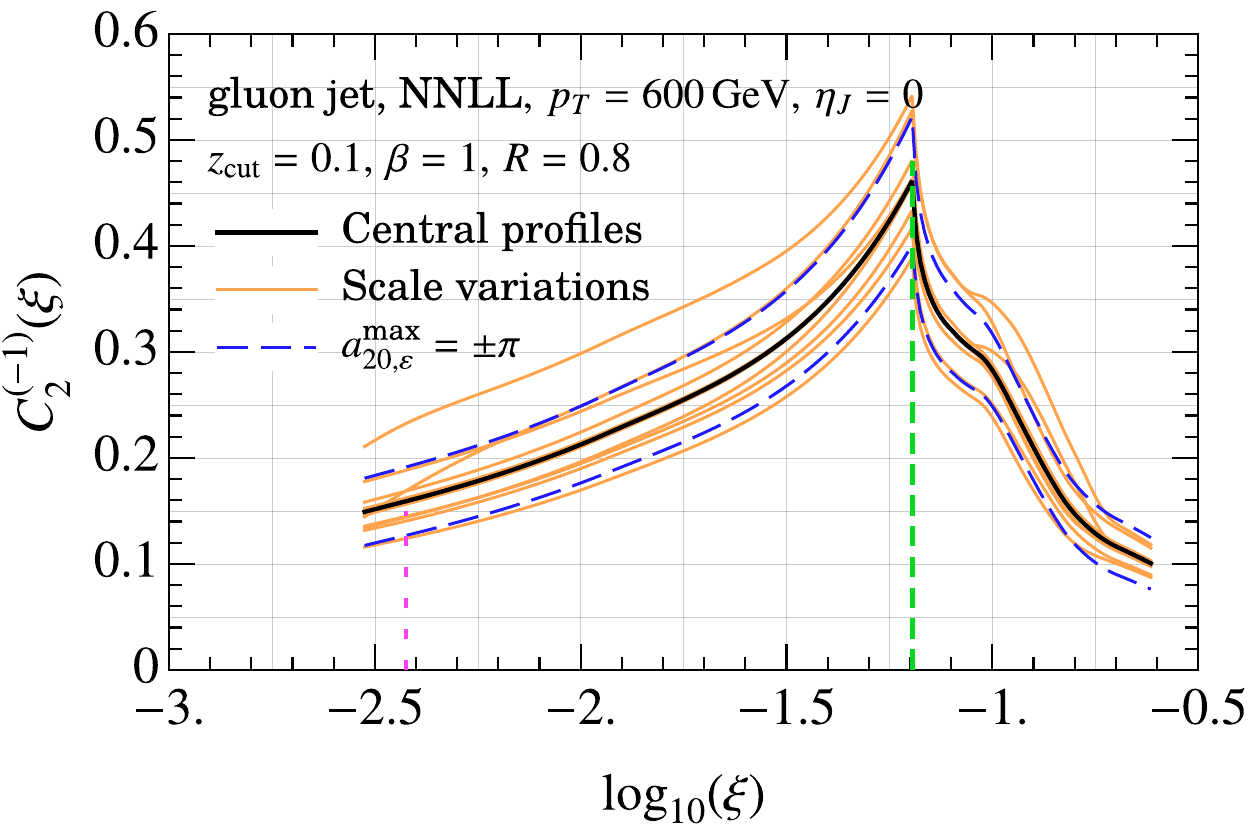}
\includegraphics[width=.48\linewidth]{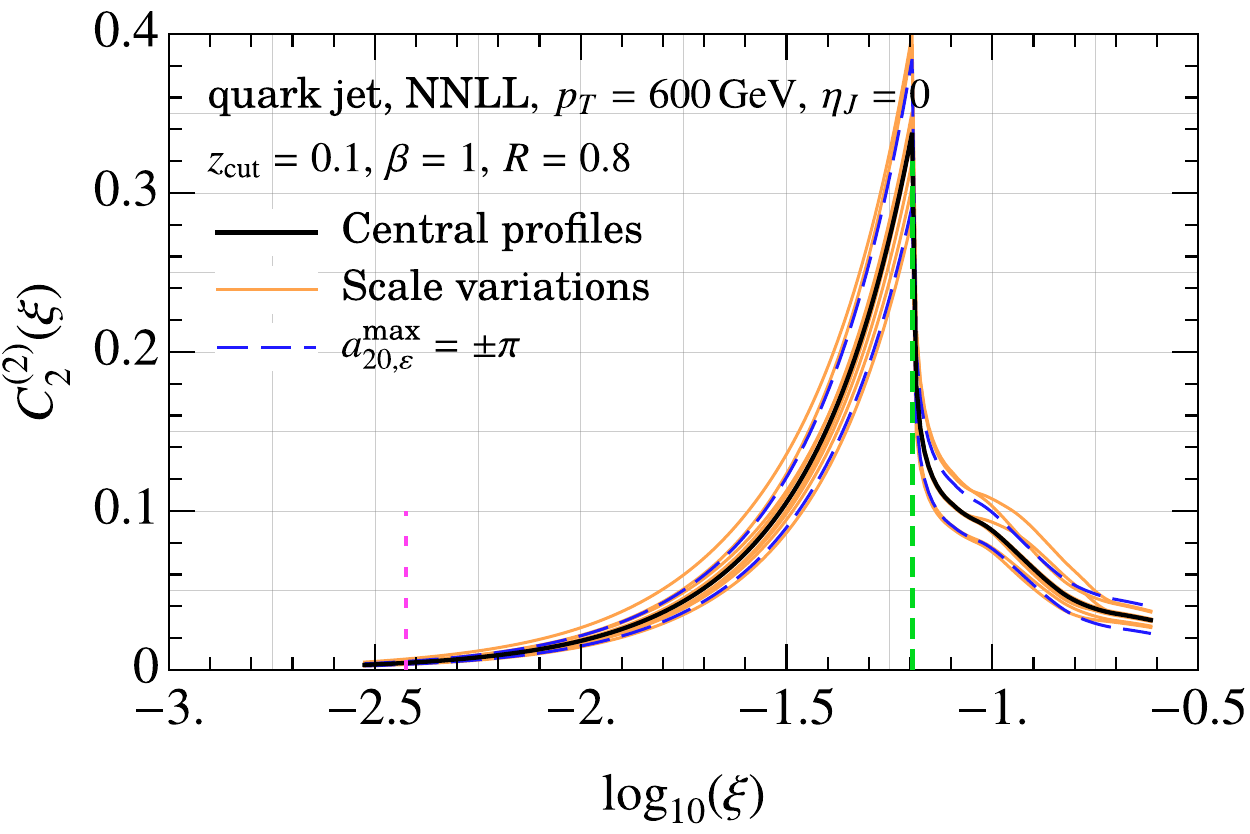}
\includegraphics[width=.48\linewidth]{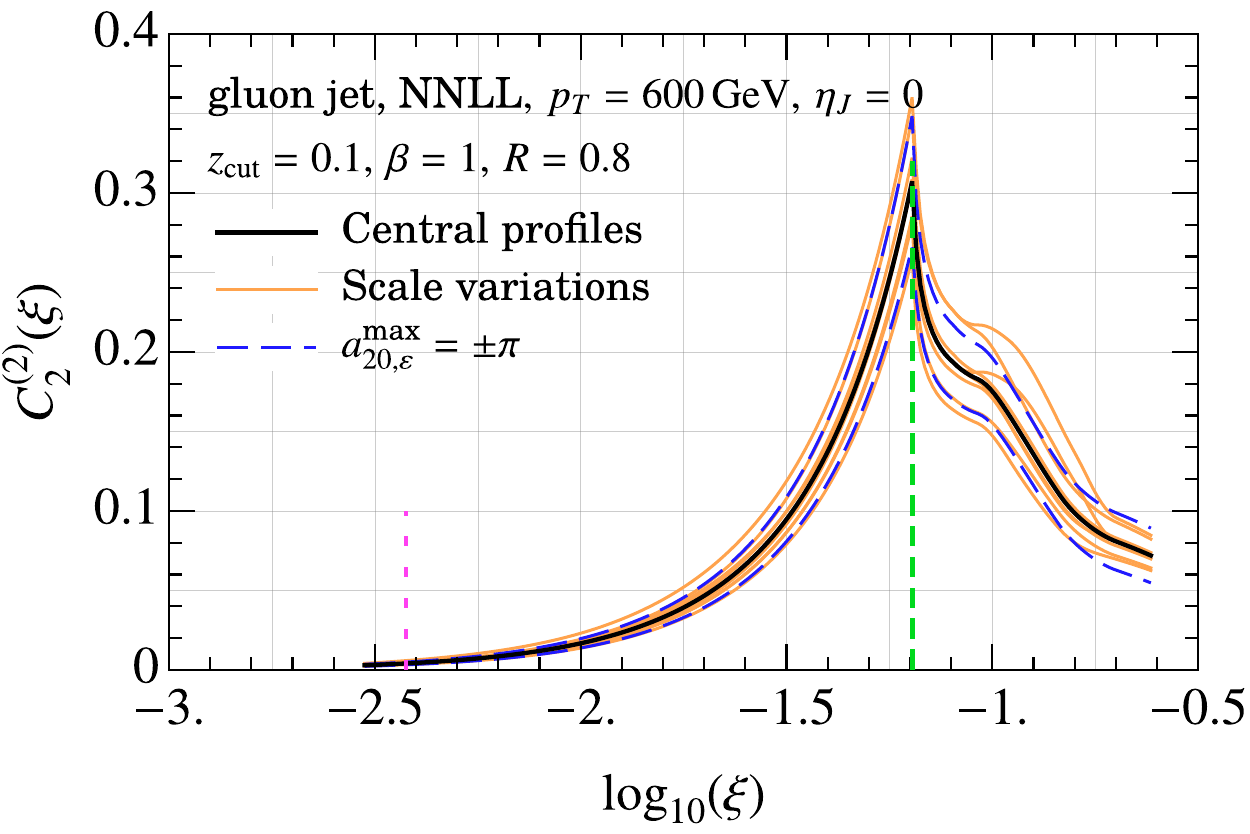}
\caption{$C_{2\kappa}^{(n)}$ for $n = -1$ (top) and $n =2$ (bottom) for quark and gluon jets. The vertical lines denote the extent of the SDOE region.}
\label{fig:c2vary}
\end{figure}

In \fig{c2match} we demonstrate how the matching for $C_{2\kappa}^{(-1)}$ works. We see that the overlap piece completely cancels the soft drop resummed cross section in the plain jet mass region and the plain jet mass cross section in the soft drop resummation region.
In \fig{c2vary} we show the impact of scale variations and the variation of the nuisance parmaeter $a_{20,\veps}^{\rm max}$ for two-loop non-logarithmic pieces for $C_{2\kappa}^{(-1,2)}$ moments. Unlike the doubly differential cross section considered in the previous section the boundary cross section is not directly related to single differential jet mass cross section. As a result we find a significant impact of scale variation left in the ratio, which dominates the nuisance parmaeter variation. The variations for $C_{2\kappa}^{(2)}$ are somewhat smaller than those of the $C_{2\kappa}^{(-1)}$ moment. This is because for positive moments $n = 2$, the contribution from the int-$R_g$ and min-$R_g$ regions is naturally power suppressed. The scale variations for small $r_g$ tend to be larger due to smaller scales being probed.

\section{Comparison with previous results}
\label{sec:compare}

\begin{figure}[t]
\centering
\includegraphics[width=\linewidth]{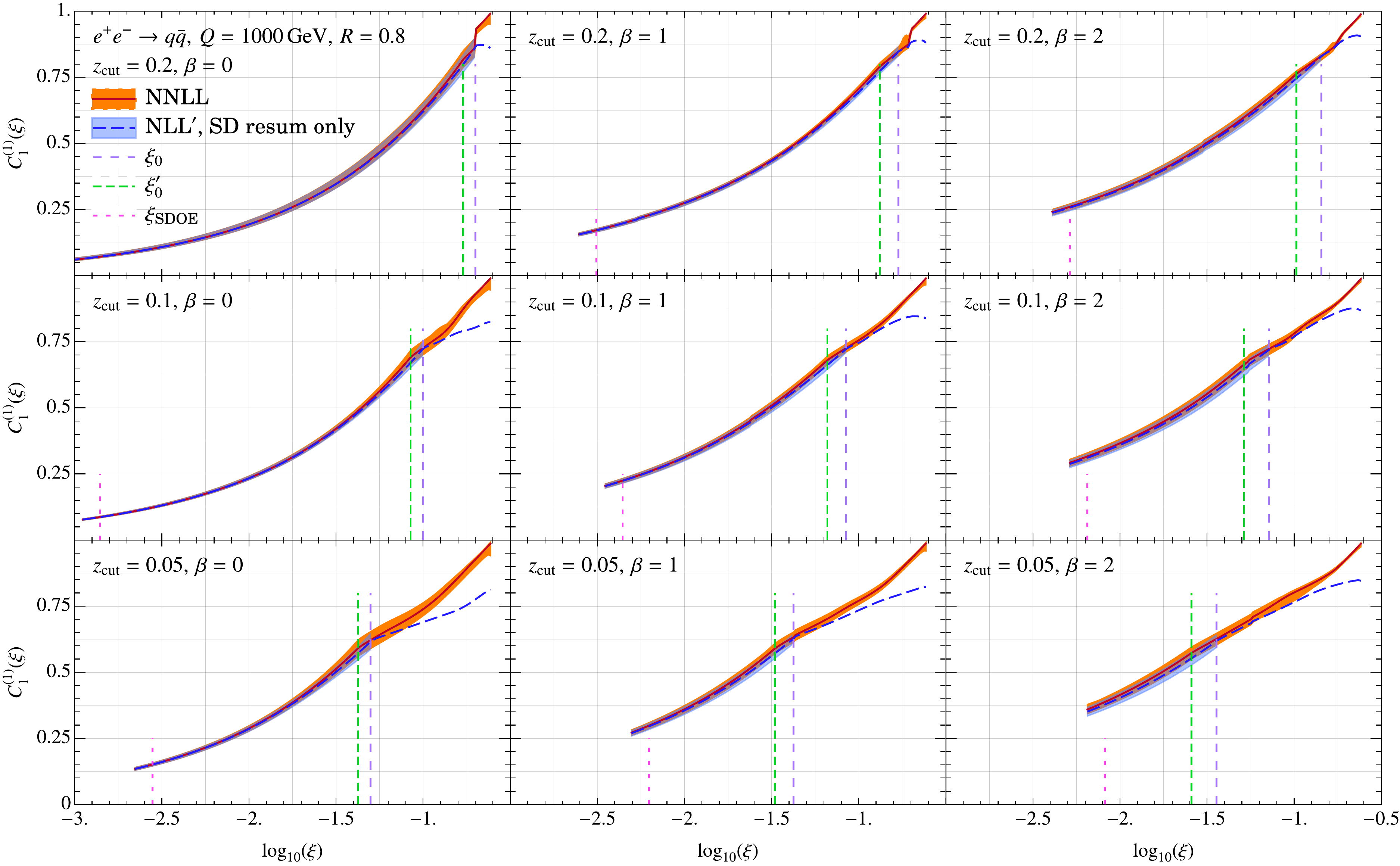}
\includegraphics[width=\linewidth]{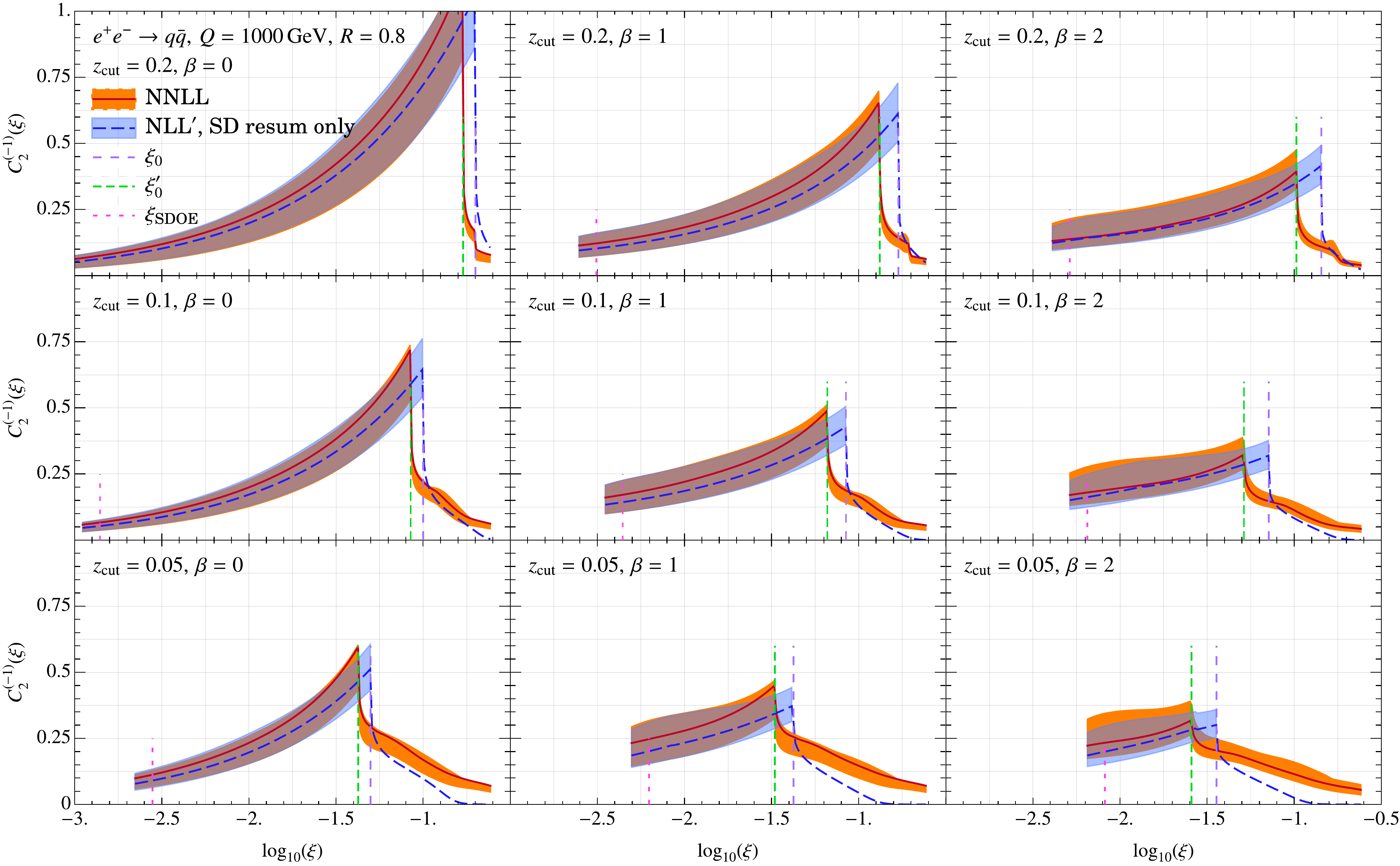}
\caption{Comparison of NNLL prediction of $C_1^{(1)}$ and $C_2^{(-1)}$ against the NLL$'$ prediction from \Refcite{Pathak:2020iue} in the soft drop resummation region.}
\label{fig:c1c2eelegacy}
\end{figure}

Having discussed the NNLL computation of the Wilson coefficients we now investigate the impact of improvements achieved in this work from matching to the plain jet mass region.
In \fig{c1c2eelegacy} we show the comparison of NNLL results for $C_{1q}^{(1)}$ and $C_{2q}^{(1)}$ and the previous result derived in \Refcite{Pathak:2020iue}, taking the $\ee\ra q \bar q$ process for illustration. A comparison with LL and NLL results, as well as with a computation in the coherent branching framework can be found in \Refcite{Pathak:2020iue}.
Crucially, in the NLL$'$ calculation in \Refcite{Pathak:2020iue} the power corrections associated with the max-$R_g$ regime in the soft drop resummation region shown in \eq{MaxPC2} were systematically dropped. Hence, the result is not valid close to the soft drop cusp and beyond. For this reason we only show the central curve in \fig{c1c2eelegacy} obtained by simply switching from soft drop to plain jet mass resummation profiles. Our NNLL calculation reveals that the power corrections from matching to the plain jet mass region are not large in the region $\xi < \xi_0' < \xi_0$ for typical grooming parameters. For both $C_{1\kappa}^{(1)}$ and $C_{2\kappa}^{(-1)}$ the difference in the region where the two results overlap, however, becomes significant for low grooming cases such as $\zcut = 0.05$ and $\beta = 2$.
On the other hand, we see a significant impact on $C_{2\kappa}^{(-1)}$ from the shift of the soft drop
cusp from the NLL location, $\xi_0$, to NNLL location, $\xi_0'$.
We now investigate the impact of this shift on the cusp location on the numerical size of the power
corrections.

\begin{figure}[t]
\centering
\includegraphics[width=.48\linewidth]{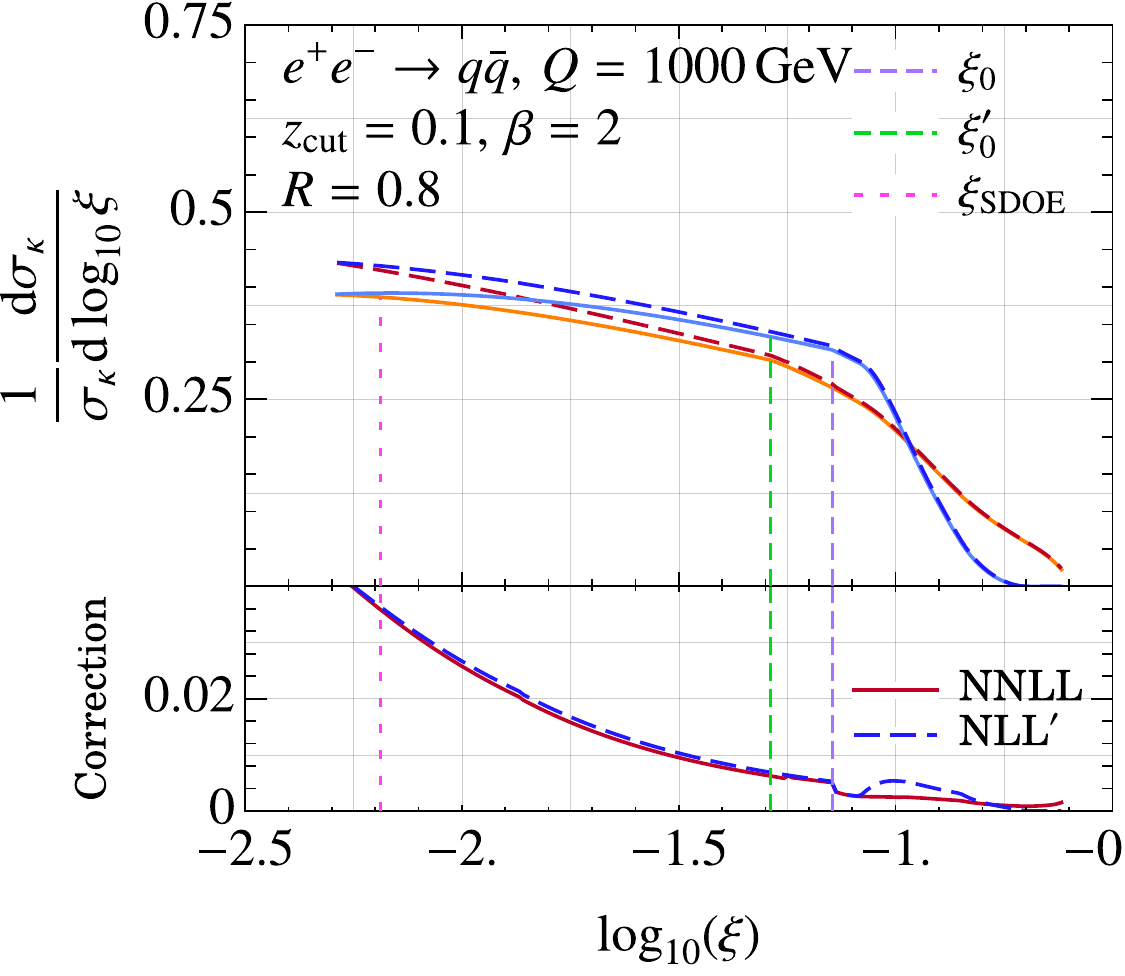}
\includegraphics[width=.48\linewidth]{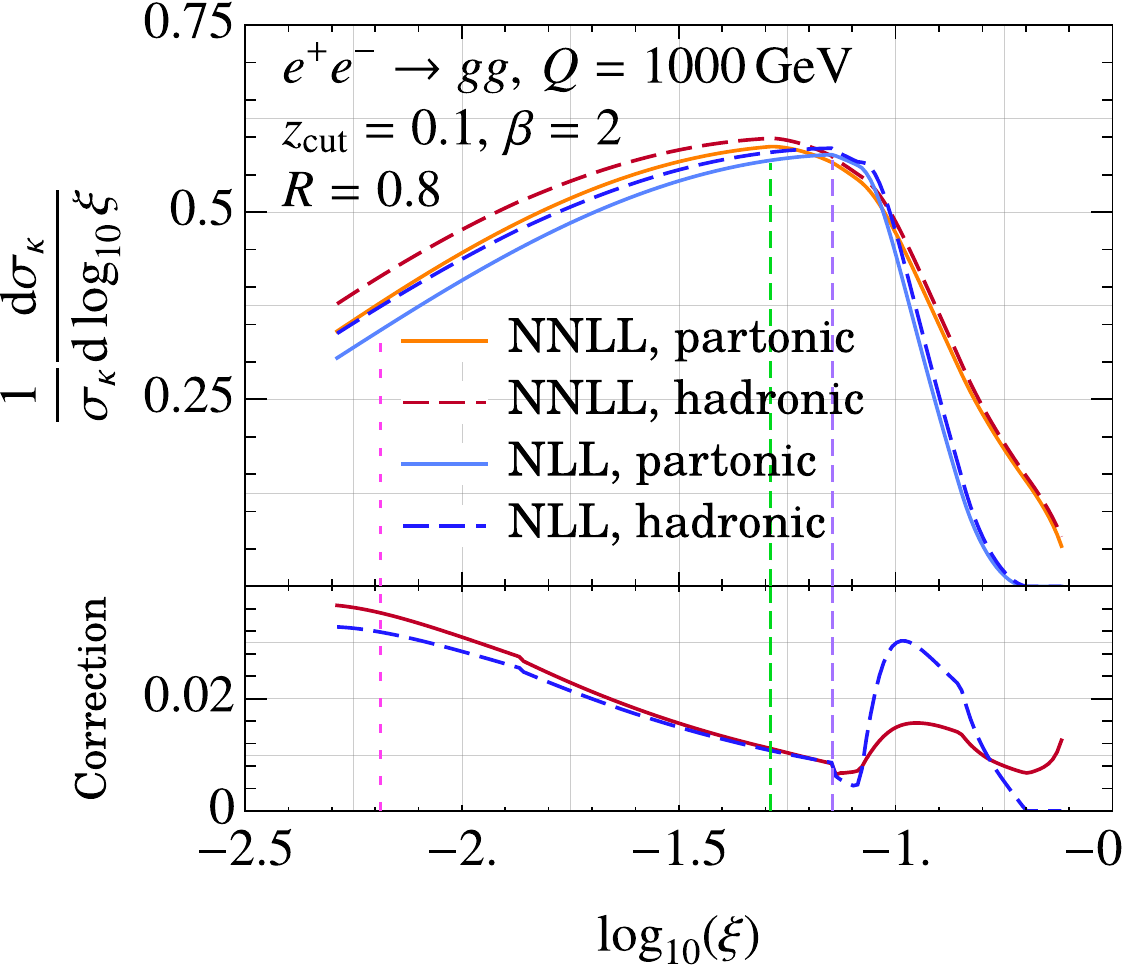}
\caption{Hadronization corrections in the jet mass spectrum by combining the Wilson coefficients and
the corresponding moments following \eq{NP}. The insets show correction to the parton level curve for
two different calculations of perturbative weights. The same sets of legends apply to both plots and the
corresponding insets.}
\label{fig:mJhad}
\end{figure}
We first assess the improvements in the perturbative weights achieved in this work by considering the impact on the jet mass spectrum.
We note that the NLL jet mass spectrum exhibits cusp at $\xi_0$, whereas the NNLL (or NLL$'$) spectrum at $\xi_0'$. To include hadronization corrections consistently we must ensure that this is also the case for the Wilson coefficients.
In \fig{mJhad} we combine the NLL$'$ results for perturbative weights from \Refcite{Pathak:2020iue} without $(\xi/\xi_0)^{1/(2+\beta)}$ soft drop power corrections with the NLL jet mass spectrum,
and the NNLL results derived in this work (including these power corrections) with the NNLL jet mass spectrum.
Here, as a reference, we used $\{\Oq,\Uqa,\Uqb\} = \{0.55,-0.57,1.06\}$ obtained from fitting to \Pythiaxx
hadronization model in \Refcite{Ferdinand:2023vaf}.\footnote{These values are chosen purely for
illustration purposes and we do not claim to have determined these parameters by fitting to \Pythiaxx.}
As can be seen in the top panel, \fig{mJhad}, the nonperturbative corrections start to grow as we reduce
the jet masses past the respective cusp location shown in dashed vertical lines.
We find, however, that the correction to the parton level for two different calculations of perturbative
weights shown in the bottom inset differs significantly only beyond the cusp at $\xi_0$, with the NNLL
calculation being the reliable one in this
region.

\begin{figure}[t]
\centering
\includegraphics[width=.45\linewidth]{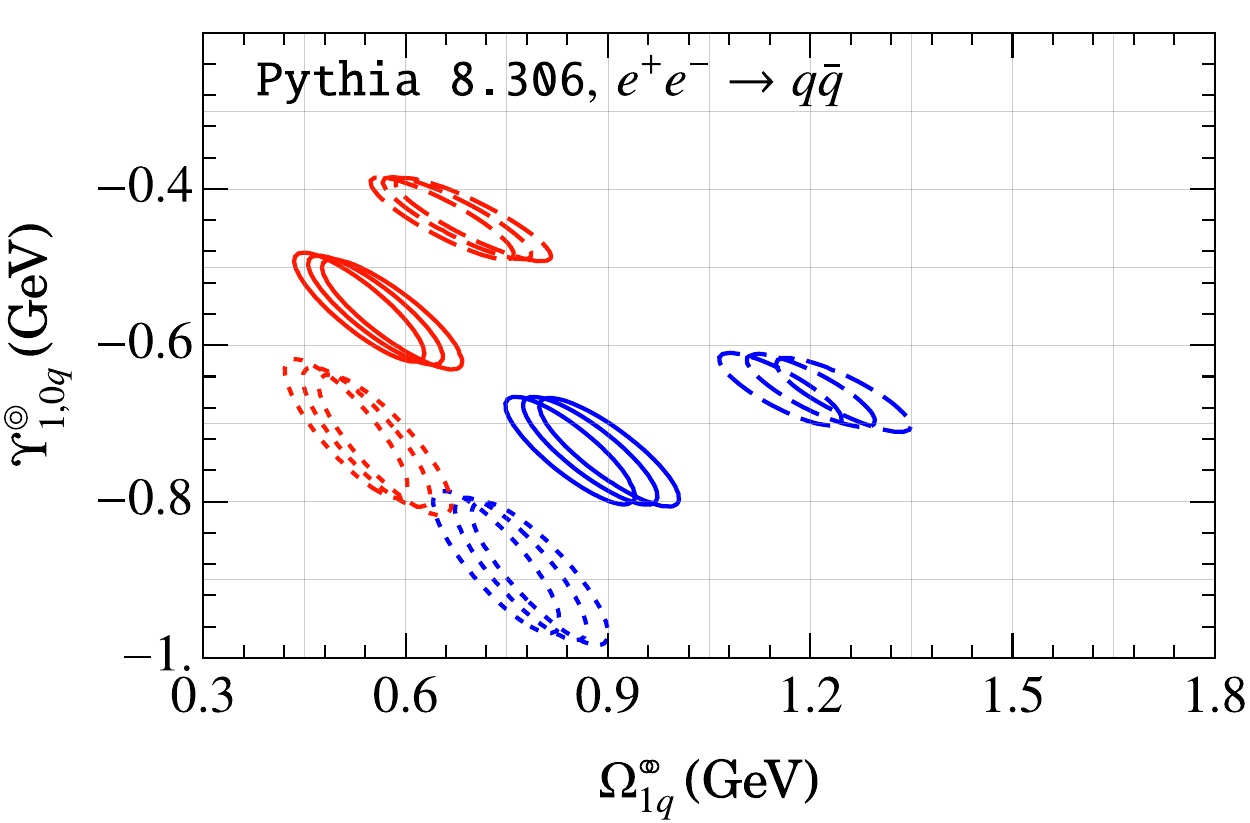}
\includegraphics[width=.45\linewidth]{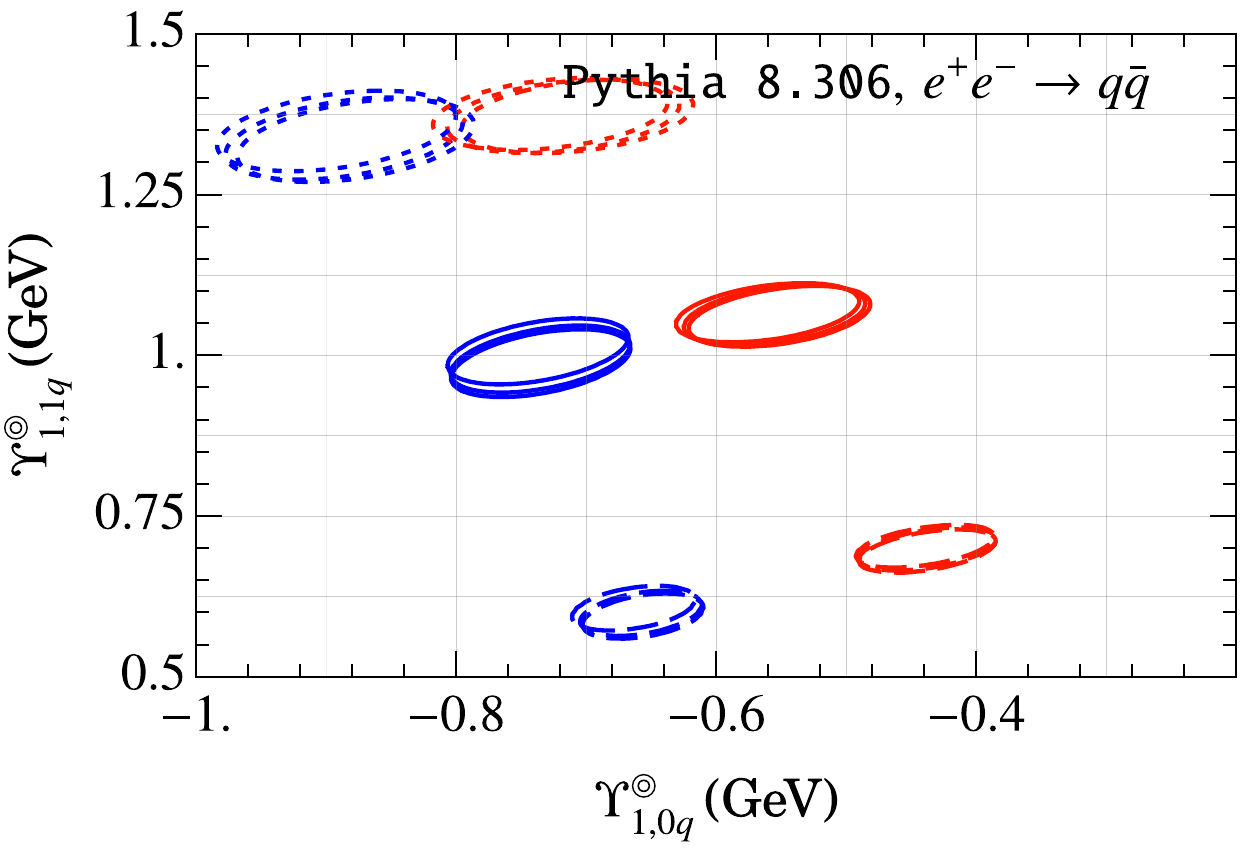}
\includegraphics[width=.55\linewidth]{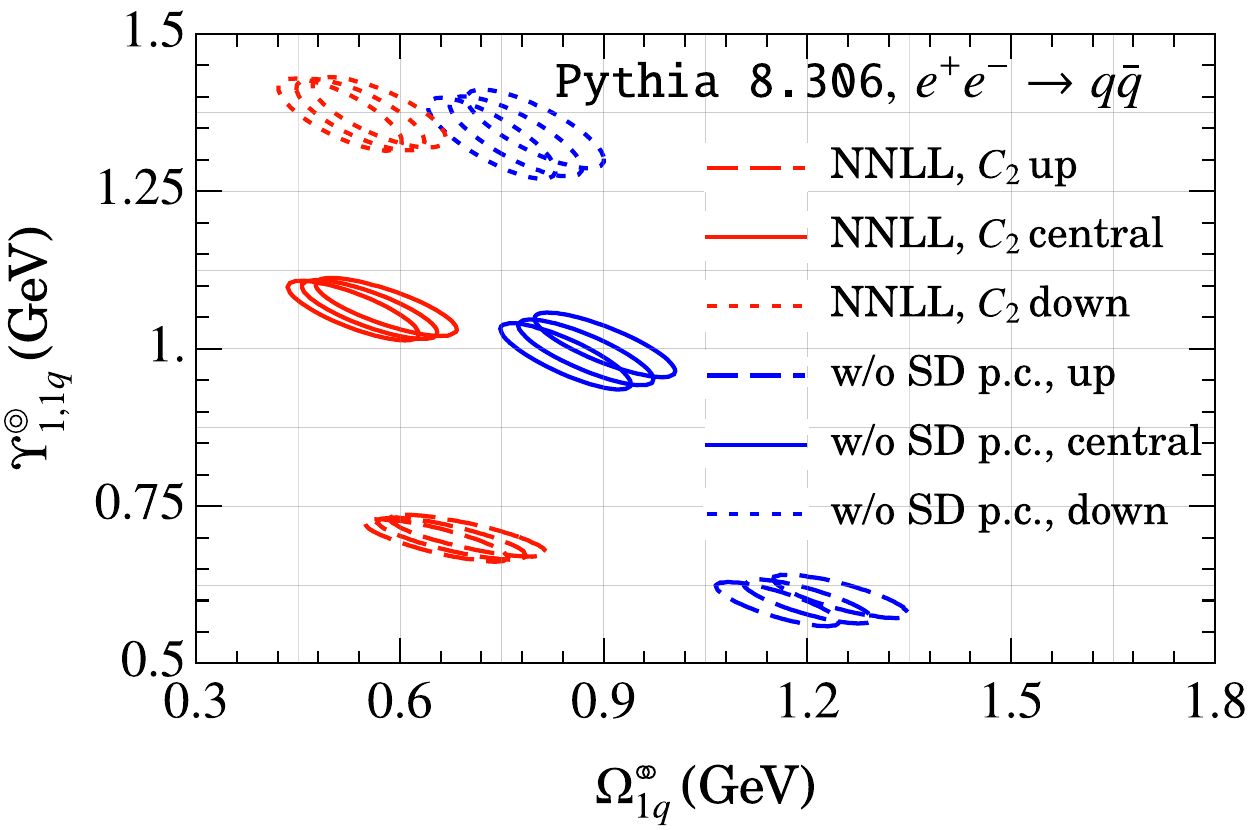}
\caption{Results of fitting to \Pythiaxx using NNLL $C_\kappa^{(n)}$ coefficients and with dropping soft drop power corrections.}
\label{fig:parthad}
\end{figure}

We can further investigate this effect by carrying out simple Monte Carlo fitting analysis along the lines of \Refcite{Ferdinand:2023vaf}. More specifically, we treat the parton shower prediction in the MC as the perturbative calculation and fit for the nonperturbative parameters $\{\Oq,\Uqa,\Uqb\}$ in $\ee \ra q\bar q$ process by comparing it with the hadron level curve and using analytical calculations of the Wilson coefficients.
We consider two different cases where
calculation of Wilson coefficients in \Refcite{Pathak:2020iue} without soft drop power corrections and the
NNLL calculation of this work is employed. When fitting using the NNLL result of this work, we take the fit
region to be $\xi \in (\xi_{\rm SDOE}, \xi_0')$, where $\xi_{\rm SDOE}$ was defined above in \eq{xiSDOE}
and with $\rho$ in \eq{xiSDOE} chosen to be 4.5~\cite{Ferdinand:2023vaf}. On the other hand, since in
the NLL$'$ calculation of \Refcite{Pathak:2020iue} the soft drop power corrections were systematically
dropped, it is natural to choose the fit range to be $\xi \in (\xi_{\rm SDOE}, \xi_0)$. Additionally, following
the procedure of \Refcite{Ferdinand:2023vaf}, we interpret the uncertainty in the extractions of these
moments by varying the Wilson coefficients within their respective uncertainty bands shown in
\fig{c1c2eelegacy}. In \fig{parthad} we show the results of this fitting procedure.
The dominant uncertainty comes from variations in $C_{2\kappa}^{(-1)}$ which are shown as dotted and dashed ellipses, whereas the uncertainty in $C_{1\kappa}^{(1)}$ being small leads to tighter variations. From the spread of the ellipses we can gauge that the two fit results are compatible. However, the fit results for NLL$'$ computation, where soft drop power corrections are also dropped, lead to larger uncertainty (and a slightly lower $\chi^2$ value). Thus, the improved calculation with matching in the plain jet mass region enables a more precise calibration of hadronization models in terms of soft drop nonperturbative parameters. At the same time, through this comparison it is reassuring to see that by including power corrections, we do not significantly alter the outcome the determination of these parameters.
A discussion on how the choice of lower end of the fit range $\xi_{\rm SDOE}$ impacts the fit results can be found in \Refcite{Ferdinand:2023vaf}.

Finally, we note that in \fig{mJhad} we have not truncated the hadronization corrections in the ungroomed region, and have incorporated as shown in \eq{NP} also for jet masses larger than the respective cusp locations. While the prescription of \eq{NP} is strictly valid only in the SDOE region, this approach nevertheless captures the main features of transition from the groomed to ungroomed region. This is because the $C_{2\kappa}^{(-1)}$ coefficient which is specific to effects of soft drop automatically becomes small in the ungroomed region for $\xi > \xi_0, \xi_0'$, whereas the $C_{1\kappa}^{(1)}$ coefficient tends to $1$, the value at the jet radius boundary. This is consistent with the behavior expected in the ungroomed region as described in \eq{NPUngroomed}. However, because of the jet boundary, the dipole geometry valid in the SDOE region gets distorted, and as such $\Omega_{1\kappa}^{\figeight}$ is a distinct moment from $\Omega_\kappa^{(1)}$ for the ungroomed region. Nevertheless, the power corrections are very small in this region and numerically the differences between these two moments may be ignored.
Hence, our NNLL result valid throughout the jet mass spectrum provides a starting point for matching the nonperturbative power corrections across the two regions.

\section{Comparison with simulations}
\label{sec:mc}

In this section we show a comparison of the numerical implementation of the NNLL calculation of the
moments above against simulations of quarks and gluon jets in $\ee$ and $\pp$ collisions in \Pythiaxx and
\Herwigxx. For \Pythiaxx simulations we employ the default $p_T$-ordered shower and for
\Herwigxx the default angular-ordered shower. Both parton showers are simulated with LO
hard process. Below we only show results for parton-level predictions and do not turn on
hadronization and underlying event models.
In the simulations we reconstruct anti-$k_T$~\cite{Cacciari:2008gp} jets using
\texttt{Fastjet}~\cite{Cacciari:2011ma}, and analyze them using jet analysis software
\texttt{JETlib}~\cite{jetlib}. We note that since the parton shower results are $\sim$ LL accurate, we do not
anticipate a perfect agreement with the more precise NNLL results. Nevertheless, we will see below that
the parton shower results are largely in agreement with analytical calculations within perturbative
uncertainty.

\subsection{Results for $C_1$}

\begin{figure}[t]
\centering
\includegraphics[width=\linewidth]{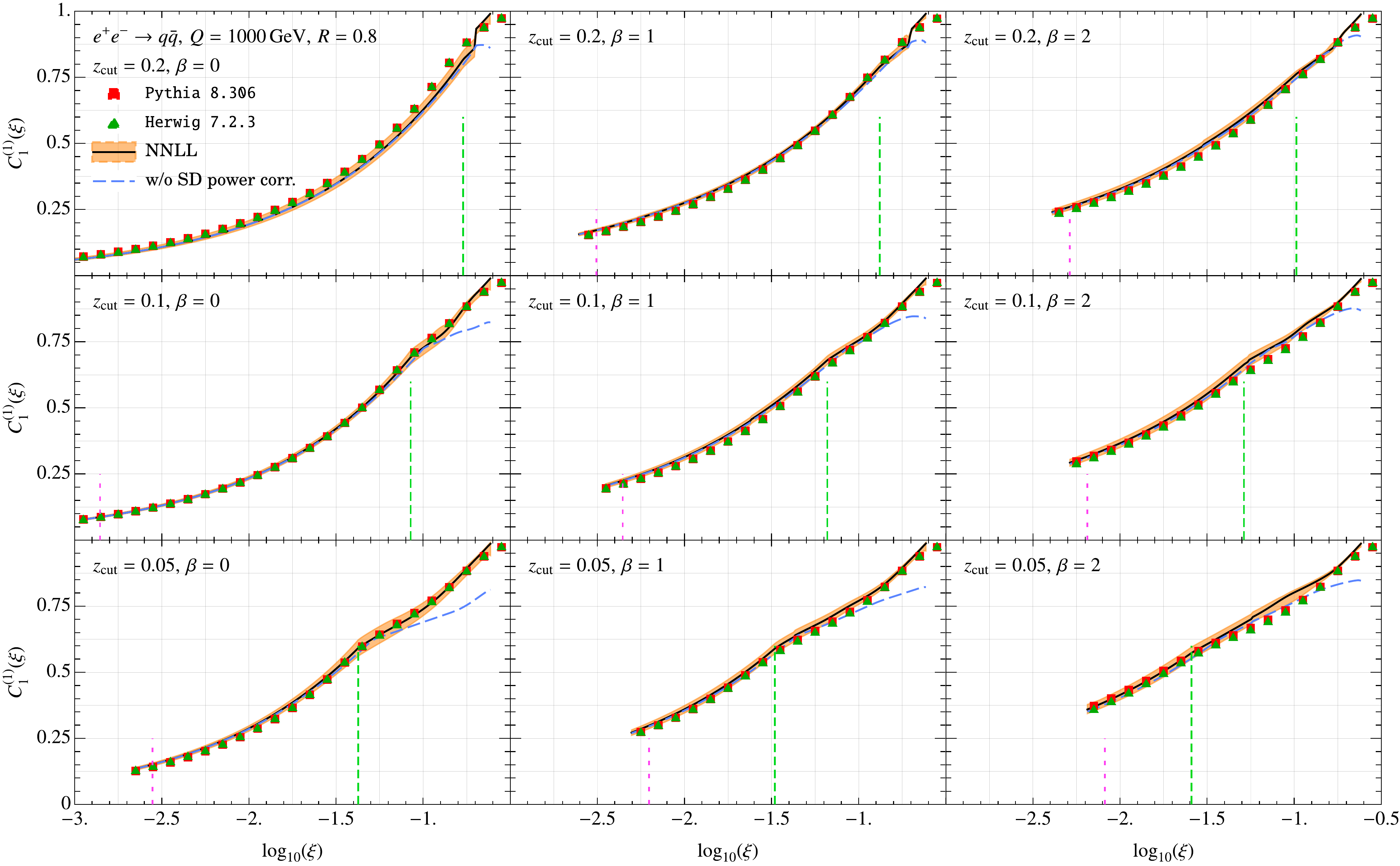}
\includegraphics[width=\linewidth]{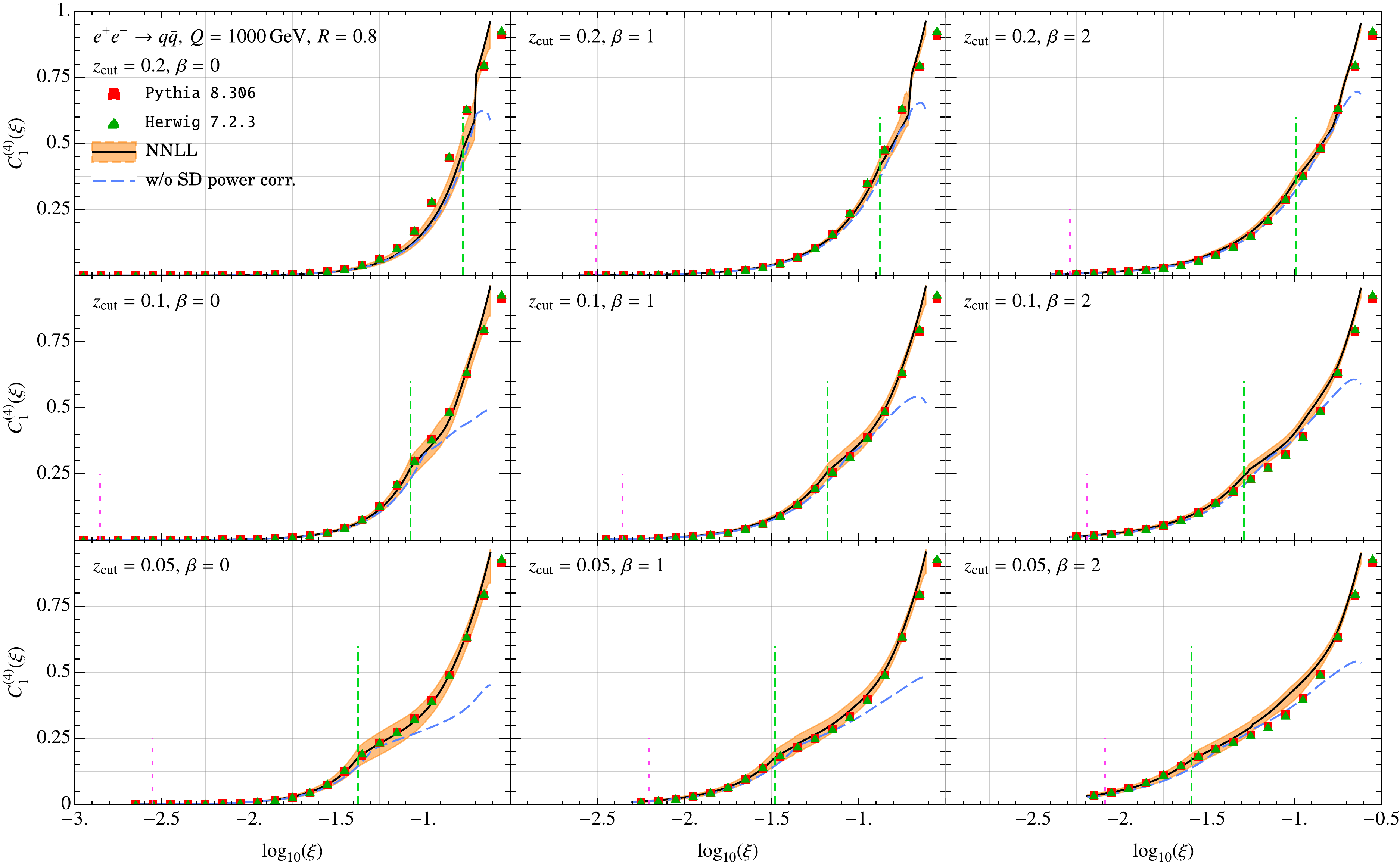}
\caption{Comparison of NNLL prediction of $C_1^{(1)}$ (top) and $C_1^{(4)}$ (bottom) against parton level simulations of quark jets in $\ee$ collisions in \Pythiaxx and \Herwigxx.}
\label{fig:c1eeqq}
\end{figure}

\begin{figure}[t]
\centering
\includegraphics[width=\linewidth]{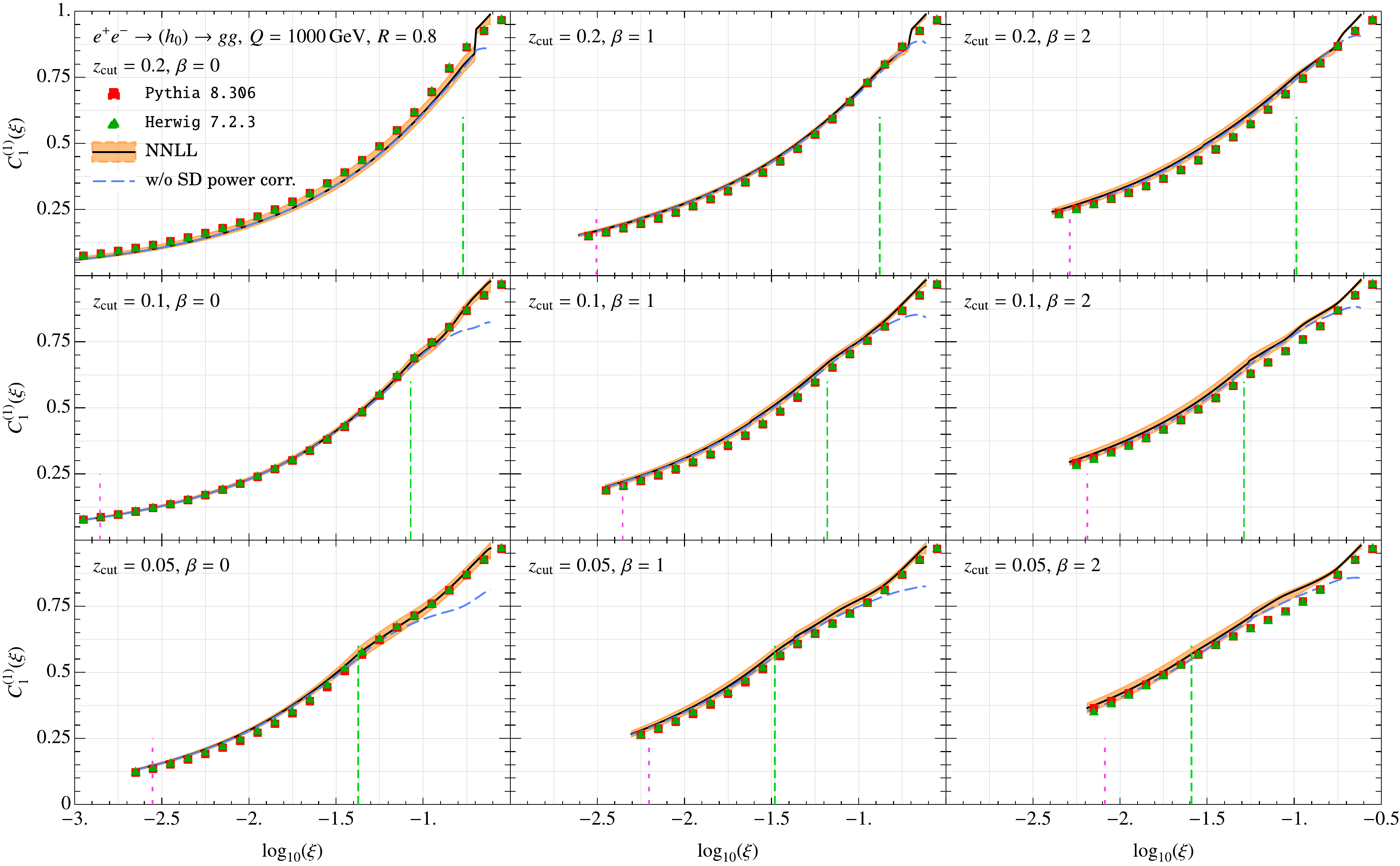}
\includegraphics[width=\linewidth]{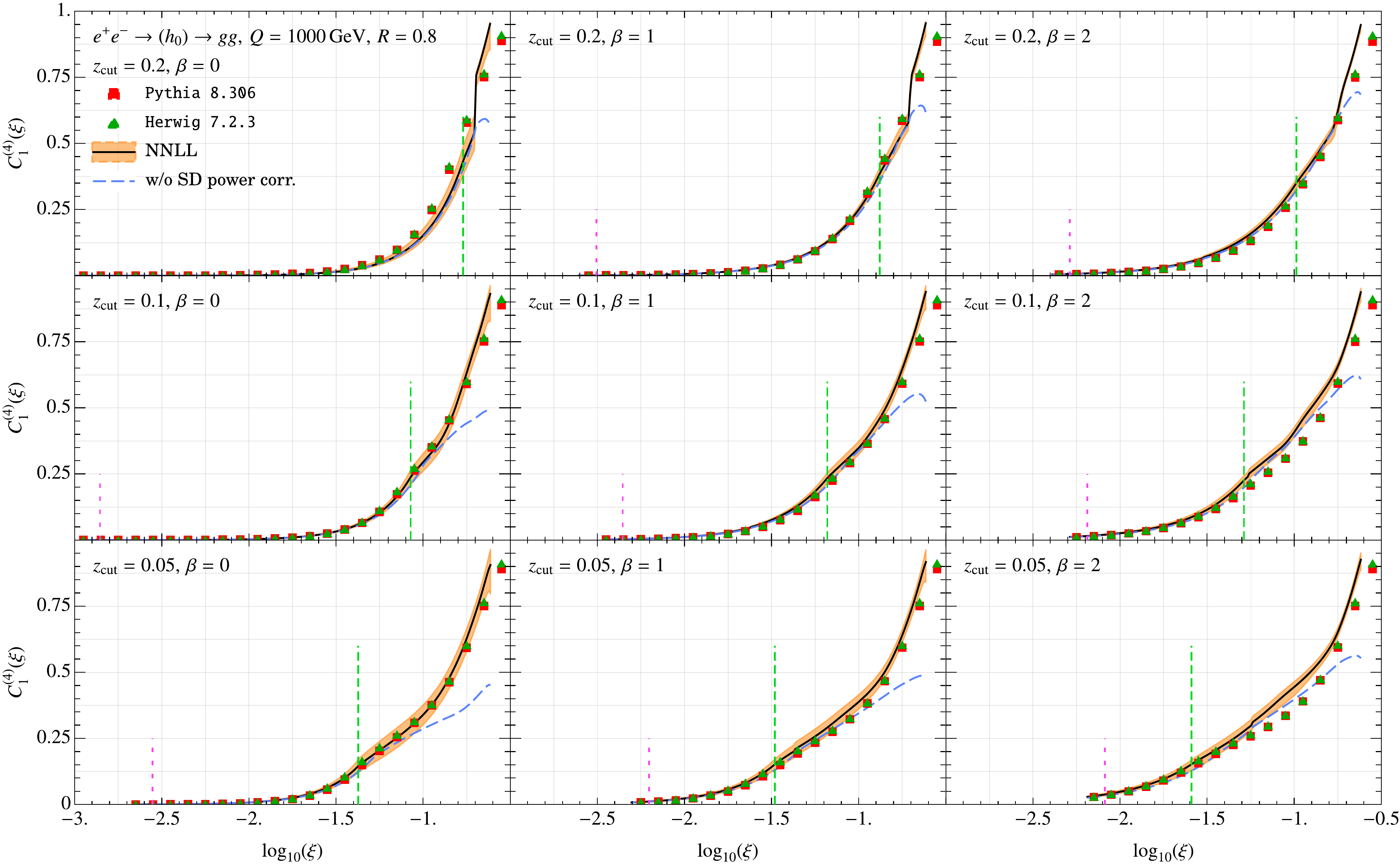}
\caption{Comparison of NNLL prediction of $C_1^{(1)}$ and $C_1^{(4)}$ against parton level simulations of gluon jets in isolation in \Pythiaxx and \Herwigxx.}
\label{fig:c1eegg}
\end{figure}

In \fig{c1eeqq} we show compare the analytical prediction for $C_{1\kappa}^{(n)}$ for phenomenologically relevant cases of $n = 1,4$ for quark jets in $\ee \ra q\bar q$ parton level process simulated at leading order for all combinations of $\zcut \in \{0.05, 0.1, 0.2\}$ (rows) and $\beta \in \{0,1,2\}$ (columns).
We will stick to leading order treatment of the hard scattering since the main goal of this work is the hadronization model and its interface with the parton shower which can be well isolated with the help of soft drop in the SDOE region.
Within the SDOE region shown between the vertical lines we find the MC extraction to agree very well with the analytical result, and we expect this to work even when higher order hard matching corrections are included.
Beyond the cusp, however, we expect that the prediction for $\xi_0 < \xi \lesssim 1$ to be modified with more precise matching to account for additional jets, though
continue to find a very good agreement for $\ee$ collisions for our LO simulations. We have also shown the result for the central curve of NLL$'$ calculation without soft drop power corrections in dashed line.
We also simulate the $\ee \ra h_0 \ra gg$ process to study gluon jets in isolation shown in \fig{c1eegg} and find similar results as quark jets.

\begin{figure}[t]
\centering
\includegraphics[width=\linewidth]{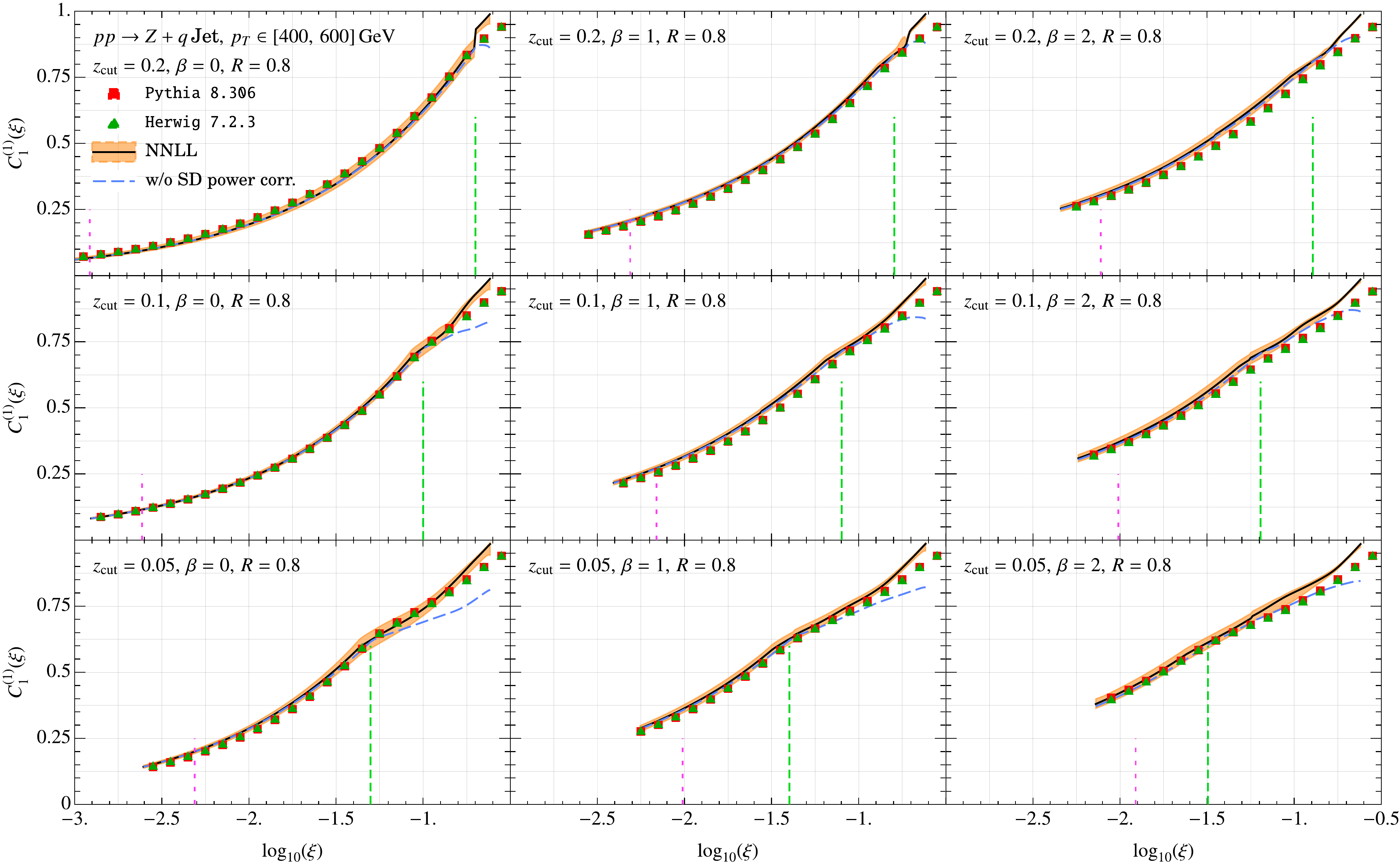}
\includegraphics[width=\linewidth]{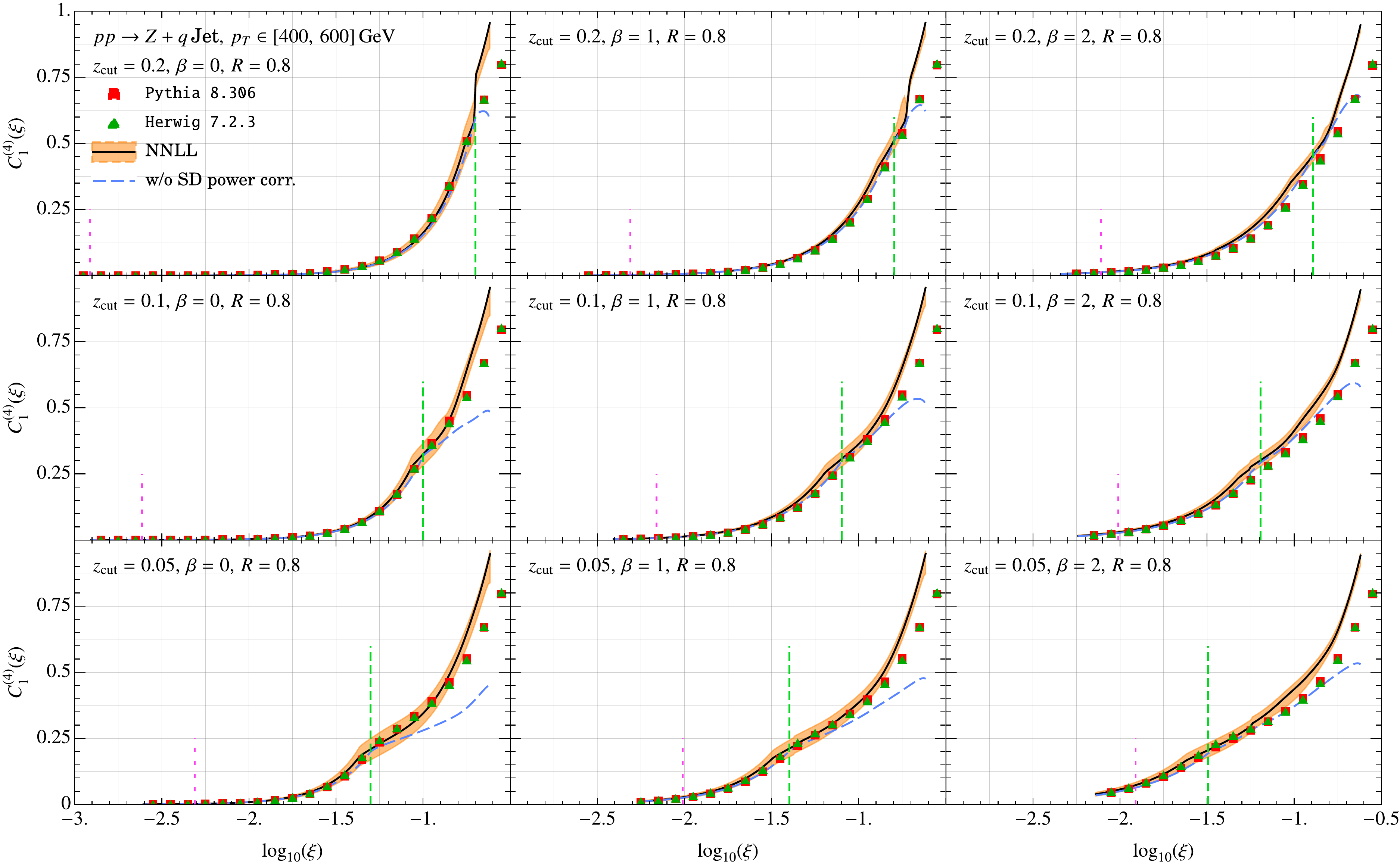}
\caption{Comparison of NNLL prediction of $C_1^{(1)}$ (top) and $C_1^{(4)}$ (bottom) against parton level simulations of quark jets in $pp$ collisions in \Pythiaxx and \Herwigxx.}
\label{fig:c1ppq}
\end{figure}

In \figs{c1ppq}{c1ppg} we compare the analytical prediction for quarks and gluon jets in $pp$ collisions against simulations. Here we simulate $pp \ra Z + q/g$ jet process and sample the leading jet.
We see that both MC simulations yield almost identical results and agree with the NNLL result in the SDOE region, though there are some small noticeable differences for $\beta = 2$ case.
We find that despite the differences in the theory setup for inclusive jets and the jet selection criteria in simulations they remain close to the analytical result in the plain jet mass region for many of the $\zcut, \beta$ combinations.

\begin{figure}[t]
\centering
\includegraphics[width=\linewidth]{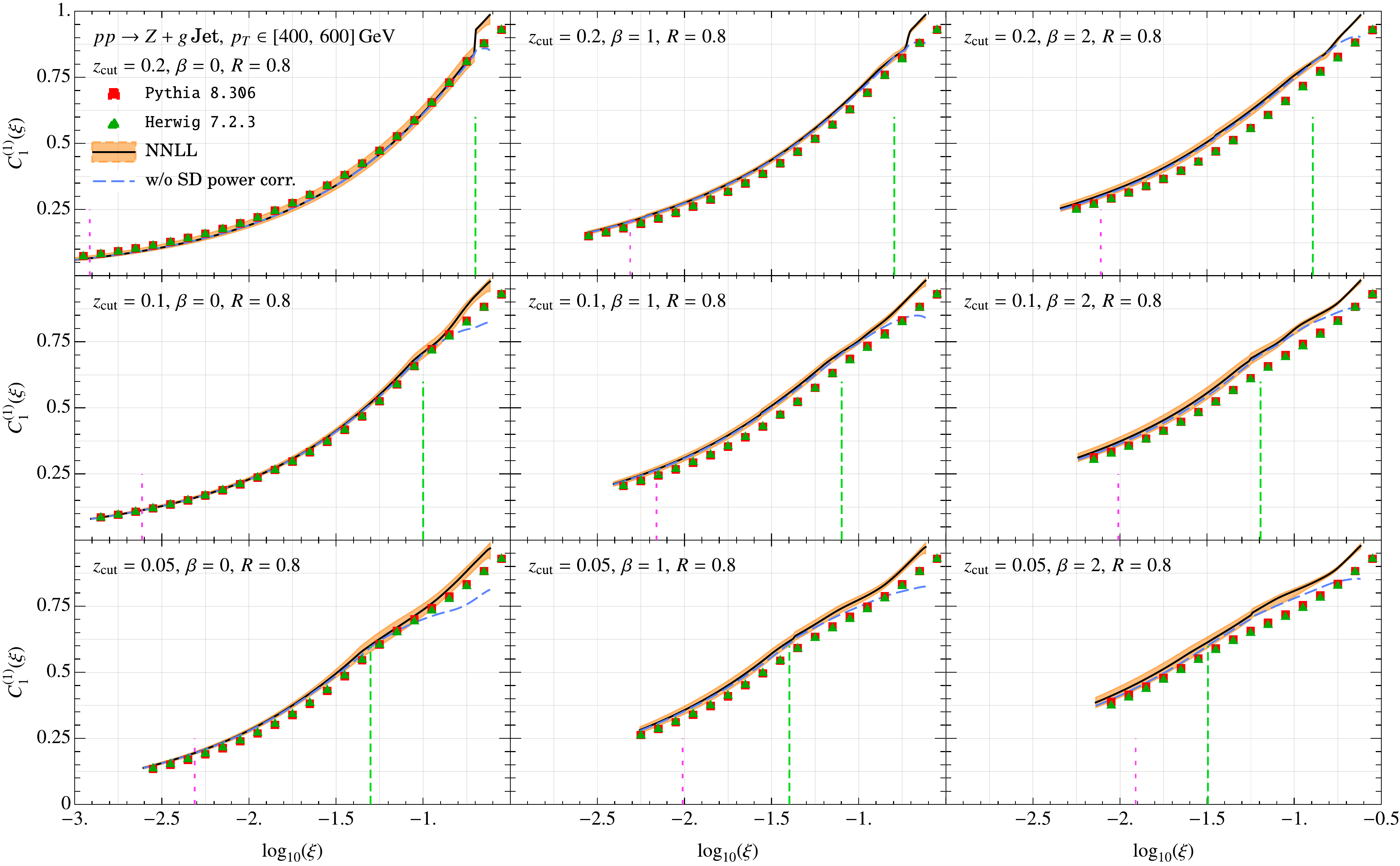}
\includegraphics[width=\linewidth]{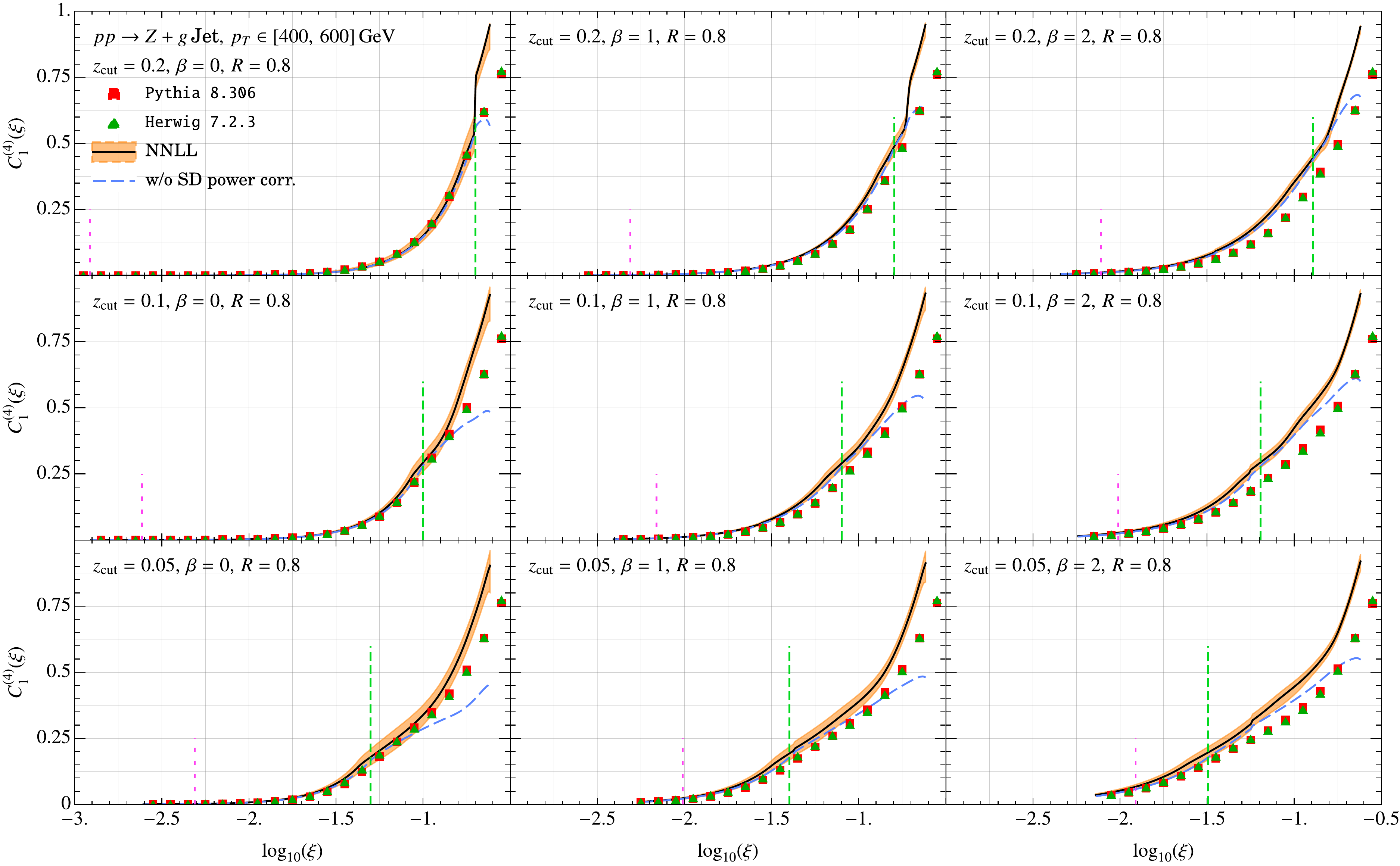}
\caption{Comparison of NNLL prediction of $C_1^{(1)}$ and $C_1^{(4)}$ against parton level simulations of gluon jets in $pp$ collisions in \Pythiaxx and \Herwigxx.}
\label{fig:c1ppg}
\end{figure}
\subsection{Results for $C_2$}
Next, we show a comparison of NNLL results for $C_{2\kappa}^{(n)}$ for $n = -1, 2$ and parton level simulations for the same processes as considered above.
Here we use an improved method for extracting of $C_{2\kappa}^{(n)}$ moments from parton showers than what was originally employed in Ref.~\cite{Hoang:2019ceu}. In Ref.~\cite{Hoang:2019ceu} the soft drop condition was shifted by a small amount $\veps/r_g$ to result in a new constraint $\Theta(z - \xi_0 r_g^\beta + \veps/r_g)$
which upon numerically differentiating with respect to $\veps$ resulted in the $C_{2\kappa}^{(-1)}$ moment. This approach however fails to work for positive moments that we consider because for small angles and $n >0$ the shift $\veps r_g^n$ becomes too small for numerical differentiation. Instead, we calculate $C_{2\kappa}^{(n)}$ here via a weighted cross-section given by
\begin{align}\label{eq:C2MC}
C_{2\kappa}^{{\rm MC}, (n)}(\xi) = \frac{\xi}{\df \hat \sigma_\kappa^{\rm MC}/\df \xi} \int \df r_g \: \sum_{\veps_i} \frac{a^{\rm fd}_i}{\big(r_g^{\rm max}(\xi)\big)^\beta \veps } \, r_g^{n} w_{{\rm sd}\ra {\rm plain}} (\xi,r_g) \frac{\df \hat \sigma^{\rm MC}_\kappa(\veps_i )}{\df r_g \df \xi}
\end{align}
where $\frac{\df \hat \sigma^{\rm MC}_\kappa(\veps_i )}{\df r_g \df \xi}$ is obtained by differentiating the jet with the shifted soft drop condition
\begin{align}
\Theta_{\rm sd}^{\rm MC} (\veps) \equiv \Theta \bigg(z - \xi_0 r_g^\beta + \veps \big(r_g^{\rm max}(\xi)\big)^\beta\bigg) \, .
\end{align}
for a range of uniformly spaced $\{\veps_i\}$ choices and the corresponding finite difference coefficients $\{a_i^{\rm fd}\}$ to build in numerical derivative.
Doing so ensure that the shifted term remains commensurate in size with the $\xi_0r_g^\beta$ term.

\begin{figure}[t]
\centering
\includegraphics[width=\linewidth]{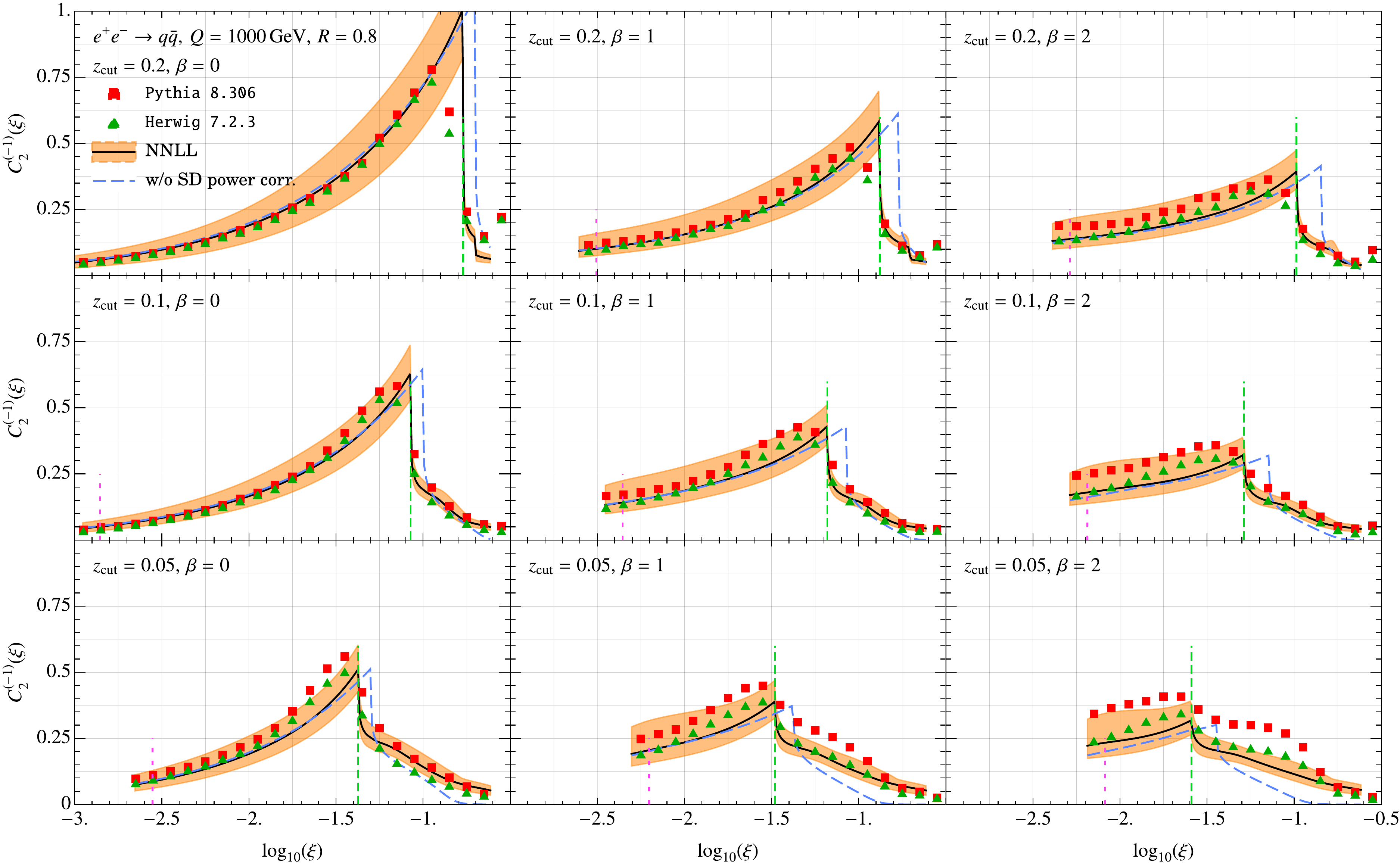}
\includegraphics[width=\linewidth]{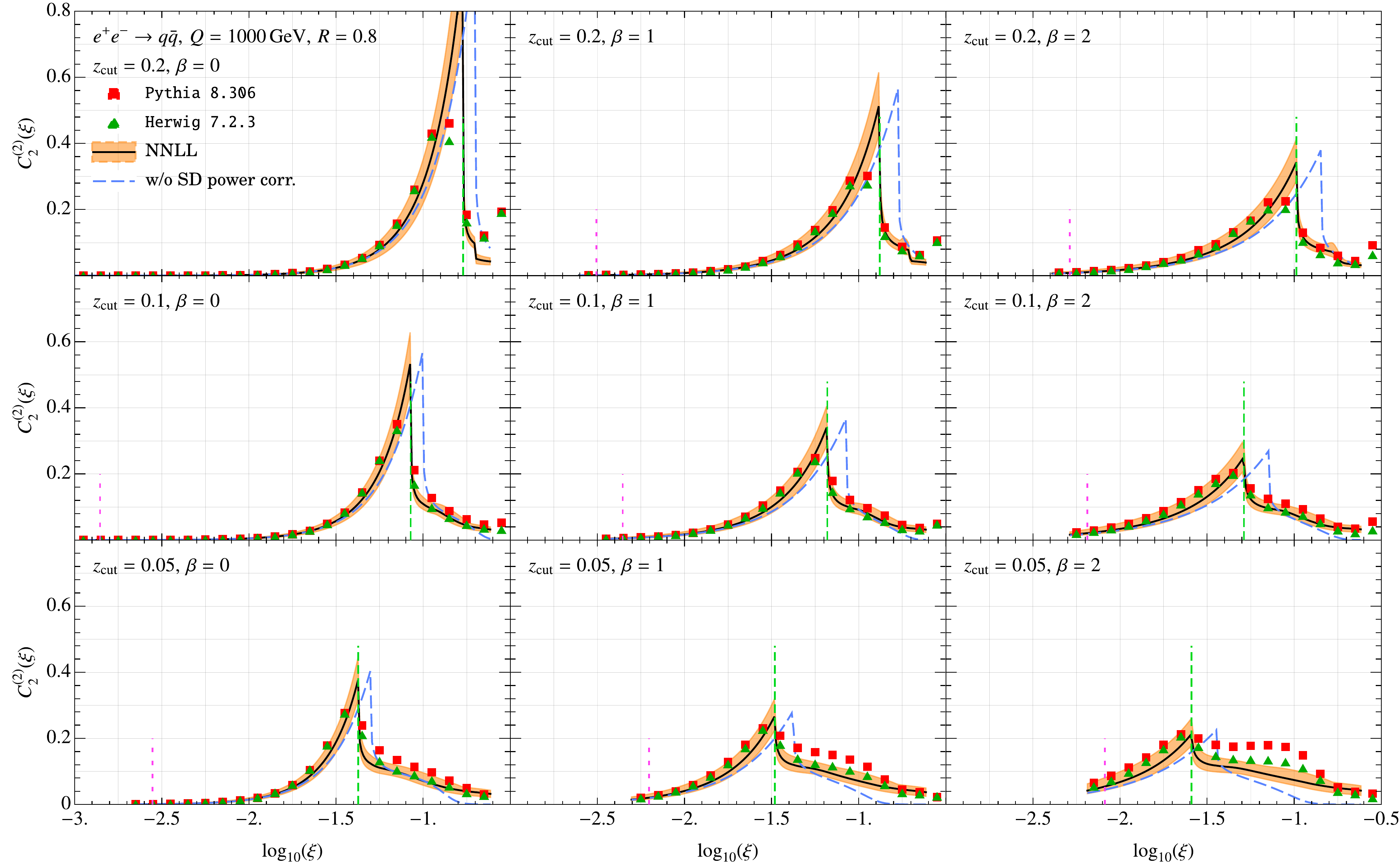}
\caption{Comparison of NNLL prediction of $C_{2\kappa}^{(-1)}$ and $C_{2\kappa}^{(2)}$ against parton level simulations in \Pythiaxx and \Herwigxx.}
\label{fig:c2eeqq}
\end{figure}

\begin{figure}[t]
\centering
\includegraphics[width=\linewidth]{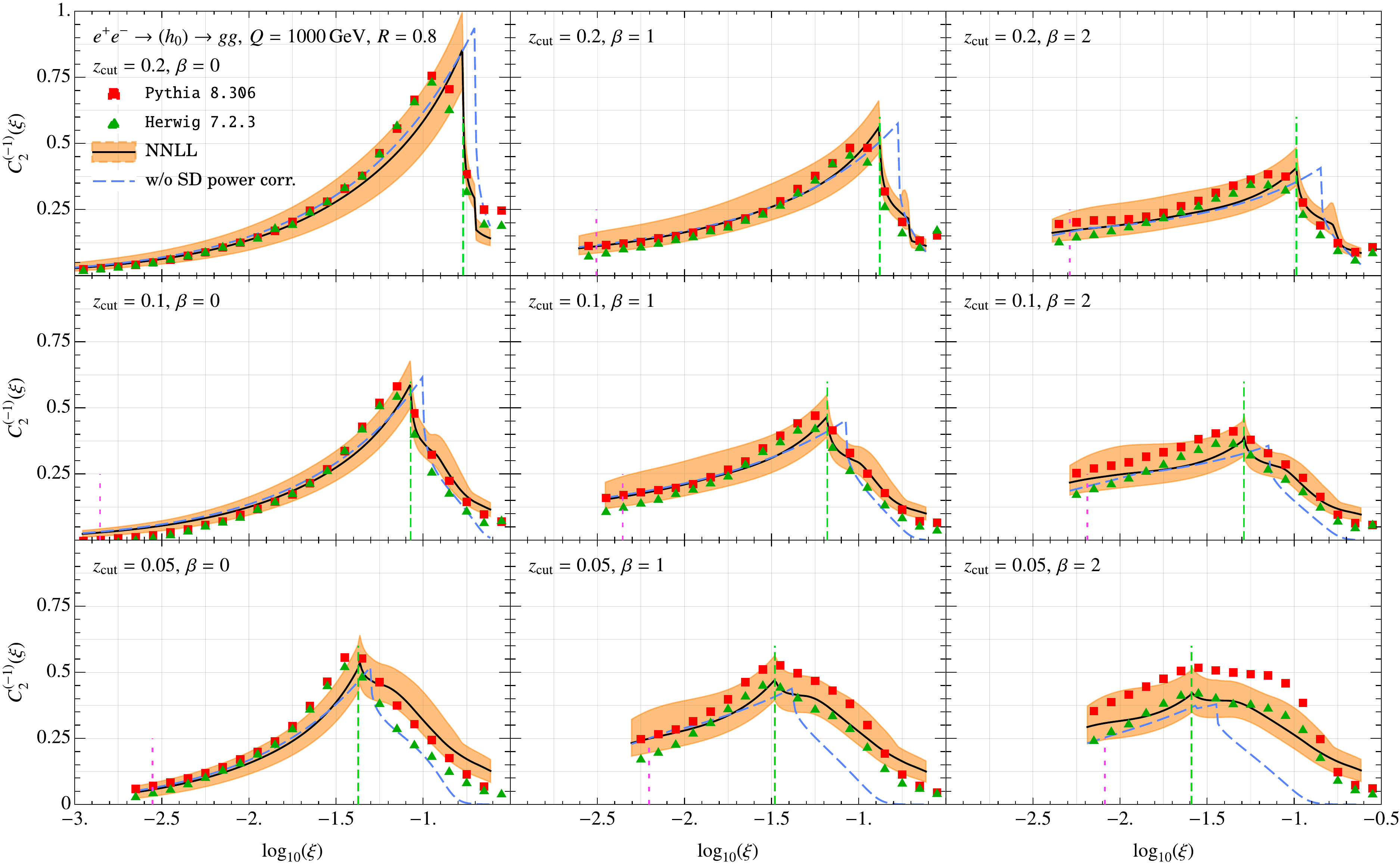}
\includegraphics[width=\linewidth]{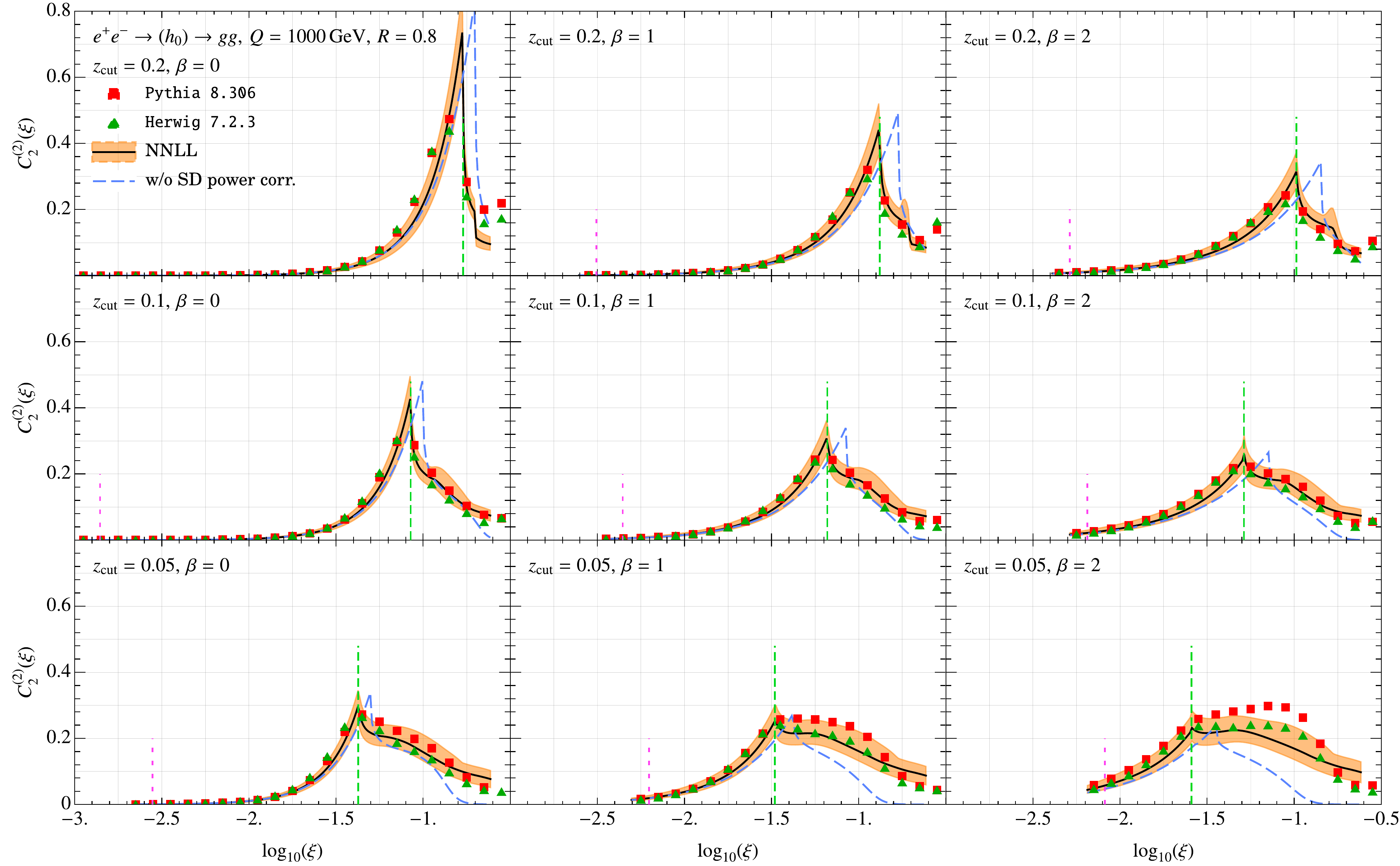}
\caption{Comparison of NNLL prediction of $C_{2\kappa}^{(-1)}$ and $C_{2\kappa}^{(2)}$ against parton level simulations of gluon jets in isolation in \Pythiaxx and \Herwigxx.}
\label{fig:c2eegg}
\end{figure}

\begin{figure}[t]
\centering
\includegraphics[width=.95\linewidth]{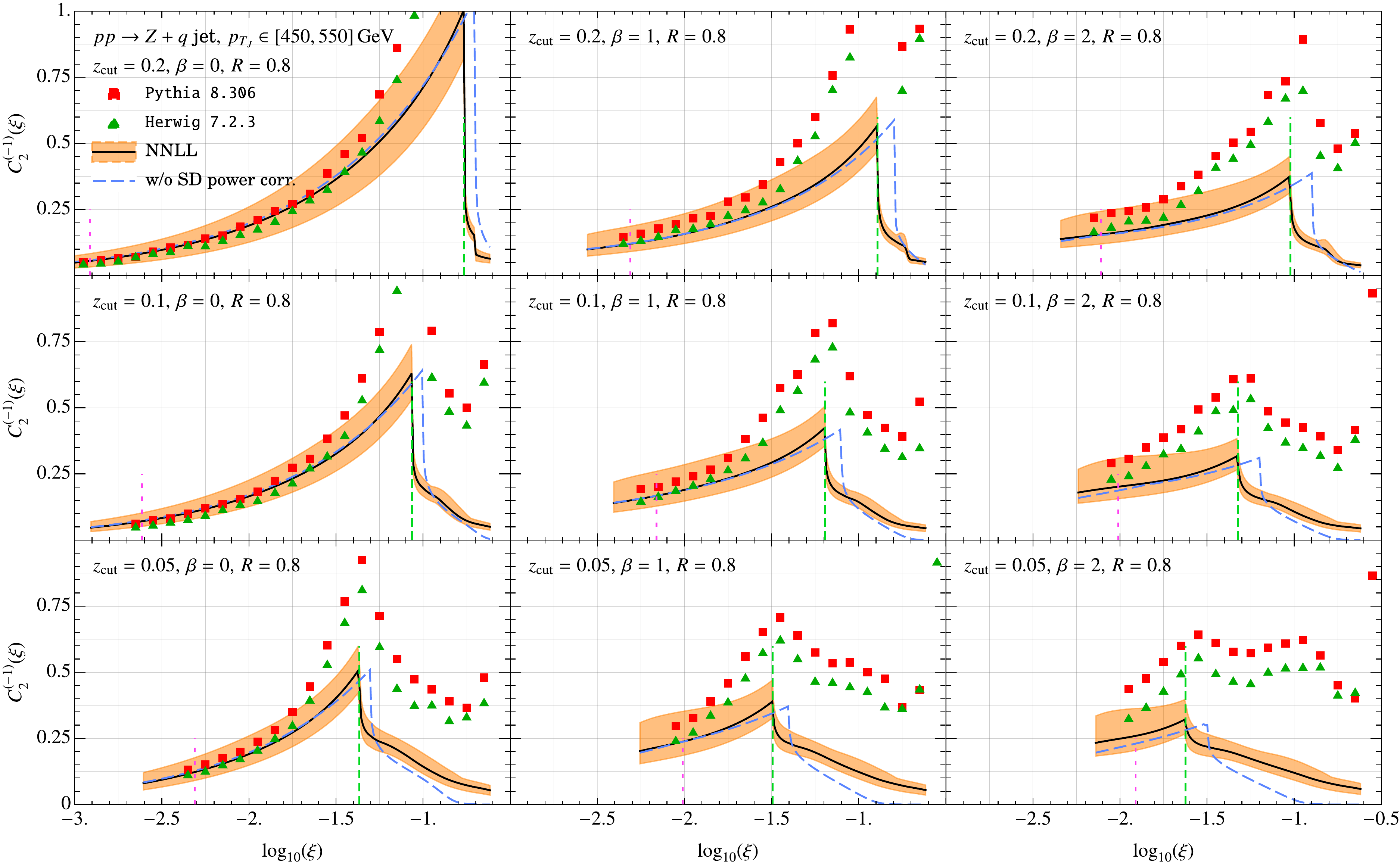}
\includegraphics[width=.95\linewidth]{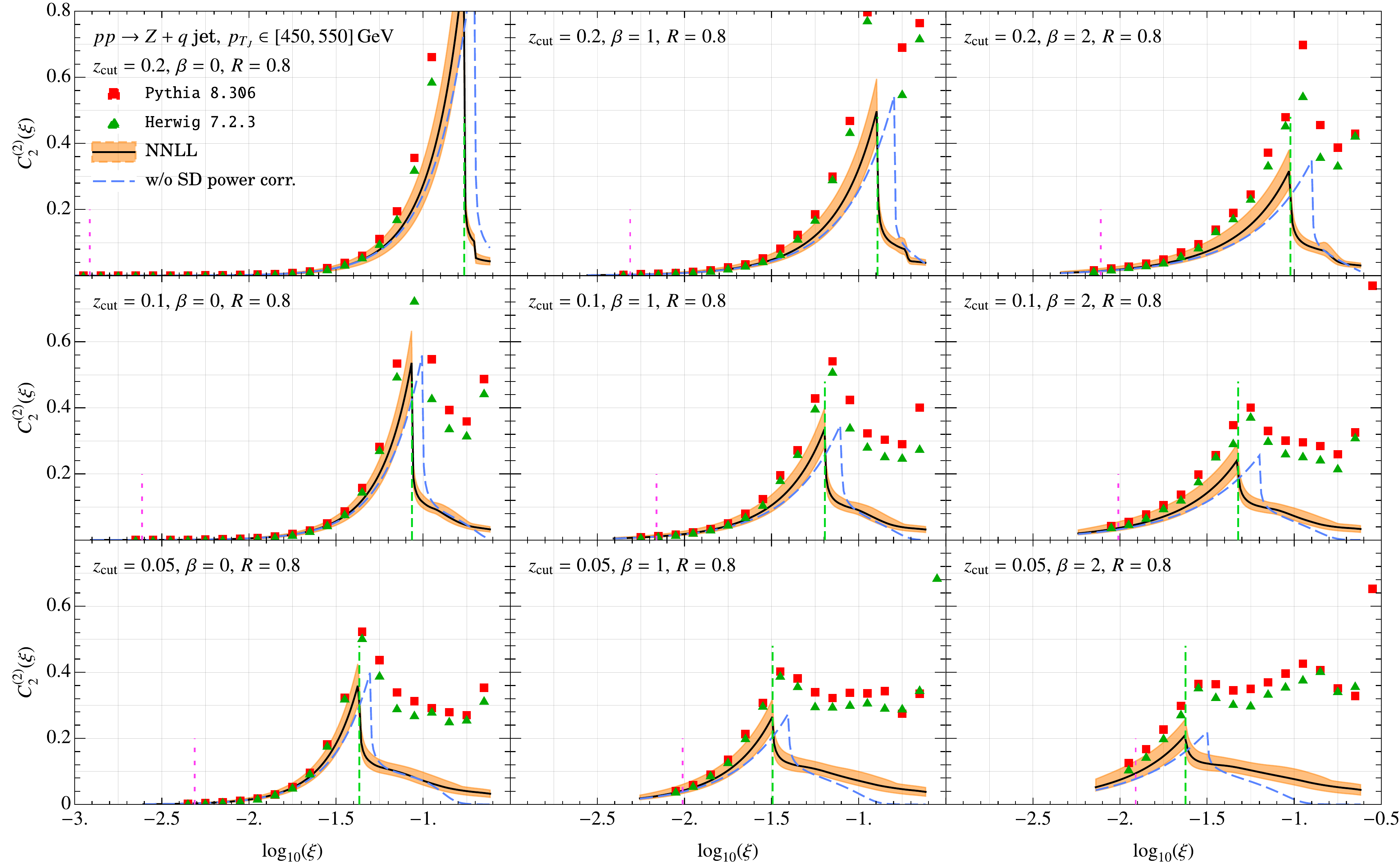}
\caption{Comparison of NNLL prediction of $C_{2\kappa}^{(-1)}$ and $C_{2\kappa}^{(2)}$ against parton level simulations of quark jets in \Pythiaxx and \Herwigxx in $pp$ collisions. The additional contribution beyond the cusp region arises from ungroomed initial state radiation.}
\label{fig:c2ppqq}
\end{figure}

\begin{figure}[t]
\centering
\includegraphics[width=.95\linewidth]{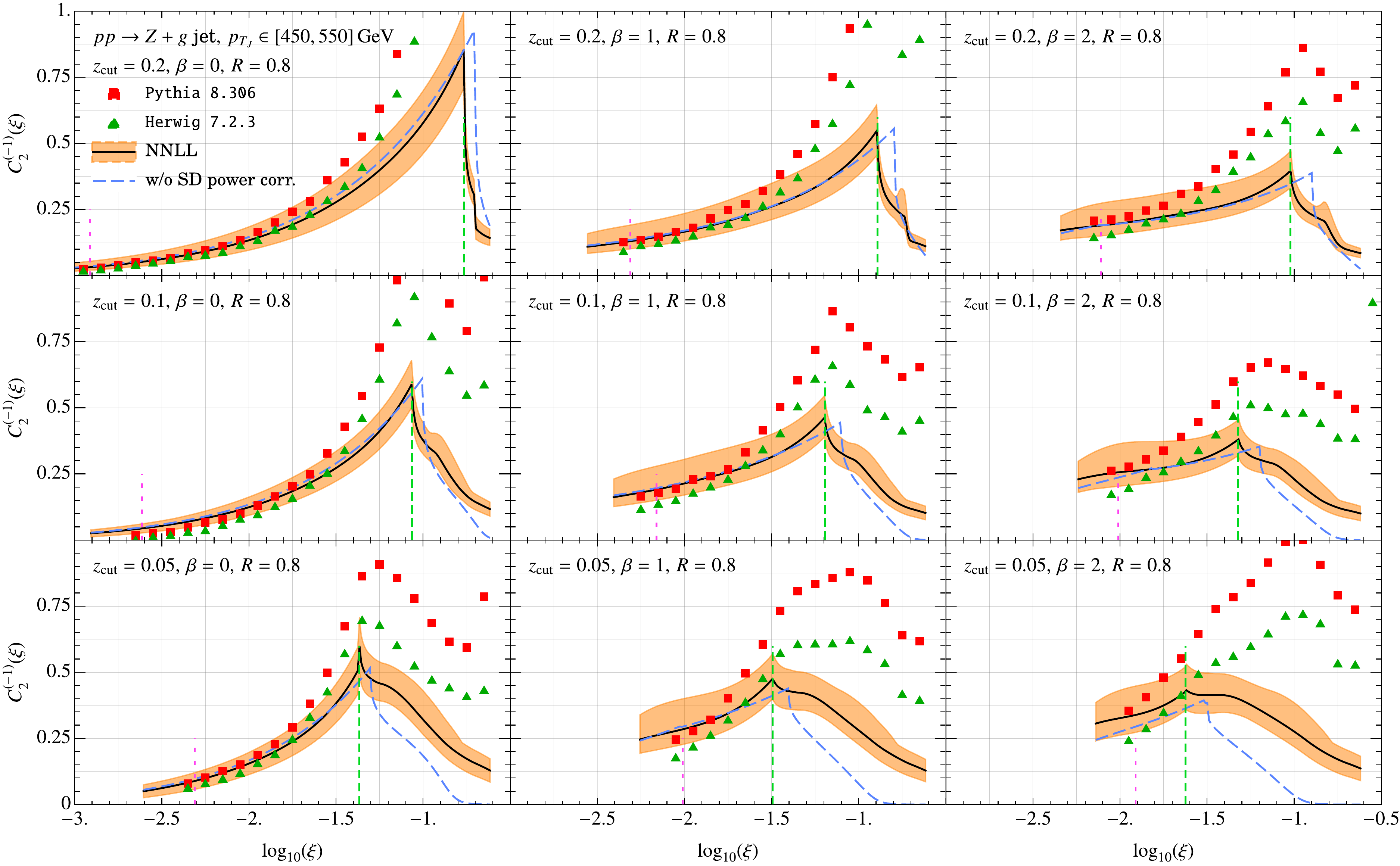}
\includegraphics[width=.95\linewidth]{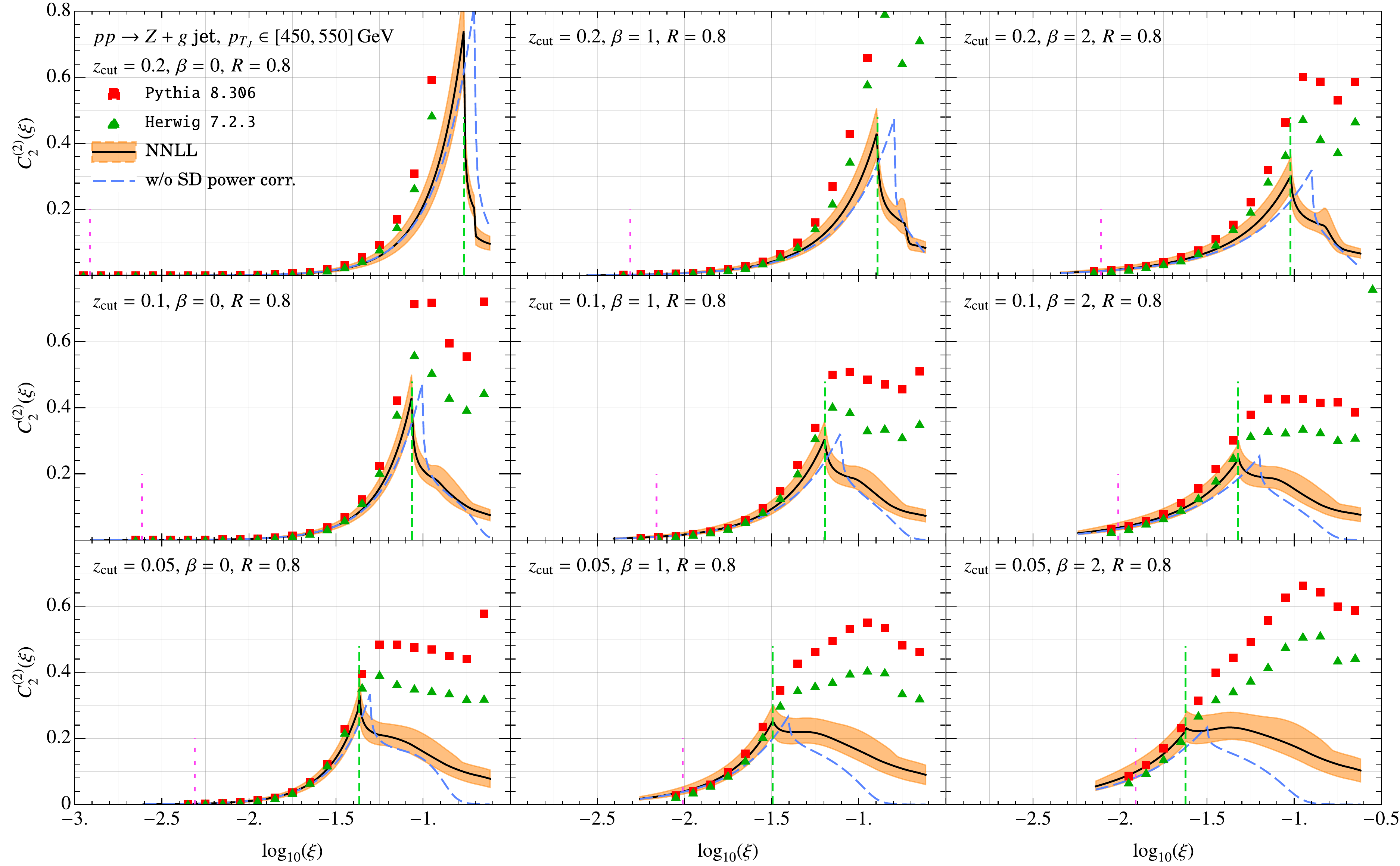}
\caption{Comparison of NNLL prediction of $C_{2\kappa}^{(-1)}$ and $C_{2\kappa}^{(2)}$ against parton level simulations of gluon jets in \Pythiaxx and \Herwigxx in $pp$ collisions. The additional contribution beyond the cusp region arises from ungroomed initial state radiation.}
\label{fig:c2ppgg}
\end{figure}

The results for $\ee \ra q \bar q$ process are shown in \fig{c2eeqq}.
Unlike in the case of $C_{1\kappa}^{(n)}$, the simulations in \Pythia and \Herwig do not agree with each other for $\beta > 0$.
Nevertheless, for all combinations of $\zcut$ and $\beta$ shown the simulations agree with analytical calculations within the NNLL uncertainty band, with the \Herwig result being closer to the central curve.
Comparing $n = -1$ and $n =2$ we see a better agreement between the analytical results and simulations for $n = 2$.
This is due to the fact that for $n = -1$ there is larger weight associated with smaller angles in \eq{C2MC} which we need to manually eliminate using the weight function to capture only the max-$R_g$ regime contribution.
We have also displayed for comparison the NLL$'$ result from Ref.~\cite{Pathak:2020iue}. We see that including soft drop power corrections results in a more accurate location of the cusp and agrees better with simulations. This is particularly noticeable for $\beta = 1,2$.
In \fig{c2eegg} we show the results for gluon jets simulated in $\ee$ collisions and can draw similar conclusions as the quark case.

In \figs{c2ppqq}{c2ppgg} comparison of analytical results with simulations in $pp$ collisions for quark and gluon jets is shown. Contrary to the $\ee$ case the simulations differ significantly in the plain jet mass region due to additional inevitable contribution from the initial state radiation. In the SDOE region, the analytical result agrees with the simulations, except however for the $n = -1$ moment, there is a disagreement for jet masses closer to the cusp.
These differences for $pp$ collisions between our analytical results and MC simulations are however not troublesome
since we expect the contribution of ISR to nonperturbative effects to be proportional to $C_{2\kappa}^{(2)}$, which is suppressed relative to that of the collinear-soft radiation within the jet.
Thus, as far as the nonperturbative corrections associated with boundary soft drop are concerned, the same moment $C_{2\kappa}^{(-1)}$ describes both $\ee$ and $pp$ scenario. For $n = 2$ moment, we however find a comparatively better agreement between MC and analytical calculation in the SDOE region.

\section{Conclusion}

\label{sec:conclusion}

In this paper we have calculated the key ingredients required to describe hadronization and underlying event corrections to the groomed jet mass spectrum in a model independent framework. These ingredients are differential jet mass cross sections weighted by certain moments of groomed jet radius and occur in combination with universal nonperturbative $\cO(\LQCD)$ constants for hadronization corrections and analogous parameters for underlying event contribution to the groomed jet mass.
A precise knowledge of these perturbative weights is crucial for precision measurements with soft drop jet mass as higher order calculations become available. The results for the moments associated with hadronization derived in this paper were used in Ref.~\cite{Hannesdottir:2022rsl} to ascertain the ultimate precision possible on the strong coupling constant determination when all the nonperturbative power corrections are left unconstrained.
At the same time, this field-theory based formalism imposes highly nontrivial constraints on the form and functional dependence of power corrections on jet kinematics and grooming parameters. This property of nonperturbative corrections can be used as a benchmark tool for improving modeling of hadronization and its interface with parton showers in event generators.
The NNLL predictions for the perturbative weights calculated in this paper were employed in Ref~\cite{Ferdinand:2023vaf} for a detailed calibration of hadronization models.

We calculated the perturbative weights associated with power corrections by computing the NNLL result for the doubly differential cross section and the boundary soft drop cross section. We improved upon the earlier calculation in Ref.~\cite{Pathak:2020iue} and extended the previous result into the ungroomed region and performed a more accurate treatment of the prediction near soft drop cusp.
We also accounted for the $\cO(\as)$ non-singular corrections and parameterized the effect of two-loop non-logarithmic pieces via nuisance parameters.
We compared the phenomenologically relevant $R_g$-moments of these cross sections against parton level extractions from \Pythiaxx and \Herwigxx simulations. The parton shower extractions of these weights, though less precise, agreed well within the uncertainty bands for $\ee \ra q \bar q$ and $\ee \ra gg$ processes within the SDOE region.
We saw some disagreement near the cusp for $C_{2\kappa}^{(-1)}$ moment for $pp$ collisions due to inevitable contribution from the initial state radiation in simulations. On the other hand, the moments that are relevant for ISR and the underlying event were found to be in better agreement in the SDOE region for simulations of jets in $pp$ collisions.

In comparing the results of this work with the previous computation in \Refcite{Pathak:2020iue} we found that the results in the soft drop resummation region did not differ significantly. However, by including the ungroomed region, we ensure a more comprehensive treatment of the problem. This matching also described the significant shift in the soft drop cusp arising at NNLL accuracy, that cannot be systematically accounted for otherwise. Additionally, in the region soft drop resummation region, the fact that the new calculation falls within tthe uncertainty of the previous results can be seen as a consistency check, giving us increased confidence in the validity of our theoretical framework. It is also reassuring to see that the new calculation, though more intricate, does not introduce any unexpected large power corrections in the soft drop resummation region that were previously neglected. Finally, the result of the perturbative weights extended into the ungroomed region provides a starting point for equivalently extending the power corrections beyond the soft drop cusp.

In summary, we believe that this paper provides motivation for carrying out analyses with unfolded data
from colliders.
With the prospects of high quality data to be delivered from high-luminosity phase of the LHC, it will be
very exciting to use the approach presented in this paper to further our understanding of hadronization
effects as well as carry out precision measurements using groomed observables.

\section*{Acknowledgements}
\enlargethispage{20pt}

I am grateful to Anna Ferdinand and Kyle Lee for their collaboration during initial stages. I especially thank Anna Ferdinand for cross checking the results of the computations against Monte Carlo simulations, and Kyle Lee for discussions on the effective field theory regions and modes, and cross checking various intermediate calculations.
I immensely grateful to Simon Pl\"atzer for technical support with \Herwig.
A numerical implementation of the NNLL calculation described above in \texttt{C++} building on core classes of \texttt{SCETlib}~\cite{scetlib} will be made available as a part of the \texttt{scetlib::sd} module~\cite{scetlibSD}. I thank Johannes Michel for support with above mentioned software.
I acknowledge support from DESY (Hamburg, Germany), a member of the Helmholtz Association HGF. I was previously a member of the Lancaster-Manchester-Sheffield Consortium for Fundamental Physics, which is supported by the UK Science and Technology Facilities Council (STFC) under grant number ST/T001038/1.

\appendix

\section{Anomalous dimensions}

\label{app:anomdim}

In this appendix we consolidate the anomalous dimensions up to $\cO(\as^2)$ for NNLL resummation of the various factorization functions appearing in our analysis as well as results of certain functions in Laplace space. We refer the reader to App.~A of Ref.~\cite{Pathak:2020iue} for a details of the notation for anomalous dimension and RG evolution kernels.

The cusp anomalous dimensions are given by
\begin{align}
\Gamma_{N_{\rm incl}^\kappa} [\alpha_s]&= -2 \, C_\kappa \Gamma^{\text{cusp}}[\alpha_s] \,,\\
\Gamma_{J^\kappa} [\alpha_s]&= +2 \, C_\kappa \Gamma^{\text{cusp}}[\alpha_s] \,, \nn\\
\Gamma_{\cC^\kappa} [\alpha_s]&= +2 \, C_\kappa \Gamma^{\text{cusp}}[\alpha_s] \,, \nn\\
\Gamma_{S_{c_m}^\kappa}[\alpha_s] &= -2 C_\kappa \Gamma^{\rm cusp} [\alpha_s] \, , \nn\\
\Gamma_{S_G^\kappa} [\alpha_s] &= + 2 C_\kappa \Gamma^{\rm cusp} [\alpha_s] \, , \nn \\
\Gamma_{S_{c_g}^\kappa} [\alpha_s] &= - 2 C_\kappa \Gamma^{\rm cusp} [\alpha_s] \, , \nn \\
\Gamma_{S_{c}^\kappa}[\alpha_s] &= -2 C_\kappa \Gamma^{\rm cusp} [\alpha_s] \, , \nn
\end{align}
where with the convention
\begin{align}\label{eq:GammaCusp}
\Gamma^{\rm cusp} [\alpha_s] = \sum_{n = 0}^\infty \Gamma^{\rm cusp}_n\Big(\frac{\alpha_s}{4\pi}\Big)^{n+1} \, ,
\end{align}
up to NNLL we have~\cite{Korchemsky:1987wg,Moch:2004pa,Henn:2019swt}
\begin{align}
\Gamma^{\rm cusp}_0 &= 4 \, , \\
\Gamma^{\rm cusp}_1 &= 8\bigg[\Big(\frac{67}{18} - \frac{\pi^2}{6}\Big)C_A - \frac{5}{9} n_f\bigg] \nn \\
\Gamma^{\rm cusp}_2&= 16 \bigg[\Big(\frac{245}{24} - \frac{67}{54} + \frac{11\pi^4}{180} + \frac{11}{6}\Big)C_A^2 + \Big(-\frac{209}{108} + \frac{5\pi^2}{27} - \frac{7}{3}\zeta_3\Big)C_A n_f \nn\\
&\quad \qquad+ \Big(\frac{-55}{27} + 2\zeta_3\Big) C_F n_f -\frac{1}{27} n_f^2 \bigg] \, .\nn
\end{align}

Next, we state the non-cusp anomalous dimensions following the same series expansion as in \eq{GammaCusp}. The non-cusp anomalous dimensions of the normalization factor $N_{\rm incl}^\kappa$ for quark and gluon jets are given by
\begin{align}
\gamma_0^{N^q} = - 6 C_F \,, \qquad
\gamma_0^{N^g} = -2 \beta_0 \, ,
\end{align}
\begin{align}
\gamma_1^{N^q} &= \big(-3 + 4\pi^2 - 48 \zeta_3) C_F^2
+ \bigg(-\frac{961}{27} - \frac{11\pi^2}{3} + 52 \zeta_3 \bigg) C_F C_A + \bigg(\frac{260}{27} + \frac{4\pi^2}{3}\bigg) C_F n_f T_F \, , \nn \\
\gamma_1^{N_g} &= \bigg(-\frac{118}{9} + 4 \zeta_3\bigg) C_A^2 + \bigg( -\frac{38}{9} + \frac{\pi^2}{3}
\bigg) C_A \beta_0 - 4 \beta_1 \, .
\end{align}
For the jet function we have
\begin{align}
\gamma_0^{J_q} &= 6 C_F \, ,\qquad
\gamma_0^{J_g} = 2 \beta_0 \, ,
\end{align}
\begin{align}
\gamma_1^{J_q} & = C_F \left[ C_F \left(3-4 \pi^2+48 \zeta_3 \right) + C_A \left( \frac{1769}{27} + \frac{22 \pi^2}{9} - 80 \zeta_3 \right) + T_R n_f \left(-\frac{484}{27} - \frac{8 \pi^2}{9} \right) \right] \,,\nn
\\
\gamma_1^{J_g} & = C_A^2 \left( \frac{2192}{27} - \frac{22 \pi^2}{9} - 32 \zeta_3 \right) + C_A T_R n_f \left(-\frac{736}{27}+ \frac{8 \pi^2}{9} \right) - 8 C_F T_R n_f \,.
\end{align}
The one-loop non-cusp anomalous dimension of all the soft functions are zero (with the exception of the boundary soft drop non-cusp anomalous dimension for $\beta = 0$ in \eq{gamma0eps}). For $S_{c_m}^\kappa$ (and $S^\kappa_{\rm plain}$), we have
\begin{align}
\gamma_1^{S_{c_m}^\kappa} &= C_\kappa \bigg( C_A \Big(-\frac{808}{27} + \frac{11 \pi^2}{9} + 28 \zeta_3\Big)
+ n_f T_F \Big(\frac{224}{27} - \frac{4\pi^2}{9}\Big)
\bigg) \, .
\end{align}
At two-loops a numerical approximation of the global soft anomalous dimension reads~\cite{Bell:2018vaa,Hannesdottir:2022rsl}:
\begin{align}
\gamma_1^{S_G^\kappa}(\beta) = \frac{C_\kappa}{1 +\beta} \bigg( \gamma_{C_F}^{S_G^\kappa} (\beta)+n_f \gamma_{T_F}^{S_G^\kappa}(\beta) + \gamma_{C_A}^{S_G^\kappa}(\beta)
\bigg) \, ,
\end{align}
where
\begin{align}\label{eq:gamma1SG}
\gamma_{C_F}^{S_G^\kappa} (\beta)
&= C_F \big( 0.00563338 \beta^3 - 0.621462 \beta^2 - 1.11337 \beta + 16.9974\big) \, ,
\\
\gamma_{T_F}^{S_G^\kappa}(\beta)
&= T_F \big(- 0.26041 \beta^3 + 2.01765 \beta^2 + 3.48117 \beta - 10.9341\big) \, , \nn
\\
\gamma_{C_A}^{S_G^\kappa}(\beta)
&= C_A \big(
+0.640703 \beta^3 +3.37308 \beta^2 +3.68876 \beta - 20.4351
\big) \, . \nn
\end{align}

The anomalous dimension of the hard-collinear function $\cC^\kappa(\xi,r_g, Q,\mu)$ is simply the negative of that of the normalization factor $N_{\rm incl}^\kappa$. Likewise the anomalous dimension of $S_{c_g}^\kappa$ is negative of that of the global soft function:
\begin{align}
\gamma_{\cal C^{\kappa}}[\alpha_s] &= - \gamma_{N^\kappa}[\alpha_s] \, , \\
\gamma_{S_{c_g}^\kappa}[\alpha_s] &= - \gamma_{S_G^\kappa}[\alpha_s] \, . \nn
\end{align}
Finally, the non-cusp anomalous dimension of the collinear soft function $S_c^\kappa$ can be obtained using RG consistency of the max-$R_g$ cross section:
\begin{align}
\gamma_1^{S_c^\kappa}(\beta) = - \gamma_1^{N^\kappa} - \gamma_1^{J_\kappa} - \gamma_1^{S_G^\kappa}(\beta) \, .
\end{align}

\section{Profile functions}
\label{app:prof}

Here we discuss the profile functions that incorporate the appropriate canonical scales for various factorization functions as well as enable us to estimate perturbative uncertainty through their varaitions. Here we simply summarize the final formulae of original implementation discussed in great detail in Ref.~\cite{Pathak:2020iue} in the soft drop resummation region, which was further extended into the plain jet mass region in Ref.~\cite{Hannesdottir:2022rsl}.

Firstly, the hard scale and global-soft scales are defined as
\begin{align}\label{eq:muNgs}
\mu_N &\equiv e_N Q \,, \qquad
\mu_{gs} \equiv e_{N} \qcut \, , \qquad e_N \in [0.5, 2] \,,
\end{align}
where the parameter $e_N$ is varied in the range shown. In the fixed-order cross section we use the same $\mu_N$ scale.
We vary the two scales with the same parameters so as to be consistent with matching at the soft drop cusp by ensuring that $\mu_{gs}/\mu_N = \xi_0$.
\subsection{Plain jet mass profiles}
\label{app:profPlain}
Next, we summarize the profile scales and their variations associated with the plain jet mass cross section as described in Ref.~\cite{Hannesdottir:2022rsl}:
\begin{align}\label{eq:profPlain}
\tilde \mu_s^{\rm plain} (\xi ; \lambda ) &\equiv \mu_N \Big[f_{\rm vary}^{\rm plain}(\xi)\Big]^\lambda
f_{\rm run}^{\rm plain}(\xi) \, , \\
\mu_s^{\rm plain} (\xi ; \lambda) &\equiv f_{\rm freeze}\big[\tilde \mu_{s}^{\rm plain}(\xi;\lambda)\big] \nn\, , \\
\mu_J^{\rm plain}(\xi ; \lambda, \gamma ) &\equiv
\mu_N^{\frac{1}{2}+\gamma} \big(\mu^{\rm plain}_s(\xi ; \lambda)\big)^{\frac{1}{2} -\gamma} \, . \nn
\end{align}
The jet and soft scales being proportional to $\mu_N$ inherit the $e_N$ norm-variation in \eq{muNgs}. The interpolation of the canonical ungroomed soft scales between various regions is governed by $f_{\rm run}^{\rm plain}(\xi)$ given by
\begin{align}
\label{eq:frunmJ}
f_{\rm run}^{\rm plain}(\xi)
&\equiv \left\{\begin{array}{ll}
x_0 \big(1+ \frac{\xi^2}{4 x_0^2}\big) & ~~~~~~~~~~\xi \leq 2 x_0\\[2pt]
\xi & ~~~~~~~~~~ 2 x_0 < \xi \leq x_1\\[2pt]
\xi + \frac{(2 - x_2 - x_3)(\xi - x_1)^2}{2(x_2 - x_1)(x_3 - x_1)} & ~~~~~~~~~~ x_1 < \xi \leq x_2\\[2pt]
1- \frac{(2 - x_1- x_2)(\xi - x_3)^2}{2(x_3 - x_1)(x_3 - x_2)} & ~~~~~~~~~~ x_2 < \xi \leq x_3\\[2pt]
1 & ~~~~~~~~~~x_3 < \xi \leq 1\\[2pt]
\end{array}
\right.\;.
\end{align}
The regions $\xi < x_0$, $2x_0 < \xi < x_1$ and $x_3 < \xi < 1$ respectively correspond to the ungroomed-nonperturbative, ungroomed-resummation and fixed order region. $x_0$ is determined by the point where the soft scale freezes to a nonperturbative scale,
\begin{align}\label{eq:x0}
x_0 &= \frac{n_0}{(\mu_N/1 {\rm GeV})} \, ,
\end{align}
where we take the default value of $n_0 = 1$. Additionally, varying this parameter tests sensitivity of the cross section to nonperturbative effects:
\begin{align}
&\text{Vary $\as$ freezing scale:}& &n_0 \in [0.75, 1.25] \,.&
\end{align}
The other three parameters take different values depending on the transition point $\xi_0$. Firstly, $x_3$ is given by
\begin{align}
\label{eq:x123}
&\xi_0 < 0.2&&:& &x_3 = 0.2 \, ,& \\
&0.2\leq \xi_0< 0.25&&:& &x_3 = \frac{0.25 + \xi_0}{2} \, ,& \nn\\
&0.25 \leq \xi_0 &&:& &x_3 = \min \{\xi_0 + 0.1, 1\} \, , & \nn
\end{align}
and $x_1$ and $x_2$ are given by
\begin{align}
x_1 = \Theta(x_3 - 1.15 \xi_0 ) 1.15 \xi_0 + \Theta(1.15\xi_0 - x_3) \xi_0\, ,\qquad x_2 = \frac{x_1+x_3}{2} \, .
\end{align}

Having defined the canonical soft scale $\tilde \mu_s(\xi)$ we implement its variation in the plain jet mass resummation region alone, different than the overall normalization probed by $e_N$, using the trumpet function
\begin{align}
f^{\rm plain}_{\rm vary}(\xi)
\equiv \left\{ \begin{array}{l l}
1 &\qquad 0 < \xi \leq \xi_0 \\
\zeta\big(\xi, \xi_0, x_{\rm mid}; 1 , 2\big)\,, & \qquad \xi_0 \leq \xi < x_{\rm mid} \\
\zeta \big(\xi, x_{\rm mid}, x_3; 2, 1\big)\,, & \qquad x_{\rm mid} \leq \xi \leq x_3 \\
1 & \qquad x_3 < \xi \leq 1
\end{array}
\right. .
\end{align}
where
\begin{align}\label{eq:xmid}
x_{\rm mid} \equiv \frac{\xi_0 + x_3}{2} \, ,
\end{align}
and
\begin{align}\label{eq:zetaProfDef}
\zeta \big(\xi, x_{\rm start}, x_{\rm end}; a_1, a_2\big) \equiv
\left\{ \begin{array}{l l}
a_1 &\qquad \xi \leq x_{\rm start} \\
a_1 + \frac{2(a_2 - a_1)(\xi - x_{\rm start})^2}{(x_{\rm end} - x_{\rm start})^2}
\,,
& \qquad x_{\rm start} \leq \xi < \frac{x_{\rm start} + x_{\rm end}}{2} \\
a_2 - \frac{2(a_2 - a_1)(x_{\rm end} - \xi)^2}{(x_{\rm end} - x_{\rm start})^2}
\,,
& \qquad \frac{x_{\rm start} + x_{\rm end}}{2} \leq \xi \leq x_{\rm end} \\
a_2 & \qquad x_{\rm end} < \xi
\end{array}
\right. .
\end{align}
This variation is controlled via the parameter $\lambda$:
\begin{align}
&\text{Trumpet variation in plain jet mass region:}& &\lambda \in [-1, 0.3] \,.&
\end{align}
Finally we avoid the scale varying below the nonperturbative scale $n_0$ by re-freezing via the function
\begin{equation}\label{eq:ffreeze}
f_{\rm freeze}[\mu]
\equiv \left\{ \begin{array}{l r}
\mu\,, & \qquad \mu \geq 2n_0 \\
n_0 \Big(1 +\frac{\mu^2}{4n_0^2}\Big)\,, & \qquad \mu < 2n_0
\end{array}
\right. .
\end{equation}

The jet scale in \eq{profPlain} is derived from the ungroomed soft scale using the canonical see-saw relation with $\gamma = 0$. Breaking this canonical relation defines another variation:
\begin{align}\label{eq:varyGamma}
&\text{Break jet-hard-soft see-saw canonical relation:}&&\gamma \in [-0.1,0.1] \,,&
\end{align}
We have written a ``plain'' subscript on the jet scale as we use a sightly different prescription for freezing in the nonperturbative region for the jet scale in the soft drop resummed cross section below.

\subsection{Soft drop profiles}

We first summarize min-$R_g$ profiles:
\begin{align}\label{eq:profMin}
\tilde \mu_{cs_g} (r_g; \alpha) &\equiv \big (f^{\rm sd\,res.}_{\rm vary}(r_g^{1+\beta})\big)^\alpha \,\mu_{gs}\,f^{\rm sd\,res.}_{\rm run}\big(r_g^{1+\beta}\big) \, , \\
\mu_{cs_g} (r_g; \alpha ) &\equiv f_{\rm freeze}\big[\tilde \mu_{cs_g}(r_g;\alpha)\big] \, , \nn \\
\mu_{\cC}(r_g; \alpha, \gamma) &\equiv \mu_N \bigg(\frac{\mu_{cs_g}(r_g;\alpha)}{\mu_{gs}}\bigg)^{\frac{1-\gamma}{1+\gamma}\frac{1}{1+\beta}}
\, , \nn
\end{align}
Here the soft scale depends on the groomed jet radius $r_g$ instead of the jet mass, and the interpolating function is defined to be
\begin{align}
\label{eq:frunRg}
f^{\rm sd\,res.}_{\rm run}(r_g)
&\equiv \left\{\begin{array}{ll}
y_0 \Big(1+ \frac{r_g^2}{4y_0^2}\Big) & ~~~~~~~~~~r_g \leq 2 y_0\\[4pt]
r_g & ~~~~~~~~~~ 2 y_0 < r_g \leq 1
\end{array}
\right.\;.
\end{align}
Here, the parameter governing the transition into the (soft drop) nonperturbative region is given by
\begin{align}\label{eq:y0}
y_0 \equiv \frac{n_0}{(\mu_{gs}/1\, {\rm GeV})}
>\frac{\Lambda_{\rm QCD}}{\qcut}
\, .
\end{align}
As above, the trumpet variation that vanishes at the end points is governed by
\begin{equation}\label{eq:fVary}
f^{\rm sd\,res.}_{\rm vary}(r_g)
= \left\{ \begin{array}{l r}
2(1 - r_g^2)\,, & \qquad r_g < 0.5 \\
1 + 2(1- r_g)^2\,, & \qquad 0.5 \leq r_g \leq 1
\end{array}
\right. ,
\end{equation}
With $\alpha = 0$ in \eq{profMin} the new variation we have is given by
\begin{align}
&\text{Trumpet variation in the soft drop resummation region:}&
&\alpha \in [-1,1] \,.&
\end{align}
As in the case of jet scale above, the hard-collinear scale $\mu_{\cC}$ is derived from the $\mu_{cs_g}$ scale as shown in \eq{profMin} using the canonical relation for $\gamma = 0$. To ensure that the jet scale in the max-$R_g$ and int-$R_g$ region merges with the hard-collinear scale for $\gamma \neq 0$, we are required to use the same parameter for the two and vary them together as in \eq{varyGamma}.

Next, we summarize the implementation of profiles for max-$R_g$ cross section in soft drop resummation region:
\begin{align}\label{eq:profMax}
\tilde \mu^{\rm sd\,res.}_s (\xi)
&\equiv \mu_{gs} \:\Bigg[ f^{\rm sd\,res.}_{\rm run}\bigg( \Big(\frac{\xi}{\xi_0}\Big)^{\frac{1+\beta}{2+\beta}}\bigg)\Bigg]^{\frac{2+\beta}{1+\beta}} \,,\\
\tilde \mu_{cs} (\xi; \alpha , \rho) &\equiv
\Bigg[ f^{\rm sd\,res.}_{\rm vary}\bigg ( \Big(\frac{\xi}{\xi_0}\Big)^{\frac{1+\beta}{2+\beta}}\bigg) \Bigg]^{\alpha} \big(\tilde \mu^{\rm sd\,res.}_s(\xi)\big)^{\frac{1+\beta + \rho}{2+\beta}} \big(\mu_{gs}\big)^{\frac{1-\rho}{2+\beta}}
\,, \nn \\
\mu_{cs} (\xi; \alpha,\rho ) &\equiv f_{\rm freeze}\big[\tilde \mu_{cs}(\xi;\alpha,\rho)\big] \nn\, ,
\\
\mu_J (\xi;\alpha, \gamma,\rho) &\equiv \mu_N^{\frac{1}{2}+\gamma} \Bigg[ \mu_{cs}(\xi; \alpha,\rho) \: \bigg(\frac{\mu_{cs}(\xi; \alpha,\rho )}{\mu_{gs}}\bigg)^{\frac{1}{1+\beta}}\Bigg]^{\frac{1}{2}-\gamma} \, . \nn
\end{align}
In addition to the functions and variation parameters discussed already above, we have a new canonical relation between the ungroomed soft scale, collinear-soft scale and global-soft scale. This relation allows us to define the c-soft scale in terms of the other two, whereas the auxiliary ``ungroomed-soft scale'' $\tilde \mu^{\rm sd\,res.}_s (\xi)$ itself is derived using the $\mu_{cs_g}$ scale. Thus, the variation of $\rho$ with default value $\rho = 0$ now probes the effect of breaking this canonical relation:
\begin{align}
&\text{Break $\mu_{gs}$, $\mu_{cs}$ and plain soft see-saw canonical relation:}&
&\rho \in [-0.1,0.1] \,.&
\end{align}
Lastly, in the intermediate region, we use the $\mu_J, \mu_{gs}, \mu_{cs_g}$ and the hard scale above, and use the soft drop canonical see-saw relation to define the $\mu_{cs_m}$ scale:
\begin{align}\label{eq:profInt}
\mu_{cs_m}(\xi, r_g; \alpha, \rho )
&\equiv \mu_{cs}(\xi; \alpha,\rho) \: \bigg(\frac{\mu_{cs}(\xi; \alpha,\rho )}{\mu_{gs}}\bigg)^{\frac{1}{1+\beta}}\bigg(\frac{\mu_{cs_g}(r_g;\alpha)}{\mu_{gs}}\bigg)^{\frac{-(1-\rho)}{1+\rho + \beta }}
\,.
\end{align}
This concludes the summary of profile scales implementation and their variations.

\section{Weight functions}

\label{app:weight}

Here we describe how the weight functions constructed in \Refcite{Pathak:2020iue} are extended into the plain jet mass resummation region. We first list down the relevant canonical profile scales
\begin{align}
&\mu_{J}^{\rm can.} = Q \sqrt{\xi}
\,,&
&\mu_{gs}^{\rm can.} = \qcut \, ,&
&\mu_{cs_m}^{\rm can.} = Q \xi/r_g \, ,&
&\mu_{cs_g}^{\rm can.} = \qcut r_g^{1+\beta} \,,&
\end{align}
Using these scales and the size of power corrections in the intermediate-$R_g$ regime in \eq{IntPC} we can define parameters that control transition from intermediate to min and max-$R_g$ regimes. For transition from intermediate to min-$R_g$ regime, we define
\begin{align}
\lambda_{\rm min} \equiv \frac{\mu_{cs_m}}{\mu_J} = \frac{\sqrt{\xi}}{r_g} \, .
\end{align}
In transitioning to the max-$R_g$ regime, we define two parameters for cases $\xi < \xi_0'$ and $\xi > \xi_0'$:
\begin{align}
&\lambda_{\rm max}^{\rm sd\,res.} \equiv \frac{\mu_{cs_g}}{\mu_{cs_m}} = \frac{\xi_0}{\xi}r_g^{2+\beta} \,,&
&\lambda_{\rm max}^{\rm plain} \equiv \frac{\mu_{cs_g}}{\mu_{gs}}= r_g^{1+\beta} \, .&
\end{align}
Next, to determine whether or not the intermediate regime is required, we calculate the angle $r_{g,t}$ for which $\lambda_{\rm min} = \lambda_{\rm max}$:
\begin{align}
r_{g,t}^{\rm sd\,res.} (\xi) \equiv \Big[\frac{\xi}{\xi_0} \sqrt{\xi}\Big]^{\frac{1}{3+\beta}} \, , \qquad
r_{g,t}^{\rm plain} (\xi) \equiv \big(\sqrt{\xi}\big)^{\frac{1}{2+\beta}} \, .
\end{align}
which can be combined as
\begin{align}
r_{g,t} (\xi) \equiv \Big(\frac{\xi}{\xi_0}\Big)^{a_{\rm sd}(\log_{10}\xi)} (\sqrt{\xi})^{a_{\rm plain}(\log_{10}\xi)} \, ,
\end{align}
where the exponents are modified as a function of $\xi$, and are given by
\begin{align}
a_{\rm sd} (\log_{10}\xi) &\equiv a \Big(\log_{10}(\xi); \: \frac{1}{3+\beta}, 0\Big) \, , \\
a_{\rm plain}(\log_{10}\xi) &\equiv a \Big(\log_{10}(\xi); \: \frac{1}{3+\beta}, \frac{1}{2+\beta} \Big) \, , \nn
\end{align}
where $a(\log_{10}(\xi); x,y)$ is a function that smoothly transitions from the value $x$ to $y$ as $\xi$ is increased past $\xi = \xi_0^\prime$, and is defined as
\begin{align}
a(\log_{10}(\xi);\: x,y) &\equiv \zeta \big(\log_{10}\xi ,\: \big[\log_{10}(\xi_0^\prime) - \delta \xi\big], \: \big[\log_{10}(\xi_0^\prime) + \delta \xi\big]; \: x, y\big ) \, ,
\end{align}
where the function $\zeta(\xi,x_{\rm start}, x_{\rm end}; a_1 , a_2)$ was defined above in \eq{zetaProfDef} and we set $\delta \xi = 0.75$.

We will take $\lambda = 1/3$ as the reference power correction to determine the validity of the intermediate regime; i.e. if $\lambda_{\rm max}, \lambda_{\rm min} < \lambda$ then we implement resummation in the intermediate region defined by
\begin{align}\label{eq:rgPC}
r_{g,\rm PC}^{\rm min}(\xi,\lambda) \leq r_g \leq r_{g,\rm PC}^{\rm max}(\xi,\lambda) \, ,
\end{align}
where
\begin{align}
r_{g,\rm PC}^{\rm min}(\xi,\lambda) \equiv \frac{\sqrt{\xi}}{\lambda} \, ,
\end{align}
Now for the max-$R_g$ regime, we have two cases:
\begin{align}
r^{\rm sd\,res.}_{g,\rm PC}(\xi, \lambda) \equiv \Big(\frac{\xi}{\xi_0} \lambda\Big)^{\frac{1}{2+\beta}}
\, ,
\qquad
r^{\rm plain}_{g,\rm PC}(\xi,\lambda) \equiv \lambda^{\frac{1}{1+\beta}} \, ,
\end{align}
the following combination of which defines $r_{g,\rm PC}^{\rm max}$ in \eq{rgPC}:
\begin{align}
r_{g,\rm PC}^{\rm max} (\xi, \lambda) \equiv
\Big(\frac{\xi}{\xi_0}\Big)^{a_{\rm sd}^{\rm PC}(\log_{10}\xi)} \lambda^{a_{\rm plain}^{\rm PC}(\log_{10}\xi)}
\, ,
\end{align}
where
\begin{align}
a_{\rm sd}^{\rm PC} (\log_{10}\xi) &\equiv a \Big (\log_{10}\xi ; \:\frac{1}{2+\beta}, 0 \Big ) \, ,
\nn \\
a_{\rm plain}^{\rm PC} (\log_{10}\xi) &\equiv a (\log_{10}\xi ; \: \frac{1}{2+\beta} , \frac{1}{1+\beta} \Big) \, .
\end{align}
We finally introduce a transition function $X(r_g,r_{g,t})$ for a given value of $\xi$:
\begin{align}
X(r_g, r_{g,t}) = \frac{1}{2} \bigg(1 + \tanh \Big(x_t \frac{r_g - r_{g,t}}{r_g^{\rm max}(\xi) - r_g^{\rm min}(\xi)}\Big)\bigg) \, , \qquad x_t = 20 \, ,
\end{align}
and design the weight functions for the three EFT regimes as
\begin{align}
&\text{2-EFT:}& &w_{\rm max} = X (r_g, r_{g,t} (\xi))\,, & &w_{\rm int} = 0 \, ,& \\
&\text{3-EFT:}& &w_{\rm max } = X (r_g, r_{g,\rm PC}^{\rm max} (\xi, \lambda)) \, ,&
&w_{\rm int} = \big[1 - X(r_g, r_{g,\rm PC}^{\rm max}(\xi,\lambda ))\big] X(r_g, r_{g,\rm PC}^{\rm min}(\xi, \lambda)) \, ,& \nn
\end{align}
In either of these cases $w_{\rm min} = 1 - w_{\rm int} - w_{\rm max}$.
With the above construction, these weight functions turn on in their respective regime.

\bibliographystyle{JHEP}
\bibliography{sd}

\end{document}

%% file: commands.tex
\usepackage{amsmath,amssymb,mathtools} 
\usepackage{bm}
\usepackage{graphicx}
\usepackage{xspace}
\usepackage{wrapfig}
\usepackage[dvipsnames]{xcolor}
\usepackage{slashed}
\usepackage[normalem]{ulem}
\usepackage[utf8]{inputenc}
\usepackage{mathtools}
\usepackage{stmaryrd}
\usepackage{dsfont}
\usepackage{appendix}
\usepackage{xspace}
\usepackage{upgreek}
\usepackage{multirow}
\usepackage{verbatim}
\usepackage{todonotes}

\usepackage{scrextend}

\usepackage{tabularx}
\usepackage{tensor}

\usepackage{marginnote}

\definecolor{darkgreen}{rgb}{0.0, 0.4, 0.0}

\newcommand{\df}{\mathrm{d}}

\newcommand{\bn}{{\bar{n}}}

\newcommand{\ra}{\rightarrow}

\newcommand{\as}{\alpha_s}

\newcommand{\eps}{\epsilon}

\newcommand{\cL}{{\mathcal L}}

\newcommand{\nn}{\nonumber}

\newcommand{\cut}{\mathrm{cut}}
\renewcommand{\max}{\mathrm{max}}

\newcommand{\zcut}{z_{\rm cut}}
\newcommand{\qcut}{Q_{\rm cut}}

\newcommand{\Pythiaxx}{\texttt{Pythia\xspace8.3}\xspace}
\newcommand{\Pythia}{\texttt{Pythia}\xspace}

\newcommand{\Herwig}{\texttt{Herwig}\xspace}
\newcommand{\Herwigxx}{\texttt{Herwig\xspace7.2}\xspace}
\newcommand{\Sherpa}{\texttt{Sherpa}\xspace}

\def\ln{\textrm{ln}}
\def\df{\textrm{d}}
\def\nn{\nonumber}
\def\MS{\overline{\rm MS}}

\def\LQCD{\Lambda_{\rm QCD}}

\def\bndry{\varocircle}
\def\figeight{{\circ\!\!\circ}}

\newcommand{\ee}{{e^+e^-}}
\newcommand{\pp}{{pp}}

\newcommand{\Ok}{\Omega_{1\kappa}^{\figeight}}
\newcommand{\Uka}{\Upsilon_{1,0\kappa}^{\bndry}}
\newcommand{\Ukb}{\Upsilon_{1,1\kappa}^{\bndry}}
\newcommand{\Oq}{\Omega_{1q}^{\figeight}}
\newcommand{\Uqa}{\Upsilon_{1,0q}^{\bndry}}
\newcommand{\Uqb}{\Upsilon_{1,1q}^{\bndry}}

\def\veps{\varepsilon}
\allowdisplaybreaks[2]

\newcommand{\cG}{\mathcal{G}}
\newcommand{\cF}{\mathcal{F}}
\newcommand{\cR}{\mathcal{R}}
\newcommand{\cC}{\mathcal{C}}
\newcommand{\cO}{\mathcal{O}}
\newcommand{\cQ}{\mathcal{Q}}

\newcommand{\tcG}{\tilde{\mathcal{G}}}
\newcommand{\tcGk}{\tilde{\mathcal{G}}_\kappa}

\newcommand{\cH}{\mathcal{H}}
\newcommand{\cJ}{\mathcal{J}}
\newcommand{\cS}{\mathcal{S}}

\newcommand{\mC}{\textcolor{blue}{C}}
\newcommand{\mcC}{\textcolor{Purple}{\mathcal{C}}}
\newcommand{\mSp}{\textcolor{JungleGreen}{S_{\rm plain}}}
\newcommand{\mSG}{\textcolor{OliveGreen}{S_{G}}}
\newcommand{\mCS}{\textcolor{Magenta}{{\rm CS}}}
\newcommand{\mCSg}{\textcolor{BurntOrange}{{\rm CS}_g}}
\newcommand{\mCSm}{\textcolor{Red}{{\rm CS}_m}}
\newcommand{\crS}{\mathscr{S}}

%% file: main.bbl
\providecommand{\href}[2]{#2}\begingroup\raggedright\begin{thebibliography}{100}

\bibitem{Larkoski:2017jix}
A.~J. Larkoski, I.~Moult and B.~Nachman, \emph{{Jet Substructure at the Large
  Hadron Collider: A Review of Recent Advances in Theory and Machine
  Learning}},  \href{https://arxiv.org/abs/1709.04464}{{\ttfamily 1709.04464}}.

\bibitem{Kogler:2018hem}
R.~Kogler et~al., \emph{{Jet Substructure at the Large Hadron Collider:
  Experimental Review}},
  \href{https://doi.org/10.1103/RevModPhys.91.045003}{\emph{Rev. Mod. Phys.}
  {\bfseries 91} (2019) 045003}
  [\href{https://arxiv.org/abs/1803.06991}{{\ttfamily 1803.06991}}].

\bibitem{Lee:2006fn}
C.~Lee and G.~F. Sterman, \emph{{Universality of nonperturbative effects in
  event shapes}}, {\emph{eConf} {\bfseries C0601121} (2006) A001}
  [\href{https://arxiv.org/abs/hep-ph/0603066}{{\ttfamily hep-ph/0603066}}].

\bibitem{Korchemsky:1994is}
G.~P. Korchemsky and G.~F. Sterman, \emph{{Nonperturbative corrections in
  resummed cross-sections}},
  \href{https://doi.org/10.1016/0550-3213(94)00006-Z}{\emph{Nucl. Phys. B}
  {\bfseries 437} (1995) 415}
  [\href{https://arxiv.org/abs/hep-ph/9411211}{{\ttfamily hep-ph/9411211}}].

\bibitem{Korchemsky:1997sy}
G.~P. Korchemsky, G.~Oderda and G.~F. Sterman, \emph{{Power corrections and
  nonlocal operators}}, \href{https://doi.org/10.1063/1.53732}{\emph{AIP Conf.
  Proc.} {\bfseries 407} (1997) 988}
  [\href{https://arxiv.org/abs/hep-ph/9708346}{{\ttfamily hep-ph/9708346}}].

\bibitem{Korchemsky:1999kt}
G.~P. Korchemsky and G.~F. Sterman, \emph{{Power corrections to event shapes
  and factorization}},
  \href{https://doi.org/10.1016/S0550-3213(99)00308-9}{\emph{Nucl. Phys. B}
  {\bfseries 555} (1999) 335}
  [\href{https://arxiv.org/abs/hep-ph/9902341}{{\ttfamily hep-ph/9902341}}].

\bibitem{Belitsky:2001ij}
A.~V. Belitsky, G.~P. Korchemsky and G.~F. Sterman, \emph{{Energy flow in QCD
  and event shape functions}},
  \href{https://doi.org/10.1016/S0370-2693(01)00899-1}{\emph{Phys. Lett. B}
  {\bfseries 515} (2001) 297}
  [\href{https://arxiv.org/abs/hep-ph/0106308}{{\ttfamily hep-ph/0106308}}].

\bibitem{Dasgupta:2007wa}
M.~Dasgupta, L.~Magnea and G.~P. Salam, \emph{{Non-perturbative QCD effects in
  jets at hadron colliders}},
  \href{https://doi.org/10.1088/1126-6708/2008/02/055}{\emph{JHEP} {\bfseries
  02} (2008) 055} [\href{https://arxiv.org/abs/0712.3014}{{\ttfamily
  0712.3014}}].

\bibitem{Stewart:2014nna}
I.~W. Stewart, F.~J. Tackmann and W.~J. Waalewijn, \emph{{Dissecting Soft
  Radiation with Factorization}},
  \href{https://doi.org/10.1103/PhysRevLett.114.092001}{\emph{Phys. Rev. Lett.}
  {\bfseries 114} (2015) 092001}
  [\href{https://arxiv.org/abs/1405.6722}{{\ttfamily 1405.6722}}].

\bibitem{Ellis:2009me}
S.~D. Ellis, C.~K. Vermilion and J.~R. Walsh, \emph{{Recombination Algorithms
  and Jet Substructure: Pruning as a Tool for Heavy Particle Searches}},
  \href{https://doi.org/10.1103/PhysRevD.81.094023}{\emph{Phys. Rev.}
  {\bfseries D81} (2010) 094023}
  [\href{https://arxiv.org/abs/0912.0033}{{\ttfamily 0912.0033}}].

\bibitem{Cacciari:2014gra}
M.~Cacciari, G.~P. Salam and G.~Soyez, \emph{{SoftKiller, a particle-level
  pileup removal method}},
  \href{https://doi.org/10.1140/epjc/s10052-015-3267-2}{\emph{Eur. Phys. J. C}
  {\bfseries 75} (2015) 59} [\href{https://arxiv.org/abs/1407.0408}{{\ttfamily
  1407.0408}}].

\bibitem{Krohn:2009th}
D.~Krohn, J.~Thaler and L.-T. Wang, \emph{{Jet Trimming}},
  \href{https://doi.org/10.1007/JHEP02(2010)084}{\emph{JHEP} {\bfseries 02}
  (2010) 084} [\href{https://arxiv.org/abs/0912.1342}{{\ttfamily 0912.1342}}].

\bibitem{Dasgupta:2013ihk}
M.~Dasgupta, A.~Fregoso, S.~Marzani and G.~P. Salam, \emph{{Towards an
  understanding of jet substructure}},
  \href{https://doi.org/10.1007/JHEP09(2013)029}{\emph{JHEP} {\bfseries 09}
  (2013) 029} [\href{https://arxiv.org/abs/1307.0007}{{\ttfamily 1307.0007}}].

\bibitem{Larkoski:2014wba}
A.~J. Larkoski, S.~Marzani, G.~Soyez and J.~Thaler, \emph{{Soft Drop}},
  \href{https://doi.org/10.1007/JHEP05(2014)146}{\emph{JHEP} {\bfseries 05}
  (2014) 146} [\href{https://arxiv.org/abs/1402.2657}{{\ttfamily 1402.2657}}].

\bibitem{Butterworth:2008iy}
J.~M. Butterworth, A.~R. Davison, M.~Rubin and G.~P. Salam, \emph{{Jet
  substructure as a new Higgs search channel at the LHC}},
  \href{https://doi.org/10.1103/PhysRevLett.100.242001}{\emph{Phys. Rev. Lett.}
  {\bfseries 100} (2008) 242001}
  [\href{https://arxiv.org/abs/0802.2470}{{\ttfamily 0802.2470}}].

\bibitem{Frye:2017yrw}
C.~Frye, A.~J. Larkoski, J.~Thaler and K.~Zhou, \emph{{Casimir Meets Poisson:
  Improved Quark/Gluon Discrimination with Counting Observables}},
  \href{https://doi.org/10.1007/JHEP09(2017)083}{\emph{JHEP} {\bfseries 09}
  (2017) 083} [\href{https://arxiv.org/abs/1704.06266}{{\ttfamily
  1704.06266}}].

\bibitem{Dreyer:2018tjj}
F.~A. Dreyer, L.~Necib, G.~Soyez and J.~Thaler, \emph{{Recursive Soft Drop}},
  \href{https://doi.org/10.1007/JHEP06(2018)093}{\emph{JHEP} {\bfseries 06}
  (2018) 093} [\href{https://arxiv.org/abs/1804.03657}{{\ttfamily
  1804.03657}}].

\bibitem{Lee:2022ige}
K.~Lee, B.~Me\c{c}aj and I.~Moult, \emph{{Conformal Colliders Meet the LHC}},
  \href{https://arxiv.org/abs/2205.03414}{{\ttfamily 2205.03414}}.

\bibitem{Komiske:2022enw}
P.~T. Komiske, I.~Moult, J.~Thaler and H.~X. Zhu, \emph{{Analyzing N-point
  Energy Correlators Inside Jets with CMS Open Data}},
  \href{https://arxiv.org/abs/2201.07800}{{\ttfamily 2201.07800}}.

\bibitem{Holguin:2022epo}
J.~Holguin, I.~Moult, A.~Pathak and M.~Procura, \emph{{A New Paradigm for
  Precision Top Physics: Weighing the Top with Energy Correlators}},
  \href{https://arxiv.org/abs/2201.08393}{{\ttfamily 2201.08393}}.

\bibitem{Chen:2022swd}
H.~Chen, I.~Moult, J.~Thaler and H.~X. Zhu, \emph{{Non-Gaussianities in
  collider energy flux}},
  \href{https://doi.org/10.1007/JHEP07(2022)146}{\emph{JHEP} {\bfseries 07}
  (2022) 146} [\href{https://arxiv.org/abs/2205.02857}{{\ttfamily
  2205.02857}}].

\bibitem{Chen:2022jhb}
H.~Chen, I.~Moult, J.~Sandor and H.~X. Zhu, \emph{{Celestial blocks and
  transverse spin in the three-point energy correlator}},
  \href{https://doi.org/10.1007/JHEP09(2022)199}{\emph{JHEP} {\bfseries 09}
  (2022) 199} [\href{https://arxiv.org/abs/2202.04085}{{\ttfamily
  2202.04085}}].

\bibitem{Andres:2022ovj}
C.~Andres, F.~Dominguez, R.~Kunnawalkam~Elayavalli, J.~Holguin, C.~Marquet and
  I.~Moult, \emph{{Resolving the Scales of the Quark-Gluon Plasma with Energy
  Correlators}},  \href{https://arxiv.org/abs/2209.11236}{{\ttfamily
  2209.11236}}.

\bibitem{Frye:2016aiz}
C.~Frye, A.~J. Larkoski, M.~D. Schwartz and K.~Yan, \emph{{Factorization for
  groomed jet substructure beyond the next-to-leading logarithm}},
  \href{https://doi.org/10.1007/JHEP07(2016)064}{\emph{JHEP} {\bfseries 07}
  (2016) 064} [\href{https://arxiv.org/abs/1603.09338}{{\ttfamily
  1603.09338}}].

\bibitem{Frye:2016okc}
C.~Frye, A.~J. Larkoski, M.~D. Schwartz and K.~Yan, \emph{{Precision physics
  with pile-up insensitive observables}},
  \href{https://arxiv.org/abs/1603.06375}{{\ttfamily 1603.06375}}.

\bibitem{Hoang:2017kmk}
A.~H. Hoang, S.~Mantry, A.~Pathak and I.~W. Stewart, \emph{{Extracting a Short
  Distance Top Mass with Light Grooming}},
  \href{https://arxiv.org/abs/1708.02586}{{\ttfamily 1708.02586}}.

\bibitem{Cal:2021fla}
P.~Cal, K.~Lee, F.~Ringer and W.~J. Waalewijn, \emph{{The soft drop momentum
  sharing fraction $z_g$ beyond leading-logarithmic accuracy}},
  \href{https://arxiv.org/abs/2106.04589}{{\ttfamily 2106.04589}}.

\bibitem{Larkoski:2017bvj}
A.~Larkoski, S.~Marzani, J.~Thaler, A.~Tripathee and W.~Xue, \emph{{Exposing
  the QCD Splitting Function with CMS Open Data}},
  \href{https://doi.org/10.1103/PhysRevLett.119.132003}{\emph{Phys. Rev. Lett.}
  {\bfseries 119} (2017) 132003}
  [\href{https://arxiv.org/abs/1704.05066}{{\ttfamily 1704.05066}}].

\bibitem{Gutierrez-Reyes:2019msa}
D.~Gutierrez-Reyes, Y.~Makris, V.~Vaidya, I.~Scimemi and L.~Zoppi,
  \emph{{Probing Transverse-Momentum Distributions With Groomed Jets}},
  \href{https://doi.org/10.1007/JHEP08(2019)161}{\emph{JHEP} {\bfseries 08}
  (2019) 161} [\href{https://arxiv.org/abs/1907.05896}{{\ttfamily
  1907.05896}}].

\bibitem{Makris:2018npl}
Y.~Makris and V.~Vaidya, \emph{{Transverse Momentum Spectra at Threshold for
  Groomed Heavy Quark Jets}},
  \href{https://doi.org/10.1007/JHEP10(2018)019}{\emph{JHEP} {\bfseries 10}
  (2018) 019} [\href{https://arxiv.org/abs/1807.09805}{{\ttfamily
  1807.09805}}].

\bibitem{Chien:2019osu}
Y.-T. Chien and I.~W. Stewart, \emph{{Collinear Drop}},
  \href{https://arxiv.org/abs/1907.11107}{{\ttfamily 1907.11107}}.

\bibitem{Cal:2020flh}
P.~Cal, K.~Lee, F.~Ringer and W.~J. Waalewijn, \emph{{Jet energy drop}},
  \href{https://doi.org/10.1007/JHEP11(2020)012}{\emph{JHEP} {\bfseries 11}
  (2020) 012} [\href{https://arxiv.org/abs/2007.12187}{{\ttfamily
  2007.12187}}].

\bibitem{Stewart:2022ari}
I.~W. Stewart and X.~Yao, \emph{{Pure quark and gluon observables in collinear
  drop}}, \href{https://doi.org/10.1007/JHEP09(2022)120}{\emph{JHEP} {\bfseries
  09} (2022) 120} [\href{https://arxiv.org/abs/2203.14980}{{\ttfamily
  2203.14980}}].

\bibitem{KunnawalkamElayavalli:2017hxo}
R.~Kunnawalkam~Elayavalli and K.~C. Zapp, \emph{{Medium response in JEWEL and
  its impact on jet shape observables in heavy ion collisions}},
  \href{https://doi.org/10.1007/JHEP07(2017)141}{\emph{JHEP} {\bfseries 07}
  (2017) 141} [\href{https://arxiv.org/abs/1707.01539}{{\ttfamily
  1707.01539}}].

\bibitem{Andrews:2018jcm}
H.~A. Andrews et~al., \emph{{Novel tools and observables for jet physics in
  heavy-ion collisions}},  \href{https://arxiv.org/abs/1808.03689}{{\ttfamily
  1808.03689}}.

\bibitem{Mehtar-Tani:2016aco}
Y.~Mehtar-Tani and K.~Tywoniuk, \emph{{Groomed jets in heavy-ion collisions:
  sensitivity to medium-induced bremsstrahlung}},
  \href{https://doi.org/10.1007/JHEP04(2017)125}{\emph{JHEP} {\bfseries 04}
  (2017) 125} [\href{https://arxiv.org/abs/1610.08930}{{\ttfamily
  1610.08930}}].

\bibitem{Casalderrey-Solana:2019ubu}
J.~Casalderrey-Solana, G.~Milhano, D.~Pablos and K.~Rajagopal,
  \emph{{Modification of Jet Substructure in Heavy Ion Collisions as a Probe of
  the Resolution Length of Quark-Gluon Plasma}},
  \href{https://arxiv.org/abs/1907.11248}{{\ttfamily 1907.11248}}.

\bibitem{Brewer:2021hmh}
J.~Brewer, Q.~Brodsky and K.~Rajagopal, \emph{{Disentangling jet modification
  in jet simulations and in Z+jet data}},
  \href{https://doi.org/10.1007/JHEP02(2022)175}{\emph{JHEP} {\bfseries 02}
  (2022) 175} [\href{https://arxiv.org/abs/2110.13159}{{\ttfamily
  2110.13159}}].

\bibitem{ALICE:2017nij}
{\scshape ALICE} collaboration, S.~Acharya et~al., \emph{{First measurement of
  jet mass in Pb\textendash{}Pb and p\textendash{}Pb collisions at the LHC}},
  \href{https://doi.org/10.1016/j.physletb.2017.11.044}{\emph{Phys. Lett. B}
  {\bfseries 776} (2018) 249}
  [\href{https://arxiv.org/abs/1702.00804}{{\ttfamily 1702.00804}}].

\bibitem{ATLAS:2012am}
{\scshape ATLAS} collaboration, G.~Aad et~al., \emph{{Jet mass and substructure
  of inclusive jets in $\sqrt{s}=7$ TeV $pp$ collisions with the ATLAS
  experiment}}, \href{https://doi.org/10.1007/JHEP05(2012)128}{\emph{JHEP}
  {\bfseries 05} (2012) 128} [\href{https://arxiv.org/abs/1203.4606}{{\ttfamily
  1203.4606}}].

\bibitem{ATLAS:2012nnf}
{\scshape ATLAS} collaboration, G.~Aad et~al., \emph{{ATLAS Measurements of the
  Properties of Jets for Boosted Particle Searches}},
  \href{https://doi.org/10.1103/PhysRevD.86.072006}{\emph{Phys. Rev. D}
  {\bfseries 86} (2012) 072006}
  [\href{https://arxiv.org/abs/1206.5369}{{\ttfamily 1206.5369}}].

\bibitem{ATLAS:2017zda}
{\scshape ATLAS} collaboration, M.~Aaboud et~al., \emph{{Measurement of the
  Soft-Drop Jet Mass in pp Collisions at $\sqrt{s} = 13$ TeV with the ATLAS
  Detector}}, \href{https://doi.org/10.1103/PhysRevLett.121.092001}{\emph{Phys.
  Rev. Lett.} {\bfseries 121} (2018) 092001}
  [\href{https://arxiv.org/abs/1711.08341}{{\ttfamily 1711.08341}}].

\bibitem{ATLAS:2019dty}
{\scshape ATLAS} collaboration, G.~Aad et~al., \emph{{Measurement of the jet
  mass in high transverse momentum $Z(\rightarrow b\overline{b})\gamma$
  production at $\sqrt{s}= 13$ TeV using the ATLAS detector}},
  \href{https://doi.org/10.1016/j.physletb.2020.135991}{\emph{Phys. Lett. B}
  {\bfseries 812} (2021) 135991}
  [\href{https://arxiv.org/abs/1907.07093}{{\ttfamily 1907.07093}}].

\bibitem{ATLAS:2019mgf}
{\scshape ATLAS} collaboration, G.~Aad et~al., \emph{{Measurement of soft-drop
  jet observables in $pp$ collisions with the ATLAS detector at $\sqrt {s}$ =13
  TeV}}, \href{https://doi.org/10.1103/PhysRevD.101.052007}{\emph{Phys. Rev. D}
  {\bfseries 101} (2020) 052007}
  [\href{https://arxiv.org/abs/1912.09837}{{\ttfamily 1912.09837}}].

\bibitem{CMS:2017tdn}
{\scshape CMS} collaboration, C.~Collaboration, \emph{{Measurement of the
  differential jet production cross section with respect to jet mass and
  transverse momentum in dijet events from pp collisions at $\sqrt{s}$ = 13
  TeV}}, .

\bibitem{CMS:2018ypj}
{\scshape CMS} collaboration, A.~M. Sirunyan et~al., \emph{{Measurement of jet
  substructure observables in $\mathrm{t\overline{t}}$ events from
  proton-proton collisions at $\sqrt{s}=$ 13TeV}},
  \href{https://doi.org/10.1103/PhysRevD.98.092014}{\emph{Phys. Rev. D}
  {\bfseries 98} (2018) 092014}
  [\href{https://arxiv.org/abs/1808.07340}{{\ttfamily 1808.07340}}].

\bibitem{CMS:2019fak}
{\scshape CMS} collaboration, A.~M. Sirunyan et~al., \emph{{Measurement of the
  Jet Mass Distribution and Top Quark Mass in Hadronic Decays of Boosted Top
  Quarks in $pp$ Collisions at $\sqrt{s} =$ TeV}},
  \href{https://doi.org/10.1103/PhysRevLett.124.202001}{\emph{Phys. Rev. Lett.}
  {\bfseries 124} (2020) 202001}
  [\href{https://arxiv.org/abs/1911.03800}{{\ttfamily 1911.03800}}].

\bibitem{CMS:2018fof}
{\scshape CMS} collaboration, A.~M. Sirunyan et~al., \emph{{Measurement of the
  groomed jet mass in PbPb and pp collisions at $ \sqrt{s_{\mathrm{NN}}}=5.02 $
  TeV}}, \href{https://doi.org/10.1007/JHEP10(2018)161}{\emph{JHEP} {\bfseries
  10} (2018) 161} [\href{https://arxiv.org/abs/1805.05145}{{\ttfamily
  1805.05145}}].

\bibitem{ALICE:2021njq}
{\scshape ALICE} collaboration, S.~Acharya et~al., \emph{{Measurements of the
  groomed and ungroomed jet angularities in pp collisions at $\sqrt{s} = 5.02$
  TeV}},  \href{https://arxiv.org/abs/2107.11303}{{\ttfamily 2107.11303}}.

\bibitem{CMS:2021iwu}
{\scshape CMS} collaboration, A.~Tumasyan et~al., \emph{{Study of quark and
  gluon jet substructure in Z+jet and dijet events from pp collisions}},
  \href{https://doi.org/10.1007/JHEP01(2022)188}{\emph{JHEP} {\bfseries 01}
  (2022) 188} [\href{https://arxiv.org/abs/2109.03340}{{\ttfamily
  2109.03340}}].

\bibitem{ATLAS:2019kwg}
{\scshape ATLAS} collaboration, M.~Aaboud et~al., \emph{{Measurement of
  jet-substructure observables in top quark, $W$ boson and light jet production
  in proton-proton collisions at $\sqrt{s}=13$ TeV with the ATLAS detector}},
  \href{https://doi.org/10.1007/JHEP08(2019)033}{\emph{JHEP} {\bfseries 08}
  (2019) 033} [\href{https://arxiv.org/abs/1903.02942}{{\ttfamily
  1903.02942}}].

\bibitem{Bachu:2020nqn}
B.~Bachu, A.~H. Hoang, V.~Mateu, A.~Pathak and I.~W. Stewart, \emph{{Boosted
  top quarks in the peak region with NL3L resummation}},
  \href{https://doi.org/10.1103/PhysRevD.104.014026}{\emph{Phys. Rev. D}
  {\bfseries 104} (2021) 014026}
  [\href{https://arxiv.org/abs/2012.12304}{{\ttfamily 2012.12304}}].

\bibitem{ATLAS:2021urs}
{\scshape ATLAS} collaboration, \emph{{A precise interpretation for the top
  quark mass parameter in ATLAS Monte Carlo simulation}},
  {\emph{ATL-PHYS-PUB-2021-034} (2021) }.

\bibitem{Marzani:2019evv}
S.~Marzani, D.~Reichelt, S.~Schumann, G.~Soyez and V.~Theeuwes, \emph{{Fitting
  the Strong Coupling Constant with Soft-Drop Thrust}},
  \href{https://doi.org/10.1007/JHEP11(2019)179}{\emph{JHEP} {\bfseries 11}
  (2019) 179} [\href{https://arxiv.org/abs/1906.10504}{{\ttfamily
  1906.10504}}].

\bibitem{Hannesdottir:2022rsl}
H.~S. Hannesdottir, A.~Pathak, M.~D. Schwartz and I.~W. Stewart,
  \emph{{Prospects for strong coupling measurement at hadron colliders using
  soft-drop jet mass}},
  \href{https://doi.org/10.1007/JHEP04(2023)087}{\emph{JHEP} {\bfseries 04}
  (2023) 087} [\href{https://arxiv.org/abs/2210.04901}{{\ttfamily
  2210.04901}}].

\bibitem{Kardos:2020gty}
A.~Kardos, A.~J. Larkoski and Z.~Tr\'ocs\'anyi, \emph{{Groomed jet mass at high
  precision}},
  \href{https://doi.org/10.1016/j.physletb.2020.135704}{\emph{Phys. Lett. B}
  {\bfseries 809} (2020) 135704}
  [\href{https://arxiv.org/abs/2002.00942}{{\ttfamily 2002.00942}}].

\bibitem{Sjostrand:2007gs}
T.~Sjostrand, S.~Mrenna and P.~Z. Skands, \emph{{A Brief Introduction to PYTHIA
  8.1}}, \href{https://doi.org/10.1016/j.cpc.2008.01.036}{\emph{Comput. Phys.
  Commun.} {\bfseries 178} (2008) 852}
  [\href{https://arxiv.org/abs/0710.3820}{{\ttfamily 0710.3820}}].

\bibitem{Gleisberg:2008ta}
T.~Gleisberg, S.~Hoeche, F.~Krauss, M.~Schonherr, S.~Schumann, F.~Siegert
  et~al., \emph{{Event generation with SHERPA 1.1}},
  \href{https://doi.org/10.1088/1126-6708/2009/02/007}{\emph{JHEP} {\bfseries
  02} (2009) 007} [\href{https://arxiv.org/abs/0811.4622}{{\ttfamily
  0811.4622}}].

\bibitem{Bahr:2008pv}
M.~Bahr et~al., \emph{{Herwig++ Physics and Manual}},
  \href{https://doi.org/10.1140/epjc/s10052-008-0798-9}{\emph{Eur. Phys. J.}
  {\bfseries C58} (2008) 639}
  [\href{https://arxiv.org/abs/0803.0883}{{\ttfamily 0803.0883}}].

\bibitem{Marzani:2017mva}
S.~Marzani, L.~Schunk and G.~Soyez, \emph{{A study of jet mass distributions
  with grooming}}, \href{https://doi.org/10.1007/JHEP07(2017)132}{\emph{JHEP}
  {\bfseries 07} (2017) 132}
  [\href{https://arxiv.org/abs/1704.02210}{{\ttfamily 1704.02210}}].

\bibitem{Reichelt:2021svh}
D.~Reichelt, S.~Caletti, O.~Fedkevych, S.~Marzani, S.~Schumann and G.~Soyez,
  \emph{{Phenomenology of jet angularities at the LHC}},
  \href{https://doi.org/10.1007/JHEP03(2022)131}{\emph{JHEP} {\bfseries 03}
  (2022) 131} [\href{https://arxiv.org/abs/2112.09545}{{\ttfamily
  2112.09545}}].

\bibitem{Marzani:2017kqd}
S.~Marzani, L.~Schunk and G.~Soyez, \emph{{The jet mass distribution after Soft
  Drop}}, \href{https://doi.org/10.1140/epjc/s10052-018-5579-5}{\emph{Eur.
  Phys. J. C} {\bfseries 78} (2018) 96}
  [\href{https://arxiv.org/abs/1712.05105}{{\ttfamily 1712.05105}}].

\bibitem{Dokshitzer:1995qm}
Y.~L. Dokshitzer, G.~Marchesini and B.~R. Webber, \emph{{Dispersive approach to
  power behaved contributions in QCD hard processes}},
  \href{https://doi.org/10.1016/0550-3213(96)00155-1}{\emph{Nucl. Phys. B}
  {\bfseries 469} (1996) 93}
  [\href{https://arxiv.org/abs/hep-ph/9512336}{{\ttfamily hep-ph/9512336}}].

\bibitem{Dokshitzer:1997iz}
Y.~L. Dokshitzer, A.~Lucenti, G.~Marchesini and G.~P. Salam,
  \emph{{Universality of 1/Q corrections to jet-shape observables rescued}},
  \href{https://doi.org/10.1016/S0550-3213(97)00650-0}{\emph{Nucl. Phys. B}
  {\bfseries 511} (1998) 396}
  [\href{https://arxiv.org/abs/hep-ph/9707532}{{\ttfamily hep-ph/9707532}}].

\bibitem{Dokshitzer:1998pt}
Y.~L. Dokshitzer, A.~Lucenti, G.~Marchesini and G.~P. Salam, \emph{{On the
  universality of the Milan factor for 1 / Q power corrections to jet shapes}},
  \href{https://doi.org/10.1088/1126-6708/1998/05/003}{\emph{JHEP} {\bfseries
  05} (1998) 003} [\href{https://arxiv.org/abs/hep-ph/9802381}{{\ttfamily
  hep-ph/9802381}}].

\bibitem{Dasgupta:2009tm}
M.~Dasgupta and Y.~Delenda, \emph{{On the universality of hadronisation
  corrections to QCD jets}},
  \href{https://doi.org/10.1088/1126-6708/2009/07/004}{\emph{JHEP} {\bfseries
  07} (2009) 004} [\href{https://arxiv.org/abs/0903.2187}{{\ttfamily
  0903.2187}}].

\bibitem{Dasgupta:2009an}
M.~Dasgupta and Y.~Delenda, \emph{{Hadronisation corrections to jets in the
  $k_t$ algorithm}},  in \emph{{17th International Workshop on Deep-Inelastic
  Scattering and Related Subjects}}, (Berlin, Germany), pp.~67--73, Science
  Wise Publ., 2009, \href{https://arxiv.org/abs/0906.5463}{{\ttfamily
  0906.5463}}.

\bibitem{Salam:2001bd}
G.~P. Salam and D.~Wicke, \emph{{Hadron masses and power corrections to event
  shapes}}, \href{https://doi.org/10.1088/1126-6708/2001/05/061}{\emph{JHEP}
  {\bfseries 05} (2001) 061}
  [\href{https://arxiv.org/abs/hep-ph/0102343}{{\ttfamily hep-ph/0102343}}].

\bibitem{Collins:1981ta}
J.~C. Collins and G.~F. Sterman, \emph{{Soft Partons in {QCD}}},
  \href{https://doi.org/10.1016/0550-3213(81)90370-9}{\emph{Nucl. Phys. B}
  {\bfseries 185} (1981) 172}.

\bibitem{Collins:1985ue}
J.~C. Collins, D.~E. Soper and G.~F. Sterman, \emph{{Factorization for Short
  Distance Hadron - Hadron Scattering}},
  \href{https://doi.org/10.1016/0550-3213(85)90565-6}{\emph{Nucl. Phys. B}
  {\bfseries 261} (1985) 104}.

\bibitem{Collins:1988ig}
J.~C. Collins, D.~E. Soper and G.~F. Sterman, \emph{{Soft Gluons and
  Factorization}},
  \href{https://doi.org/10.1016/0550-3213(88)90130-7}{\emph{Nucl. Phys. B}
  {\bfseries 308} (1988) 833}.

\bibitem{Collins:1989gx}
J.~C. Collins, D.~E. Soper and G.~F. Sterman, \emph{{Factorization of Hard
  Processes in QCD}},
  \href{https://doi.org/10.1142/9789814503266_0001}{\emph{Adv. Ser. Direct.
  High Energy Phys.} {\bfseries 5} (1989) 1}
  [\href{https://arxiv.org/abs/hep-ph/0409313}{{\ttfamily hep-ph/0409313}}].

\bibitem{Collins:2011zzd}
J.~Collins, \emph{{Foundations of perturbative QCD}}. Cambridge University
  Press, 2013.

\bibitem{Bauer:2000ew}
C.~W. Bauer, S.~Fleming and M.~E. Luke, \emph{{Summing Sudakov logarithms in $B
  \to X_s \gamma$ in effective field theory}},
  \href{https://doi.org/10.1103/PhysRevD.63.014006}{\emph{Phys. Rev.}
  {\bfseries D63} (2000) 014006}
  [\href{https://arxiv.org/abs/hep-ph/0005275}{{\ttfamily hep-ph/0005275}}].

\bibitem{Bauer:2000yr}
C.~W. Bauer, S.~Fleming, D.~Pirjol and I.~W. Stewart, \emph{{An Effective field
  theory for collinear and soft gluons: Heavy to light decays}},
  \href{https://doi.org/10.1103/PhysRevD.63.114020}{\emph{Phys. Rev.}
  {\bfseries D63} (2001) 114020}
  [\href{https://arxiv.org/abs/hep-ph/0011336}{{\ttfamily hep-ph/0011336}}].

\bibitem{Bauer:2001yt}
C.~W. Bauer, D.~Pirjol and I.~W. Stewart, \emph{{Soft collinear factorization
  in effective field theory}},
  \href{https://doi.org/10.1103/PhysRevD.65.054022}{\emph{Phys. Rev.}
  {\bfseries D65} (2002) 054022}
  [\href{https://arxiv.org/abs/hep-ph/0109045}{{\ttfamily hep-ph/0109045}}].

\bibitem{Bauer:2001ct}
C.~W. Bauer and I.~W. Stewart, \emph{{Invariant operators in collinear
  effective theory}},
  \href{https://doi.org/10.1016/S0370-2693(01)00902-9}{\emph{Phys. Lett.}
  {\bfseries B516} (2001) 134}
  [\href{https://arxiv.org/abs/hep-ph/0107001}{{\ttfamily hep-ph/0107001}}].

\bibitem{Bauer:2002nz}
C.~W. Bauer, S.~Fleming, D.~Pirjol, I.~Z. Rothstein and I.~W. Stewart,
  \emph{{Hard scattering factorization from effective field theory}},
  \href{https://doi.org/10.1103/PhysRevD.66.014017}{\emph{Phys. Rev.}
  {\bfseries D66} (2002) 014017}
  [\href{https://arxiv.org/abs/hep-ph/0202088}{{\ttfamily hep-ph/0202088}}].

\bibitem{Abbate:2010xh}
R.~Abbate, M.~Fickinger, A.~H. Hoang, V.~Mateu and I.~W. Stewart, \emph{{Thrust
  at N$^3$LL with Power Corrections and a Precision Global Fit for
  $\alpha_s$($m_Z$)}},
  \href{https://doi.org/10.1103/PhysRevD.83.074021}{\emph{Phys. Rev.}
  {\bfseries D83} (2011) 074021}
  [\href{https://arxiv.org/abs/1006.3080}{{\ttfamily 1006.3080}}].

\bibitem{Abbate:2012jh}
R.~Abbate, M.~Fickinger, A.~H. Hoang, V.~Mateu and I.~W. Stewart,
  \emph{{Precision Thrust Cumulant Moments at $N^3$LL}},
  \href{https://doi.org/10.1103/PhysRevD.86.094002}{\emph{Phys. Rev. D}
  {\bfseries 86} (2012) 094002}
  [\href{https://arxiv.org/abs/1204.5746}{{\ttfamily 1204.5746}}].

\bibitem{Hoang:2015hka}
A.~H. Hoang, D.~W. Kolodrubetz, V.~Mateu and I.~W. Stewart, \emph{{Precise
  determination of $\alpha_s$ from the $C$-parameter distribution}},
  \href{https://doi.org/10.1103/PhysRevD.91.094018}{\emph{Phys. Rev.}
  {\bfseries D91} (2015) 094018}
  [\href{https://arxiv.org/abs/1501.04111}{{\ttfamily 1501.04111}}].

\bibitem{Hoang:2019ceu}
A.~H. Hoang, S.~Mantry, A.~Pathak and I.~W. Stewart, \emph{{Nonperturbative
  Corrections to Soft Drop Jet Mass}},
  \href{https://arxiv.org/abs/1906.11843}{{\ttfamily 1906.11843}}.

\bibitem{Pathak:2020iue}
A.~Pathak, I.~W. Stewart, V.~Vaidya and L.~Zoppi, \emph{{EFT for Soft Drop
  Double Differential Cross Section}},
  \href{https://doi.org/10.1007/JHEP04(2021)032}{\emph{JHEP} {\bfseries 04}
  (2021) 032} [\href{https://arxiv.org/abs/2012.15568}{{\ttfamily
  2012.15568}}].

\bibitem{Chen:2021uws}
Y.~Chen et~al., \emph{{Jet energy spectrum and substructure in e$^{+}$e$^{-}$
  collisions at 91.2 GeV with ALEPH Archived Data}},
  \href{https://doi.org/10.1007/JHEP06(2022)008}{\emph{JHEP} {\bfseries 06}
  (2022) 008} [\href{https://arxiv.org/abs/2111.09914}{{\ttfamily
  2111.09914}}].

\bibitem{Larkoski:2015zka}
A.~J. Larkoski, I.~Moult and D.~Neill, \emph{{Non-Global Logarithms,
  Factorization, and the Soft Substructure of Jets}},
  \href{https://doi.org/10.1007/JHEP09(2015)143}{\emph{JHEP} {\bfseries 09}
  (2015) 143} [\href{https://arxiv.org/abs/1501.04596}{{\ttfamily
  1501.04596}}].

\bibitem{Benkendorfer:2021unv}
K.~Benkendorfer and A.~J. Larkoski, \emph{{Grooming at the Cusp: All-Orders
  Predictions for the Transition Region of Jet Groomers}},
  \href{https://arxiv.org/abs/2108.02779}{{\ttfamily 2108.02779}}.

\bibitem{Kang:2019prh}
Z.-B. Kang, K.~Lee, X.~Liu, D.~Neill and F.~Ringer, \emph{{The soft drop
  groomed jet radius at NLL}},
  \href{https://arxiv.org/abs/1908.01783}{{\ttfamily 1908.01783}}.

\bibitem{Cacciari:2008gn}
M.~Cacciari, G.~P. Salam and G.~Soyez, \emph{{The Catchment Area of Jets}},
  \href{https://doi.org/10.1088/1126-6708/2008/04/005}{\emph{JHEP} {\bfseries
  04} (2008) 005} [\href{https://arxiv.org/abs/0802.1188}{{\ttfamily
  0802.1188}}].

\bibitem{Ferdinand:2023yyy}
A.~E. Ferdinand and A.~Pathak, \emph{{On impact of underlying event on groomed
  jet mass}},  2023.

\bibitem{Cal:2019gxa}
P.~Cal, D.~Neill, F.~Ringer and W.~J. Waalewijn, \emph{{Calculating the angle
  between jet axes}},  \href{https://arxiv.org/abs/1911.06840}{{\ttfamily
  1911.06840}}.

\bibitem{Bell:2018vaa}
G.~Bell, R.~Rahn and J.~Talbert, \emph{{Two-loop anomalous dimensions of
  generic dijet soft functions}}, {\emph{Nucl. Phys. B} {\bfseries 936} (2018)
  520} [\href{https://arxiv.org/abs/1805.12414}{{\ttfamily 1805.12414}}].

\bibitem{Ferdinand:2023vaf}
A.~Ferdinand, K.~Lee and A.~Pathak, \emph{{Field-Theoretic Analysis of
  Hadronization Using Soft Drop Jet Mass}},
  \href{https://arxiv.org/abs/2301.03605}{{\ttfamily 2301.03605}}.

\bibitem{Ellis:1996mzs}
R.~K. Ellis, W.~J. Stirling and B.~R. Webber, \emph{{QCD and collider
  physics}}, vol.~8. Cambridge University Press, 2, 2011.

\bibitem{Kang:2016mcy}
Z.-B. Kang, F.~Ringer and I.~Vitev, \emph{{The semi-inclusive jet function in
  SCET and small radius resummation for inclusive jet production}},
  \href{https://doi.org/10.1007/JHEP10(2016)125}{\emph{JHEP} {\bfseries 10}
  (2016) 125} [\href{https://arxiv.org/abs/1606.06732}{{\ttfamily
  1606.06732}}].

\bibitem{Kang:2018jwa}
Z.-B. Kang, K.~Lee, X.~Liu and F.~Ringer, \emph{{The groomed and ungroomed jet
  mass distribution for inclusive jet production at the LHC}},
  \href{https://doi.org/10.1007/JHEP10(2018)137}{\emph{JHEP} {\bfseries 10}
  (2018) 137} [\href{https://arxiv.org/abs/1803.03645}{{\ttfamily
  1803.03645}}].

\bibitem{Cal:2019hjc}
P.~Cal, F.~Ringer and W.~J. Waalewijn, \emph{{The jet shape at NLL$'$}},
  \href{https://doi.org/10.1007/JHEP05(2019)143}{\emph{JHEP} {\bfseries 05}
  (2019) 143} [\href{https://arxiv.org/abs/1901.06389}{{\ttfamily
  1901.06389}}].

\bibitem{Bauer:2003pi}
C.~W. Bauer and A.~V. Manohar, \emph{{Shape function effects in $B \to X_s
  \gamma$ and $B \to X_u \ell \bar \nu$ decays}},
  \href{https://doi.org/10.1103/PhysRevD.70.034024}{\emph{Phys. Rev.}
  {\bfseries D70} (2004) 034024}
  [\href{https://arxiv.org/abs/hep-ph/0312109}{{\ttfamily hep-ph/0312109}}].

\bibitem{Bosch:2004th}
S.~W. Bosch, B.~O. Lange, M.~Neubert and G.~Paz, \emph{{Factorization and shape
  function effects in inclusive B meson decays}},
  \href{https://doi.org/10.1016/j.nuclphysb.2004.07.041}{\emph{Nucl. Phys. B}
  {\bfseries 699} (2004) 335}
  [\href{https://arxiv.org/abs/hep-ph/0402094}{{\ttfamily hep-ph/0402094}}].

\bibitem{Cacciari:2008gp}
M.~Cacciari, G.~P. Salam and G.~Soyez, \emph{{The anti-k$_T$ jet clustering
  algorithm}}, \href{https://doi.org/10.1088/1126-6708/2008/04/063}{\emph{JHEP}
  {\bfseries 04} (2008) 063} [\href{https://arxiv.org/abs/0802.1189}{{\ttfamily
  0802.1189}}].

\bibitem{Cacciari:2011ma}
M.~Cacciari, G.~P. Salam and G.~Soyez, \emph{{FastJet User Manual}},
  \href{https://doi.org/10.1140/epjc/s10052-012-1896-2}{\emph{Eur. Phys. J. C}
  {\bfseries 72} (2012) 1896}
  [\href{https://arxiv.org/abs/1111.6097}{{\ttfamily 1111.6097}}].

\bibitem{jetlib}
A.~E. Ferdinand, A.~Pathak et~al., \emph{{\texttt{JETlib}: A C++ Package for
  Efficient Analyses with Monte Carlo Event Generators}},  2019.

\bibitem{scetlib}
M.~A. Ebert, J.~K.~L. Michel, F.~J. Tackmann et~al., \emph{{SCETlib: A C++
  Package for Numerical Calculations in QCD and Soft-Collinear Effective
  Theory}}, {\emph{DESY-17-099} (2018) }.

\bibitem{scetlibSD}
A.~Pathak, \emph{\texttt{scetlib::sd}: A {C}++ library for for soft drop
  resummed observables in \texttt{SCETlib}},  2022.

\bibitem{Korchemsky:1987wg}
G.~P. Korchemsky and A.~V. Radyushkin, \emph{{Renormalization of the Wilson
  Loops Beyond the Leading Order}},
  \href{https://doi.org/10.1016/0550-3213(87)90277-X}{\emph{Nucl. Phys. B}
  {\bfseries 283} (1987) 342}.

\bibitem{Moch:2004pa}
S.~Moch, J.~A.~M. Vermaseren and A.~Vogt, \emph{{The Three loop splitting
  functions in QCD: The Nonsinglet case}},
  \href{https://doi.org/10.1016/j.nuclphysb.2004.03.030}{\emph{Nucl. Phys.}
  {\bfseries B688} (2004) 101}
  [\href{https://arxiv.org/abs/hep-ph/0403192}{{\ttfamily hep-ph/0403192}}].

\bibitem{Henn:2019swt}
J.~M. Henn, G.~P. Korchemsky and B.~Mistlberger, \emph{{The full four-loop cusp
  anomalous dimension in $\mathcal{N}=4$ super Yang-Mills and QCD}},
  \href{https://arxiv.org/abs/1911.10174}{{\ttfamily 1911.10174}}.

\end{thebibliography}\endgroup
